\documentclass[12pt,a4paper,twoside]{book}
\usepackage[T1]{fontenc}
\usepackage[table]{xcolor}
\usepackage[utf8]{inputenc}
\usepackage{mathptmx} 
\usepackage{fancyhdr} 
\usepackage{longtable}
\usepackage{multicol}
\usepackage{sectsty}
\usepackage[ngerman,english]{babel}
\usepackage{makeidx} 
\usepackage{subcaption} 
\usepackage[nottoc,notindex]{tocbibind} 
\usepackage{tocloft}
\usepackage{ifthen} 
\usepackage{substr} 
\usepackage[final]{pdfpages}
\usepackage{graphicx}
\usepackage[printonlyused]{acronym}
\usepackage{changepage}  

\usepackage[normalem]{ulem}

\usepackage{tcolorbox}

\usepackage{array}
\usepackage{booktabs}
\usepackage{colortbl}

\usepackage[hang,splitrule]{footmisc} 
\usepackage{titlesec}
\usepackage{geometry}

\usepackage{amssymb}
\usepackage{pifont}
\usepackage{tabularx}
\newcolumntype{L}[1]{>{\raggedright\arraybackslash}p{#1}} 
\newcolumntype{C}[1]{>{\centering\arraybackslash}p{#1}} 
\newcolumntype{R}[1]{>{\raggedleft\arraybackslash}p{#1}} 

\usepackage{subscript}
\usepackage{rotating}
\usepackage{multirow}

\usepackage{marvosym}
\usepackage{amsmath}
\usepackage{amsfonts}   
\usepackage{wrapfig}
\usepackage{afterpage}
\usepackage{csquotes}

\usepackage{blindtext}

\usepackage{pgfplots}
\pgfplotsset{compat=1.12}
\usepackage{tikz}
\usetikzlibrary{arrows, decorations.pathmorphing, decorations.markings, backgrounds, positioning, fit, petri}

\usepackage[gen]{eurosym}

\usepackage{wrapfig}

\label{Color Definition}
\usepackage{xcolor}
\definecolor{platinum}{HTML}{E5E4E2}
\colorlet{diagrambg}{platinum}
\usepackage{color}
\usepackage{colortbl}

\usepackage{mdframed}

\usepackage{csvsimple}

\usepackage{pifont}
\makeindex

\usepackage{pdfpages}

\usepackage[square,numbers,sort]{natbib}
\usepackage{doi}

\usepackage{hyperref}
\hypersetup {
pdfpagemode = {UseNone},
pdftitle = {Interactive Tools for Reproducible Science. Understanding, Supporting, and Motivating Reproducible Science Practices},
pdfauthor = {Sebastian Feger},
pdflang = {en-US},
bookmarks=true,bookmarksopen=true,colorlinks=true,
    linkcolor=black,citecolor=black,filecolor=darkmagenta,
    menucolor=darkred,
    urlcolor=darkcyan,plainpages=false
}

\usepackage{makecell} 
\usepackage[math]{cellspace}

\usepackage{longtable}

\usepackage{chngcntr}
\counterwithout{footnote}{chapter}

\newcommand{\mytitle}{Interactive Tools for Reproducible Science}
\newcommand{\mysubtitle}{Understanding, Supporting, and Motivating \protect\\ Reproducible Science Practices}
\newcommand{\myname}{Sebastian Stefan Feger}

\usepackage{natbib}

\label{Quotes}

\label{Footnotes}


\makeatletter
\renewenvironment{theindex}
  {\if@twocolumn
      \@restonecolfalse
   \else
      \@restonecoltrue
   \fi
   \setlength{\columnseprule}{0pt}
   \setlength{\columnsep}{20pt}
   \begin{multicols}{2}[\chapter*{\indexname}]
   \markboth{\MakeUppercase\indexname}%
            {\MakeUppercase\indexname}%
   \thispagestyle{plain}
   \setlength{\parindent}{0pt}
   \setlength{\parskip}{0pt plus 0.3pt}
   \relax
   \let\item\@idxitem}%
  {\end{multicols}\if@restonecol\onecolumn\else\clearpage\fi}
\makeatother

\AtBeginDocument{%
	\definecolor{gray}{rgb}{0.8,0.8,0.8}
	\definecolor{darkgray}{RGB}{102,102,102}
	\definecolor{lightyellow}{RGB}{250,250,210}
}

\newcommand{\clearemptydoublepage}{%
  \ifthenelse{\boolean{@twoside}}{\newpage{\pagestyle{empty}\cleardoublepage}}%
  {\clearpage}}

\usepackage{ifpdf}

\ifpdf
  \usepackage{color}
  \definecolor{darkred}{rgb}{.25,0,0}
  \definecolor{darkgreen}{rgb}{0,.2,0}
  \definecolor{darkmagenta}{rgb}{.2,0,.2}
  \definecolor{darkcyan}{rgb}{0,.15,.15}
  \definecolor{headings}{rgb}{0,0,.3}
  \definecolor{primaryColor}{HTML}{A2C2DE}
  \definecolor{secondaryColor}{HTML}{E1EAF4}
  \definecolor{tertiaryColor}{HTML}{cbd6e5}
  \definecolor{MyGray}{rgb}{0.86,0.87,0.88}
  
  \definecolor{ReferenceGray}{rgb}{0.91,0.91,0.91}
  \definecolor{darkgray}{rgb}{.4,.4,.4}
  \definecolor{lightgray}{rgb}{.85,.85,.85}
  
  \usepackage{graphicx}
	
  \renewcommand{\\}{}
    \hypersetup {
    pdfpagemode = {UseOutlines}, 
    pdftitle = {\mytitle},
    pdfsubject = {Dissertation of \myname},
    pdfauthor = {\myname},
    pdfdisplaydoctitle = true
  }

\else
\fi

\graphicspath{{../../Figures/General}}

\usepackage[scaled=0.85]{berasans}
\makeatletter

\label{Chapter Head}
\def\@makechapterhead#1{%
\thispagestyle{empty}
  \vspace*{-70\p@}%
  {\parindent \z@ \raggedright \normalfont
    \ifnum \c@secnumdepth >\m@ne
      \if@mainmatter
      		\par\vspace*{\fill}
      		\IfSubStringInString{: }{#1}{%
			\begin{flushright}
			\fontsize{50}{40}\fontfamily{qhv}\textbf{\thechapter }\par\nobreak
			\vspace{10pt}
			\fontsize{20}{80}\fontfamily{qhv}\textbf{\BeforeSubString{: }{#1}}\par\nobreak
			\fontsize{20}{80}\fontfamily{qhv}\textbf{\BehindSubString{: }{#1}}\par\nobreak
			  \end{flushright}
		}{
				
				\begin{flushright}
				\fontsize{50}{40}\fontfamily{qhv}\textbf{\thechapter }\par\nobreak
				\vspace{10pt}
				\fontsize{20}{80}\fontfamily{qhv}\textbf{#1}\par\nobreak
			     \end{flushright}
				\par\nobreak
        }
      \fi
  
        \vskip 30\p@
    
  }}
  
\label{Chapter Synopsis}

\label{Subsections}
\renewcommand\section{\@startsection {section}{1}{\z@}%
                                   {-3.5ex \@plus -1ex \@minus -.2ex}%
                                   {2.3ex \@plus.2ex}%
                                   {\fontfamily{qhv}\Large\bfseries}}
\renewcommand\subsection{\@startsection{subsection}{2}{\z@}%
                                     {-3.25ex\@plus -1ex \@minus -.2ex}%
                                     {1.5ex \@plus .2ex}%
                                     {\fontfamily{qhv}\large\bfseries}}
\renewcommand\subsubsection{\@startsection{subsubsection}{3}{\z@}%
                                     {-1.75ex\@plus -1ex \@minus -.2ex}%
                                     {0.1ex \@plus .1ex}%
                                     {\fontfamily{qhv}\normalsize\bfseries}}
\renewcommand\paragraph{\@startsection{paragraph}{4}{\z@}%
                                    {1.75ex \@plus1ex \@minus.2ex}%
                                    {0.1ex \@plus .1ex}%
                                    {\normalfont\normalsize\itshape\rmfamily}}
\renewcommand\subparagraph{\@startsection{subparagraph}{5}{\parindent}%
                                       {3.25ex \@plus1ex \@minus .2ex}%
                                       {-1em}%
                                      {\normalfont\normalsize\sffamily}}

\let\oldtabular\tabular
\renewcommand{\tabular}{\small\oldtabular}

\def\@part[#1]#2{%
    \ifnum \c@secnumdepth >-2\relax
      \refstepcounter{part}%
      \addcontentsline{toc}{part}{\thepart\hspace{1em}#1}%
    \else
      \addcontentsline{toc}{part}{#1}%
    \fi
    \markboth{}{}%
    {\centering
     \interlinepenalty \@M
     \normalfont
     \ifnum \c@secnumdepth >-2\relax
       \textcolor{gray}{\fontsize{240}{288}\mdseries\textrm{\thepart}}%
       \par
     \fi
     \centering \normalfont
     \fontsize{30}{36}\mdseries\scshape\textrm{#2}\par}%
    \@endpart}
\def\@spart#1{%
    {\centering
     \interlinepenalty \@M
     \normalfont \Huge \bfseries \SS@parttitlefont {#1}\par}%
    \@endpart}

\label{Header}
\makeatletter
\def\@fancyhead#1#2#3#4#5{#1\hbox to\headwidth{\fancy@reset
  \@fancyvbox\headheight{\hbox
   {\rlap{\parbox[b]{\headwidth}{\fontsize{10}{14}\rmfamily\mdseries\raggedright#2}}\hfill
      \parbox[b]{\headwidth}{\fontsize{12}{14}\rmfamily\mdseries\centering#3}\hfill
      \llap{\parbox[b]{\headwidth}{\fontsize{10}{14}\rmfamily\mdseries\raggedleft#4}}}\headrule}}#5}
\makeatother









\newcommand{\publicationsbegin}{ 
	\begin{list}{\raise .5ex\hbox{\scriptsize$\bullet$}}
		{ \setlength{\itemsep}{5pt}      \setlength{\parsep}{3pt} 
			\setlength{\topsep}{6pt}       \setlength{\partopsep}{0pt}
			\setlength{\leftmargin}{2.5em} \setlength{\labelwidth}{1em}
			\setlength{\labelsep}{0.7em} } }
	
	\newcommand{\publicationsend}{
\end{list}  }

\makeatletter
\long\def\@makecaption#1#2{%
  \vskip\abovecaptionskip
  \vskip2mm
  \bfseries
  \centering
  \sbox\@tempboxa{#1: #2}%
	\ifdim \wd\@tempboxa >0.95\textwidth
	  \begin{minipage}{0.95\textwidth}
    	\bfseries\small{#1:} \normalfont\small #2\par
    \end{minipage}
  \else
    \global \@minipagefalse
    \bfseries\small{#1:} \normalfont\small #2\par
  \fi
  \vskip\belowcaptionskip}
\makeatother

\usepackage{bibentry}
\usepackage{paralist} 
\usepackage{epigraph}
\usepackage{multirow, tabularx}
\usepackage{rotating}
\usepackage{txfonts}
\usepackage{mathptmx}
\usepackage{booktabs}
\usepackage{ccicons}
\counterwithout{footnote}{chapter}
\usepackage{wrapfig} 
\usepackage{array}
\usepackage{eqnarray}
\usepackage{booktabs}
\usepackage{color}
\usepackage{colortbl}
\usepackage{verbatim} 
\usepackage[hang,splitrule]{footmisc} 
\usepackage{titletoc} 
\usepackage{enumitem} 
\usepackage{overpic}
\usepackage{units}
\usepackage{longtable,array,ragged2e}

\makeatletter
\newcommand*{\rom}[1]{\expandafter\@slowromancap\romannumeral #1@}

\def\@makechapterhead#1{%
	\vspace*{-35\p@}%
	{\parindent \z@ \raggedright \normalfont
		\ifnum \c@secnumdepth >\m@ne
		\if@mainmatter
		\fontsize{20}{24}\mdseries\textsf{\@chapapp}\space\fontsize{80}{96}\mdseries\textsf{\thechapter}
		\hrulefill
		\par\nobreak
		\vskip 30\p@
		\fi
		\else
		
		\fi
		\interlinepenalty\@M
		
		\IfSubStringInString{? }{#1}{%
			\fontsize{36}{50}\mdseries\textsf{\BeforeSubString{? }{#1}}?\par\nobreak
			\fontsize{22}{26}\mdseries\textsf{\BehindSubString{? }{#1}}\par\nobreak
		}{%
			\fontsize{26}{40}\bfseries\textsf{#1}\par\nobreak
		}
		
		\vskip 40\p@
}}

\label{Page Heading}
\renewcommand{\chaptermark}[1]{\markboth{ #1}{}}
\renewcommand{\sectionmark}[1]{\markright{ #1}}
\newcommand{\phstyle}{\fontfamily{qhv}\fontsize{10}{14}\bfseries}
\author{\myname\\\small mariam.hassib@ifi.lmu.de}
\title{\mytitle\\
\mysubtitle\\}
\date{\svndate\ \small(rev. \svnrevision)}

\newmdenv[
  topline=false,
  rightline=false,
  bottomline=false,
  skipabove=\topsep,
  skipbelow=\topsep,
  linewidth=2pt,
  linecolor=darkgray
]{definitionbar}

\nobibliography*
\makeatletter

	{\end{multicols}\if@restonecol\onecolumn\else\clearpage\fi}
\makeatother

\newcommand*{\compress}{\@minipagetrue}
\makeatother

\label{Anecdote}

\label{Shortcuts}
\usepackage{xspace}

\newcommand\CommitmentModel{Stage-Based Model of Personal RDM Commitment}
\newcommand\CommitmentModelTrailingSpace{Stage-Based Model of Personal RDM Commitment }

\newcommand\ReproDefinition{Reproducibility in data-driven scientific discovery concerns the \textbf{ease of access} to scientific resources, as well as their \textbf{completeness}, to the degree required for \textbf{efficiently} and \textbf{effectively} \textbf{\textit{interacting}} with scientific work.}

\newcommand\rowincludegraphics[2][]{\raisebox{-0.45\height}{\includegraphics[#1]{#2}}}

  {\begin{list}{}%
          {\setlength{\leftmargin}{#1}}%
          \item[]%
  }
  {\end{list}}
  
\def\sf#1{\textcolor{blue}{$\leftarrow$ [sf: \emph{#1}]}}

\begin{document}
	
    \frontmatter

	{

		\cleardoublepage
		\clearpage
		
		\pagestyle{empty}

		\addtolength{\oddsidemargin}{11mm}
		\addtolength{\evensidemargin}{11mm}
		\addtolength{\topmargin}{11mm}

		\newgeometry{inner=30pt,outer=30pt,top=30pt}
		\noindent\begin{minipage}{188mm} 
			
			\begin{picture}(0,0)
			\put(258.7,-858.8){\includegraphics[width=140mm]{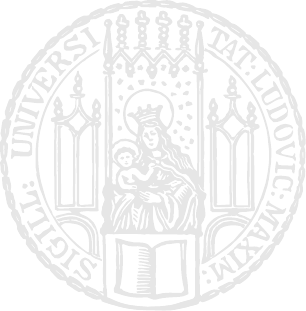}} 
			\end{picture}
			
			\setlength{\fboxsep}{0mm}%
			\pagestyle{empty}%
			\noindent

			\vspace*{23mm}
			
			\begin{center}
				\vspace*{20mm}
				\rule{\textwidth}{0.5pt}
				\vskip 5mm
				\fontsize{28}{34}\rmfamily\scshape\mytitle
				\par
				\vskip 15pt
				\fontsize{22}{24}\rmfamily\scshape\mysubtitle
				\par
				\vskip 5mm

				\rule{\textwidth}{0.5pt}
				\vskip 20mm
				\huge \textbf{Dissertation}\\ 
				\vspace*{5mm}

				\normalfont\Large an der Fakultät für Mathematik, Informatik und Statistik\\
				der Ludwig-Maximilians-Universität München\\
				\vspace*{10mm}
				vorgelegt von\\
				\huge{\scshape{\textbf{\myname}}}\\
				\normalfont\Large M.Sc.~in Informatik\\
				\vspace*{20mm}
				\normalfont\Large München, den 12. März 2020 
			\end{center}
		\end{minipage}
		
		\cleardoublepage
		
		\clearpage
	}
	\pagestyle{fancy}%

	\renewcommand{\chaptermark}[1]{%
		\markboth{\thechapter~~#1}{}}
	
	
	\begin{picture}(0,0)%
	\put(181.7,-773.8){\includegraphics[width=140mm]{./img/general/lmu-siegel-grey.pdf}} 
	\end{picture}
	
	\vspace*{70mm}
	{\large\noindent
		\centering{
			\begin{tabular}{ll}
				\large{Erstgutachter:}	& \large{Prof.\ Dr.\ Albrecht Schmidt}\\[1ex]
				\large{Zweitgutachter:} & \large{Prof.\ Dr.\ Pawe{\l} W. Wo{\'z}niak}\\[1ex]
				\large{Drittgutachter:}	& \large{Prof.\ Dr.\ Barry Brown}
			\end{tabular}
		}\\
	}

	\vspace*{25mm}
	\begin{center}
	{\large\noindent
	Datum der Disputation: 17. Juli 2020}
	\end{center}

	\cleardoublepage
	
	\setlength{\parindent}{0pt} 
	\setlength{\parskip}{7pt plus 2pt minus 1pt}
	
\selectlanguage{english}

\chapter*{Preface}\markboth{PREFACE}{}
\addcontentsline{toc}{chapter}{Preface}

This thesis presents research which I conducted between 2017 and 2020 at CERN, the European Organization for Nuclear Research. My doctoral research was financed through the Wolfgang Gentner Scholarship. This scholarship is funded by the \ac{BMBF} and integrated into the general CERN Doctoral Student Programme. The Wolfgang Gentner Scholarship mandates close collaboration between CERN researchers and German universities. Throughout my doctoral research at CERN, Albrecht Schmidt was my primary supervisor. My doctoral research was further supervised by Paweł Woźniak and Sünje Dallmeier-Tiessen. I was enrolled at the University of Stuttgart (2017 and 2018) and at LMU Munich (2018 -- 2020). In close collaboration with my university and CERN supervisors, we published results from my doctoral research at different venues. I added references to these publications in the beginning of related chapters and sections. To reflect the collaborative nature of this research, I decided to use the scientific plural throughout this thesis.

\clearemptydoublepage
	
\selectlanguage{english}
\markboth{Abstract}{Abstract}
\section*{\LARGE\rmfamily\bfseries\scshape{Abstract}}

Reproducibility should be a cornerstone of science. It plays an essential role in research validation and reuse. In recent years, the scientific community and the general public became increasingly aware of the \textit{reproducibility crisis}, i.e. the wide-spread inability of researchers to reproduce published work, including their own. The reproducibility crisis has been identified in most branches of data-driven science. The effort required to document, clean, preserve, and share experimental resources has been described as one of the core contributors to this irreproducibility challenge. Documentation, preservation, and sharing are key reproducible research practices that are of little perceived value for scientists, as they fall outside the traditional academic reputation economy that is focused on novelty-driven scientific contributions. 

Scientific research is increasingly focused on the creation, observation, processing, and analysis of large data volumes. On one hand, this transition towards \textit{computational} and \textit{data-intensive science} poses new challenges for research reproducibility and reuse. On the other hand, increased availability and advances in computation and web technologies offer new opportunities to address the reproducibility crisis. A prominent example is the World Wide Web (WWW), which was developed in response to researchers’ needs to quickly share research data and findings with the scientific community. The WWW was invented at the European Organization for Nuclear Research (CERN). CERN is a key laboratory in High Energy Physics (HEP), one of the most data-intensive scientific domains. This thesis reports on research connected in the context of \ac{CAP}, a Research Data Management (RDM) service tailored to CERN's major experiments. We use this scientific environment to study the role and requirements of interactive tools in facilitating reproducible research.

In this thesis, we build a wider understanding of researchers' interactions with tools that support research documentation, preservation, and sharing. From an \ac{HCI} perspective the following aspects are fundamental: (1) Characterize and map requirements and practices around research preservation and reuse. (2) Understand the wider role and impact of RDM tools in scientific workflows. (3) Design tools and interactions that promote, motivate, and acknowledge reproducible research practices.

Research reported in this thesis represents the first systematic application of HCI methods in the study and design of interactive tools for reproducible science. We have built an empirical understanding of reproducible research practices and the role of supportive tools through research in HEP and across a variety of scientific fields. We designed prototypes and implemented services that aim to create rewarding and motivating interactions. We conducted mixed-method evaluations to assess the \ac{UX} of the designs, in particular related to usefulness, suitability, and persuasiveness. We report on four empirical studies in which 42 researchers and data managers participated.

In the first interview study, we asked HEP data analysts about RDM practices and invited them to explore and discuss CAP. Our findings show that tailored preservation services allow for introducing and promoting meaningful rewards and incentives that benefit contributors in their research work. Here, we introduce the term \textit{secondary usage forms} of RDM tools. While not part of the core mission of the tools, secondary usage forms motivate contributions through meaningful rewards. We extended this research through a cross-domain interview study with data analysts and data stewards from a diverse set of scientific fields. Based on the findings of this cross-domain study, we contribute a Stage-Based Model of Personal RDM Commitment Evolution that explains how and why scientists commit to open and reproducible science.

To address the motivation challenge, we explored if and how gamification can motivate contributions and promote reproducible research practices. To this end, we designed two prototypes of a gamified preservation service that was inspired by CAP. Each gamification prototype makes use of different underlying mechanisms. HEP researchers found both implementations valuable, enjoyable, suitable, and persuasive. The gamification layer improves visibility of scientists and research work and facilitates content navigation and discovery. Based on these findings, we implemented six \textit{tailored science badges} in CAP in our second gamification study. The badges promote and reward high-quality documentation and special uses of preserved research. Findings from our evaluation with HEP researchers show that tailored science badges enable novel forms of research repository navigation and content discovery that benefit users and contributors. We discuss how the use of tailored science badges as an incentivizing element paves new ways for interaction with research repositories.

Finally, we describe the role of HCI in supporting reproducible research practices. We stress that tailored RDM tools can improve content navigation and discovery, which is key in the design of secondary usage forms. Moreover, we argue that incentivizing elements like gamification may not only motivate contributions, but further promote secondary uses and enable new forms of interaction with preserved research. Based on our empirical research, we describe the roles of both HCI scholars and practitioners in building interactive tools for reproducible science. Finally, we outline our vision to transform computational and data-driven research preservation through ubiquitous preservation strategies that integrate into research workflows and make use of automated knowledge recording.

In conclusion, this thesis advocates the unique role of HCI in supporting, motivating, and transforming reproducible research practices through the design of tools that enable effective RDM. We present practices around research preservation and reuse in HEP and beyond. Our research paves new ways for interaction with RDM tools that support and motivate reproducible science.

\clearemptydoublepage

\markboth{Zusammenfassung}{Zusammenfassung}
\section*{\LARGE\rmfamily\bfseries\scshape{Zusammenfassung}}
\selectlanguage{ngerman}

Reproduzierbarkeit sollte ein wissenschaftlicher Grundpfeiler sein, da sie einen essenziellen Bestandteil in der Validierung und Nachnutzung von Forschungsarbeiten darstellt. Verfügbarkeit und Vollständigkeit von Forschungsmaterialien sind wichtige Voraussetzungen für die Interaktion mit experimentellen Arbeiten. Diese Voraussetzungen sind jedoch oft nicht gegeben. Zuletzt zeigten sich die Wissenschaftsgemeinde und die Öffentlichkeit besorgt über die Reproduzierbarkeitskrise in der empirischen Forschung. Diese Krise bezieht sich auf die Feststellung, dass Forscher oftmals nicht in der Lage sind, veröffentlichte Forschungsergebnisse zu validieren oder nachzunutzen. Tatsächlich wurde die Reproduzierbarkeitskrise in den meisten Wissenschaftsfeldern beschrieben. Eine der Hauptursachen liegt in dem Aufwand, der benötigt wird, um Forschungsmaterialien zu dokumentieren, vorzubereiten und zu teilen. Wissenschaftler empfinden diese Forschungspraktiken oftmals als unattraktiv, da sie außerhalb der traditionellen wissenschaftlichen Belohnungsstruktur liegen. Diese ist zumeist ausgelegt auf das Veröffentlichen neuer Forschungsergebnisse.

Wissenschaftliche Forschung basiert zunehmend auf der Verarbeitung und Analyse großer Datensätze. Dieser Übergang zur rechnergestützten und daten-intensiven Forschung stellt neue Herausforderungen an Reproduzierbarkeit und Forschungsnachnutzung. Die weite Verbreitung des Internets bietet jedoch ebenso neue Möglichkeiten, Reproduzierbarkeit in der Forschung zu ermöglichen. Die Entwicklung des World Wide Web (WWW) stellt hierfür ein sehr gutes Beispiel dar. Das WWW wurde in der Europäischen Organisation für Kernforschung (CERN) entwickelt, um Forschern den weltweiten Austausch von Daten zu ermöglichen. CERN ist eine der wichtigsten Großforschungseinrichtungen in der Teilchenphysik, welche zu den daten-intensivsten Forschungsbereichen gehört. In dieser Arbeit berichten wir über unsere Forschung, die sich auf CERN Analysis Preservation (CAP) fokussiert. CAP ist ein Forschungsdatenmanagement-Service (FDM-Service), zugeschnitten auf die größten Experimente von CERN. 

In dieser Arbeit entwickeln und kommunizieren wir ein erweitertes Verständnis der Interaktion von Forschern mit FDM-Infrastruktur. Aus Sicht der Mensch-Computer-Interaktion (MCI) sind folgende Aspekte fundamental: (1) Das Bestimmen von Voraussetzungen und Praktiken rund um FDM und Nachnutzung. (2) Das Entwickeln von Verständnis für die Rolle und Auswirkungen von FDM-Systemen in der wissenschaftlichen Arbeit. (3) Das Entwerfen von Systemen, die Praktiken unterstützen, motivieren und anerkennen, welche die Reproduzierbarkeit von Forschung vorantreiben.

Die Forschung, die wir in dieser Arbeit beschreiben, stellt die erste systematische Anwendung von MCI-Methoden in der Entwicklung von FDM-Systemen für Forschungsreproduzierbarkeit dar. Wir entwickeln ein empirisches Verständnis von Forschungspraktiken und der Rolle von unterstützenden Systemen durch überwiegend qualitative Forschung in Teilchenphysik und darüber hinaus. Des Weiteren entwerfen und implementieren wir Prototypen und Systeme mit dem Ziel, Wissenschaftler für FDM zu motivieren und zu belohnen. Wir verfolgten einen Mixed-Method-Ansatz in der Evaluierung der Nutzererfahrung bezüglich unserer Prototypen und Implementierungen. Wir berichten von vier empirischen Studien, in denen insgesamt 42 Forscher und Forschungsdaten-Manager teilgenommen haben.

In unserer ersten Interview-Studie haben wir Teilchenphysiker über FDM-Praktiken befragt und sie eingeladen, CAP zu nutzen und über den Service zu diskutieren. Unsere Ergebnisse zeigen, dass die mensch-zentrierte Studie von speziell angepassten FDM-Systemen eine besondere Blickweise auf das Entwerfen von Anreizen und bedeutungsvollen Belohnungen ermöglicht. Wir führen den Begriff secondary usage forms (Zweitnutzungsformen) in Bezug auf FDM-Infrastruktur ein. Hierbei handelt es sich um Nutzungsformen, die Forschern sinnvolle Anreize bieten, ihre Arbeiten zu dokumentieren und zu teilen. Basierend auf unseren Ergebnissen in der Teilchenphysik haben wir unseren Forschungsansatz daraufhin auf Wissenschaftler und Forschungsdatenmanager aus einer Vielzahl verschiedener und diverser Wissenschaftsfelder erweitert. In Bezug auf die Ergebnisse dieser Studie beschreiben wir ein zustandsbasiertes Modell über die Entwicklung individueller Selbstverpflichtung zu FDM. Wir erwarten, dass dieses Modell designorientierte Denk- und Methodenansätze in der künftigen Implementierung und Evaluation von FDM-Infrastruktur beeinflussen wird.

Des Weiteren haben wir einen Forschungsansatz zu Spielifizierung (Gamification) verfolgt, in dem wir untersucht haben, ob und wie Spielelemente FDM-Praktiken motivieren können. Zunächst haben wir zwei Prototypen eines spielifizierten FDM-Tools entwickelt, welche sich an CAP orientieren. Obwohl die beiden Prototypen auf sehr unterschiedlichen Entwurfskonzepten beruhen, fanden Teilchenphysiker beide angemessen und motivierend. Die Studienteilnehmer diskutierten insbesondere verbesserte Sichtbarkeit individueller Forscher und wissenschaftlicher Arbeiten. Basierend auf den Ergebnissen dieser ersten Studie zu Spielifizierung in FDM haben wir im nächsten Schritt sechs speziell zugeschnittene Forschungs-Abzeichen (tailored science badges) in CAP implementiert. Die Abzeichen bewerben das ausführliche Dokumentieren sowie besondere Nutzen der auf dem Service zugänglichen Forschungsarbeiten. Die Ergebnisse unserer Evaluierung mit Teilchenphysikern zeigen, dass die speziell zugeschnittenen Forschungs-Abzeichen neue und effektivere Möglichkeiten bieten, Forschungsmaterialien systematisch zu durchsuchen und zu entdecken. Hierdurch profitieren sowohl Nutzer als auch Forschungsdaten-Beisteuernde. Basierend auf den Ergebnissen diskutieren wir, wie die Forschungs-Abzeichen neue Formen der Interaktion mit großen Forschungsrepositorien ermöglichen.

Zum Schluss heben wir die besondere Rolle von MCI in der Entwicklung unterstützender FDM-Infrastruktur hervor. Wir betonen, dass speziell an Forschungspraktiken angepasste Systeme neue Ansätze in der Interaktion mit wissenschaftlichen Arbeiten ermöglichen. Wir beschreiben zwei Modelle und unsere Erwartung, wie MCI die Entwicklung künftiger FDM-Systeme nachhaltig beeinflussen kann. In diesem Zusammenhang präsentieren wir auch unsere Vision zu ubiquitären Strategien, die zum Ziel hat, Forschungsprozesse und Wissen systematisch festzuhalten.

\clearemptydoublepage
\selectlanguage{english}  
	\cleardoublepage

\selectlanguage{english}
\markboth{Acknowledgements}{Acknowledgements}
\section*{\LARGE\rmfamily\bfseries\scshape{Acknowledgements}}




I am very grateful that I was given the opportunity to conduct my doctoral research in such exciting and stimulating environments. I am even more grateful for learning to know so many outstanding people who became friends over the past years — both, professional and personal. Fairly acknowledging the contributions of everyone might very well make this part the most complicated to write.

\begin{center}
---
\end{center}

I want to thank \textbf{Albrecht Schmidt} for his greatly inspiring supervision of my doctoral research. From our first meeting, to the submission of this thesis, Albrecht always introduced a well-balanced mix of thought-provoking visions and pragmatic solutions. His calm and supportive nature enabled me to conduct my research with confidence and joy. I am equally grateful to \textbf{Paweł Woźniak}, who not only taught me how to be successful in our field, but how to enjoy my research to the fullest. Paweł’s enthusiasm for research affects and inspires the people around him. His knowledge about various threads of HCI motivated me to explore and enjoy diverse research topics and strategies. I am just as thankful to \textbf{Sünje Dallmeier-Tiessen}. None of this would have been possible without Sünje's vision and initiative towards human-centered design of RDM tools. I am grateful to Sünje for her supportive, demanding, and genuinely curious way of supervising my research at CERN. Notably, I am not only grateful to Sünje for improving my research skills, but also for helping me become a semi-advanced table tennis player.

I thank \textbf{Barry Brown} for agreeing to review this thesis. I am looking forward to meet Barry and to discuss my research with him and the committee. Further, I want to thank \textbf{Michael Hauschild} for his initiative, passion, and commitment towards the Wolfgang Gentner scholarship, as well as the exciting physics insights he discussed at coffee meetings. I am glad for the discussions I shared with \textbf{Tibor Šimko} and for his calm and patient way of listening to new ideas. 

During the past years, many of you saw me frequently visit my home institute in Stuttgart and Munich to discuss my research plan and progress. Equally important was re-connecting with the great people there. I learned from you, I wrote papers with some of you, and I shared great moments with all of you. I thank \textbf{Jakob Karolus} for becoming a good friend and travel companion who spontaneously jumps into any adventure, from Bommerlunder to statistical analysis; \textbf{Thomas Kosch} for being a Macarons-smashing friend who accepted me even without a thoroughly documented profile when we first met. I thank \textbf{Florian Lang} for being a friend who tolerates spontaneous room parties, for sharing culinary experiences, and for being a reliable craftsman who returns tools borrowed. The last one is a rare quality. Further, I thank \textbf{Matthias Hoppe} for preparing me to survive in jungles; \textbf{Pascal Knierim} for ensuring our safety on the slopes; \textbf{Jasmin Niess} for her support and many rounds of Kakerlakensalat; \textbf{Lars Lischke}, \textbf{Francisco Kiss}, \textbf{Lewis Chuang}, \textbf{Fiona Draxler}, and \textbf{Rufat Rzayev} for fun moments and great discussions; \textbf{Passant El.Agroudy} for creative culinary ideas; \textbf{Klaudia Greif} for her great hospitality in Łódź; \textbf{Anja Mebus} for helping me select suitable dates for visiting the institute; \textbf{Tonja Machulla} for taking the time to talk about goals and interests; and \textbf{Bastian Pfleging} for teaching creative skiing exercises.

I thank \textbf{Pascal Oser} for being a good friend with whom I could talk about research and life, as well as for sharing his private pool; \textbf{Stephanie van de Sandt} for countless discussions about our research, intercultural differences, and Feuerzangenbowle; \textbf{Sebastian Bott} for countless Squash matches; \textbf{Felix Ehm} for his curiosity and for taking me along to one of the coolest trips ever; \textbf{Marina Savino} for organizing all those official trips; \textbf{Achintya Rao} for sharing his interesting stories and for talking about PhD research; \textbf{Ana Trisovic} for being a source of ideas; \textbf{Salvatore Mele} for welcoming me to the team; \textbf{Ania Trzcińska} for sabotaging my achievements playing Munchkin; \textbf{Pamfilos Fokianos} for ` \textit{It's ok}' when things go bad; \textbf{Giannis Tsanaktsidis} for talking about motorbikes; \textbf{Alex Kohls} for providing guidance in selecting vacation destinations suitable for post-thesis submission; \textbf{Kamran Naim} for his interest in my research and for organizing the best parties; \textbf{Artemis Lavasa}, \textbf{Robin Dasler}, \textbf{Xiaoli Chen}, \textbf{Harri Hirvonsalo}, \textbf{Ilias Koutsakis}, \textbf{Antonios Papadopoulos}, \textbf{Jan Okraska}, \textbf{Marcos Oliveira}, \textbf{Jennifer Dembski}, \textbf{Stella Christodoulaki}, \textbf{Jelena Brankovic}, \textbf{Diego Rodriguez}, \textbf{Rokas Maciulaitis}, \textbf{Jens Vigen}, and \textbf{Micha Moskovic} for countless conversations and for making work and life at CERN pleasurable.

Further, I am grateful to \textbf{Sebastian Suchanek}, \textbf{Fabienne Kirschner}, \textbf{Philipp Lacatusu}, \textbf{Melanie and Benjamin Maier}, \textbf{Laura Comella}, and \textbf{Yves Fischer} for their impact and support.

I am grateful beyond words to my family. My parents \textbf{Simone and Gerhard Feger} provided me with love and the environment needed to grow to the person that I am today. I thank \textbf{Günther Brommer} for encouraging me to achieve more; \textbf{Irene Burkhardt} for countless enjoyable visits in Achern; \textbf{Stefan Brommer} for being a great inspiration; \textbf{Oliver, Antje, and Leon} for their curiosity; \textbf{Karin Mayer}, \textbf{Maria and Gottfried Feger}, and \textbf{Helga Stadler} for all their contributions to my life. Last but not least, I am extremely grateful to \textbf{Sara Marconi} for her love and support, for making me happy, and for sharing exciting plans for our future.

\textit{Thank you all.}

\textit{Danke.}

\clearemptydoublepage
	
	\cleardoublepage
	
	\makeatletter
	\setlength{\cftbeforepartskip}{7ex}
	\setlength{\cftbeforechapskip}{5.5ex}
	\setlength{\cftbeforesecskip}{0.75ex}
	\setlength{\cftbeforesubsecskip}{0.1ex}
	\setlength{\cftbeforetoctitleskip}{-5mm}
	\setlength{\cftbeforeloftitleskip}{-5mm}
	\setlength{\cftbeforelottitleskip}{-5mm}
	\renewcommand{\cfttoctitlefont}{
		\LARGE\usefont{OT1}{ptm}{b}{sc}\selectfont
	}
	\renewcommand{\cftloftitlefont}{
		\LARGE\usefont{OT1}{ptm}{b}{sc}\selectfont
	}
	\renewcommand{\cftlottitlefont}{
		\LARGE\usefont{OT1}{ptm}{b}{sc}\selectfont
	}
	\renewcommand{\cftpartfont}{
		\fontsize{14}{18}\usefont{OT1}{ptm}{b}{sc}\selectfont
	}
	\renewcommand{\cftchapfont}{
		\fontsize{13}{17}\usefont{OT1}{ptm}{b}{n}\selectfont
	}	

	\setcounter{tocdepth}{2}
	\setcounter{secnumdepth}{2}
	\pdfbookmark[1]{Table of Contents}{table}
	\markboth{Table of Contents}{Table of Contents}
	\settocname{Table of Contents}
	\tableofcontents
	\cleardoublepage	
	
	
	\mainmatter
	
	\part{Introduction and Background}\label{part:intro}


\chapter{Introduction }
\label{ch:introduction}

Reproducibility is an essential tool of the scientific method that enables research validation and reuse. Discussions about values and challenges of reproducibility date back several hundred years. In the 17$^{th}$ century, Robert Boyle designed an air pump to generate and study vacuum. At the time, the study and existence of vacuum caused disputes in the scientific world. The air pump itself was a complex device that was inaccessible for most researchers. This made it difficult to reproduce Boyle's findings. Thus, the air pump and related discoveries represent an early example of the values and challenges of reproducibility in validating scientific findings. And science still faces some of those challenges today.

Experimental verification is the core topic in the \textit{science-in-fiction} novel \textit{Cantor's Dilemma}~\cite{djerassi2012cantor}. The novel was written by Carl Djerassi, a chemist who was instrumental in the development of the oral contraceptive pill \cite{ball2015carl}. In \textit{Cantor's Dilemma}, Djerassi illustrated how scientists work and addressed some grand challenges and topics of discussion in science, including the fair reflection of contributions and the verification of experimental research. The novel introduces the fictional Professor Isidore Cantor. Professor Cantor makes a breakthrough in cancer research that is expected to favor him in the race for the Nobel Prize. To test his theory, he designs an experiment that his fictional research assistant Dr.~Jeremiah~Stafford conducts. Since the experiment is successful, Cantor and Stafford publish the theory and the experimental results and wait for an independent verification of their findings. The novel highlights the value of independent verification of results, addresses issues of trust in science, and illustrates how missing information in the experimental documentation and protocol hinder research validation. 

The importance of independent research validation is also reflected in recent reports related to the potential discovery of a fundamental \textit{fifth force of nature}. Those reports gained attention in both scientific and popular media \cite{cockburn_2019}, showing some parallels to Djerassi's novel. Krasznahorkay et al. \cite{krasznahorkay2019new} described new experimental evidence for which they ``are expecting more, independent experimental results [...] in the coming years.'' Given successful replication of the results, Jonathan Feng, a professor of physics and astronomy, stated that ``this would be a no-brainer Nobel Prize.'' \cite{prior_2019}

Issues related to documentation, as described in Cantor's Dilemma, are reflected in the research that we present in this thesis. And also inaccessibility of experimental resources still poses serious challenges to reproducibility in science today. While preserving and sharing research are basic requirements for reproducibility ~\cite{Bechhofer2013, Wilkinson2016, FAIR}, they require substantial efforts to clean and document resources ~\cite{Borgman:1297241}. Yet, the traditional academic reputation economy is focused on rewarding novel contributions, rather than reproductions. Thus, it has been argued, that the scientific culture not only lacks support for systematic reproducibility, but even impairs reproducible research practices \cite{Begley2012,Collaboration2012,fecher2017reputation}.

In a large-scale survey by Baker \cite{Baker2016}, ${90\%}$ of the 1,576 participating researchers agreed that there was a reproducibility crisis. More than half of the researchers who took the questionnaire agreed that there was even a \textit{significant} crisis. Most researchers who participated in that study reported that they tried to reproduce work in the past, but ultimately failed to do so. Notably, this is true both for work published by someone else, as well as the researchers' own work. Based on the survey, Baker found that factors related to competition and time pressure contribute strongly to irreproducible research. The unavailability of methods, code, and raw data was referred to as factors contributing to irreproducibility by around ${80\%}$ of the participants. Approximately the same number of participants agreed that \textit{Incentives for better practice} could boost scientific reproducibility.

The findings from Baker's survey study are based on responses from scientists working in a wide range of scientific fields, including chemistry, physics, engineering, biology, medicine, and environmental science. Definitions, practices, and requirements for reproducible research differ between scientific fields \cite{Feitelson2015, Schmidt2009}. Yet, what they have in common is the ongoing transformation of research practices through the wide-spread use of information technology in scientific research. In fact, \textit{computational science} is referred to as the 3$^{rd}$ paradigm of science \cite{bell2009beyond}. This digital transformation provides opportunities for more efficient and effective preservation and sharing of experimental material. However, even though barriers for sharing digital resources are low, availability of research material remains a major concern ~\cite{echtler2018open, Stodden2014}. Concerns related to irreproducibility have been voiced even in modern computational fields like Artificial Intelligence \cite{hutson2018artificial, gundersen2018state}. This is alarming, as data volumes in science continue to grow rapidly. In fact, \textit{data-intensive science} has been described as an evolving 4$^{th}$ paradigm of science \cite{bell2009beyond}. 

\textit{E-Science}, ``the application of computer technology to the undertaking of modern scientific investigation'' \cite{bohle2013science}, is strongly related to the notion of data-intensive science. Large data volumes, grid computing, and distributed collaboration are some of its defining features. E-Science does not only create new opportunities for global collaboration. It is also expected to enable systematic sharing and reuse in science \cite{Jirotka2006, Karasti2006}. It is mainly through today's availability and access to online technologies that we see new opportunities for the development of tools that support scientists in their \ac{RDM} \cite{Pasquetto:2016:ODS:2858036.2858543, Worden2017}. 

Two completely different types of supportive RDM tools are emerging: general data repositories and community-tailored services \cite{Wallis2013}. General data repositories, like Dryad\footnote{https://datadryad.org/stash} and Zenodo\footnote{https://zenodo.org/}, enable submission, preservation, and sharing of any kind of digital data, making such repositories suitable for all scientific fields. Instead, community-tailored services map research workflows of a target domain. This mapping enables more targeted preservation, discovery, and reuse through domain-tailored language \cite{chen2016cern, Jackson2013}. However, the design, implementation, and maintenance of tailored tools is more difficult and expensive \cite{schwartz2010data, cragin2010data}. Overall, we need to further our understanding of researchers' requirements for supportive RDM tools and interconnections between preservation, sharing, tools, and knowledge lifecycles \cite{Jirotka2006}. As Jackson and Barbrow pointed out, we ``need to supplement or replace generic, tool-centered, and aspirational accounts of cyberinfrastructure development with approaches that start from the individual histories of practice and value in specific scientific fields'' \cite{Jackson2013}.

The systematic study of infrastructure development must not only focus on easing RDM practices, but also on motivating them. In fact, it has been argued that only minimizing the effort required to follow reproducible research practices might not be sufficient to engage scientists at large \cite{Borgman2006}. Thus, the study of requirements for the development and adoption of RDM infrastructure must reflect the role of policies \cite{Pasquetto:2016:ODS:2858036.2858543}, in particular those issued by publishers \cite{Belhajjame2014, Stodden2014} and research funders \cite{russell2013if, Rosenblatt336ed5}. While enforcement will always play a role in ensuring compliance, our research focused on understanding how the design of supportive technology can create meaningful motivation for researchers to follow core reproducible practices. The work of Rowhani-Farid et al. stresses the importance of this research approach \cite{Rowhani-Farid2017}. They conducted a systematic literature review of incentives in the medical domain and found that although ``data is the foundation of evidence-based health and medical research, it is paradoxical that there is only one evidence-based incentive to promote data sharing.'' The authors referred to \ac{OSB} \cite{Kidwell2016}, issued by the Center for Open Science\footnote{https://cos.io/our-services/open-science-badges/}. They concluded that ``more well-designed studies are needed in order to increase the currently low rates of data sharing.'' The research reported in this thesis responds to the call for studying incentives through a systematic application of HCI methods.




%
%






%
%
\section{Research Context}

This thesis is based on research conducted primarily at the \ac{CERN}. CERN\footnote{https://home.cern/} is a leading laboratory in \ac{HEP}, located at the border between France and Switzerland, close to Geneva. The research work was supported by the CERN \ac{SIS}\footnote{http://library.cern/}, and in particular S\"unje Dallmeier-Tiessen. CERN is an international organization that is publicly funded. It has 23 member states and seven associate member states\footnote{Retrieved October 2, 2019. https://home.cern/about/who-we-are/our-governance/member-states}. CERN is best know for its research on the \ac{LHC}, the world's largest particle accelerator \cite{Evans2008}. The LHC\footnote{https://home.cern/science/accelerators/large-hadron-collider} consists of a 27-kilometer underground ring, designed to collide particles in four locations. Four main detectors measure particle collisions at these collision points, which make up CERN's four largest research collaborations: ALICE, ATLAS, CMS, and LHCb \cite{Gustafsson2006}.

HEP in general, and CERN in particular, represent ideal environments to study practices, needs, and requirements of reproducible science. Our research profited from five defining characteristics. First, in terms of challenges for reproducibility, parallels can be identified between the LHC and Boyle's air pump. While the air pump was accessible to very few researchers in the 17$^{th}$ century, the LHC and its detectors are unique research apparatus. In order to validate findings, the LHC collaborations perform their research mostly independent from other collaborations. This is especially true for the two general-purpose detectors ATLAS and CMS \cite{Cho1564}. Effective RDM that enables internal and external reproducibility is needed to establish trust in the results and the responsible use of unique data recorded by publicly founded research experiments.

Second, HEP is one of the most data-intensive branches of science. In 2017, CERN reported having permanently archived more than 200 petabytes of data \cite{Melissa:2276551}, making CERN a shining example of data-intensive science \cite{bell2009beyond, kouzes2009changing}. This vast amount of data poses extreme challenges to accessibility and reproducibility. Yet, findings from our research are expected to inform the design of supportive tools far beyond CERN and HEP. In fact, we consider as third defining characteristic CERN's demonstrated ability to advance the overall development of computing technology --- both within and outside of science. In response to the challenges posed by large data volumes, the \ac{WWW} was invented at CERN in 1989, to enable rapid sharing of data, codes, and findings with the scientific community \cite{Berners-Lee1992,birth:1998446, bentley1995supporting}. Furthermore, CERN played an instrumental role in the advancement of grid computing \cite{segal2000grid}. Those are examples that show how CERN's unique requirements informed the design of technology far beyond the scope of physics research. Similarly, we expect that our findings will benefit RDM and reproducible practices beyond particle physics.

Fourth, the openness in HEP represents a strength of this scientific domain. Scholarly communication in HEP is characterized by the \textit{preprint server} culture, which enables physicists to freely and immediately share resources and ideas \cite{Gentil-Beccot2010,doi:10.1177/0306312716659373}. Velden \cite{Velden2013} illustrated this openness in her ethnographic study which involved experimental physics groups at shared radiation facilities. She found that those groups shared information despite competition.

Finally, the size and distributed organization of the four major LHC collaborations provide a highly valuable framework for studying collaborative data science practices, with particular regard to reproducibility and reuse. In fact, the LHC collaborations involve hundreds of institutes worldwide\footnote{https://greybook.cern.ch/greybook/researchProgram/detail?id=LHC}, making them a shining example of data-intensive, collaborative e-Science \cite{newman2003data}. ATLAS and CMS are the two largest LHC collaborations. While ``ATLAS comprises about 3000 scientific authors from 183 institutions around the world''\footnote{Retrieved February 6, 2020. https://atlas.cern/discover/collaboration}, ``CMS has over 4000 particle physicists, engineers, computer scientists, technicians and students from around 200 institutes and universities''\footnote{Retrieved February 6, 2020. https://cms.cern/collaboration}. While globally distributed, the work of the collaborations is focused on data collected locally within the experiments' detectors. The special structure of CERN's collaborations attracted much attention among social scientists. Merali \cite{Merali2010} stressed that responsibility within the collaborations is distributed among the highly specialized teams, rather than mandated top-down. In Merali's study, a spokesperson noted that ``in industry, if people don't agree with you and refuse to carry out their tasks, they can be fired, but the same is not true in the LHC collaborations.'' Because ultimately, ``physicists are often employed by universities, not by us.'' This lack of a strong top-down structure provides for an interesting environment, as it makes enforcing RDM practices difficult. Instead, it calls for solutions that make data management meaningful, rewarding, and motivating.

In this thesis, we argue repeatedly that our findings are likely to hold value in a broader scientific context. As science becomes increasingly data-intensive, we believe that our findings will become influential in guiding RDM infrastructure design well beyond HEP. Yet, we need to emphasize that most studies reported in this thesis represent empirical research conducted with HEP researchers. In addition to the HEP studies, we conducted one cross-domain study involving researchers and research data managers from a wide variety of diverse scientific fields. Still, with the focus on HEP, we advise practitioners and researchers working in other scientific domains to carefully validate applicability of our findings and guidelines.

%
%

\section{Research Questions}

Large data volumes, open scholarly communication practices, and a highly collaborative character of research work make computation-driven HEP a shining example of data-intensive e-Science \cite{newman2003data, bell2009beyond, kouzes2009changing}, which provides new opportunities for systematic and effective RDM \cite{Jirotka2006, Karasti2006}. Thus, CERN provides an ideal environment to address our primary \ac{RQ}: \textbf{How to design interactive tools that support and improve reproducible research practices?}

Based on our primary RQ, calls for field-specific investigations \cite{Jackson2013}, the characterized lack of meaningful incentives \cite{Rowhani-Farid2017}, and our own intermediate findings, we identified four secondary RQs. First, we investigate the role of technology in supporting reproducible practices (RQ1). Second, we study how preservation tools impact current practices around RDM, reproducibility, and reuse --- both within and outside of HEP (RQ2). Based on those findings and related work \cite{Kidwell2016, Rowhani-Farid2017}, we pose RQ3: How can gamification and motivational design stimulate preservation and sharing in science? Based on the sum of our findings, we aim to gain a holistic understanding of how HCI and its methods can impact reproducible science. Thus, we address RQ4: How does HCI contribute to open and reproducible science? Table \ref{tab:rqs} presents all research questions and refers to corresponding chapters and sections.

\cellspacetoplimit 2pt
\cellspacebottomlimit 2pt

\begin{table}
  \centering
  \begin{tabular}{Sl Sl Sc Sc}
    {\small\textit{}}
    & {\textbf{Research Question}}
    & {\textbf{Part}}
    & {\textbf{Chapter}}\\
    \toprule
    RQ & \makecell[l]{How to design interactive tools that support and improve \\ reproducible research practices?}  & & \\ \hline
    
    RQ1 & \makecell[l]{What is the role of technology in supporting reproducible \\ research practices?} & \hyperref[part:intro]{\ref{part:intro}}, \ref{part:requirements} & \hyperref[ch:open]{\ref{section:open}}, \ref{ch:requirements}, \ref{ch:cross_domain} \\
    \hline
    
    RQ2 & \makecell[l]{How do preservation tools impact practices around RDM, \\ reproducibility, and reuse?} & \hyperref[part:requirements]{\ref{part:requirements}} & \\
    RQ2.1 & \makecell[l]{What are practices and challenges of resource sharing and \\ reuse in HEP data analysis?} & \hyperref[part:requirements]{\ref{part:requirements}} & \hyperref[ch:requirements]{\ref{ch:requirements}} \\ 
    RQ2.2 & \makecell[l]{How can research preservation tools support HEP \\ workflows and incentivize contributions?} &
    \hyperref[part:requirements]{\ref{part:requirements}} & \hyperref[ch:requirements]{\ref{ch:requirements}} \\
    RQ2.3 & \makecell[l]{How do findings from HEP compare to other, diverse \\ fields of science?} & \hyperref[part:requirements]{\ref{part:requirements}} & \hyperref[ch:cross_domain]{\ref{ch:cross_domain}} \\
    \hline
    
    RQ3 & \makecell[l]{How can gamification and motivational design stimulate \\ preservation and sharing in science?} & \hyperref[part:gamification]{\ref{part:gamification}} & \\
    RQ3.1 & \makecell[l]{What are requirements and perceptions of gamification \\ in the context of reproducible science?} & \hyperref[part:gamification]{\ref{part:gamification}} & \hyperref[ch:gamification_requirements]{\ref{ch:gamification_requirements}} \\
    RQ3.2 & \makecell[l]{How does the implementation of game design elements \\ impact research practices and interaction with preservation tools?} & \hyperref[part:gamification]{\ref{part:gamification}} & \hyperref[ch:tailored_badges]{\ref{ch:tailored_badges}} \\
    \hline
    
    RQ4 & \makecell[l]{How does HCI contribute to open and reproducible science?} & \hyperref[part:conclusion]{\ref{part:conclusion}} & 
    \hyperref[ch:hci_role]{\ref{ch:hci_role}}\\
    
    \bottomrule
  \end{tabular}
  \caption{Overview of the addressed research questions.}~\label{tab:rqs}
\end{table}

%
%
\section{Methodology and Evaluation}
\label{section:method}

Our research is based on \ac{UCD} \cite{norman1986user, abras2004user}. We conducted interview studies with scientists to map practices, understand technology interaction, describe design requirements, and evaluate interactive tools. Based on our understanding of user needs, we designed preservation service prototypes and a gamified research data management tool. We conducted mixed-method evaluation studies with HEP researchers and derived design implications from our findings. To the best of our knowledge, this represents the first systematic application of HCI methods in studying and designing interactive tools for research reproducibility.

\begin{figure}
  \centering
  \includegraphics[width=0.9\columnwidth]{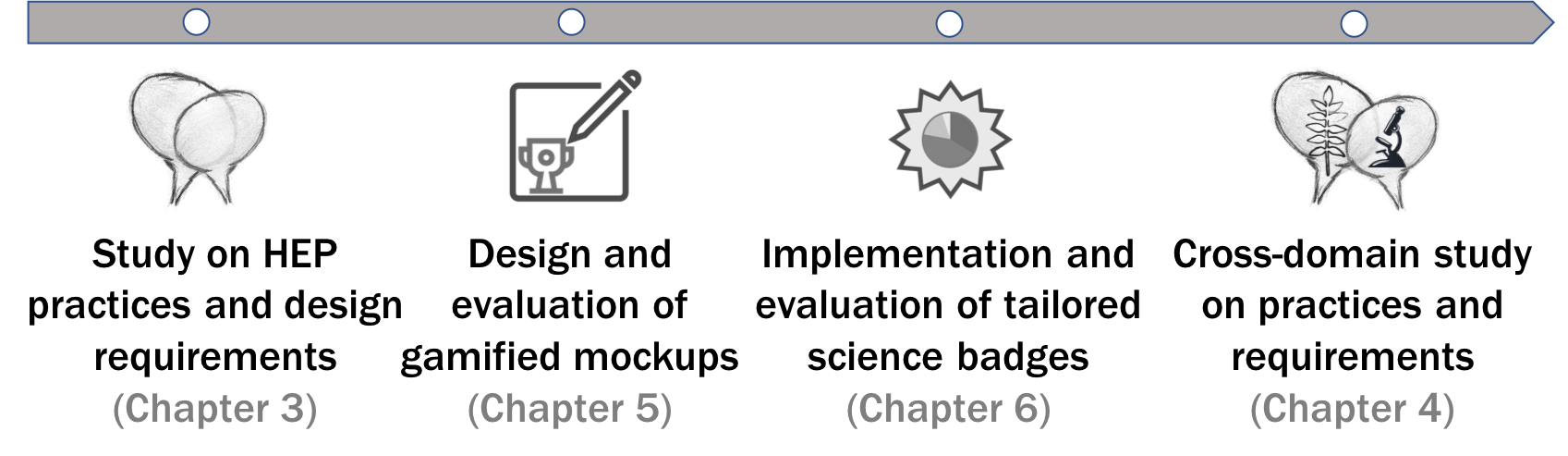}
  \caption{High-level overview of the research process.}~\label{fig:research_process}
\end{figure}

The research and conclusions reported in this thesis are based on four empirical studies. Figure \ref{fig:research_process} depicts our sequential research process. First, we conducted an interview study at CERN to map practices around RDM, reproducibility and reuse in HEP (see Chapter \hyperref[ch:requirements]{\ref{ch:requirements}}). As part of this study, we asked physics researchers to explore the CAP prototype service and to discuss its value, concerns, and challenges. This allowed us to investigate design requirements and present design implications. In particular, we found that researchers needed strong incentives to contribute to a preservation service. Based on those findings, we investigated opportunities for motivational design. In particular, we studied applications of gamification in the scientific context. Gamification, the ``use of game design elements in non-game contexts'' \cite{Deterding2011}, has proven to create motivation and engagement across a wide variety of different applications \cite{Ibanez2014, knaving2018understanding, Oprescu2014}. But, applications and research in the science context were mostly limited to participation of the general public in scientific processes (citizen science)~\cite{Eveleigh2013}. We designed prototypes of two contrasting gamified research preservation services that we evaluated in a mixed-method study with HEP data analysts (see Chapter \ref{ch:gamification_requirements}). Based on the findings from this evaluation, we, third, implemented and evaluated tailored science badges in CAP (see Chapter \ref{ch:tailored_badges}). Finally, we conducted a cross-domain study on practices around RDM, reproducibility, and reuse with researchers and data managers from a wide variety of scientific fields (see Chapter \ref{ch:cross_domain}). For this study, we designed a generic preservation service prototype that is inspired by CAP. The goal of this cross-domain investigation was to relate our findings from HEP to practices in other domains and to inform the design of supportive tools beyond physics and natural sciences.

In total, we conducted 45 interviews and evaluation sessions with 42 distinct participants. In order to create a most thorough understanding of the data, I transcribed 34 of those sessions myself non-verbatim. The remaining 11 recordings were transcribed by a professional transcription service. In total, we recorded around 29 hours of interview and evaluation sessions. To understand how interactive tools impact current practices around RDM, reproducibility, and reuse, we conducted a total of 24 semi-structured interviews with researchers and data managers from within and outside of HEP (see Part \ref{part:requirements}). Our studies on requirements and effects of gamification and motivational design focused on 21 mixed-method evaluation sessions (see Part \ref{part:gamification}). We recruited highly trained and skilled participants for all our studies. All participants held an academic degree. Out of the 42 participants, 7 were PhD students, and 30 participants held a doctoral degree. Out of those 30 PhD holders, we identified seven participants with a particularly senior role (e.g. member of the upper management, professor, team leader). We provide details of all recruited participants in the corresponding sections.

Regarding the chosen methodology, we want to emphasize the qualitative focus of our research as both a strength and limitation of our work. We acknowledge that questions concerning reproducibility, rigour, and transparency of qualitative research have been raised \cite{mays2020quality, meyrick2006good, tsai2016promises}. In this context, we would like to stress that we made several resources available to the reviewers of this PhD thesis, conference and journal reviewers, and openly as supplementary material accompanying our publications. Those include the study protocols and data analysis reports. We would like to further stress that a focus on qualitative methods was needed to ensure that the tools we design are suitable for supporting and motivating comprehensive RDM and that novel interactive tools do not risk to alienate early open science adopters. In our research, the qualitative study of practices around RDM and requirements for tool design allowed to build a thorough understanding of the delicate interplay between various drivers for RDM commitment and the design of supportive RDM tools. We argue that a rigorous qualitative approach was needed to build a solid foundation for future systems design. In particular, it allowed to describe a fundamental change in service design paradigms (see ``secondary usage forms'' in Part \ref{part:requirements}) and enabled the systematic mapping of perceptions and description of design guidelines regarding the use of gamification in highly skilled environments (see Part \ref{part:gamification}). 

%
%

\section{Contributing Publications}

A substantial part of this thesis is based on research reported in peer-reviewed publications, or research that has been submitted to international peer-reviewed venues. One section is based on a MetaArXiv preprint. The decision to make our vision of ubiquitous research preservation freely and immediately available in a preprint reflects the scholarly communication practice in HEP. With the exception of one, I am the first author of all the publications. The exception is a \textit{Nature Physics} paper for which I had been involved in the discussion from concept to publication. However, I have not made major contributions to that manuscript. I list this publication here and base Section \hyperref[section:open]{\ref{section:open}} on it, as the paper motivates the need for technology in supporting reproducible research practices in HEP. Thus, it represents a key motivation for the PhD research reported in this thesis. All other publications profited from close collaboration and discussions with the co-authors, but they were primarily written by myself. Table \ref{tab:publications} provides an overview of all publications, their status, co-authors, and corresponding chapters in this thesis. 

\newpage

\cellspacetoplimit 8pt
\cellspacebottomlimit 8pt

\begin{longtable}{clc}

    \textbf{Chapter} & \textbf{Publication} & \textbf{Status} \\
    \hline
    \endhead
    \hline
    \endfoot

    & & \\

    \hyperref[ch:open]{\ref{section:open}} & \multicolumn{1}{m{10cm}}{\raggedright Open is not enough \smallskip 
    \newline 
    Xiaoli Chen, \uline{Sünje Dallmeier-Tiessen}, Robin Dasler, \textbf{Sebastian Feger}, Pamfilos Fokianos, Jose Benito Gonzalez,  Harri Hirvonsalo, Dinos Kousidis, Artemis Lavasa, Salvatore Mele, Diego Rodriguez Rodriguez, \uline{Tibor Šimko}, Tim Smith, \uline{Ana Trisovic}, Anna Trzcinska, Ioannis Tsanaktsidis, Markus Zimmermann, Kyle Cranmer, Lukas Heinrich, Gordon Watts, Michael Hildreth, Lara Lloret Iglesias, Kati Lassila-Perini \& Sebastian Neubert.
    \newline
    \textbf{Nature Physics (2018)}. 7 pages.
    \newline 
    \url{https://doi.org/10.1038/s41567-018-0342-2}}
    & Published \cite{chen2018open} \\
    
    & & \\

    \hyperref[ch:requirements]{\ref{ch:requirements}} & \multicolumn{1}{m{10cm}}{\raggedright Designing for Reproducibility: A Qualitative Study of Challenges and Opportunities in High Energy Physics \smallskip  \newline 
    \uline{\textbf{Sebastian Feger}}, Sünje Dallmeier-Tiessen, Albrecht Schmidt, and Paweł W. Woźniak 
    \newline
    In CHI Conference on Human Factors in Computing Systems Proceedings \textbf{(CHI 2019)}. 14 pages.
    \newline
    \url{https://doi.org/10.1145/3290605.3300685}} & Published \cite{Feger2019Requirements} \\

    & & \\
    
    \hyperref[ch:cross_domain]{\ref{ch:cross_domain}}, \ref{fig:rdm_model} & \multicolumn{1}{m{10cm}}{\raggedright `Yes, I comply!': Motivations and Practices around Research Data Management and Reuse across Scientific Fields \smallskip
    \newline 
    \uline{\textbf{Sebastian Feger}}, Paweł W. Woźniak, Lars Lischke, and Albrecht Schmidt
    \newline
    In Proceedings of the ACM on Human-Computer Interaction, Vol. 4, \textbf{CSCW}2, Article 141 (October 2020). ACM, New York, NY. 26 pages.
    \newline
    \url{https://doi.org/10.1145/3415212}}
    & \makecell[c]{Published \cite{Feger2020Comply}}  \\
  
    & & \\

    \hyperref[ch:gamification_motivation]{\ref{ch:gamification_requirements}} & \multicolumn{1}{m{10cm}}{\raggedright Just Not The Usual Workplace: Meaningful Gamification in Science  \smallskip
    \newline 
    \uline{\textbf{Sebastian Feger}}, Sünje Dallmeier-Tiessen, Paweł W. Woźniak, and Albrecht Schmidt
    \newline
    In: Dachselt, R. \& Weber, G. (Hrsg.), \textbf{Mensch und Computer 2018} -- Workshopband. 6 pages.
    \newline
    \url{https://doi.org/10.18420/muc2018-ws03-0366}} & Published \cite{feger2018just} \\
    
    & & \\

    \newpage
    
    & & \\

    \hyperref[ch:gamification_requirements]{\ref{ch:gamification_requirements}} & \multicolumn{1}{m{10cm}}{\raggedright Gamification in Science: A Study of Requirements in the Context of Reproducible Research \smallskip
    \newline 
    \uline{\textbf{Sebastian Feger}}, Sünje Dallmeier-Tiessen, Paweł W. Woźniak, and Albrecht Schmidt  
    \newline
    In CHI Conference on Human Factors in Computing Systems Proceedings \textbf{(CHI 2019)}. 14 pages.
    \newline
    \url{https://doi.org/10.1145/3290605.3300690}} & Published \cite{Feger2019Gamification} \\
    
    & & \\
    
    \hyperref[ch:tailored_badges]{\ref{ch:tailored_badges}} & \multicolumn{1}{m{10cm}}{\raggedright Tailored Science Badges: Enabling New Forms of Research Interaction. 12 pages. \smallskip
    \newline 
    \uline{\textbf{Sebastian Feger}}, Paweł W. Woźniak, Jasmin Niess, and Albrecht Schmidt} & \makecell[c]{Manuscript is\\being prepared\\for submission} \\

    & & \\

    \hyperref[ch:hci_role]{\ref{ch:hci_role}} & \multicolumn{1}{m{10cm}}{\raggedright More Than Preservation: A Researcher-Centered Approach to Reproducibility in Data Science \smallskip
    \newline 
    \uline{\textbf{Sebastian Feger}} and Paweł W. Woźniak
    \newline
    Accepted and presented at the \textbf{CHI 2019} Workshop on \textit{Human-Centered Study of Data Science Work Practices}. Published on CERN CDS. 4 pages.
    \newline
    \url{http://cds.cern.ch/record/2677268}} & Published \cite{Feger:2677268}\\   

    & & \\

    \hyperref[ch:hci_role]{\ref{ch:hci_role}} & \multicolumn{1}{m{10cm}}{\raggedright The Role of HCI in Reproducible Science: Understanding, Supporting and Motivating Core Practices \smallskip
    \newline 
    \uline{\textbf{Sebastian Feger}}, Sünje Dallmeier-Tiessen, Paweł W. Woźniak, and Albrecht Schmidt  
    \newline
    In Extended Abstracts of the CHI Conference on Human Factors in Computing Systems \textbf{(CHI 2019)}. 6 pages.
    \newline
    \url{https://doi.org/10.1145/3290607.3312905}} & Published \cite{Feger2019RoleHCI}\\    
    
    & & \\
    
    \hyperref[ch:hci_role]{\ref{ch:hci_role}} & \multicolumn{1}{m{10cm}}{\raggedright More than preservation: Creating motivational designs and tailored incentives in research data repositories \smallskip
    \newline 
    \uline{\textbf{Sebastian Feger}}, Sünje Dallmeier-Tiessen, Pamfilos Fokianos, Dinos Kousidis, et al. 
    \newline
    Peer-reviewed, accepted presentation proposal for a full talk at \textbf{Open Repositories 2019}. Published on CERN CDS. 5 pages.
    \newline
    \url{https://cds.cern.ch/record/2691945}} & Published \cite{Feger:2691945} \\
    
    & & \\

    \newpage
    
    & & \\
    
    \hyperref[section:urp]{\ref{section:urp}} & \multicolumn{1}{m{10cm}}{\raggedright Ubiquitous Research Preservation: Transforming Knowledge Preservation in Computational Science. \smallskip
    \newline 
    \uline{\textbf{Sebastian Feger}}, Sünje Dallmeier-Tiessen, Pascal Knierim, Passant El.Agroudy, Paweł W. Woźniak, and Albrecht Schmidt   
    \newline
    MetaArXiv Preprint. 4 pages.
    \newline
    \url{https://doi.org/10.31222/osf.io/qmkc9}
    } & Published \cite{feger2020urp}  \\ 

    & & \\

  \caption{Overview of publications that contribute to this thesis.}~\label{tab:publications}
\end{longtable}

Table \ref{table:personal_contributions} details my personal contributions to the publications and corresponding studies listed in Table \ref{tab:publications}.

\cellspacetoplimit 8pt
\cellspacebottomlimit 8pt

\begin{longtable}{p{.48\textwidth} p{.48\textwidth}} 
\textbf{Publication} & \textbf{Personal Contributions} \\
    \hline
    \endhead
    \hline
    \endfoot
    
Open is not enough \cite{chen2018open} & My PhD research contributed to the conceptual design of this publication. I made suggestions for minor improvements in the manuscript. \\
    
    \midrule
    
    Designing for Reproducibility: A Qualitative Study of Challenges and Opportunities in High Energy Physics \cite{Feger2019Requirements} \newline \newline ‘Yes, I comply!’: Motivations and Practices around Research Data Management and Reuse across Scientific Fields \cite{Feger2020Comply} & I drafted the study designs and protocols, conducted all interviews, transcribed most of the interviews, analysed all transcriptions, and made the most contributions to all sections of the final manuscripts. \\
    
    \midrule
    
    Gamification in Science: A Study of Requirements in the Context of Reproducible Research \cite{Feger2019Gamification} \newline \newline Tailored Science Badges: Enabling New Forms of Research Interaction & I drafted the study designs and protocols, designed and implemented the gamification prototypes, conducted all mixed-method evaluation sessions, transcribed all of the interviews, analysed all recorded data, and made the most contributions to all sections of the final manuscripts. \\
    
    \midrule
    
    Just Not The Usual Workplace: Meaningful Gamification in Science \cite{feger2018just} \newline \newline More Than Preservation: A Researcher-Centered Approach to Reproducibility in Data Science \cite{Feger:2677268} \newline \newline The Role of HCI in Reproducible Science: Understanding, Supporting and Motivating Core Practices \cite{Feger2019RoleHCI} \newline \newline More than preservation: Creating motivational designs and tailored incentives in research data repositories \cite{Feger:2691945} & I conceptually designed the various publications and made the most contributions to all sections of the final manuscripts. \\
    
    \midrule
    
\caption{Detailed description of my personal contributions to the publications referred to in this PhD thesis.} 
\label{table:personal_contributions}
\end{longtable}

%
%
\section{Thesis Outline}

\begin{figure}
  \centering
  \includegraphics[width=1.0\columnwidth]{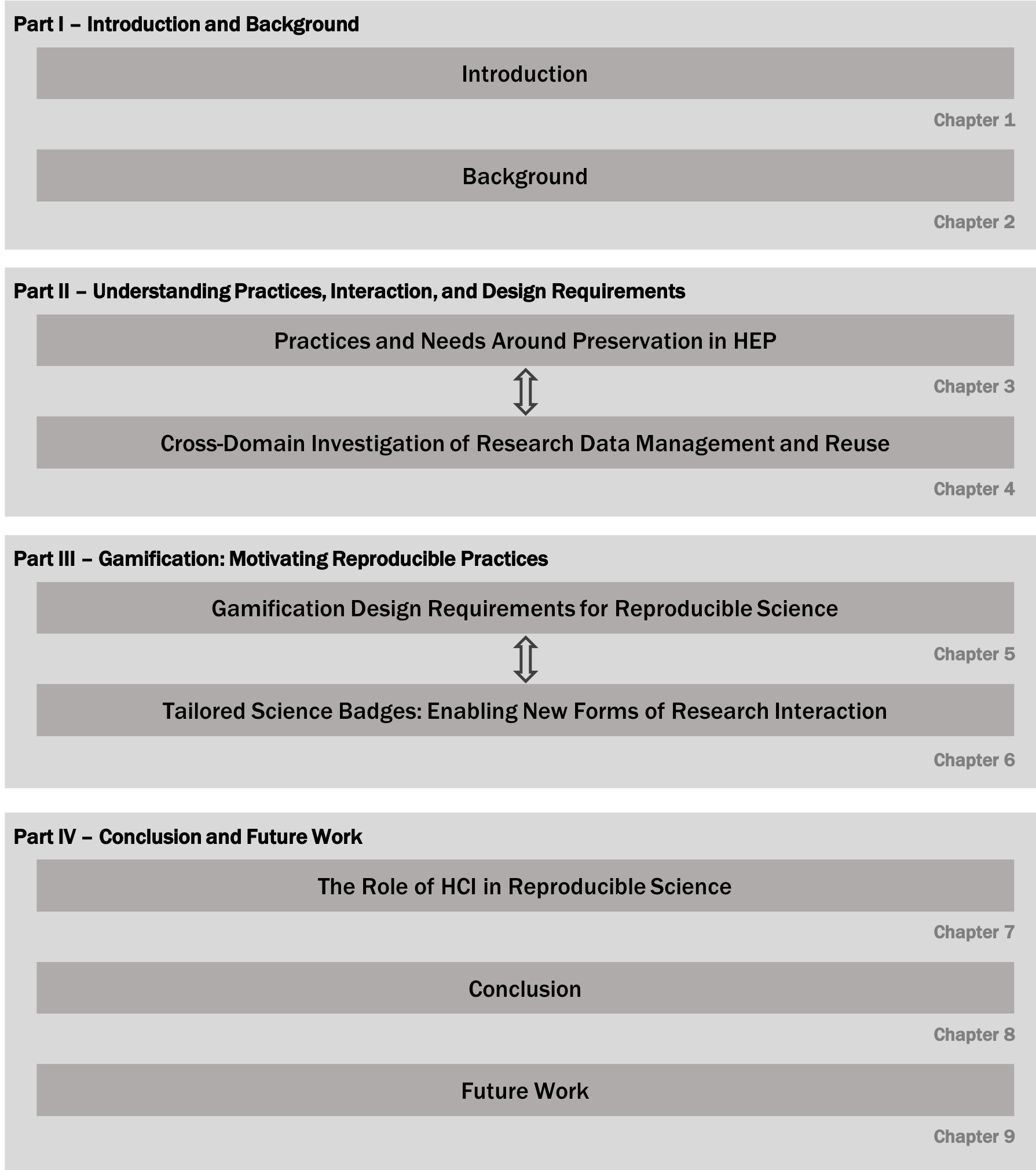}
  \caption{Outline of this thesis.}~\label{fig:outline}
\end{figure}

This thesis is organized into four parts and nine chapters. Figure \ref{fig:outline} provides a detailed overview of the structure of this thesis and the interconnections and dependencies between different chapters. In the following, we provide an overview of the various parts and chapters. Some of those descriptions are partially based on the abstracts of related publications that a chapter is based on.

\begin{flushright}
\textit{Part \ref{part:intro}: Introduction and Background}
\end{flushright}

\textbf{Chapter \ref{ch:introduction} - Introduction}

We motivate our research focus and discuss current challenges and opportunities related to RDM, reproducibility, and reuse in science. We introduce and describe HEP, CERN, and CERN's largest collaborations as our research context. We present the research questions that guide our work and discuss the methodology used to address those questions. We further detail the publications that contributed to this thesis and outline how the thesis is structured.

\textbf{Chapter \ref{ch:background} - Background}

In this chapter, we reflect on definitions of the term \textit{reproducibility}, discussions around the value of replication in HCI, and the needs and requirements of reproducible research. Based on related work and our findings, we introduce the first researcher-centered definition of reproducibility and state that we expect this definition to impact future design thinking of supportive tools for reproducible science. Next, we discuss a paper that we published in Nature Physics: Open is not enough. The paper stresses the importance of technology support in following reproducible practices and motivates our research. In this context, we also detail the CAP service and the service infrastructure at CERN. We then introduce gamification and review requirements for designing meaningful gamified tools. Gamification plays an important role in our research, as we investigate its potential in motivating reproducible science practices. Finally, we reflect on related work investigating the production, processing, and reuse of scientific data and software. We describe how our research was influenced by previous findings and stress the unique perspective of our research on reproducible science practices.

\begin{flushright}
\textit{Part \ref{part:requirements}: Understanding Practices, Interaction, and Design Requirements}
\end{flushright}

\textbf{Chapter \ref{ch:requirements} - Practices and Needs Around Preservation in HEP}

This chapter is based on our first HEP study, which focused on understanding how to design services and tools that support documentation, preservation, and sharing. We report on our interview study with 12 experimental physicists, studying requirements and opportunities in designing for research preservation and reproducibility. In this study, we asked HEP data analysts about RDM practices and invited them to explore and discuss CAP. They reported concerns, hopes, and challenges related to the adoption of the service. The findings highlight the value of a tailored preservation service in lowering efforts for documenting and sharing research. Yet, participants stressed that only lowering efforts was not enough. Our findings suggest that we need to design for motivation and rewards in order to stimulate contributions and to address the observed scalability challenge. Therefore, researchers' attitudes towards communication, uncertainty, collaboration, and automation need to be reflected in design. Based on our findings, we present a systematic view of user needs and constraints that define the design space of systems which support reproducible practices in HEP research.

\textbf{Chapter \ref{ch:cross_domain} - Cross-Domain Investigation of Research Data Management and Reuse}

We report on our cross-domain study that expands on the findings from the previous interview study in HEP. In order to understand practices and needs of data science workers in relation to documentation, preservation, sharing, and reuse, we conducted an interview study with 15 scientists and data managers from diverse scientific domains. Our findings relate to human data management interventions across five core concepts: Practice, Adoption, Barriers, Education, and Impact. We contribute an analysis of the technology and infrastructure components involved within those components of data management. Our work increases the understanding of how to design systems that support data management, promote reproducibility, and enable reuse.

\newpage

\begin{flushright}
\textit{Part \ref{part:gamification}: Gamification: Motivating Reproducible Practices}
\end{flushright}

\textbf{Chapter \ref{ch:gamification_requirements} - Gamification Design Requirements for Reproducible Science}

In this chapter, we reflect on how gamification could motivate reproducible practices in science. We stress that while the application of gamification in corporate work environments has received significant research attention, little focus has been placed on gamification of tools employed in the scientific workplace. We report on our first empirical study of gamification in the context of reproducible research. In particular, we explored possible uses of gamification to support reproducible practices in HEP. We designed two interactive prototypes of a research preservation service that use contrasting gamification strategies. The evaluation of the prototypes showed that gamification needs to address core scientific challenges, in particular the fair reflection of quality and individual contribution. Through thematic analysis, we identified four themes which describe perceptions and requirements of gamification in research: Contribution, Metrics, Applications, and Scientific practice. Based on these, we discuss design implications for gamification in science.

\textbf{Chapter \ref{ch:tailored_badges} - Tailored Science Badges: Enabling New Forms of Research Interaction}

To further our understanding of the impact of game design elements in highly skilled research settings, we implemented six science badges tailored to a physics research preservation service. Our mixed-method evaluation with 11 research physicists focused on assessing trust, suitability, and commitment towards the badges and their three core mechanisms: community votes, clear goals, and community usage. Our findings suggest that researchers find the tailored science badges useful, suitable, and persuasive overall, although their assessment of individual badges differed. We present design implications related to meaningful criteria, repository navigation, and content discovery. Finally, we discuss uses of game design elements beyond motivation.

\begin{flushright}
\textit{Part \ref{part:conclusion}: Conclusion and Future Work}
\end{flushright}

\textbf{Chapter \ref{ch:hci_role} - The Role of HCI in Understanding, Supporting, and Motivating Reproducible Science}

In this chapter, we describe HCI's role in reproducible science. In particular, we introduce two models: a \CommitmentModel, and a conceptual model of UCD in reproducible science. Based on those, we describe the role of both HCI researchers and practitioners in understanding, supporting, and motivating reproducible research practices. Finally, we envision HCI's role in transforming RDM strategies through ubiquitous forms of knowledge and research preservation.

\textbf{Chapter \ref{ch:conclusion} - Conclusion}

We summarize our research contributions. We further comment on the role of replication in HCI and discuss limitations of our work.

\textbf{Chapter \ref{ch:future_work} - Future Work}

Based on our findings, we present opportunities and challenges that should be addressed by future work. In particular, we illustrate how future HCI research could impact the transition between the various stages of the \CommitmentModel.

\chapter{Background }
\label{ch:background}

In this chapter, we discuss related work and findings relevant to the core concepts and topics addressed in this thesis. First, we reflect on definitions of reproducibility and related concepts, including repeatability, replicability, and reuse. We stress the current ambiguity of definitions and provide working definitions for HEP. We further recognize that replication is a topic of interest in the HCI community. We relate to discussions on replication in HCI and illustrate how our work contributes to replication efforts. Furthermore, we discuss needs and requirements of reproducibility and introduce \ac{RDM} in this context. Finally, we propose a general researcher-centered definition of reproducibility.

Next, we sketch the data life cycle in CERN's experiments. We introduce the service infrastructure created to support preservation, reuse, and open access at CERN. In particular, we introduce the CAP service. Related to the development of those services, we stress that availability of data and resources is not the only requirement in enabling reproducibility and reuse. Instead, we emphasize that sharing and open access strategies need to be implemented in concert with appropriate tools. That way, we outline the motivation for the research we present thereafter.

Third, we introduce gamification and describe opportunities and limitations of gamification as a design tool to motivate desired practices. In particular, we emphasize the importance of meaningful game design that creates commitment amongst users, rather than implementing game elements that do not fit the context. This understanding of design requirements is reflected in the study designs of our gamification research, presented in Part \ref{part:gamification} of this thesis.

Finally, we present related work on design requirements for tools embedded in scientific environments. Based on previous findings designing for scientific communities, we motivate our systematic user-centered design approach. We further reflect on findings related to the role of citations, funding, and policies in motivating compliance with reproducible practices. We provide an overview of related findings that emphasize the value of understanding production, sociotechnical frameworks, and uses of data and software in scientific sharing and reuse. This reflection enables us to discuss findings from our research in the wider context of incentives, enforcement, and science infrastructure design.

%
%

\section{Research Reproducibility}

Leek and Peng \cite{Leek2015} defined reproducibility ``as the ability to recompute data analytic results given an observed dataset and knowledge of the data analysis pipeline.'' While this definition fits well in the context of research in particle physics, we must mention that a wide variety of definitions of reproducibility exist. In this section, we initially provide an overview of those definitions and related concepts. In particular, we reflect on discussions on \textit{replication} in HCI and highlight how the research presented in this thesis can contribute to a better understanding of the value of replication in the HCI community. Next, we describe RDM practices that are crucial in fostering science reproducibility. Finally, we introduce our researcher-centered definition of reproducibility that is based on the findings of the research presented in this thesis.

\subsection{Overview of Definitions and Related Concepts}

There are several terms related to reproducibility. Those include replicability, repeatability, and reusability. In the scientific discourse --- and sometimes even within the same article or publication --- they are often used interchangeably. In fact, the specific meaning of the individual concepts can vary between different disciplines ~\cite{Feitelson2015}. The \ac{ACM} stated that a ``variety of research communities have embraced the goal of reproducibility in experimental science. Unfortunately, the terminology in use has not been uniform'' \cite{acmbadgesweb}. The ACM defined repeatability, replicability, and reproducibility based on the acting team and the origin of the experimental setup:

\begin{itemize}

    \item \textbf{Repeatability (Same team, same experimental setup)}: The measurement can be obtained with stated precision by the same team using the same measurement procedure, the same measuring system, under the same operating conditions, in the same location on multiple trials. For computational experiments, this means that a researcher can reliably repeat her own computation.
    
    \item \textbf{Reproducibility (Different team, same experimental setup)}: The measurement can be obtained with stated precision by a different team using the same measurement procedure, the same measuring system, under the same operating conditions, in the same or a different location on multiple trials. For computational experiments, this means that an independent group can obtain the same result using the author's own artifacts.

    \item \textbf{Replicability (Different team, different experimental setup)}: The measurement can be obtained with stated precision by a different team, a different measuring system, in a different location on multiple trials. For computational experiments, this means that an independent group can obtain the same result using artifacts which they develop completely independently.

\end{itemize}

These definitions are particularly valuable in the context of our research, as they target computational experiments and our research at CERN is focused on data-intensive computational science. Based on those ACM definitions, and terminology introduced by Goble \cite{goble_repro} and Barba \cite{barba2018terminologies}, we referred to an extended set of terms based on their \textit{purpose} in a Nature Physics paper \cite{chen2018open}. Those terms and corresponding descriptions are listed in Table \ref{tab:definitions}. In the same paper, we introduced our own definitions of the same terms from the angle of particle physics, as shown in Table \ref{tab:definitions_hep}.

\begin{table}[]
    \centering
    \begin{tabular}{l l l}
         \textbf{Term} & \textbf{Purpose} & \textbf{Description}  \\
    \midrule
         Rerun & Robust & Variations on experiment and set-up, conducted in the same lab \\
         Repeat	& Defend & Same experiment, same set-up, same lab \\
         Replicate & Certify & Same experiment, same set-up, independent lab \\
         Reproduce & Compare &Variations on experiment and set-up, independent labs \\
         Reuse & Transfer & Different experiment
    \end{tabular}
    \caption{Science reproducibility terminology introduced by ACM, Goble, and Barba. Based on Chen et al. \cite{chen2018open}.}
    \label{tab:definitions}
\end{table}

\begin{table}
\includegraphics[width=1.0\columnwidth]{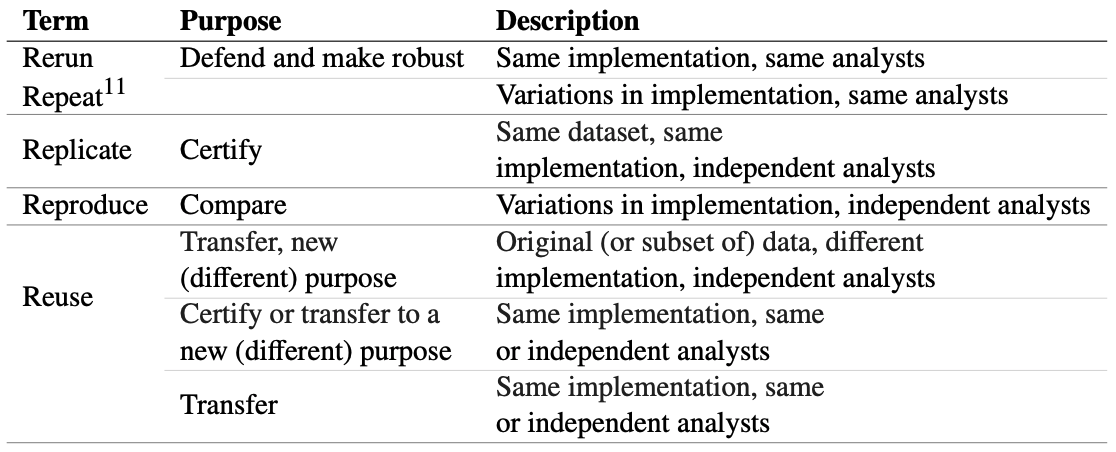}
\caption{Terminology related to science reproducibility in particle physics research. Based on Chen et al. \cite{chen2018open}.} 
\label{tab:definitions_hep} 
\end{table}



Feitelson's characterization of reproducibility \cite{Feitelson2015} fits well with the proposed definitions. He referred to the ``reproduction of the gist of an experiment: implementing the same general idea, in a similar setting, with newly created appropriate experimental apparatus.'' This definition relates well to particle physics where data analyses are enriched by adding later observational data. As this happens over the course of several years, the notion of reproducibility applies as well: analyses are not just re-executed, but rather enriched by new observations. This enrichment represents a type of reuse that is further reflected in the expanded definition of the reuse concept in Table \ref{tab:definitions_hep}. Thus, our work on reproducible research practices is closely connected to the re-usability of experimental resources. In fact, by referencing reproducible practices, ``we aim generally at environments in which researchers are encouraged to describe, preserve and share their work, in order to make resources \textit{re-usable} in the future.'' We introduced this working definition of research reproducibility in our CHI 2019 paper on HEP practices and design requirements \cite{Feger2019Requirements}, which guided the research reported in this thesis.

%
%

\subsection{Replication in HCI}

In HCI, it is lively to refer to research \textit{replication}. In our community, the role of replicability is discussed as well. In fact, Greiffenhagen and Reeves \cite{Greiffenhagen2013} asked: '\textit{Is replication important for HCI?}' They stressed that we need to investigate \textit{aims and motivations} for replication in HCI research. The authors argued that this discussion should distinguish between research that is \textit{replicable} and research that is \textit{replicated}:

\begin{itemize}
    \item \textit{Replicable} refers to research that can, \textit{in principle}, be replicated.
    \item \textit{Replicated} acknowledges reserach that \textit{has been} replicated.
\end{itemize}

Greiffenhagen and Reeves stressed that this formal distinction impacts the very core of HCI's role in science, similar to ``psychology's own debates around its status as a science (that) are also consonant with these foundational concerns of 'being replicable'.'' They stated that ``to focus the discussion of replication in HCI, it would be very helpful if one could gather more examples from different disciplines, from biology to physics, to see whether and how replications are valued in these.'' In this thesis, we report on practices around reproducibility and reuse in particle physics and well beyond. Thus, we expect that our findings will contribute to discussions on the role of replication in HCI.

Two RepliCHI\footnote{http://www.replichi.com/} workshops at CHI 2013 and CHI 2014 represent some of the most structured early efforts towards investigating and advocating the role of replication in HCI. In their 2013 workshop abstract, Wilson et al. \cite{Wilson2013} stressed that HCI researchers ``have almost no drive and barely any reason to consider replicating the work of other HCI researchers.'' They argued that the novelty-driven publication model prevents publishing research replication attempts. They further highlighted rapidly changing technology and its social acceptance as a barrier for structured replication in HCI. The authors described four forms of replication in HCI that we present in an abbreviated form:

\begin{itemize}
    \item \textbf{Direct Replication.} Direct Replication consists of attempting to entirely replicate a study or system, using the same format and with the same tools, and experimental protocol. The aim of direct replications is often to replicate a specific finding. Direct Replication is often driven by the aspirations of strong science to confirm that results are true, are not created by an unseen bias, or that they apply in different contexts (geographic, cultural, topic, task) to the original study \cite{mullet1997your}.
    
    \item \textbf{Conceptual Replication.} Conceptual Replications are systems and studies that focus on a certain principle or phenomenon and confirm findings using alternative methods. Of the three approaches, this is most common in HCI, in that multiple studies demonstrate the principles of direct manipulation. Many instances, however, are post-hoc reflections of their findings in the context of prior work. Through this approach we surmise heuristics about best practices for design or for evaluation.
    
    \item \textbf{Replicate \& Extend.} Replicate+Extend is a common research method in which people first reach the level of prior research before investigating it further. This may involve reproducing a phenomenon before specifically investigating it further, or by building on the findings of the study.
    
    \item \textbf{Applied Case Studies.} One common form of replication is application --- a special instance of conceptual replication. If HCI research produces a finding, and its application in real world contexts confirms it, then case studies are a form of replication.
\end{itemize}

Based on those four forms of replication, Wilson et al. discussed various benefits. The authors highlighted that ``an archive of research findings that reflect directly on prior work would be highly valuable for our community.'' In addition, replication is expected to increase confidence in research findings, and the replication of studies is a valuable method in teaching HCI practices. Cockburn, Gutwin, and Dix \cite{Cockburn:2018:HNM:3173574.3173715} advocated for experimental preregistration in HCI as a means to increase transparency and confidence in study results. In the CHI 2014 RepliCHI II workshop abstract \cite{Wilson:2014:RWI:2559206.2559233}, Wilson et al. reflected on the outcome of the previous RepliCHI workshop. Based on case studies of replication attempts, position papers, and experience reports, they presented an evolved understanding of replication, highlighting the importance of understanding \textit{why} replication should be attempted. The authors ``recommend that people identify clear motivations and reasons to investigate prior work, and to identify areas where contributions will be made.'' As part of this evolved understanding, the RepliCHI organizers stressed that in order to extend prior work, its findings need to be recreated first. However, they stated that ``because it is impossible to completely replicate research, we conclude that by revisiting work, we cannot prove that the original work was wrong or right, but only that we can or cannot find further evidence.''

The RepliCHI workshops touched upon the lack of incentives for researchers to replicate work or to make work replicable, by stressing that our publication system is driven by novelty. Connected to this motivation issue, ACM introduced six badges that are designed to promote and acknowledge sharing and reproducible research\footnote{https://www.acm.org/publications/policies/artifact-review-badging}: \textit{Artifacts Evaluated --- Functional}, \textit{Artifacts Evaluated --- Reusable}, \textit{Artifacts Available}, \textit{Results Replicated}, and \textit{Results Reproduced} \cite{acmbadgesweb}. ACM conferences and journals can adopt those badges which are expected to increase the discoverability of publications in the ACM Digital Library. In Part \ref{part:gamification} of this thesis, we expand on requirements for gamification in science and relate to the use of badges in navigating research repositories. In this context, we highlight the value and importance of discussing adoption of those badges within the ACM \ac{SIGCHI}.

The Transparent Statistics in HCI Working Group\footnote{https://transparentstatistics.org/} concluded a \ac{SIG} at CHI 2016 \cite{kay2016special}, a workshop at CHI 2017 \cite{Kay:2017:MTS:3027063.3027084}, and a SIG at CHI 2018 \cite{Wacharamanotham:2018:SIG:3170427.3185374}. At the time of writing, the working group proposed nine guiding principles\footnote{Retrieved March 1, 2020. https://transparentstats.github.io/guidelines/principles.html\#guiding-principles}. Amongst others, they advocated for experimental pre-registration \cite{Cockburn:2018:HNM:3173574.3173715, nosek2018preregistration} in order to ensure \textit{Process Transparency}. Preregistration is an important Open Science practice that effectively deals with issues of HARKing (Hypothesising After the Results are Known). The guiding principle \textit{Material Availability} corresponds particularly to our research. According to this principle, ``sharing as much study material as possible is a core part of transparent statistics, as it greatly facilitates peer scrutiny and replication. Being able to run the experimental software and examine what participants saw (the techniques, tasks, instructions, and questions asked) is essential in order for other researchers to understand the details of a study.'' Echtler and Häußler \cite{echtler2018open} investigated sharing practices in the HCI community. They analyzed all papers, notes, and extended abstracts published at CHI 2016 and CHI 2017 and found that source code was released for less than three percent of those papers. Wacharamanotham et al. \cite{wacharamanotham2019transparency} surveyed authors of papers accepted at CHI 2018 and CHI 2019. They investigated sharing practices and confirmed that sharing is uncommon. The authors identified several reasons, including data and privacy protection concerns, lack of participants' consent, but also a lack of motivation and knowledge about effective sharing practices.

The community efforts impact policies and practice at CHI. For CHI 2020, two official guides were adapted based on the collaborative efforts between members of the Transparent Statistics group and conference responsibles: \textit{Guide to A Successful CHI Submission}\footnote{Retrieved March 1, 2020. https://chi2020.acm.org/authors/papers/guide-to-a-successful-submission/} and \textit{Guide to Reviewing Papers}\footnote{Retrieved March 1, 2020. https://chi2020.acm.org/guide-to-reviewing-papers/}. Amongst others, changed instructions emphasize sharing of research materials. An overview of changes is available on the webpage of the Transparent Statistics in HCI working group\footnote{Retrieved March 1, 2020. https://transparentstatistics.org/2019/08/01/updates-to-chi-submission-and-reviewing-guides/}. In this context, it should be noted that we made a wide set of research materials available as supplementary material for all full papers submitted to SIGCHI conferences.

\subsection{Needs and Requirements}
\label{section:needs_req_repro_background}

To foster and enable reproducibility and reuse, scientists must follow comprehensive RDM practices \cite{darlington2010principles, de10aspects, strasser2015research, qin2016metadata}. RDM is referred to as ``the organisation of data, from its entry to the research cycle through to the dissemination and archiving of valuable results'' \cite{whyte2011making}. The FAIR data principles\footnote{https://www.force11.org/group/fairgroup/fairprinciples} \cite{Wilkinson2016, FAIR} demand data to be \textit{findable, accessible, interoperable, and reusable.} In the context of data-intensive computational science, it is clear that \textit{data} refers to any sort of experimental resources, from datasets and code scripts, to run-time information \cite{echtler2018open, hurlin2014runmycode, sufi2014software}. \textit{Preserving} and \textit{sharing} all those resources are core reproducible practices that require efforts to describe, prepare, and document them ~\cite{Borgman:1297241}.

\begin{figure}
    \centering
    \includegraphics[width=1.0\columnwidth]{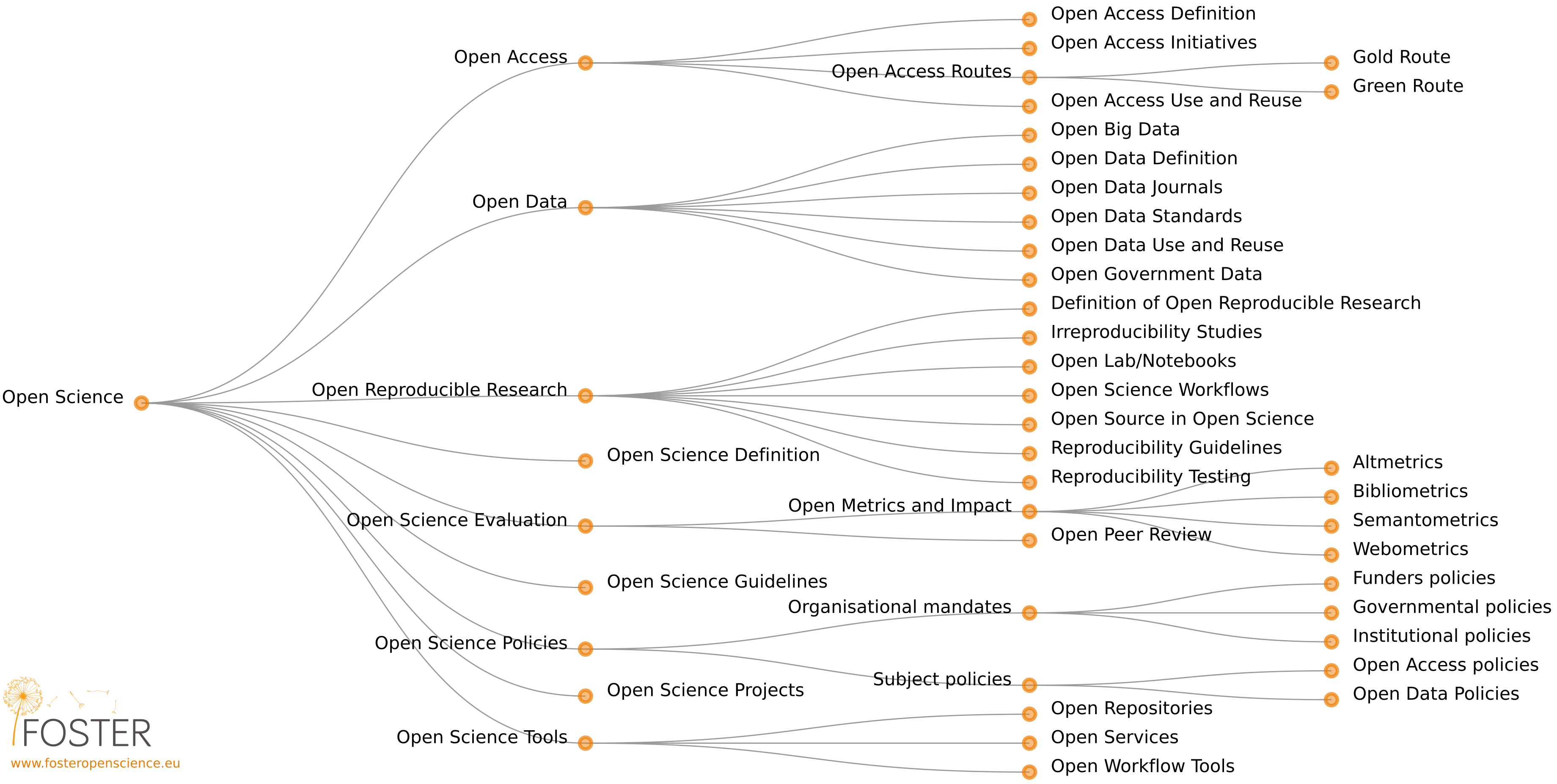}
    \caption[Open Science taxonomy described by FOSTER Plus.]{Open Science taxonomy described by FOSTER Plus.\usebox0}
    \label{fig:os_taxonomy}
\end{figure}

The notion of \textit{sharing} as a requirement in research reproducibility is closely linked to the \ac{OS} movement. OS has been characterized as ``transparent and accessible knowledge that is shared and developed through collaborative networks'' \cite{vicente2018open}. In their well-received \textit{Manifesto for reproducible science}, Munafò et al. \cite{munafo2017manifesto} referred to OS as ``the process of making the content and process of producing evidence and claims transparent and accessible to others. Transparency is a scientific ideal, and adding `open' should therefore be redundant. In reality, science often lacks openness: many published articles are not available to people without a personal or institutional subscription, and most data, materials and code supporting research outcomes are not made accessible, for example, in a public repository.'' This statement hints towards the multiple facets of OS: from open publication access, to the sharing of research artifacts. FOSTER\footnote{https://www.fosteropenscience.eu}, a project funded by the \ac{EU}, developed an OS taxonomy to formally reflect the various aspects related to OS \cite{pontika2015fostering}. Figure \ref{fig:os_taxonomy} shows that the first instance of their OS taxonomy refers to nine key terms: Open Access, Open Data, Open Reproducible Research, Open Science Definition, Open Science Evaluation, Open Science Guidelines, Open Science Policies, Open Science Projects, and Open Science Tools. The value of creating and providing a detailed taxonomy is also reflected in the findings from Konkol et al. \cite{konkol2019computational} who found that researchers have different understandings of what a term like \textit{Open Reproducible Research} means.

Online services such as Zenodo and the \ac{OSF}\footnote{https://osf.io} cover several of these aspects as they provide means to openly share scientific data and software, as well as papers and preprints. Organizations like the \ac{COS} advocate the importance of pre-registration in science.\footnote{https://cos.io/prereg/} Pre-registration is an important mechanism to improve research transparency and requires scientists only to publish their study plan and hypotheses before data collection takes place. Several web services support pre-registration, including OSF and AsPredicted\footnote{https://aspredicted.org}. 

\sbox0{\footnotemark}

\begin{figure}
    \centering
    \includegraphics[width=1.0\columnwidth]{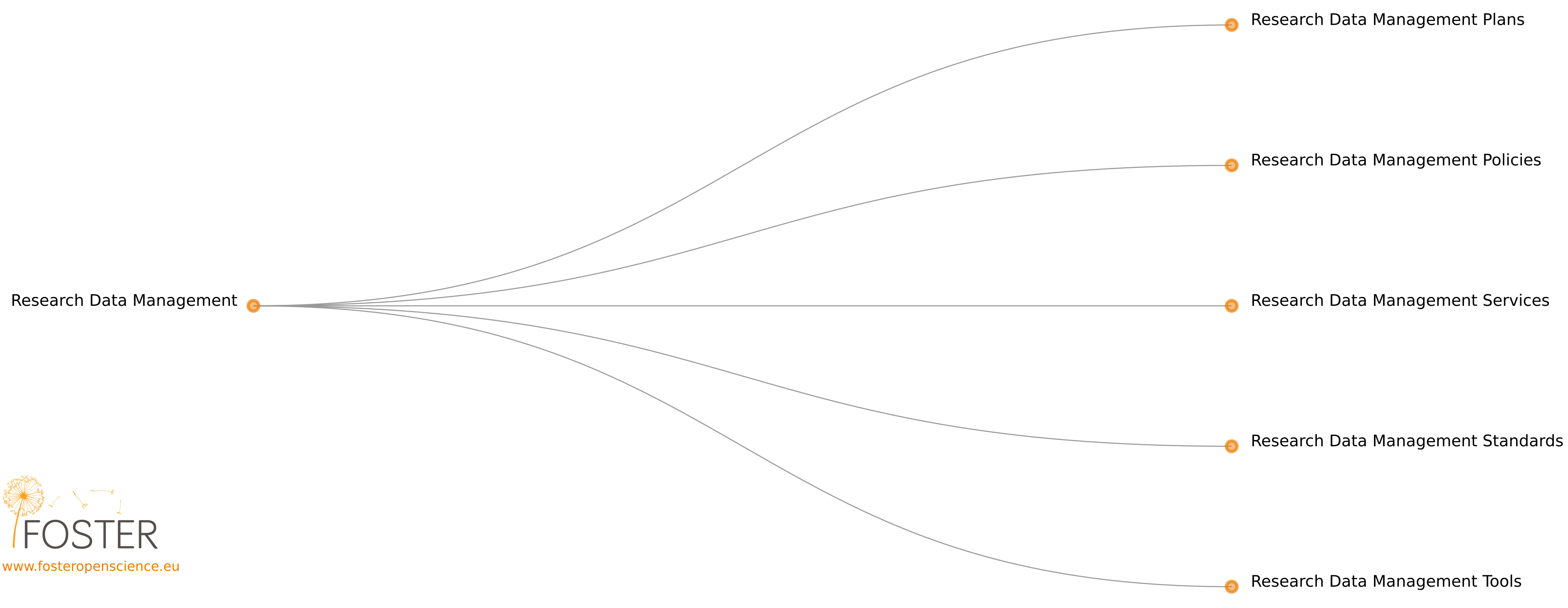}
    \caption[RDM taxonomy described by FOSTER Plus.]{RDM taxonomy described by FOSTER Plus.\usebox1}
    \label{fig:rdm_taxonomy}
\end{figure}

Our research ultimately relates to all of the first-level OS taxonomy terms --- implicitly or explicitly --- and in particular to \textit{Open Reproducible Research}, \textit{Open Metrics and Impact (Open Science Evaluation)}, \textit{Open Science Policies}, and \textit{Open Science Tools}. This is also reflected in our description of the needed integration into the wider ecosystem of RDM tools in Section \ref{section:role_practitioners}. In general, our findings and design implications focus on supporting and motivating RDM practices, which FOSTER described as closely connected to OS \cite{pontika2015fostering}. They introduced a RDM taxonomy, depicted in Figure \ref{fig:rdm_taxonomy}, to reflect the importance of RDM in open and reproducible science. We identify our main research contributions in relation to the design of supportive and rewarding \textit{Research Data Management Tools.}

\footnotetext{Published under a Creative Commons Attribution 4.0 International License (\url{https://creativecommons.org/licenses/by/4.0/}) by FOSTER Plus \cite{pontika2015fostering}: \url{https://www.fosteropenscience.eu/taxonomy/term/104}.}

Transparency and Openness are key values of OS that are clearly reflected in the two taxonomies. A committee formed by representatives from journals and funding agencies, as well as disciplinary experts, developed standards for open practices across scientific journals. Those \ac{TOP} guidelines comprise eight standards with three levels of stringency \cite{nosek2016transparency}. The eight standards are: Citation Standards, Data Transparency, Analytic Methods (Code) Transparency, Research Materials Transparency, Design and Analysis Transparency, Study Preregistration, Analysis Plan Preregistration, and Replication. An overview of the standards and their three levels of stringency is available on the web page of the COS.\footnote{https://cos.io/top/} The HCI community also began to investigate applicability and suitability of the TOP Guidelines in HCI research \cite{chuang2018transparency}.
 
\begin{figure}
    \centering
    \includegraphics[width=0.65\columnwidth]{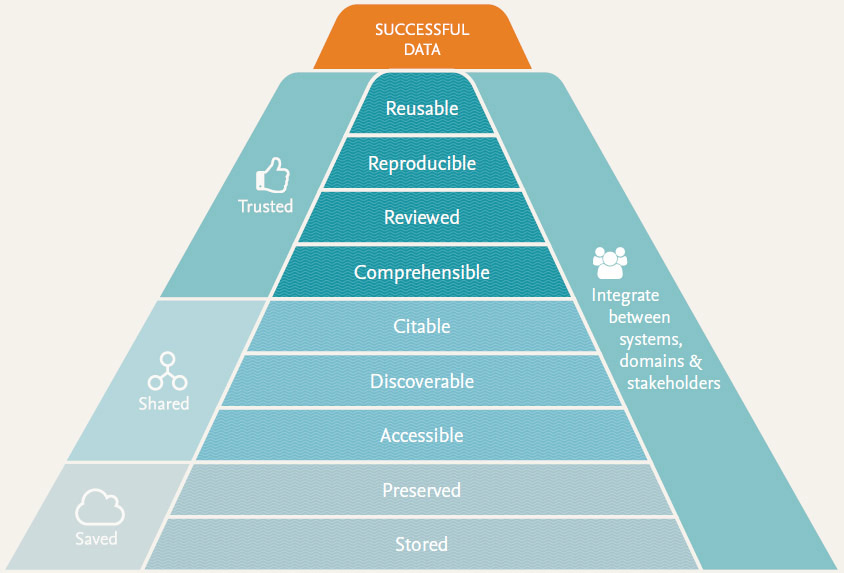}
    \caption[Pyramid of effective data aspects as proposed by de Waard et al.]{Pyramid of effective data aspects as proposed by de Waard et al.\usebox2 }
    \label{fig:rdm_pyramid}
\end{figure}

De Waard et al. \cite{de10aspects} described a pyramid of ten aspects of effective data management that is based on four key factors: \textit{Saved} (Stored and Preserved), \textit{Shared} (Accessible, Discoverable, and Citable), \textit{Trusted} (Comprehensible, Reviewed, Reproducible, and Reusable), and \textit{Successful Data}. As illustrated in Figure \ref{fig:rdm_pyramid}, the authors argued that the ten aspects and concepts must be integrated between systems, domains and stakeholders, in order to build a foundation for effective data. In the context of this integration, the authors stressed that ``in building systems for data reuse or data citation, the practices of current systems for storing and sharing data need to be taken into account.'' This closely relates to our research on sharing practices in HEP and beyond.

\sbox1{\footnotemark}

\footnotetext{Published under a Creative Commons Attribution 4.0 International License (\url{https://creativecommons.org/licenses/by/4.0/}) by FOSTER Plus \cite{pontika2015fostering}: \url{https://www.fosteropenscience.eu/themes/fosterstrap/images/taxonomies/rdmanagement.png}.}

\sbox2{\footnotemark}

\footnotetext{Published under a Creative Commons Attribution 4.0 International License (\url{https://creativecommons.org/licenses/by/4.0/}) by Waard et al. \cite{de10aspects}: \url{https://www.elsevier.com/connect/10-aspects-of-highly-effective-research-data}.}
Chard et al. ~\cite{Chard2015} discussed the value of dedicated data publication systems for data-intensive science. The authors stressed that sharing on basic, connected storage services, like Dropbox, is not sufficient. They argued that dedicated systems are needed to ensure that data are identifiable, described, and findable. Stodden and Miguez \cite{Stodden2014} also described the value of infrastructure in following best practices in computational science. In particular, they referred to the ability to deal with very large data and highlighted that dedicated systems provide features related to citations and versioning.


\subsection{Towards a Researcher-Centered Definition of Reproducibility}

\label{section:definition}

The reflection on various definitions of reproducibility and related concepts in this section underlined the ambiguity and often interchangeable use of those terms. Key elements used to characterize reproducibility relate to the performing team (i.e. same team / different team), the experimental setup (i.e. same setup / different setup), and the purpose. We understand that those characteristics can be very important in the scientific discourse. Being able to state clearly and comprehensively what has been done (e.g. repeated, replicated, or reproduced) is of great value in communicating scientific progress and validation. Even if this clear understanding is limited to a specific scientific domain. For example, a clear and consistent wording is certainly important in communicating verification of Cantor's experiment or the discovery of a fifth fundamental force. However, we argue that formal definitions and differences between various concepts often do not reflect practical concerns in the day-to-day work of most researchers. In this section, we move towards a researcher-centered definition of reproducibility that reflects common concerns of researchers. This approach is in line with findings from van de Sandt et al. \cite{van2019definition} who analyzed differences between \textit{use} and \textit{reuse} in science. They concluded that differences in those terminologies often do not reflect scientific realities and proposed to refer only to scientific \textit{use}.

Based on our research in HEP and across various scientific domains, we find that researchers are mainly concerned with three aspects of reproducibility:

\begin{enumerate}
    \item \textbf{Access.} In Part \ref{part:requirements}, we report our research of practices and requirements around preservation and reuse in HEP and across various scientific domains. We find that researchers' main concern when referring to \textit{reproducibility} lies in gaining access to resources they need. Vines et al. \cite{vines2014availability} found that the availability of research data decreased rapidly with the age of published articles. They contacted authors, requesting data for a reproducibility study. Investigating data availability from 516 studies, with article ages ranging from 2 to 22 years, the authors found that ``the odds of the data being extant decreased by 17\% per year.''
    
    Based on our research, we find that access needs depend on the complexity and gain of the resources' use. For example, the independent verification of findings from an experiment might only require a very thorough experimental protocol. It might additionally require the raw datasets and metadata describing their recording in case of a unique information source, like HEP experiments or medical data of patients \cite{rolland2013beyond}. Instead, it will require the complete set of computational resources in case colleagues want to re-run an experiment to change the visualization of a plot. The notion of ``all about getting the plots'' is reflected in the work of Howison and Herbsleb \cite{howison2011scientific}, as well as our research in HEP. Participant P9 of the study reported in Chapter \ref{ch:requirements} described the need for access related to the simple creation of plots: 
    
    \begin{quote}
        It happens that a summary plot that gets shown at conferences and everywhere gets obsolete and needs to be updated. And in the best case you figure out just where to change a number. In other cases, you have to change the structure of the plot. Because there is a qualitatively new information that has to enter. So, you have to re-format the plot. And there is that gray area... I mean before it gets convenient to just put it in the trash bin and rewrite from scratch, there is that gray area by recycling the old macro written by someone else. And the person who wrote the macro disappears again. This happens a lot of times.
    \end{quote}

    \item \textbf{Effort to gain access.} The final part of the former quote relates to a concern that we describe throughout this thesis: the balance between the potential gain in re-using scientific resources and the effort needed to gain access to those artifacts. Little effort might need to be invested when resources are stored in accessible repositories and colleagues and mentors even point to them. Engaging in personal communication to request resources from colleagues is already more demanding, although very common \cite{huang2013meanings} (see Chapter \ref{ch:requirements}). Substantially more effort might need to be invested in case former colleagues left the institution, left research altogether, or are reluctant to share information. Those efforts might even be futile.
    
    \item \textbf{Ease-of-use.} Our research and related work showed that access to scientific resources is only one requirement in science reuse. Successful reuse depends on trust \cite{mayernik2008whose, faniel2010reusing, zimmerman2007not} and resource documentation \cite{rolland2013beyond}. Findings in Part \ref{part:requirements} emphasize the growing importance of automated analysis workflows and executability of computational environments in data-intensive science. Chapter \ref{ch:tailored_badges} reports on HEP researchers' appreciation of game design elements that reward re-executable analyses and provide new interaction forms for discovering and navigating these resources. 
\end{enumerate}

Based on those characteristics, we introduce a researcher-centered definition of reproducibility that reflects described characteristics. We expect that such a definition can reshape and broaden our understanding of the challenges involved in motivating reproducible science practices, and impact the design of supportive science infrastructure.

 \begin{tcolorbox}[
colframe=gray!70,
colback=blue!1,
coltitle=gray!10!black,
title = Our Researcher-Centered Definition of Reproducibility] 
   
\ReproDefinition

\end{tcolorbox}

%
%

\section{Open is Not Enough: Infrastructure Needs in HEP}
\label{section:open}

In the previous section, we related reproducibility and connected concepts to sharing and accessibility of research. In this section, we detail the life cycle of research data in CERN's experiments and depict the infrastructure developed to support reproducible research practices at different stages of the research life cycle. In particular, we introduce the CAP service. We further stress that openness is not enough to enable reproducible research. Instead, we motivate the development of tools, designed to support openness and accessibility of resources, that are appropriate for the specific environment and goal.

\begin{tcolorbox}[title = This section is based on the following publication.]
Xiaoli Chen, \uline{Sünje Dallmeier-Tiessen}, Robin Dasler, Sebastian Feger, Pamfilos Fokianos, Jose Benito Gonzalez,  Harri Hirvonsalo, Dinos Kousidis, Artemis Lavasa, Salvatore Mele, Diego Rodriguez Rodriguez, \uline{Tibor Šimko}, Tim Smith, \uline{Ana Trisovic}, Anna Trzcinska, Ioannis Tsanaktsidis, Markus Zimmermann, Kyle Cranmer, Lukas Heinrich, Gordon Watts, Michael Hildreth, Lara Lloret Iglesias, Kati Lassila-Perini \& Sebastian Neubert. 2018. Open is not enough. Nature Physics, 15(2), 113–119. 
\newline\url{https://doi.org/10.1038/s41567-018-0342-2}

\color{darkgray}
------------------------------------------------------------------------------------------------------\newline
The author list is presented in alphabetical order. The three main authors are underlined.

\end{tcolorbox}

\subsection{Data Life Cycle and Reuse in HEP}

In Table \ref{tab:definitions}, we referred to the descriptions of reproducibility and related terms by Goble \cite{goble_repro} and Barba \cite{barba2018terminologies}. These concepts assume a research environment in which multiple labs have the equipment necessary to duplicate an experiment, which essentially makes the experiments portable. In the particle physics context, however, the immense cost and complexity of the experimental set-up essentially make the independent and complete replication of HEP experiments unfeasible and unhelpful. HEP experiments are set up with unique capabilities, often being the only facility or instrument of their kind in the world; they are also constantly being upgraded to satisfy requirements for higher energy, precision and level of accuracy. The experiments at the LHC are prominent examples. It is this uniqueness that makes the experimental data valuable for preservation so that it can be later reused with other measurements for comparison, confirmation or inspiration.

Our considerations in HEP begin after gathering the data. This means that we are more concerned with repeating or verifying the computational analysis performed over a given dataset rather than with data collection. Therefore, in Table \ref{tab:definitions_hep}, we presented a variation of these definitions that takes into account a research environment in which `experimental set-up' refers to the implementation of a computational analysis of a defined dataset, and a `lab' can be thought of as an experimental collaboration or an analysis group.

In the case of computational processes, physics analyses themselves are intrinsically complex due to the large data volume and algorithms involved \cite{brun2012web}. In addition, the analysts typically study more than one physics process and consider data collected under different running conditions. Although comprehensive documentation on the analysis methods is maintained, the complexity of the software implementations often hides minute but crucial details, potentially leading to a loss of knowledge concerning how the results were obtained \cite{pasquier2017if}.

\begin{figure}
    \centering
    \includegraphics[width=1.0\columnwidth]{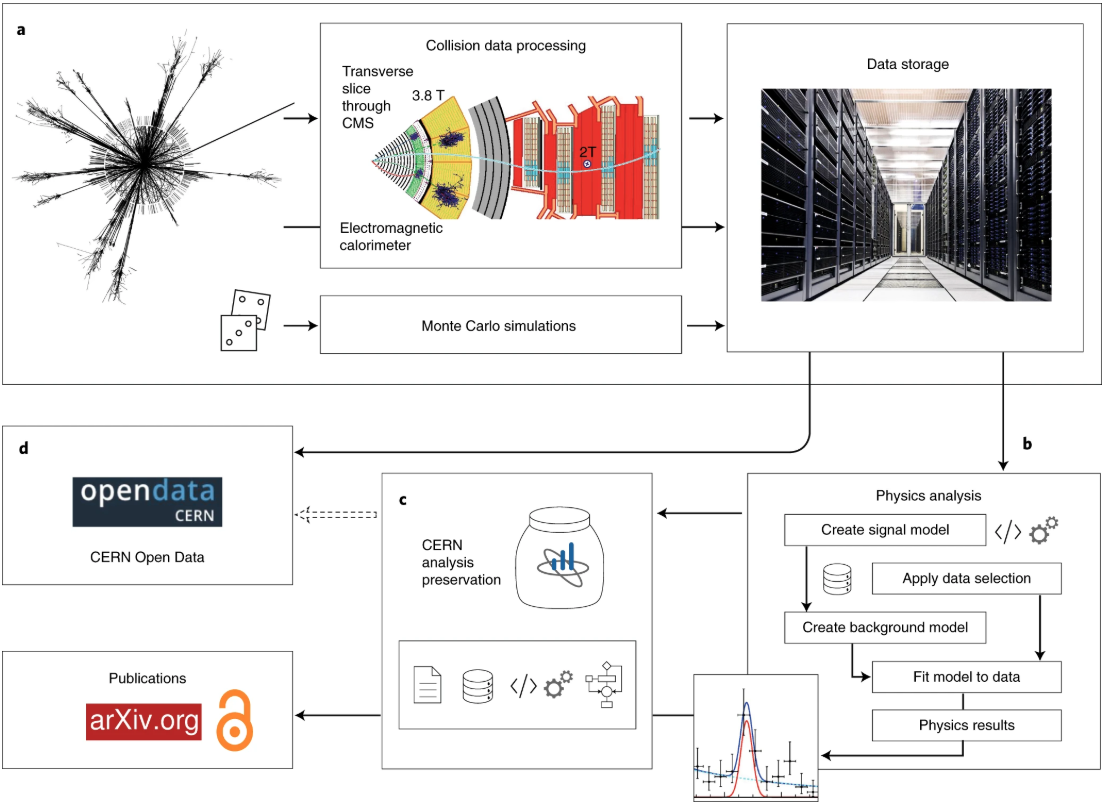}
    \caption[Data continuum in LHC experiments.]{Data continuum in LHC experiments.\usebox3}
    \label{fig:lhc_data}
\end{figure}

In absence of solutions for analysis capture and preservation, knowledge of specific methods and how they are applied to a given physics analysis might be lost. To tackle these community-specific challenges, a collaborative effort (coordinated by CERN, but involving the wider community) has emerged. Figure \ref{fig:lhc_data} depicts the data continuum from proton-proton collisions in the LHC (a) to public data releases (d):

\begin{itemize}
    \item \textbf{a.} The experimental data from proton–proton collisions in the Large Hadron Collider are being collected by particle detectors run by the experimental collaborations ALICE, ATLAS, CMS and LHCb. The raw experimental data is further filtered and processed to give the collision dataset formats that are suitable for physics analyses. In parallel, the computer simulations are being run in order to provide necessary comparison of experimental data with theoretical predictions.
    
    \item \textbf{b.} The stored collision and simulated data are then released for individual physics analyses. A physicist may perform further data reduction and selection procedures, which are followed by a statistical analysis on the data. Physics results are derived taking into account statistical and systematic uncertainties. The results often summarize which theoretical models have predictions that are consistent with the observations once background estimates have been included. The analysis assets being used by the individual researcher include the information about the collision and simulated datasets, the detector conditions, the analysis code, the computational environments, and the computational workflow steps used by the researcher to derive the histograms and the final plots as they appear in publications.
    
    \item \textbf{c.} The CERN Analysis Preservation (CAP) service captures all the analysis assets and related documentation via a set of `push' and `pull' protocols, so that the analysis knowledge and data are preserved in a trusted long-term digital repository for preservation purposes.
    
    \item \textbf{d.} The CERN Open Data service publishes selected data as they are released by the LHC collaborations into the public domain after an embargo period of several years depending on the collaboration data management plans and preservation policies.
    
\end{itemize}

\sbox3{\footnotemark}

\footnotetext{Published under a Creative Commons Attribution 4.0 International License (\url{https://creativecommons.org/licenses/by/4.0/}) by Chen et al. \cite{chen2018open}: \url{https://www.nature.com/articles/s41567-018-0342-2}. Credit: CERN (a); Dave Gandy (b,c, code icon); SimpleIcon (b,c, gear icon); Andrian Valeanu (b,c, data icon); Umar Irshad (c, paper icon); Freepik (c, workflow icon).}

In the next section, we detail the CAP and COD services, as well as the \ac{REANA}\footnote{http://www.reanahub.io/} platform that is closely connected to CAP.

\subsection{CERN Analysis Preservation and Reuse Framework}
\label{section:cap_framework}

In the case of particle physics, it may be true that openness, in the sense of unfettered access to data by the general public, is not necessarily a prerequisite for the reproducibility of the research. We can take the LHC collaborations as an example: while they generally strive to be open and transparent in both their research and their software development \cite{foundation2017roadmap, elmer2017strategic}, analysis procedures and the previously described challenges of scale and data complexity mean that there are certain necessary reproducibility use cases that are better served by a tailored tool rather than an open data repository.

Such tools need to preserve the expertise of a large collaboration that flows into each analysis. Providing a central place where the disparate components of an analysis can be aggregated at the start, and then evolve as the analysis gets validated and verified, will fill this valuable role in the community. Confidentiality might aid this process so that the experts can share and discuss in a protected space before successively opening up the content of scrutiny to ever larger audiences, first within the collaboration and then later via peer review to the whole HEP community. Cases in point are CAP and REANA, which we describe next.

\subsubsection{CERN Analysis Preservation (CAP)}

The CERN Analysis Preservation (CAP) service is a digital repository instance dedicated to describing and capturing analysis assets. The service uses a flexible metadata structure conforming to \ac{JSON} schemas that describe the analysis in order to help researchers identify, preserve and find information about components of analyses. These JSON components define everything from experimental configurations to data samples and from analysis code to links to presentations and publications. By assembling such schemas, we are creating a standard way to describe and document an analysis in order to facilitate its discoverability and reproducibility.

\begin{figure}
  \centering
  \includegraphics[width=0.75\columnwidth]{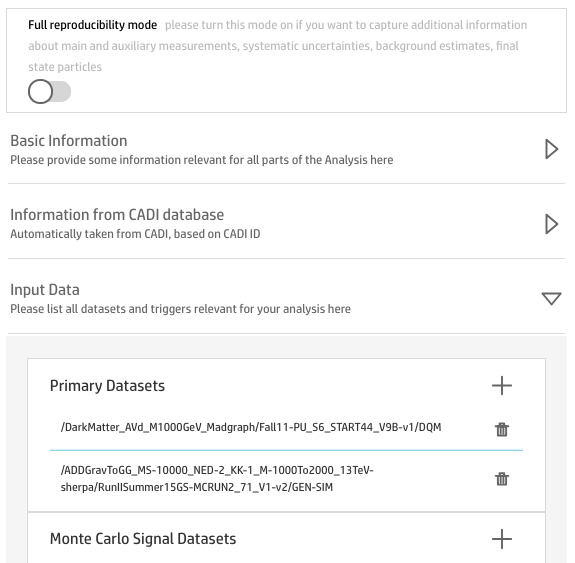}
  \caption{CAP supports documentation and preservation through tailored templates.}~\label{fig:cap_template}
\end{figure}

The design of CAP is based on feedback from the four major LHC collaborations ALICE, ATLAS, CMS, and LHCb. Based on that feedback, the CERN developers published Use Cases in 2015 \cite{dallmeier_tiessen_use_cases}. The service entered Beta development phase in November 2018, and is still in this phase today. Access to the service is limited to members of the collaborations, as the release of experimental data in the LHC collaborations is subject to embargo periods. However, the software code is freely available on Github\footnote{https://github.com/cernanalysispreservation/}. The service is based on the open source Invenio\footnote{https://invenio-software.org/} framework for large-scale digital repositories. Invenio represents the core technology for numerous services, including\footnote{https://invenio-software.org/showcase/} Zenodo and United Nations digital libraries. CAP is designed with the goal of supporting researchers in preserving and sharing their work, and easing collaborative research and analysis retrieval \cite{chen2016cern}.

\begin{figure}
  \centering
  \includegraphics[width=0.4\columnwidth]{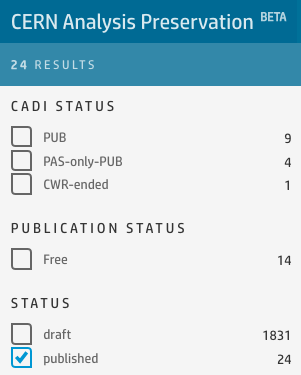}
  \caption{Screenshot of CAP facets designed to meet search and reuse needs.}~\label{fig:cap_facets}
\end{figure}

The research description templates are at the core of the CAP service. Figure \ref{fig:cap_template} details part of the template that maps research in CMS. Members of the collaborations can review analysis information within this template structure for analyses that they have access rights to. And they can enter and edit information of analyses for which they have edit rights. The analysts can freely create and describe analyses on the service, assign edit rights, and make analyses available to all members of their collaboration. An overview of analyses is available on the dashboard. Researchers can use a search box to look for preserved work on the service. A set of community- and collaboration-tailored search facets helps to navigate the repository and the search results, as depicted in Figure \ref{fig:cap_facets}.

\begin{figure}
  \centering
  \includegraphics[width=1.0\columnwidth]{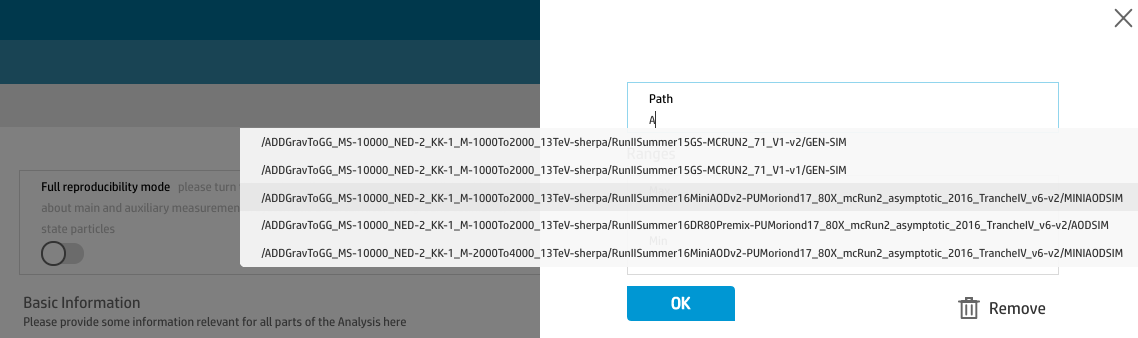}
  \caption{Auto-suggest and auto-complete mechanisms ease documentation on CAP.}~\label{fig:cap_supportive}
\end{figure}

The experiment-tailored design of CAP allows implementing supportive mechanisms. Those include domain-specific auto-suggest and auto-complete features. Figure \ref{fig:cap_supportive} shows an example of the auto-completion for input datasets.

\subsubsection{REANA}

We argue that physics analyses ideally should be automated from inception in such a way that they can be executed with a single command. Automating the whole analysis while it is still in its active phase permits to both easily run the `live' analysis process on demand as well as to preserve it completely and seamlessly once it is over and the results are ready for publication. Thinking of restructuring a finished analysis for eventual reuse after its publication is often too late. Facilitating future reuse starts with the first commit of the analysis code. This is the purpose served by the Reusable Analyses service, REANA: a standalone component of the framework dedicated to instantiating preserved research data analyses on the cloud. While REANA was born from the need to rerun analyses preserved in the CERN Analysis Preservation framework, it can be used to run ‘active’ analyses before they are published and preserved.

Using information about the input datasets, the computational environment, the software framework, the analysis code and the computational workflow steps to run the analysis, REANA permits researchers to submit parameterized computational workflows to run on remote compute clouds. REANA leverages modern container technologies to encapsulate the runtime environment necessary for various analysis steps. REANA supports several different container technologies, compute clouds, shared storage systems, and structured workflow specifications.

\subsubsection{CERN Open Data}

The \ac{COD}\footnote{http://opendata.cern.ch/} portal was released in 2014 amid a discussion as to whether the primary particle physics data, due to its large volume and complexity, would find any use outside of the LHC collaborations. In 2017, Thaler and colleagues \cite{larkoski2017exposing, tripathee2017jet} confirmed their jet substructure model predictions using the open data from the CMS experiment that were released on the portal in 2014, demonstrating that research conducted outside of the CERN collaborations could indeed benefit from such open data releases.

From its creation, the CERN Open Data service has disseminated the open experimental collision and simulated datasets, the example software, the virtual machines with the suitable computational environment, together with associated usage documentation that were released to the public by the HEP experiments. The CERN Open Data service is implemented as a standalone data repository on top of the Invenio digital repository framework. It is used by the public, by high school and university students, and by general data scientists.

Exploitation of the released open content has been demonstrated both on the educational side and for research purposes. A team of researchers, students and summer students reproduced parts of published results from the CMS experiment using only the information that was released openly on the CERN Open Data portal. This shows that the CERN Open Data service fulfils a different and complementary use case to the CERN Analysis Preservation framework. The openness alone does not sufficiently address all the required use cases for reusable research in particle physics that is naturally born `closed' in experimental collaborations before the analyses and data become openly published.



%
%
\section{Gamification}
\label{section:gamification_background}

Gamification, commonly referred to as the ``use of game design elements in non-game contexts'' \cite{Deterding2011}, is a valuable tool to create user engagement and to encourage desired behaviours \cite{Cavusoglu2015, hamari2013transforming}. Gamification has been implemented and investigated across a wide range of domains, including enterprise applications \cite{ruhi2015level, Oprescu2014, Stanculescu:2016:WPE:2818048.2820061}, education \cite{Ibanez2014, Gooch:2016:UGM:2858036.2858231, Denny:2013:EVA:2470654.2470763}, and sports \cite{Zhao:2017:KUE:3025453.3025982, Kappen:2017:GTA:3116595.3116604}. However, research on gamification in science has mostly been limited to \textit{citizen science}, trying to encourage the general public to contribute to scientific processes \cite{feger2018just, Bowser2014, Ponti:2015:SGO:2793107.2810293}. Despite very promising early indications of the positive impact of game elements on sharing in science \cite{Kidwell2016}, a wider understanding of requirements for gamification design in highly skilled scientific environments was missing.

We reflect on prior gamification research in the context of work environments and scientific practice in more detail in Part \ref{part:gamification}, where we present our research on requirements and impact of gamification in reproducible science. In the following, we lay the foundation for our research, as we relate to three fundamental components of gamification: the theoretical foundation, gamification design processes, and the spectrum of game design elements.

\subsection{Theoretical Foundation}

\textit{Flow} \cite{nakamura2009flow, mihaly1990flow} is a theory and process that has been used to inform and explain gamification design. A person who is in a \textit{flow} state is fully immersed in an activity which is considered enjoyable and fulfilling. The following dimensions are commonly described that --- in combination --- create a \textit{flow} experience \cite{hamari2014measuring, mihaly1990flow}:  Challenge-skill-balance, clear goals, control, feedback, loss of self-consciousness, autotelic experience, time transformation, concentration, and merging action-awareness. While Brühlmann et al. \cite{bruhlmann2013gamification} argued that other theories might be more suitable to explain components of motivation, they found that ``flow seems to be a very well applicable concept in the process of designing for usability''. In Part \ref{part:gamification}, we make use of several of the described dimensions in the design of gamified prototypes and tailored science badges.

Play, fun, and motivation are concepts that seem closely related \cite{knaving2013designing}.  Fontijn and Hoonhout \cite{fontijn2007functional} described three core sources of \textit{fun} in the context of playful learning: sense of \textit{accomplishment}, \textit{discovery}, and \textit{bonding}. They related \textit{discovery} to one's curiosity and exploration, and \textit{bonding} to recognition and affirmation. Those sources bear resemblance to \ac{BPNT}, which represents one of the six mini-theories of the \ac{SDT} \cite{ryan2000self}. BPNT describes three basic psychological needs: \textit{competence}, \textit{autonomy}, and \textit{relatedness}. Environments that support those needs promote psychological wellbeing and intrinsic motivation. Recently, Tyack and Mekler \cite{tyackself} found that BPNT is the most described SDT mini-theory in full paper publications at CHI and CHI PLAY.

\begin{figure}
    \centering
    \includegraphics[width=\columnwidth]{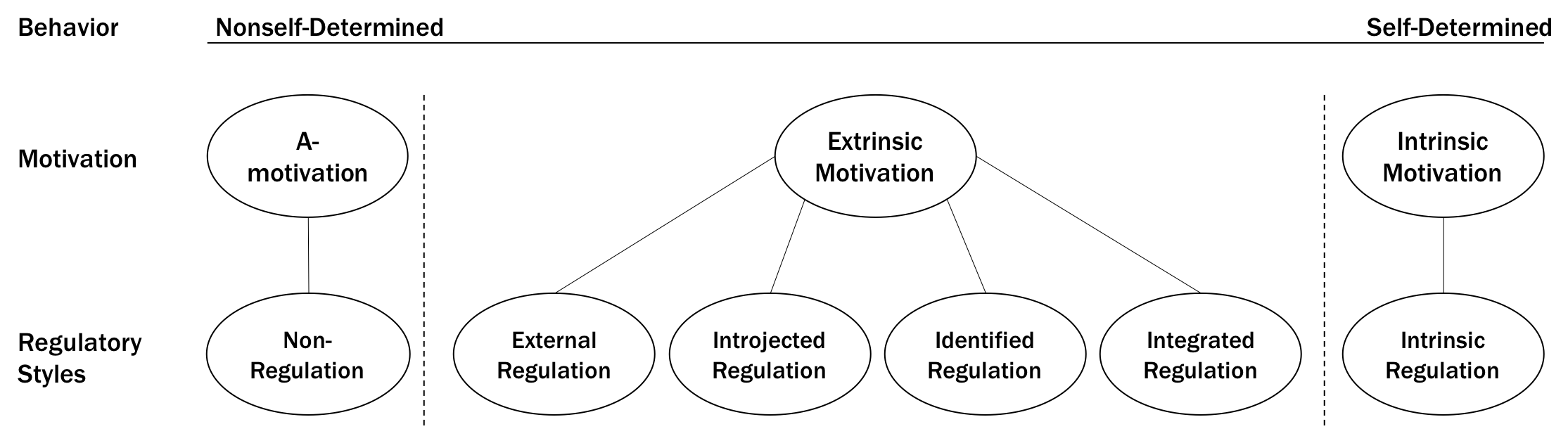}
    \caption{The Self-Determination Continuum. Adapted from \cite{ryan2000self}.}
    \label{fig:sdt_oit}
\end{figure}

SDT by Ryan and Deci \cite{ryan2000self} is a broad psychological framework and macro-theory. In their widely cited survey of gamification literature, Seaborn and Fels \cite{seaborn2015gamification} found that gamification's ``primary theoretical constructs are intrinsic and extrinsic motivation as grounded in self-determination theory (SDT).'' As illustrated by the self-determination continuum in Figure \ref{fig:sdt_oit}, SDT distinguishes between intrinsic motivation, various forms of extrinsic motivation, and amotivation. While intrinsic motivation refers to activities that are perceived personally rewarding, extrinsic motivation is created through extrinsic rewards like promotions and financial incentives. According to \ac{OIT}, another SDT mini-theory, different regulatory styles provide a basis for distinguishing between more or less self-determined forms of extrinsic motivation \cite{deci1985toward}. In particular, OIT ``recognizes that some behavioral regulations are experienced as relatively alien to the self, or imposed and heteronomous, whereas others can be very much being autonomous and self-endorsed'' \cite{ryan2009self}. The process of determining how a stimulant is internally valued is referred to as \textit{internalization}. OIT distinguishes between regulatory styles of extrinsic motivation. \textit{External Regulation} is the least self-determined form of extrinsic motivation in the self-determination continuum. It refers to situations where motivation is based on the desire to avoid punishment or obtain a reward (e.g. salary). \textit{Introjected Regulation} describes behaviours that the individual internalised partially to feel better about one's actions or to avoid self-disapproval and shame for non-compliance. \textit{Identified Regulation} describes a regulatory style characterized by further internalization where the individual \textit{identifies} herself with an activity or finds it genuinely important. When that activity becomes further aligned with one's personal values, it is said to be \textit{integrated}. We recognize that it is important to understand internalization of reproducible science practices in the design of interactive tools in general, and gamification in particular. Thus, we carefully investigate and recognize different regulatory styles and motivations in our research studies. In Future Work (Chapter \ref{ch:future_work}), we further envision the development of a standard scale designed to systematically assess regulatory styles involved in open science.

In contrast to OIT, \ac{CET} concerns intrinsic motivation. The mini-theory addresses how social contexts and rewards support or hinder intrinsic motivation. CET posits that extrinsic rewards undermine the development of intrinsic motivation towards an activity. \textit{Causality Orientation Theory} recognizes that differences in personality traits impact satisfaction of basic needs. Finally, \ac{GCT} and \ac{RMT} are both concerned with personal well-being. RMT emphasizes that relationships are essential as they satisfy the basic need for relatedness, as well as the autonomy and competence needs. GCT posits that intrinsic goals (e.g. community and relationships) contribute to well-being, while extrinsic goals are associated with lower wellness. In our research, we place particular emphasis on socio-technical contexts and the role of the scientific community in the interaction with interactive tools for reproducible science.

\subsection{Gamification Design Processes}

We know that gamification designs should focus on the wider implementation context \cite{Richards:2014:BDM:2658537.2658683} and appeal to intrinsic motivations of users \cite{Brito2015,Dale2014}. Understanding design requirements for gamification is key, as implementations of game design elements and gamified concepts that are not suitable to the task or to the users may not only lack motivational effects, but even alienate users \cite{Nicholson2015}. Systematic user-centered designs are needed \cite{werbach2012win, kumar2013gamification}. In the following, we reflect on design processes that are expected to create meaningful and motivating designs. We further relate those processes to common \ac{UCD} approaches that motivated our in-depth researcher-centered design strategies reported in Part \ref{part:gamification}.

Werbach and Hunter \cite{werbach2012win} described six steps in their gamification design process: (1) Define business objectives; (2) Delineate target behaviors; (3) Describe your players; (4) Devise activity cycles; (5) Don't forget the fun!; and (6) Deploy the appropriate tools. They emphasized that meaningful gamification requires a profound understanding of the context. Thus, they devoted the first three steps to research which satisfies the information needs. Steps four and five target the game dynamics and game mechanics. Gamification dynamics include emotions and relationships that are provoked, as players take part in the experience \cite{Robson2015}. Werbach and Hunter stressed that they can never be entered directly in the game. Instead, they are fine-tuned by the game mechanics \cite{Hunicke2004} which define the setup and rules. Game components, like badges and leaderboards, are only considered in the last stage.

Brito et al. \cite{Brito2015} developed a gamification design framework called G.A.M.E. They proposed four phases: \textit{\textbf{G}athering} of collaboration software data to understand the scenario; \textit{\textbf{A}nalysis} of collected data in order to study the collaboration problem and specify a direction; \textit{\textbf{M}odeling} the collaboration software's gameful experience that encourages the specified direction; \textit{\textbf{E}xecution} of implementation and testing of the collaboration software's gamification plan. The authors assigned the first three phases to the umbrella term \textit{Planning}. They foresee repeating phases two, three, and four as needed.

Similarities can be identified by looking at the five steps of Player-Centered Design, as described by Kumar and Herger \cite{kumar2013gamification}: (1) Know your player; (2) Identify the mission; (3) Understand human motivation; (4) Apply mechanics; and (5) Manage, monitor and measure. In stage 4, they considered mechanics that range from simple elements like points and badges to complex mechanics, like journeys and relationships. While most of the described phases seem self-explanatory, phase five requires a more detailed description. The authors proposed to \textit{manage} the mission, which means to check if it stays the same over time, or if it needs to be adapted. \textit{Monitor} player motivation calls for qualitative evaluations following the implementation. The aim is to understand the impact gamification has on the player's interaction and their perception of the system. Finally, \textit{Measure} effectiveness of mechanics calls for the identification and assessment of key performance indicators. 

\begin{table*}
  \centering
  \def\arraystretch{1.5}
  \begin{tabular}{c l l c}
     & Werbach six-steps guidelines & \makecell[l]{Player-Centered Design \\ by Kumar and Heger} & \makecell[l]{G.A.M.E \\ framework}\\
    \midrule
    \multirow{3}{*}{Study} & Define business objectives & Know your player & Gathering\\
 & Delineate target behaviors & Identify the mission & Analysis\\
 & Describe your players & Understand human motivation & \\ \hline
    \multirow{2}{*}{Design} & Devise activity cycles & \multirow{2}{*}{Apply mechanics} & \multirow{2}{*}{Modeling} \\
 & Don't forget the fun! \\ \hline
    Build & Deploy the appropriate tools & & \multirow{2}{*}{Execution} \\ 
    \cline{1-3}
    Evaluate & & \color{darkgray} Manage, \color{black} monitor and measure \\
  \end{tabular}
  \caption{Mapping of Gamification design models to UCD steps.}~\label{tab:table_design_models}
\end{table*}

To provide a more structured framework around those research and design models, we assigned the various steps and actions to common human-centered design activities. ISO 9241-210 refers to four activities: (1) Understand and specify the context of use; (2) Specify the user requirements; (3) Produce design solutions to meet user requirements; and (4) Evaluate against the requirements. In literature and amongst practitioners, various modified research and design cycles can be found. Popular models describe similar stages like Study, Design, Build, and Evaluate \cite{harper2008human} or Analysis, Design, Coding, and Quality Assurance Testing \cite{sy2007adapting}. 

In Table \ref{tab:table_design_models}, we assigned the proposed steps of the previously discussed gamification design models to a four-stage human-centered design process. The three models propose several actions within the first two stages: Study and Design. Less focus is placed on building and evaluating gamified systems. In Chapter \ref{ch:gamification_requirements} and Chapter \ref{ch:tailored_badges}, we show that this understanding of UCD in gamification informed our design and research strategy.

\subsection{Spectrum of Game Design Elements}

To understand requirements for novel application areas, we first need to build a systematic understanding of the types of game design elements and their respective constraints. Hamari et al. \cite{hamari2014does} reported on their literature review of empirical studies on gamification. They described ten motivational affordances that were tested in 24 empirical studies: Points, Leaderboards, Achievements / Badges, Levels, Story / Theme, Clear goals, Feedback, Rewards, Progress, and Challenges. Out of those, the authors noted that ``points, leaderboards, and badges were clearly the most commonly found variants.'' In this context, it needs to be noted that gamified applications that make use of points, leaderboards, and badges solely to enact business goals, likely prevents long-lasting engagement \cite{Nicholson2015}. In fact, Deterding stressed in \textit{Rethinking Gamification} \cite{fuchs2014rethinking} that ``motivational design should revolve around designing whole systems for motivational affordances, not adding elements with presumed-determined motivational effects.''

Based on study participants' self-reported preferences, Tondello et al. \cite{Tondello2017} classified gameful design elements based on 49 elements and eight groups:

\begin{itemize}
    \item \textbf{Socialization}: Social comparison or pressure, Leaderboards, Social competition, Social networks, Social status, Guilds or teams, Friend invite, Social discovery, Trading, and Scarlet letter.
    
    \item \textbf{Assistance}: Glowing choice, Beginner's luck, Signposting, Anchor juxtaposition, Power-ups or boosters, Humanity hero, Personalization, and Free lunch.
    
    \item \textbf{Immersion}: Mystery box, Easter eggs, Theme, and Narrative or story.
    
    \item \textbf{Risk / Reward}: Access, Lotteries or games of chance, Boss battles, and Challenges.
    
    \item \textbf{Customization}: Avatar, Customization, Points, and Virtual economy.

    \item \textbf{Progression}: Levels or progression, Meaning or purpose, Progress feedback, and Learning.
    
    \item \textbf{Altruism}: Knowledge sharing, Gifting, Innovation platforms, Development tools, Administrative roles, Voting mechanisms, Exploratory tasks, Creativity tools, and Meaningful choices.
    
    \item \textbf{Incentive}: Badges or achievements, Certificates, Collection Rewards or prizes, Unlockable or rare content, and Quests.
    
\end{itemize}

Tondello et al. \cite{Tondello2017} related the above gameful design elements to participants' personality traits. We describe opportunities for personality-based gamification research in science in future work (see Chapter \ref{ch:future_work}). We make extensive use of various game design elements in our requirements research on gamification in highly skilled scientific environments (see Chapter \ref{ch:gamification_requirements}). The goal of the study was to build an understanding of researchers' perceptions of the various gameful design elements. In Chapter \ref{ch:tailored_badges}, we relate the implementation of tailored science badges to gameful design elements listed above. We further reflect on gamification in the science context in the respective related work sections in Part \ref{part:gamification}.

%
%

\section{Scientific Production, Sharing, Reuse, and Tool Design}

Our research aims at understanding, supporting, and motivating core reproducible science practices with particular regard to the design and integration of suitable science infrastructure. We acknowledge that this requires a clear understanding of practices, incentives, and constraints involved in the development and sharing of scientific resources. In the first part of this section, we reflect on findings from the \ac{CSCW} literature which emphasizes that the design of cyberinfrastructure must consider practices in the creation, sharing, and reuse of scientific data and software. In this context, we stress that our research focus on motivating open and reproducible science practices paves new ways in the design of science infrastructure. In the second part of this section, we provide an overview of tools developed for science reproducibility, from model contributions to actual implementations. We further reflect on a wider set of requirements for the design of science tools.

\subsection{Understanding Production and Sharing in Science}

Jirotka et al. \cite{jirotka2013supporting} described the role of CSCW research in studying and advancing \textit{computer supported cooperative science}. They identified three areas of research challenges and opportunities. \textit{Socio-technical configurations and technologies} is one of those areas that is particularly relevant for our own research on interactive tools for reproducible science. The authors stressed that skills and practices of \textit{scientific (sub-) cultures} must be studied in order to understand how cyberinfrastructure can support new forms of collaboration. This study should consider socio-technical practices of the complete research lifecycle. Jirotka et al. further described ``the study of large-scale e-Science as virtual organisations'' where \textit{virtual} refers to distributed collaboration in a global research infrastructure. We argue that HEP research at CERN represents one of the most suitable environments to study \textit{virtual organisations}, as it relies on the collaborative effort of thousands of researchers distributed in hundreds of institutes worldwide. In Chapter \ref{ch:requirements}, we present our study on practices around preservation and reuse in HEP, which places particular attention to socio-technical requirements in the design and use of supportive cyberinfrastructure.

Data and software are key resources in empirical science today and core to any research in computational and data-intensive science. CSCW scholars studied their production, sharing, and reuse extensively. Howison and Herbsleb \cite{howison2011scientific} reported on three case studies investigating incentives and collaboration in scientific software production. Notably, one of the case studies reports on practices in HEP, including even two of CERN's experiments. The other two studies were conducted in the fields of structural biology and microbiology. They stressed that with the growing importance of software in science, scientific software work becomes increasingly subject to competition amongst scholars. In their HEP case study, they identified four types of software developed and used in this field: Analysis scripts, collaboration-specific libraries, data production software, and simulation production software. They found that analysis scripts were developed by a very small group of actual physics researchers with the goal to perform an analysis that can be reported in a publication. Instead, the other types of software are developed either by dedicated IT support staff, or members of a larger community. Such community ``service work'' provides an incentive for contributions, as it guarantees access to the collaboration's data and recognition in form of authorship lists. Discussing the findings from all three case studies, Howison and Herbsleb found that ``software is a secondary player in the world of scientific work, which is dominated by a reputation economy based on substantive scientific publications.'' They distinguished between \textit{software for academic credit} and \textit{software as supporting service}. The latter falls mostly outside the reputation economy, as it relates to commercial products or professional IT staff in large collaborations. Instead, the former software development is incentivized through the prospect of academic credit. The authors stressed that ``while academic credit shares with open source motivations the idea of reputation \cite{crowston2012free}, it is unique due to the importance of publications in that process.'' Related to academic credit, Howison and Herbsleb described a variety of challenges in attributing credit to the authors of scientific software. They concluded that ``it seems likely that significant software contributions to existing scientific software projects are not likely to be rewarded through the traditional reputation economy of science.'' As we show in this thesis, this notion of an academic reputation economy not only plays a role in the collaborative development of software but in motivating reproducible science practices in general. This understanding is also reflected in a report prepared for the \ac{EC} by the \textit{Working Group on Rewards under Open Science} \cite{o2017evaluation}. The authors argued that OS activities could be systematically encouraged and recognised through a \textit{comprehensive research career assessment}. The report refers to an \textit{Open Science Career Assessment Matrix} that lists evaluation criteria along six categories of Open Science activities: Research Output, Research Process, Service and Leadership, Research Impact, Teaching and Supervision, and Professional Experience. We repeatedly discuss findings from our research in the context of career assessment and show how the design of RDM tools can impact career perspectives of researchers who follow reproducible research practices.

Later, Howison and Herbsleb \cite{howison2013} reported on incentives and integration of improvements in science software production. They interviewed authors of software contributions made to BLAST, a key bioinformatics tool. While the findings confirmed that academic credit is a source of motivation for the production of software improvements, their integration is less likely than the integration of improvements developed through other motivations, including financial. The authors discussed several factors related to academic reputation that hinder integration into existing software projects. One challenge is the fair reflection of software contributions in publications and citations, two key mechanisms of scientific reputation. Smaller contributions to large software repositories often do not reflect in publications at all. And even if contributors are added to the author list, missing standards and awareness for software citations prevent future credit. Howison and Herbsleb related to credit in open source software development, where contributions can be tracked to individual authors at the level of single code lines. They argued that this transparency is an important aspect that needs to be considered in scientific software development. Here again, our research connects repeatedly to this notion and challenge of transparency in the context of reproducible science.

Huang et al. \cite{huang2013meanings} investigated meanings and boundaries of scientific software sharing. They report on findings from an ethnographic study conducted at a bioinformatics research center in China. The authors described ``tensions between sharing and control'' that relate to the protection of intellectual property, as well as the distribution of software based on its state and quality. They found that researchers chose different media (e.g. end-to-end email exchange and web publication) based on the state of software. The authors referred to these strategies of containment and publication as \textit{boundary management}. They discussed four types of software and stressed that sharing of each type is usually done ``within different social arrangements'': scripts and work-in-progress software is shared within small teams and through personal requests; published academic software is made available to the scientific community; commercial software is purchased for members of a team or institution; and open source software is made openly available, subject only to open source software licenses. As we point out in this thesis, these findings are valuable not only in the context of scientific software sharing, but for the design of tools supporting reproducible science in general. The authors concluded that ``what is important is not simply making more software available, but addressing issues of navigation, selection and awareness.'' Our research on gamification in (reproducible) science, reported in Part \ref{part:gamification}, shows how game design elements in general, and science badges in particular, provide new opportunities for navigating and discovering science resources.

Similar to those studies on the production and sharing of scientific software, Vertesi and Dourish \cite{vertesi2011value} studied the value of scientific data and practices around data sharing. They reported findings from their ethnographic studies with two robotic space exploration teams. They found that the teams' very different data sharing cultures stem from the way they produce their data. For example, instrument data from the \textit{Paris} project are very combinable by design. Their research questions demand use of multiple instruments. This leads to a collective understanding of data as a community-shared resource. In contrast, data from the \textit{Helen} project are not freely shared, as their production process makes them ``an expensive and hard-won (...) (resource), representing the work of independent, autonomous teams.'' The authors ``propose that data-sharing is only one set of practices in a larger data economy that encompasses production, use, and circulation.'' We considered that understanding practices around production, use, and circulation are not only valuable to the sharing of data and software, but to the wider effort of making research reproducible (see Part \ref{part:requirements}).

Birnholtz and Bietz \cite{birnholtz2003data} argued that the design of systems that aim to support data sharing in science and engineering profits from a systematic mapping of the \textit{use} of those data. They report findings from three scientific disciplines: earthquake engineering, HIV / AIDS research, and space physics. Similar to above findings from Huang et al. \cite{huang2013meanings} on \textit{software} sharing, Birnholtz and Bietz found that scientists seek fine-grained control over access to \textit{data}. They argued that ``[...] the sharing of data follows the paths established by existing social networks. Thus, one possible way to encourage data sharing behavior may be to provide facilities for communication around shared data abstractions.'' The authors discussed that doing so could profit the creator of the data in several ways. First, providing public data abstractions could attract collaborators who possess skills needed by the data creator. Second, recorded data that are a by-product of the data production and not useful to the creator might be helpful to others who have a different research perspective. Third, receiving early comments and questions related to shared preliminary data abstractions can mitigate errors at an early stage, save time, and prevent embarrassment at a later stage. Notably, in our own research, we find that providing communication mechanisms for preserved research artifacts can benefit documenting scientists through the stimulation of useful collaboration and coping with uncertainty. In Chapter \ref{ch:requirements}, we refer to \textit{secondary usage forms} of technology to describe those incentives for sharing and documentation on the analysis level.

Paine et al. \cite{paine2015examining} reported findings from their qualitative cross-domain research study of four data-intensive research groups in Atmospheric Science, Oceanography, Microbiology, and Cosmology. The focus of their study was on data processing work. The authors highlighted that besides understanding the production of data, the processing needs to be understood as part of the data and research lifecycle. The authors described three practices that are instrumental in ``transforming an initial data product in to one that is ready for scientific analysis'': data cleaning, data selection, and data transformation. Those relate to several of the \textit{human interventions in data science work practices} described later by Muller et al. \cite{Muller:2019:DSW:3290605.3300356}. We argue that understanding data processing steps in data-intensive science is important for the design of reproducible science tools, as reusing and adapting data is a common scenario of science reproducibility. Connected to the topic of data reuse, Rolland and Lee \cite{rolland2013beyond} studied scientists' data reuse practices. They conducted a qualitative study with post-doctoral researchers to understand how they (re-)use datasets in cancer epidemiology research. The authors reflected on related findings that highlight the value of support in science data sharing, with particular regards to issues of trust and reliability in using preexisting data \cite{mayernik2008whose, faniel2010reusing, zimmerman2007not}. They conducted interviews with postdocs who received access to datasets through mentors or their professional relationships, thus bypassing such issues of trust and reliability. Rolland and Lee found that the researchers required additional information about the data at different stages of the research lifecycle. They described nine types of questions that occurred repeatedly, focusing on one specific question in their paper: ``How were these data constructed?'' To answer the various questions, the postdocs employed several information seeking strategies, including conversations with their mentors and data managers. We argue that understanding communication and information exchange is crucial not only for the reuse of datasets, but even more so for complete computational research repositories. Thus, we studied information seeking strategies in the context of reproducible research in HEP, reported in Chapter \ref{ch:requirements}. We discuss design implications similar to Rolland and Lee, who concluded that ``one way to support better reuse of data is to provide better support for finding answers to this set of questions through better information management."

Overall, related work makes strong arguments for studying production, processing, use, and reuse of scientific data and software as part of the design of supportive science infrastructure. As we emphasized in this section, findings and implications from those studies greatly impacted our requirements study approach in Chapter \ref{ch:requirements}. In this context, we argue that findings and implications from our work make strong novel contributions, as they take on a new perspective of reproducibility in computational and data-intensive science that goes well beyond the simple sharing of resources, but address issues of automated analysis re-executability in big data science.

\subsection{Designing for Scientific Communities and Reproducible Science}

Today's general availability of computation and internet connectivity provides unprecedented opportunities for the systematic preservation and sharing of experimental resources. In the discussion of emerging data management tools, two key types of infrastructure need to be distinguished: \textit{general} and \textit{tailored} services \cite{Wallis2013}. \textit{General} research management services and data repositories provide support for a wide range of scientific fields. Examples of such services include Globus\footnote{https://www.globus.org/data-sharing}, Zenodo, HUBzero\footnote{https://hubzero.org/}, and Dryad. In contrast, \textit{tailored} services map practices and workflows of specific target domains, experiments, or institutes. Examples of such domain-specific repositories include the Sloan Digital Sky Survey\footnote{https://www.sdss.org/}, EarthCube\footnote{https://www.earthcube.org/}, and a number of others \cite{Stodden2014}. DesignSafe \cite{esteva2019curation} is an example of a web-based repository that focuses on specific requirements for simulation datasets. While the \textit{tailored} design approach requires extensive efforts for the implementation and maintenance, it enables a more targeted interaction with preserved research content \cite{Jackson2013}. The CAP service is a very good example of tailored infrastructure, as the tool is designed to map research workflows from the four major LHC experiments. In this section, we reflect on requirements designing tools for scientific communities in general, and reproducible science support in particular.


The research reported in this thesis focuses on the human-centered study of interaction with scientific tools and their integration into scientific practice. In fact, Oleksik et al. \cite{Oleksik:2012:BDS:2145204.2145376} stressed that in order to design and improve tools for collaborative data generation and reuse, we need to build ``a deeper understanding of the social and technological circumstances''. This is particularly important, as even small interface changes of analysis tools impact researchers behaviour \cite{Jianu:2012:ESU:2207676.2208704}. Thus, domain experts need to be involved in the design of scientific software, as Thomer et al. stressed \cite{Thomer:2016:CSS:2851581.2892549}. The need to involve domain scientists in the design and improvement of scientific tools is reflected in our research, as we recruited a total of 42 scientists and research data managers. Out of those, 30 participants held a doctoral degree and seven were PhD students. Thus, our research effectively follows calls to adopt a human-centered approach in the design of science tools, instead of focusing only on technical requirements \cite{Molin:2016:UDA:2971485.2971561}. 

Garza et al. \cite{Garza:2015:FCD:2783446.2783605} showed the impact of emphasizing ``the potential of data citations'' in a science community data system, that ``can affect researchers' data sharing preferences from private to more open.'' This is in line with related work ~\cite{2011AGUFMIN53B1628S,10.7717/peerj.175} that described citation benefits of open sharing due to improved accessibility and heightened visibility. Citations and research visibility are some of the key motivations and drivers for scientists. But, we also have to reflect the design of RDM tools in the context of strict regulations and policies. As journals and conferences started to encourage and demand resource sharing \cite{Belhajjame2014,Stodden2014}, and industry partners \cite{Rosenblatt336ed5} and funding agencies \cite{russell2013if, kervin2012research} mandate comprehensive RDM, policy compliance becomes an important aspect of tool design. In this context, Pasquetto et al. \cite{Pasquetto:2016:ODS:2858036.2858543} reported on findings from two case studies of large scientific collaborations which focused on studying relationships between policies, open data, and infrastructure requirements. Based on their findings, they confirmed that both policy rationales and compliance are closely connected to funding concerns and motivated by the goal to prove commitment to funding agencies. The authors discussed two key components of \textit{open data} definitions: the types of data referred to in the definitions, and the intended audiences. Based on their study of two different scientific settings, they discussed and confirmed differences in the focus of making resources either accessible to scientists and the general public alike \cite{leonelli2013current, boulton2011science}, or only to scientific communities \cite{wittenburg2010riding, pilat2007oecd}. 

In light of those findings, we note that the research related to the CAP service maps closely to the latter description of data accessibility amongst scientific communities. Wider scopes of data openness that include training and education of the general public are targeted by the COD portal, as described in Section \ref{section:open}. Pasquetto et al. \cite{Pasquetto:2016:ODS:2858036.2858543} highlighted that computational infrastructure is built in response to open data policies. However, they discussed a more complex dependency between policies and infrastructure design: ``while policy definitions for open data do shape scientific infrastructure, extant configurations
of available infrastructure also shape open data policies in terms of what specific types of data are covered by the policies, and how these data are to be made available., to whom, and under what conditions.'' Based on those findings, the authors confirmed ``that infrastructures are emergent, impact and are impacted by, policy, design, and practice \cite{Jirotka2006, borgman2013knowledge}.'' We note that at the time of writing, the CAP service has not yet been recognized by the LHC collaborations as a mandatory tool in the research or publication process. Thus, CAP is developed based on the initiative of certain collaboration members, the CERN SIS, and through the support of the publicly funded Freya project\footnote{https://www.project-freya.eu/en}. However, we repeatedly discuss findings from our studies in the context of data policies in this thesis.

Freya is a project funded by the \ac{EC} and a good example of Europe's centralized science infrastructure developments, further characterized by Wolfgang Kaltenbrunner \cite{kaltenbrunner2017digital}. He compared digital infrastructure projects for the humanities in Europe and the \ac{US}. Kaltenbrunner suggested ``that infrastructure actually functions as a regulatory technology, i.e. as an interface through which the different actor groups in a public science system rearticulate their mutual relations.'' Through a comparative analysis, the author described several differences in science infrastructure design between the US and Europe. The US approach is based on the expertise and leadership of digital scholars. By contrast, the centralized strategy of the EC focuses on wider transnational development and integration that seeks to prevent single research domains from taking disproportionate control over cyberinfrastructure developments. In this context, we need to emphasize that, while CERN is a partner in the FREYA project, research at CERN is mostly funded by its member states and not the EC.

Nüst et al. \cite{nust2017opening} stressed that while making relevant resources available is highly important, it often does not enable reproducibility of computational experiments. They highlighted that reproducibility in computational science requires additional information (e.g. on the runtime environment) and more systematic workflow and sharing practices. Their efforts are focused on enabling reuse of the large number of computational projects that are executed locally on researchers' computers. To this end, they introduced and discussed the \ac{ERC}. The authors described four core parts of an ERC:

\begin{itemize}
    \item \textit{Data} comprises all inputs for an analysis, ideally starting with raw measurements, for example, in form of text files, or databases.
    
    \item \textit{Software} comprises code created by a researcher and all underlying libraries or tools to reproduce the analysis in form of scripts/source code, a Dockerfile, and a Docker container.
    
    \item \textit{Documentation} comprises both instructions, such as a README file, and the actual scientific publication, e.g. in PDF format, any supplemental records, and metadata in standardized formats. The actual publication is the main output of the compendium and the core element for validation. An important metadata element are licenses for the different parts of a compendium.
    
    \item \textit{UI bindings} provide linkage between research components and user interface widgets. They can be used to attach UI widgets to static diagrams in order to make them interactive. Their representation can be stored as metadata within an ERC as part of the documentation. The resulting UI widgets open up the container and allow readers to drill deeper into results. UI bindings can unveil parameters which are required for a comprehensive understanding but are often buried in the code.

\end{itemize}

The CAP service templates provide means to submit and preserve \textit{Data, Software,} and \textit{Documentation}. However, the current CAP version does not support interaction with the research components (\textit{UI bindings}) besides downloading and uploading them. Thus, the preservation service in its current state can be seen as an advanced type of \ac{ELN} \cite{Oleksik2014, tabard2008individual} that provides a variety of supportive mechanisms. Mackay et al. \cite{Mackay2007} presented \textit{Touchstone}, which is a good example of a tool that supports replication and reuse in HCI, and that enables interaction with the experimental data. Touchstone is an experiment design platform for interaction techniques. The authors motivated the development of Touchstone, stressing that the effort needed to replicate interaction techniques makes comparisons challenging. For that reason, novel interaction techniques are often compared to very few standard techniques. On the one hand, Touchstone supports researchers in the evaluation of their experiments. On the other hand, it allows exporting and importing experiment designs described within this tool, consequently enabling and facilitating reuse and replication of research on interaction techniques. Tochstone2 \cite{Eiselmayer:2019:TIE:3290605.3300447} is available as a web application\footnote{https://beta.touchstone2.org/} that provides a direct manipulation interface for experiment designs and a declarative language that enables sharing and unambiguous communication of experimental designs.

	
	\part{Understanding Practices, Interaction, and Design Requirements}\label{part:requirements}

\chapter{Practices and Needs Around Preservation in HEP}

\label{ch:requirements}

Research repositories and data management tools are either \textit{generic}, applicable to a wide set of scientific fields, or \textit{tailored} to specific experiments, institutes, or fields of study. CAP is an excellent example of a specialized research repository. The service is tailored to CERN's four largest experiments. As the prototype matured and approached Alpha stage in 2017, we studied how particle physics researchers perceived the tool. To do so, we conducted an interview study to understand practices and needs around research preservation and reuse in HEP. As part of this study, which we report in this chapter, we introduced researchers to the CAP prototype. We invited them to explore and discuss the service, expecting that the study of a closely tailored RDM tool would benefit not only the development of CAP, but provide guidelines for the design of specialized preservation tools beyond particle physics.

In this chapter, we first detail our study design before we present the findings from the interview study. Next, we discuss design implications. In particular, we describe what \textit{secondary usage forms} are expected to motivate high-quality contributions to the preservation service.

\begin{tcolorbox}[title = This chapter is based on the following publication.]
Sebastian S. Feger, Sünje Dallmeier-Tiessen, Albrecht Schmidt, and Paweł W. Woźniak. 2019. Designing for Reproducibility: A Qualitative Study of Challenges and Opportunities in High Energy Physics. In CHI Conference on Human Factors in Computing Systems Proceedings (CHI 2019), May 4–9, 2019, Glasgow, Scotland Uk. ACM, New York, NY, USA, 14 pages.
\newline\url{https://doi.org/10.1145/3290605.3300685} 

\color{darkgray}
------------------------------------------------------------------------------------------------------\newline
Several of the study's resources are openly available as supplementary material in the ACM Digital Library.
\end{tcolorbox}

%
%
\section{Study Design}

We carried out 12 semi-structured interviews, to establish an empirical understanding of data sharing and preservation practices, as well as challenges and opportunities for systems that enable preservation and reproducibility.

\subsection{Study Participants}

In this section, we provide rich descriptions of the participants, including researchers' affiliations and experience levels. The analysts were 24 to 42 years old (average = 33, SD = 5.2). We decided not to provide information on the age of individual participants, as it would --- in combination with the additional characteristics --- allow to identify our participants. The 12 interviewees included 1 female (P8) and 11 males. The male oversampling reflects the employment structure at CERN: in 2017, between 79\% and 90\% (depending on the type of contract) of the research physicists working at CERN were male ~\cite{CERN-HR-STAFF-STAT-2017}. All interviewees were employed at CERN or at an institute collaborating with CERN. As all interviews were conducted during regular working hours, they became part of an analyst's regular work day. Accordingly, no additional remuneration was provided.

\begin{table}
  \centering
  \begin{tabular}{l c c r}
    {\small\textit{Interviewee reference}}
    & {\small \textit{Affiliation}}
    & {\small \textit{Gender}}
    & {\small \textit{Experience}}\\
    \midrule
    P1 & ATLAS & Male & Postdoc \\ 
    P2 & LHCb & Male & PhD student \\ 
    P3 & LHCb & Male & Senior researcher \\
    P4 & CMS & Male & Postdoc \\
    P5 & CMS & Male & Postdoc \\
    P6 & CMS & Male & Senior researcher \\
    P7 & CMS & Male & Senior researcher \\
    P8 & CMS & Female & PhD student \\
    P9 & CMS & Male & Convener \\
    P10 & CMS & Male & Senior researcher \\
    P11 & LHCb & Male & Convener \\
    P12 & CMS & Male & PhD student \\
  \end{tabular}
  \caption{Overview of the affiliations and professional experiences of the interviewees. 
  }~\label{tab:table_interviewees}
\end{table}

\subsubsection{Collaborations and Experience}

We interviewed data analysts working in three main LHC collaborations. Our recruitment focused on CMS and LHCb, as their preservation templates are most complex and advanced. No interviewee had a hierarchical connection to any of the authors. Table \ref{tab:table_interviewees} provides an overview of the interviewees' affiliations with the LHC collaborations.

We selected physicists with a diverse set of experiences and various roles to ensure a most complete representation of practices and perceptions. Half of the interviewees are early-stage researchers: PhD students and postdocs. The other half consists of senior researchers. As all interviewees --- except the PhD students --- held a PhD, we introduced metrics to distinguish between postdocs and senior researchers. In accordance with the maximum duration of postdoctoral fellowship contracts at CERN, we decided to consider as \textit{senior researchers} all interviewees who had worked for more than three years as postdoctoral physics researchers.

Two of the senior researchers had a convening role, or had such responsibilities within the last two years. Conveners are in charge of a working group and have a project management view. They are, however, often working on analyses themselves. Since they have this unique role within LHC collaborations, we identified them separately in Table \ref{tab:table_interviewees}.
 
\subsubsection{Cultural Diversity}

According to 2017 personnel statistics ~\cite{CERN-HR-STAFF-STAT-2017}, CERN had a total of 17,532 personnel, of which 3,440 were directly employed by the organization. CERN had 22 full member states, leading to a very diverse work environment. We decided not to list the nationalities of individual scientists, as several participants asked us not to do so and because we were concerned that participants could be identified based on the rich characterization already consisting of affiliation, experience, and gender. However, we report the nationalities involved. The participants were in alphabetical order: British, Finnish, German, Indian, Iranian, Italian, Spanish, and Swiss. The official working languages at CERN are English and French, with English being the predominant language in technical fields. All interviews were conducted in English. Working in a highly international environment at CERN, all interviewees had a full professional proficiency in English communication.

\subsection{Interview Protocol}

Initially, participants were invited to articulate questions and were asked to sign the consent form. The 12 interviews lasted on average 46 minutes (SD = 7.6). The semi-structured interviews followed the outline of the questionnaire:

Initially, questions targeted practices and experiences regarding analysis storage, sharing, access, and reproducibility. Interviewees were encouraged to talk about expectations regarding a preservation service and the value of re-using analyses. This part of the questionnaire informed the themes \textsc{Motivation} and \textsc{Communication}. Next, we provided a short demonstration of the CAP prototype. Participants were introduced to the analysis description form and to collaborative aspects of the service: sharing an analysis with the LHC collaboration and accessing shared work. Participants were asked to imagine the service as an operational tool and were invited to describe the kind of information they would want to search for.

\begin{figure}
    \centering
    \includegraphics[width=0.95\columnwidth]{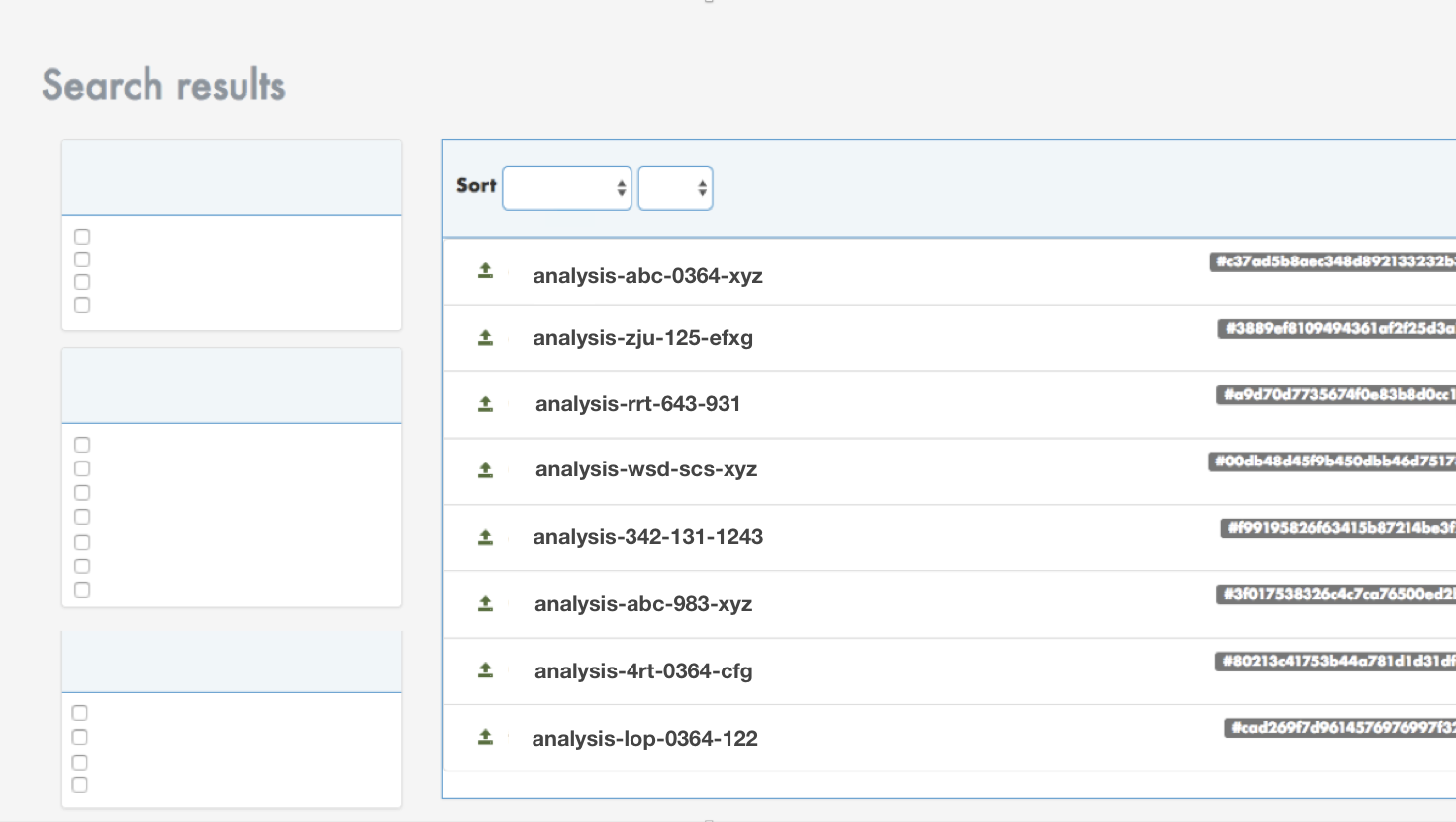}
    \caption{The search facet paper exercise.}
    \label{fig:faceting_exercise}
\end{figure}

\begin{figure}
    \centering
    \includegraphics[width=0.85\columnwidth]{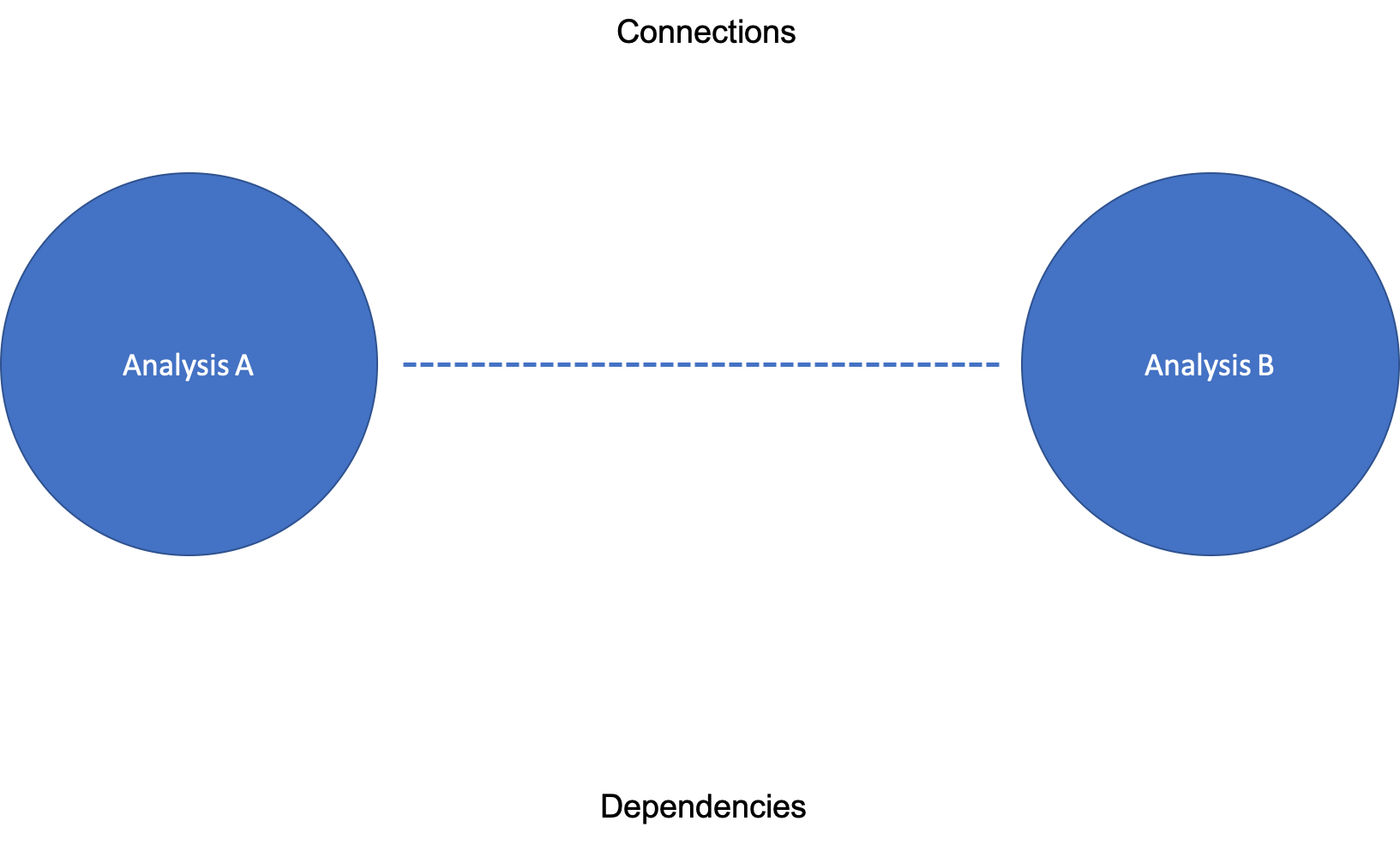}
    \caption{Analysis connections and dependencies paper exercise.}
    \label{fig:connections_dependencies}
\end{figure}

We used two paper exercises to support the effort of uncovering the underlying structure of analyses, as perceived by data analysts. In one exercise, shown in Figure \ref{fig:faceting_exercise}, participants were asked to design a faceted search for a search result page, showing a set of analyses with abstract titles. They had three empty boxes at their disposal and could enter a title and four to seven characteristics each. In the second exercise, depicted in Figure \ref{fig:connections_dependencies}, we encouraged participants to draw connections and dependencies that can exist between analyses on a printout with two circles, named \textit{Analysis A} and \textit{Analysis B}. The exercise supported us in understanding the value of a service being aware of relations between analyses. Finally, interviewees were encouraged to reflect on CAP and invited to describe how they keep aware of colleagues' ongoing analyses within their LHC collaboration.

The system-related part of the questionnaire and the paper exercises informed our results about \textsc{Uncertainty}, \textsc{Collaboration} and \textsc{Automation}.

\subsection{Data Analysis}

All interviews were transcribed non-verbatim by the principal author. We used the Atlas.ti data analysis software to organize, code, and analyze the transcriptions. Thematic analysis \cite{Blandford:2222613} was used to identify emerging themes from the interviews. We performed an initial analysis after the first six interviews were conducted. At first, we repeatedly read through the transcriptions and marked strong comments, problems, and needs. Already at this stage, it became apparent that analysts were troubled by challenges the currently employed communication and analysis workflow practices posed. After we got a thorough understanding of the kind of information contained in the transcriptions, we conducted open coding of the first six interviews. As the principal author and two co-authors discussed those initial findings, we were content to see the potential our interviews revealed: the participants already described tangible examples of how a preservation service might motivate their contribution as a strategy to overcome previously mentioned challenges. We decided not to apply any changes to the questionnaire.

As the study evolved, we proceeded with our analysis approach and revised already existing codes. We aggregated them into a total of 34 code groups that were later revised and reduced to 22 groups. The reduction was mainly due to several groups describing different approaches of communication, learning, and collaboration. For example, three smaller code groups that highlighted various aspects of e-mail communication were aggregated into one: \textit{E-Mail (still) plays key role in communication}. We continued to discuss our evolving analysis while conducting the remaining interviews. In addition, the transcript of the longest interview was independently coded by the principal author, one co-author and one external scientist, who gained expertise in thematic content analysis and was not directly involved in this study.

A late version of the paper draft was shared with the 12 interviewees and they were informed about their interviewee reference. We encouraged the participants to review the paper and to discuss any concerns with us. Eight interviewees responded (P2, P4, P5, P7, P8, P9, P11, P12), all of which explicitly approved of the paper. We did not receive critical comments regarding our work. P9 provided several suggestions, almost all of which we integrated. The CMS convener also proposed to ``argue that the under-representation of ATLAS is not a big issue, as it is likely that the attitudes in the two multi-purpose experiments are similar (the two experiments have the same goals, similar designs, and a similar number of scientists).''

%
%

\section{Findings}

Six themes emerged from our data analysis. In this section, we present each theme and our understanding of the constraints, opportunities, and implications involved.

\subsection{Motivation}

Our analysis revealed that personal motivation is a major concern in research preservation practices. In particular P1, P2, P7, P9, and P11 worried about contribution behaviors towards a preservation service. P1 further contrasted information \textit{use} and \textit{contribution}: \textit{``People may want to use information --- but we need to get them to contribute information as well.''} The analyst calls this \textit{``the most difficult task''} to be accomplished. 

Several analysts (P1, P2, P9, P11) pointed to missing incentives as the core challenge. They stressed that preserving data is not immediately rewarding for oneself, while requiring substantial time and effort. P9 highlighted that even though analysts who preserve and share their work might get slightly more citations, this is \textit{``a mild incentive. It's more motivating to start a new analysis, other than spending time encoding things--''.}

In this context, a convener critically contrasted policies with resulting preservation quality and highlighted the motivational strength of returned benefits: 

\begin{quote}
    
    If you take this extra step of enforcing all these things at this level, it's never going to get done. Because if you use this as a documentation, so I'm done, now I'm going to put these things up. If it complains, like, I don't care [...] But if there is a way of getting an extra benefit out of this, while doing your proper preservation, that is good --- that would totally work. (P11)

\end{quote}

Imagining a service that not only provides access to preserved resources, but allows systematic execution of those, the convener states that he does not \textit{``see any attitude problem anymore, because doing this sort of preservation gives you an advantage.''} Such mechanisms might also provide incentives to integrate a preservation service into the analysis workflow, which according to P9 will be crucial. The convener expects that researchers \textit{``will not adapt to data preservation afterwards. Or five percent will do.''} P2 probably falls within that category. He states: ``\textit{I want everyone else's analyses to be there and equally that means that they might want my analysis to be there.}''

\subsection{Communication}

Our analysis revealed that data analysts in HEP have a high demand for information. Yet, communication practices often depend on personal relations. All of our interviewees described the need to access code files from colleagues or highlighted how access could support them in their analysis work. Even though most analysts (P2 -- P4, P6 -- P8, P10 -- P12) explicitly stated that they share their work on repositories that provide access to their LHC collaboration, information and resource flow commonly relied on traditional methods of communication: 

\begin{quote}
The few times that I have used other people's code, I think that... I think it was sent to me by e-mail all the times. (P3) \newline
They have saved their work and then I can ask them: `where have you located this code? Can I use it?' And they might send me a link to their repository. (P8)
\end{quote}

The analysis of our interviews revealed the general practice of engaging in personal communication with colleagues in order to find resources. P4 made a common statement, i.e. colleagues pointing to existing resources:

\begin{quote}
You go to the person you know is working on that part and you ask directly: `Sorry, do you know where I can find the instructions to do that?' and he will probably point to the correct TWiki or the correct information. (P4)
\end{quote}

Personal relations are vital in this communication and information architecture. Most analysts (P1, P2, P3, P4, P6, P7, P8, P9, P11) stressed that it was important to know the right people to ask for information. P8 described the effort needed:

\begin{quote}
I mean you have to know the right people. You have to know the person who maybe was involved in 2009 in some project. And then you have to know his friend, who was doing this. And his friend and then there is somebody who did this and she can tell you how it went.
\end{quote}

But, communication and information exchange was often contained within groups and institutes. P7 stressed that for a certain technique, other groups \textit{``have better ideas. In fact, I know that they have better ideas than other groups, but they are not using them, because we are not talking to each other.''} P2 stated that \textit{``being shy and not necessarily knowing who to e-mail''} are personal reasons not to engage in communication with colleagues. The challenge to find the right colleagues to talk to is increased by the high rotation of researchers, many of them staying only few years.

Almost all analysts (P1 --- P4, P6 --- P11) in our study referred to another common issue they encounter: the lack of documentation. P6 illustrated the link between missing documentation and the need to ask for information instead:

\begin{quote}
This is really mouth-to-mouth how to do this and how to do that. I mean the problem for preservation is that at the moment it's just: ask your colleague, rather than write a documentation and then say `please read this.'
\end{quote}

Meetings and presentations are a key medium in sharing knowledge. However, the practice of considering presentations as a form of knowledge documentation makes access to information difficult:

\begin{quote}
There are cases you asked somebody: `but did they do this, actually?' And somebody says like: `I remember! Two years ago, there was this one summer meeting. We were having coffee and then they showed one slide that showed the thing.' And this slide might have never made it to the article. (P8)
\end{quote}

\subsection{Uncertainty}

\begin{figure}
\centering
  \includegraphics[width=0.85\columnwidth]{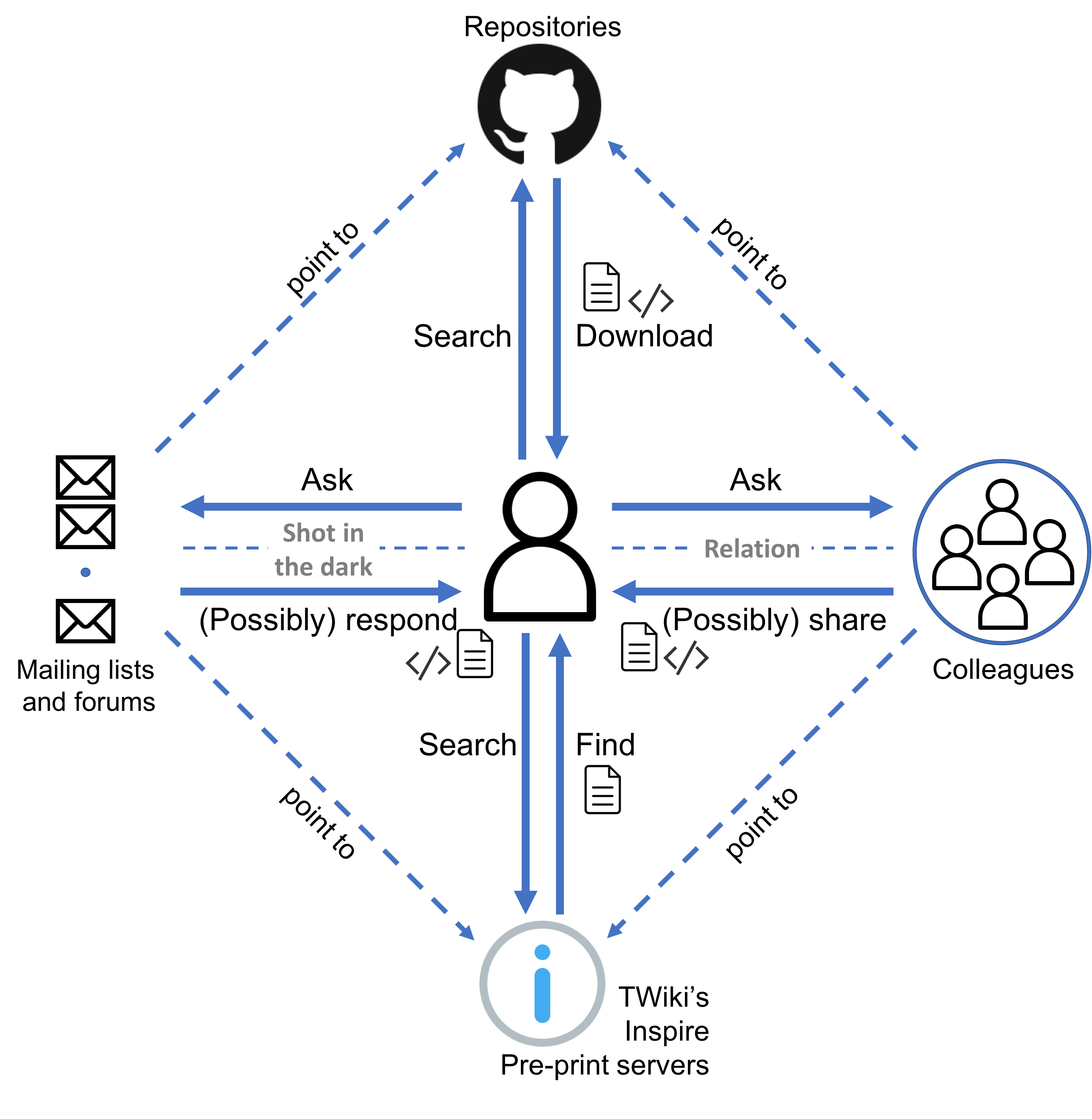}
  \caption{A visualization of information flow and communication in HEP data analysis.}~\label{fig:hep_communication}
  ~\label{fig:figure_current_flow}
\end{figure}

Our interview findings revealed that the communication and information architecture leads to two types of \textit{uncertainty}: (1) related to the \textit{accessibility} of information and resources; and (2) connected to the \textit{volatility} of data. 

\subsubsection{Accessibility}

As depicted in Figure \ref{fig:figure_current_flow}, analysts follow two principal approaches to \textit{access} information and resources: they search for them on repositories and databases or ask colleagues. The outcome of directly searching for resources contains uncertainty, as researchers might not be sure exactly what and where to search. Also, search mechanisms often represent challenges. A researcher described searching for an analysis and highlighted that \textit{``at the moment, it's sometimes hard to find even the ones that I do know exist, because I don't know whether or not they are listed maybe under the person I know. So, [name] I know that I can find... Well, actually I don't know if I can find his analysis under his GitHub user.''} (P2)

Our interviewees (P1 --- P4, P6 --- P9, P11, P12) reported that they typically contact colleagues or disseminate requests on mailing lists and forums to ask for information and resources. While mailing lists represent a shot in the dark, the success of approaching colleagues is influenced by personal relations. If successful, they receive required resources directly or are pointed to the corresponding location.

\subsubsection{Volatility}

Facing vast amounts of data and dependencies, analysts wished that a centralized preservation service helps them with uncertainty that is caused by the \textit{volatility} of data.

\textit{Analysis Integrity:} A service aware of analysis dependencies can ensure that needed resources are not deleted:

\begin{quote} 
[...] and this can be useful even while doing the analysis, because what happens is that people need to make disk space and then they say: `ah, we want to remove this and this and this dataset --- if you need it, please complain.' And if you had this in a database for example, it could be used also saying like `ah, this person is using this for this analysis' even before you would share your analysis. (P6)
\end{quote}

The analyst even highlighted the possibility to track datasets of work-in-progress that was not yet shared with the LHC collaboration. A convener also motivated the issue that comes with the removal of data and described the effort and uncertainty involved in current communication practices:

\begin{quote}
Sometimes versions get removed from disk. [...] And the physics planning group asks the conveners: `ok, is anybody still using those data?' [...] I have to send an email of which version they are using etc. [...] And at some point, if I have 30 or 40 analyses going on in my working group, it's very hard not to make a mistake in this sense if people don't answer the emails. While if I go here, I say ok, this is the data they are using --- I know what they are using --- and it takes me ten minutes and I can have a look and I know exactly. (P11)
\end{quote}

\textit{Receiving vital analysis information}: We learned that different analyses often have input datasets in common. When an analyst finds issues with a dataset, she or he draws back to the existing communication architecture:

\begin{quote} 
I present it in either one of the meetings which is to do with like that area of the detector for example. Or if it was something higher profile than maybe one of the three or four meetings which are more general, applicable to the collaboration\footnote{The interviewee is referring to the LHC collaboration.}. And from there, that would involve talking to enough people in the management and various roles...that it would then I guess propagate to...they would be again in touch with whoever they knew about that might be affected. (P2)
\end{quote}

\noindent The risk of relying on this communication flow is that one might naturally miss vital information. An analyst could be unavailable to attend the right meeting or generally not be part of it. The person sending the email might also not know about all affected analyses. This might especially be true for relevant analyses that are conducted in a working group different from the one of the analysts that are signaling the issue. A preservation service enabling researchers to signal warnings associated with a dataset or, generally, resources that are shared by various analyses, allows informing dependent analysts in a reliable manner. As being informed about discovered issues can be vital for researchers, it would be in their very interest to keep their ongoing analyses well documented in the service.

\textit{Staying Up-to-Date}: Keeping up-to-date on relevant changes can be challenging in data-intensive environments. Researchers wished for a preservation service that provides reliable dependency awareness to analysts who document their work:

\begin{quote} 
The system probably tells me: `This result is outdated. The input has changed'. Technical example. At the moment, this communication happens over email essentially. (P6)
\end{quote}

P11 told us about a concrete experience:

\begin{quote} 
He was using some number, but then at some point the new result came out and he had not realized. Nobody realized. And then, of course, when he went and presented things he was very advanced, they said `well, there is a new result --- have you used this?' `No, I have not used it.'
\end{quote}

\subsection{Collaboration}

Sharing their work openly, analysts increase their chance to engage in collaboration. Currently, useful collaboration is hindered by missing awareness of what others do. We can imagine this to be especially true outside of groups and dislocated institutes. P4 emphasized the value of collaboration:

\begin{quote} 
The nTuple production is a really time consuming part of the analysis. So, if we can produce one set of nTuples... so, one group produces them and then they can be shared by many analysis teams... this has, of course, a lot of benefits. (P4)
\end{quote}

Researchers who document their ongoing activities and interests increase their discoverability within the LHC collaboration. Thereby, they increase their chance to be asked to join an official request that might satisfy their data needs:

\begin{quote}
I want to request more simulation. [...] I would search and I would say these are the people. I would just write to them, because I want to do this few modifications. But maybe this simulation is also useful for them, so we can just get together and get something out. (P11)
\end{quote}

In fact, a convener stated that due to the size of LHC collaborations, it is difficult to be aware of other ongoing analyses: ``\textit{CMS is so big that I cannot know if someone else is already working on it. So, if this tool is intended to have also the ongoing analyses since a very early stage, this would help me if I can know who is working on that.}'' (P9)

P8 highlighted that being aware of other analyses can possibly lead to collaboration and prevent unwanted competition:

\begin{quote}
Because the issue at CMS --- and probably at whole CERN --- is that you want start working on it, but, on the other hand, it's rude if you start working on something and you publish and then you get an angry message, saying: `hey, we were just about to publish this, and you cannot do it.' [...] The rule is that everyone can study everything, but, of course, you don't want to steal anybody's subjects. So, if it wouldn't be published, you would then maybe collaborate with them. (P8)
\end{quote}

\subsection{Automation}

We see an opportunity to support researchers based on the common structure that applies to analyses: \textit{``because in the end, everybody does the same thing''} (P7). A convener characterized this theme by demanding \textit{``more and more Lego block kind analyses, keeping to a minimum the cases where you have to tailor the analysis a bit out of the path''} (P9).

\subsubsection{Templated analysis design}

As P11 articulated, the common steps and well-defined analysis structure represent an opportunity to provide checklists and templates that facilitate analysis work:

\begin{quote}
If, of course, I have some sort of checklist or some sort of template to say `what is your bookkeeping queries --- use this and that', then of course this would make my life easier. Because I would be sure I don't forget anything. (P11)
\end{quote}

The convener made two claims related to how a structured analysis description template could support researchers. First, templates help in the analysis design. Second, the service could inform about missing fragments or display warnings based on a set of defined checks. However, it is important to recognize a core challenge that comes with well-structured analysis templates: allowing for sufficient flexibility:

\begin{quote} 
Somehow these platforms tend to --- which is one of the strong points, but at the same time one of the weaknesses --- is that [...] it gives you some sort of template and makes it very easy for you to fill in the blanks. But at the same time, this makes things difficult, if you want to make very complex analyses where it's not so obvious anymore what you want to do. (P11)
\end{quote}

\subsubsection{Automate Running and Interpretation}

Several analysts (P2, P5, P7, P8, P11) expressed their wish for centralized platforms to automate tasks that they would currently have to perform manually. An interviewee stated:

\begin{quote}
So, being able to kind of see that it... might be able to submit to it and then it just goes through and runs and does everything... and I don't need to think too much about whether or not something is going to break in the middle for something that is nothing related to me, would potentially be quite nice. (P2)
\end{quote}

However, not only automating the full execution of analyses seems desirable, but also interpretation of systematics:

\begin{quote} 
And I say: `ok, now I want to know for example, which are the systematics' and you can tell me, because you know you have the information to do it by yourself. You will save a lot of time. People will be very happy I think. (P5)
\end{quote}

\subsubsection{Preventing mistakes}
P7 described how the similarity and common structure of analyses supports automated comparison and verification:

\begin{quote} 
What I would like to search is the names of the Monte Carlo samples used by other analyses. [...] the biggest mistake you can make is to forget one. Because if you forgot one, then you will see new physics, essentially. And it's a one-line mistake. (P7)
\end{quote}

Developing a feature that compares a list of dataset identifiers and that points to irregularities is trivial. Yet, as P7 continues to describe the effort needed to do the comparison at the moment, the perceived gain seems to be high:

\begin{quote} 
So, the analysis note always contains a table --- it's a PDF. Then always contains a table with a list of Monte Carlos. I often download that, look at the table and see what's missing. Copy paste things from there. But so here, I would be able to do it directly here. (P7)
\end{quote}

\subsection{Scalability}

Although not directly in the scope of the questionnaire, four interviewees (P3, P8, P9, P11) commented on the growing complexity of analysis work in HEP, stressing the importance of preservation and reproducibility. Convener P9 described issues that evolve from collecting more and more data: 

\begin{quote} 
As we collect the data, the possibility of analysis grows. In fact, we are more and more understaffed, despite of being so many in the collaboration. Because, what is interesting for the particle physics community grows as data grow. And so, we get thinner and thinner in person power in all areas that we deem crucial.
\end{quote}

The convener added that \textit{``a typical analysis cycle becomes much much longer. Typical contract duration stays the same.''} P3 detailed how the high amount of rotation and (ir-)reproducibility impact analysis duration:

\begin{quote} 
If someone goes and an analysis is not finished, it might take years. Because there was something only this person could do. I think that analysis preservation could help a lot on this. [...] But otherwise you might have to study analyses from scratch if someone important disappears. (P3)
\end{quote}

P11 agrees that \textit{``it's getting more and more complex, so I think you really need to put things together in a way that is reasonable and re-runnable in some sort of way.''} P9 coined the term \textit{orphan analyses}. It describes analyses for which no one is responsible anymore. The convener expects that \textit{``at some point it will become a crisis. Because, so far, it was a minority of cases of orphan analyses. It will become more and more frequent, unless contract durations will change. But this will not happen.''}

%
%

\section{Implications for Design}

We present challenges and opportunities in designing for research preservation and reproducibility. Our work shows that the ability to access documented and shared analyses can profit both individual researchers and groups \cite{falessi2006documenting}. Our findings hint towards what Rule et al. \cite{rule2018exploration} called ``tension between exploration and explanation in
constructing and sharing'' computational resources. Here, we primarily learned about the \textbf{need to motivate and incentivize contributions}. Based on our findings, we show how design can create motivating \textit{secondary usage forms} of the platform and its content, related to uncertainty, collaboration, and structure. While references in this section underline that the HCI community has established a long tradition of studying collaboration and communication around knowledge work, it is not yet known how to design collaborative systems that foster reproducible practices and incentivize preservation and data sharing. The following description of \textit{secondary usage forms} aims to contribute to knowledge about motivations and incentives for platforms that support research reproducibility.

\subsection{Exploit Platforms' Secondary Functions}

As described in the \textsc{Motivation} theme, getting researchers to document and preserve their work is a main concern. In this context, researchers critically commented on the impact of policies, creating little motivation to ensure the preservation quality beyond fulfilling formal requirements. Citation benefits, commonly discussed as means to encourage research sharing \cite{10.7717/peerj.175}, might also provide only a mild incentive, as time required for documentation and preservation can be spend more rewarding on novel research. This seems especially true in view of growing opportunities that result from the increasing amount of data, as described in the \textsc{Scalability} theme. Yet, researchers indicated how centralized preservation technology can uniquely benefit their work, in turn creating motivation to contribute their research. Thus, we have to \textbf{study researchers' practices, needs, and challenges in order to understand how scientists can benefit from centralized preservation technology. Doing so, we learn about the \textit{secondary function} of the platform and its content, crucial in developing powerful incentive structures}. 

\subsection{Support Coping with Uncertainty}

As we learned in the \textsc{Communication} theme, the information architecture heavily relies on personal connections and communication, leading to a high degree of \textsc{Uncertainty} related to the \textit{accessibility} and \textit{volatility} of information and data. Consequently, researchers reported encountering severe issues related to the insufficient transparency and structure that a centralized preservation service might be able to mitigate. We propose two strategies: First, a centralized preservation service can implement overviews and details of analysis dependencies not available anywhere else. Implementing corresponding features enables us to \textbf{promote preservation as effective strategy to cope with uncertainty} so that research integrity of documented dependencies can be guaranteed. Second, we further imagine documenting analyses on a dedicated, centralized service to be a powerful strategy to \textit{minimize} uncertainty towards updated dependencies and erroneous data, if the service provides awareness to researchers. In the case of data-related warnings, reliable notifications could be sent to analysts who depend on collaboration-wide resources, replacing current, less reliable communication architectures. This approach relates to uncertainties at the \textit{data layer}, as described by Boukhelifa et al. \cite{Boukhelifa2017}, who studied types of uncertainty and coping strategies of data workers in various domains. According to their work, the three main active coping strategies are: \textit{Ignore}, \textit{Understand} and \textit{Minimize}. In summary, our findings suggest that such secondary benefits might drive researchers to contribute and use the preservation tool.

\subsection{Provide Collaboration-Stimulating Mechanisms}

The \textsc{Collaboration} theme indicated the importance of cooperation in HEP. Analysts save time when they join forces with colleagues or groups with similar interests. Yet, awareness constraints resulting from the communication and information architecture often hinder further collaboration. We postulate that the preservation platform can add useful secondary benefits for theses cases. First, given the centralized interface and knowledge aggregation function of a preservation service, we see opportunities to \textbf{support locating expertise in research collaborations}. In fact, knowledge-intensive work profits from such supporting tools, as it enables sharing expertise across organizational and physical barriers \cite{Cross2004}. Ehrlich et al. \cite{ehrlich2007searching} noted that awareness of `who knows what' is indeed key to stimulating collaboration. In an organizational context, \ac{TMS} are employed to create such awareness. HEP collaborations are TMS in that the sum of knowledge is distributed among their analysts and the communication between them forms a group memory system \cite{wegner1987transactive}. Further research on the support and integration of TMS in the context of platforms for research reproducibility could increase acceptance through heightened awareness provided by such platforms. Elements of social file sharing could further stimulate discovery and exploration of relevant researchers and analyses. As noted by Shami et al. \cite{shami2011browse}, this can be particularly important in large organizations.

Second, an important benefit could be the visibility of team or project members. Taking preserved research as basis for expertise location can incentivize contributions, as scientists who document in great detail are naturally most visible, thus increasing their chances to engage in collaboration. This approach also enables us to mitigate privacy concerns, by considering only resources of analyses that have been shared with the LHC collaboration. Mining documented and shared research to provide expertise location mitigates common challenges. Typically, workplace expertise locators infer knowledge either by mining existing organizational resources like work emails \cite{Campbell2003, Gopalakrishnan2017}, or by asking employees to indicate their skills and connections within an organization \cite{Shami2007}. While automated mining of resources may cause privacy concerns, relying on users to undergo the effort of maintaining an accurate profile is slower and less complete \cite{Reichling2009}. Given increasing interdisciplinary and distributed research environments, developing such bridging mechanisms --- even though not central to the service missions --- is especially helpful.

\subsection{Support Structured Designs}

A community-tailored research preservation service can support analysts through automated mechanisms that make use of common workflow structures. Researchers pointed out that analysis work within LHC collaborations commonly follows general patterns, demanding even to further streamline processes as much as possible. We propose to \textbf{design community-tailored services that closely map research workflows to preservation templates}. That way, preservation services can provide checklists and guidance for the research and preservation process. Furthermore, automation of common workflow steps can increase efficiency. Additionally, if the preservation service is well embedded into the research workflows, it could enable supportive mechanisms like auto-suggest and auto-completion. Such steps are key to minimizing the burden of research preservation, which is of great importance as we acknowledge that the acceptance and willingness to comply with reproducible practices will always be related to the cost/benefit ratio of research preservation and sharing. Having noted the need for automation and taylorization of interfaces, we need to emphasize the significance of academic freedom when designing such services. Design has to account for all the analyses, also those that are not reflected in mainstream workflows. We have to \textbf{support creativity and novelty by leaving contributors in control}. This applies both for supportive mechanisms like auto-complete and auto-suggest, as well as for the template design.

\section{Discussion}

The study's findings and implications pointed to several relationships that are important for designing technology that enables research preservation and reproducibility. First, we have contrasted required efforts with returned benefits. It is apparent that stimuli are required to encourage researchers to conduct uninteresting and repetitive documentation and preservation tasks that in itself, and at least in the short run, are mostly unrewarding. Thus, not surprisingly, the call for policies is prominent in discussions on reproducible research. Yet, our findings hint towards the relation between preservation quality and policies, raising doubts that policies can encourage sustained commitment to documentation and preservation beyond a formal check of requirements. In this context, we argue that the relation between policies and flexibility needs to be considered. Thinking about structured description mechanisms as provided by CAP, one needs to decide on a common denominator that defines main building blocks which reflect policy requirements. However, this is likely to create two problems: (1) Lack of motivation to preserve fragments that are not part of the basic building blocks of research conducted within the hierarchical structure for which the policies apply; (2) Preservation platforms that map policies might discourage or neglect research that is not part of the fundamental building blocks.  

Facing those conflicting relationships, meaningful incentive structures could positively influence the reproducibility challenge and create a favorable shift of balance between required efforts and returned benefits. We postulate that communities dealing with the design of such systems need to invest a significant amount of time into user research to create tailored and structured designs. Further research in this area is surely needed, i.e. the evaluation of prototypes or established systems in general and with a focus on the users' exploitation of secondary benefits of the system. This call for future research in this area is particularly evident when looking at the study by Rowhani-Farid et al. \cite{Rowhani-Farid2017} who  found only one evidence-based incentive for data sharing in their systematic literature review. They conducted their study in search of incentives in the health and medical research domain, one of the branches of science that was in the focus of reproducibility discussions from the very beginning. The only reported incentive they found relates to open science badges. The authors stressed that since ``data is the foundation of evidence-based health and medical research, it is paradoxical that there is only one evidence-based incentive to promote data sharing. More well-designed studies are needed in order to increase the currently low rates of data sharing.''

Our study described secondary usage forms related to communication, uncertainty, collaboration, and automation. Described mechanisms and benefits apply not only to submissions at the end of the research lifecycle, but, rather, provide certainty and visibility for ongoing research. The significance of such contribution-stimulating mechanisms is particularly reflected in the observed scalability challenge, indicating that reproducibility in data-intensive computational science is not only a scientific ideal, but a hard requirement. This is particularly notable as the barriers to improve reproducibility through sharing of digital artefacts are rather low. Yet, it must also be noted that not all software and data can always be freely and immediately shared. The claim for reproducibility does not overrule any legal or privacy concerns.

\section{Limitations and Future Work}

We aim to foster the reproducibility of our work and to provide a base for future research. Therefore, we publicly released various study resources as part of the supplementary material of the publication on which this chapter is based. Those include the semi-structured interview questionnaire, the ATLAS.ti code group report, and the templates of the two paper exercises. As is the core idea of reproducible research, we envision future work to extend and enrich our findings and design implications by studying perceptions, opportunities, and challenges in diverse scientific fields. We can particularly profit from empirical findings in fields that are characterized by distinct scholarly communication and field practices and a differing role of reproducibility.

It should also be noted as a limitation of the study that the reference preservation service is based entirely on custom templates. While this does not reflect the majority of repositories and cloud services used today for sharing research, our findings indicate that templates are key to enable and support secondary usage forms. And even though our study focused solely on HEP, findings and implications are likely to apply to numerous fields, in particular computational and data-driven ones.

\section{Conclusion}

This chapter presented a systematic study of perceptions, opportunities, and challenges involved in designing tools that enable research preservation and reproducibility in High Energy Physics, one of the most data-intensive branches of science. The findings from our interview study with 12 experimental physicists highlight the resistance and missing motivation to preserve and share research. Given that the effort needed to follow reproducible practices can be spent on novel research --- usually perceived to be more rewarding --- we found that contributions to research preservation services can be stimulated through secondary benefits. Our data analysis revealed that contributions to a centralized preservation platform can target issues and improve efficiency related to \textsc{communication}, \textsc{uncertainty}, \textsc{collaboration}, and \textsc{automation}. Based on these findings, we presented implications for designing technology that supports reproducible research. First, we discussed how studying researchers' practices enables exploiting secondary usage forms of platforms that are expected to stimulate researchers' contributions. Centralized repositories can promote preservation as an effective strategy to cope with uncertainty, support locating expertise in research collaborations, and provide a more guided and efficient research process through preservation templates that closely map research workflows.


\chapter{Cross-Domain Investigation of Research Data Management and Reuse}

\label{ch:cross_domain}

In Chapter \ref{ch:requirements}, we reported findings from our study on practices around research preservation and reuse in HEP. We invited physics analysts to explore and discuss CAP, and found that secondary uses of RDM tools might provide meaningful benefits for contributors. In particle physics, those secondary uses relate to automation, structure, stimulation of collaborative behaviour, and coping with uncertainty. We hypothesize that the general nature of those uses make our findings applicable to research beyond particle physics. To test this hypothesis, and to expand our knowledge of practices around RDM and reuse, we conducted a cross-domain study involving 15 researchers and data managers from diverse branches of science. In this chapter, we report on the cross-domain study and our findings.

This cross-domain study also relates to the work of Muller et al. \cite{Muller:2019:DSW:3290605.3300356}. They investigated how data science workers work with data. The authors advocated the importance of better understanding data science workflows. To improve this understanding, and to inform the design of tools that improve data science workflows, they interviewed 21 data science workers at IBM. One of their core contributions is a detailed description of five ``human interventions in relation to data'': Discovery, Capture, Curation, Design, and Creation. The authors proposed that data ``wrangling operations might be mapped analytically in relation to the five interventions [...].'' This analytic mapping is expected to impact the design of tools that support data science workers. The authors noted that practitioners and researchers could profit from further investigating data science work practices beyond single organizations or enterprises. Our study relates to this by investigating data-related human interventions across scientific branches. Furthermore, our study focuses on expanding our knowledge of \textit{human interventions in relation to data management} by considering additional four interventions related to science reproducibility: Documentation, Preservation, Sharing, and Reuse.

In this chapter, we first detail our study design which is closely aligned with the HEP study, reported in Chapter \ref{ch:requirements}. Next, we present the findings from the interview study. Based on the core concepts identified in our data analysis, we introduce and discuss a \CommitmentModelTrailingSpace that we expect to inform the design of RDM tools.

\begin{tcolorbox}[title = This chapter is based on the following publication.]
Sebastian S. Feger, Paweł W. Woźniak, Lars Lischke, and Albrecht Schmidt. 2020. ‘Yes, I comply!’: Motivations and Practices around Research Data Management and Reuse across Scientific Fields. In Proceedings of the ACM on Human-Computer Interaction, Vol. 4, CSCW2, Article 141 (October 2020). ACM, New York, NY. 26 pages.
\newline\url{https://doi.org/10.1145/3415212} 

\color{darkgray}
------------------------------------------------------------------------------------------------------\newline
Several of the study's resources are openly available as supplementary material in the ACM Digital Library.
\end{tcolorbox}

%
%
\section{Study Design}

We conducted a semi-structured interview study with 15 researchers and research data managers from a diverse set of scientific fields. In this section, we detail our recruitment process and demographic data of our study participants. We further outline the structure of the interview study and detail the highly iterative and collaborative data analysis. 

\subsection{Study Participants}

\begin{table*}[t]
  \centering
  \begin{tabular}{l c c c c c c r}
    {\small\textit{Ref.}}
    & {\small \textit{Domain}}
    & {\small \textit{Role}}
    & {\small \textit{Experience}}
    & {\small \textit{Environment}}
    & {\small \textit{Gender}}\\
    \midrule
    P1 & Biology & Researcher & Postdoc & Academia & Female\\
    P2 & Meteorology & Researcher / RDM & Postdoc & Organization & Male\\
    P3 & Arts and Curation & RDM / Sen. & Master & Museum & Male\\
    P4 & Biology / Chemistry & Researcher / RDM & Postdoc & Academia & Male\\
    P5 & Physics Research & Researcher & PhD Student & Organization & Female\\
    P6 & Information Technology & Policy Officer & Master & Organization & Male\\ 
    P7 & Biology / Chemistry & Researcher & Postdoc & Academia & Male\\
    P8 & Information Technology & RDM & Master & Academia & Male\\
    P9 & Physics Research & Researcher / PM & Postdoc & Organization & Male\\
    P10 & Agricultural Research & Researcher / Sen. / PM & Master & Organization & Male\\
    P11 & Research Images Reuse & Researcher / RDM & Bachelor & Organization & Female\\
    P12 & Physics Research & Researcher & PhD Student & Organization & Male\\
    P13 & Information Science & Researcher & PhD Student & Organization & Female\\
    P14 & Geoinformatics & Researcher / RDM & PhD Student & Academia & Male\\
    P15 & Environmental Science & Researcher / RDM & Postdoc & Organization & Male\\
    \bottomrule
  \end{tabular}
  \caption{Overview of cross-domain study participants.}~\label{tab:table_interviewees_cross_domain}
\end{table*}

In the beginning of our recruitment process, we disseminated a short study abstract and call for participation among the academic circles of the authors of this work. As all authors worked in different academic organizations, we quickly recruited researchers with different backgrounds. After the first pilot interview, we discussed the interview and decided to leave the protocol for the semi-structured interviews unaltered.

Besides asking within our personal and institutional circles, we approached participants of the \ac{OR} 2019 conference. OR 2019 focused on user needs related to research repositories. As I gave a talk at this conference, we were in a good position to invite a diverse sample of researchers and research data managers to participate in our study. We studied all accepted submissions and contacted individual authors who either conducted research or worked closely with scientists on improving RDM. This approach helped us recruit several scientists who both conducted research and were in charge of RDM. We described the role of these study participants as \textit{Researcher / RDM} in Table \ref{tab:table_interviewees_cross_domain}. We further recruited participants who were solely responsible for conducting either research or RDM. In addition, we recruited one policy officer. The recruitment of participants with a diverse set of roles and responsibilities helped us map a most complete set of practices, challenges and requirements related to human interventions on research data management and scientific reuse.

The participants were 26 to 48 years old with an average age of 34 years (SD = 7.5). Eleven participants were male and four female. We conducted all interviews during regular working hours and did not provide any remuneration for the study participation. The cultural diversity amongst our study participants is the result of our conference-based recruitment strategy. Honoring the request of several interviewees, we do not list the nationalities of individual participants. However, we can list the nationalities involved in alphabetic order: Dutch, English, Finnish, German, Italian, Ugandan.  
We conducted all interviews in English.

As depicted in Table \ref{tab:table_interviewees_cross_domain}, we recruited participants from various, very different branches of science, including Biology, Chemistry, Arts, Geology, Meteorology, Physics, and Agricultural Research. The participants worked in academic institutes and research organizations. One participant worked in an arts museum which has a research department and provides access to the collection and metadata to external researchers. The participants also differed in terms of professional and academic experience. We recruited Bachelor and Master graduates, PhD Students, and Postdocs. It is worth noting that two of the Master graduates had close to or more than 15 years of professional experience. We identified them as \textit{Senior} in Table \ref{tab:table_interviewees_cross_domain}. The diverse sample enabled the study of practices around workflows in data science and requirements for supportive technology across multiple scientific domains.

\subsection{Interview Protocol}

\begin{figure}
\centering
  \includegraphics[width=0.8\columnwidth]{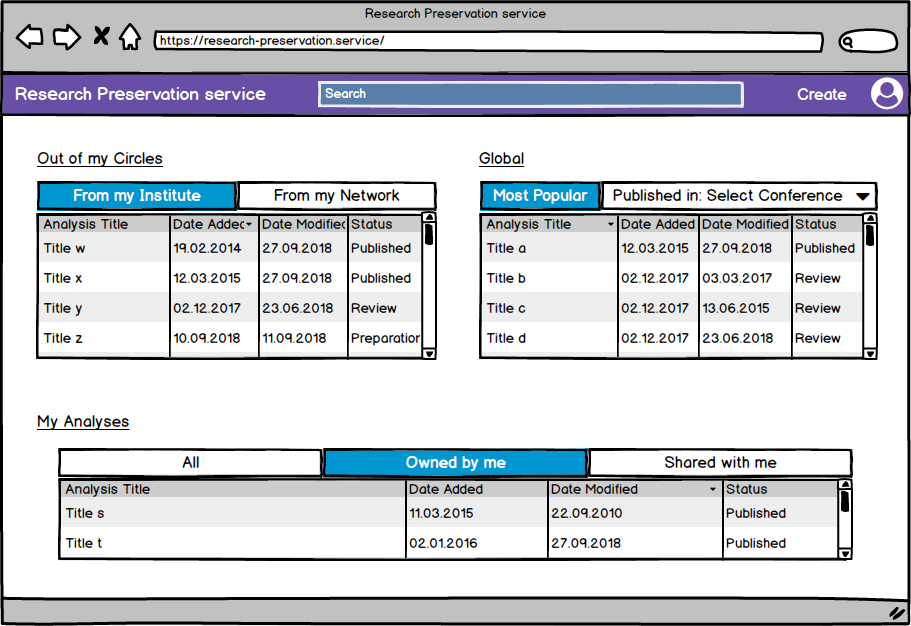}
  \caption{Dashboard of the generic preservation service used in the cross-domain study.}~\label{fig:generic_dashboard}
\end{figure}

\begin{figure}
\centering
  \includegraphics[width=0.8\columnwidth]{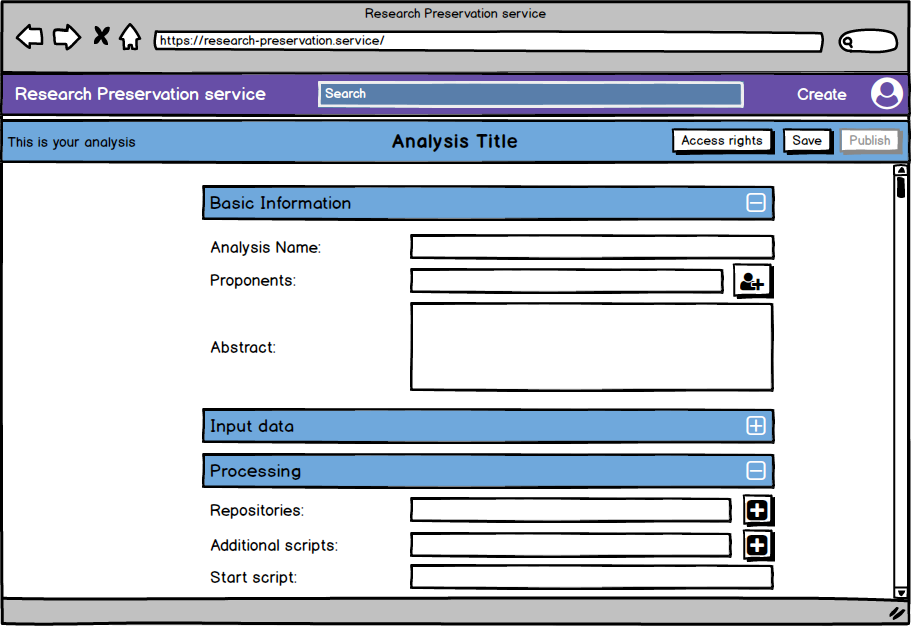}
  \caption{Template-based analysis description form of the generic preservation service.}~\label{fig:generic_templates}
\end{figure}

The interview protocol is closely aligned with the protocol we used in our HEP requirements study (see Chapter \ref{ch:requirements}). Here, we detail our interview protocol and describe modifications. The structure is as follows.

Initially, we asked the participants to briefly introduce themselves and to talk about their main responsibilities and roles. Next, we asked questions related to data used in their research field. In particular, we asked about the role of data, data provenance, and about the data life cycle. We then prompted the participants to talk about the processing of data and how processing tools are created, adapted and shared.
    
We continued by asking about practices around preservation and reuse. We were particularly interested in the storing of research artifacts and any experiences related to reuse. Based on this, we asked about their approaches to information seeking, either for research or training purposes. We concluded this part on current practices by asking about the technologies currently used to preserve, share and find research.
    
We further investigated the role of technology in RDM and reuse. In the HEP study, we asked questions related to the CAP prototype service. As this technology probe proved to be essential in our study, we designed mockups for a more general preservation service that is inspired by the design of the open source CAP service. We designed mockups for two principal views: the dashboard (Figure \ref{fig:generic_dashboard}) and the analysis page (Figure \ref{fig:generic_templates}). The analysis page is based on a research documentation template. The dashboard references research conducted or managed by the user of the service, as well as research preserved and shared by the research community. The analysis page shows a generic template with blocks for input data and processing resources.
    
We asked the study participants to tell us about their perceptions of the service and to compare such a tool to systems available in their environment. We further invited them to tell us about their search needs and to imagine what they would want to look for in the system if it was operational and contained a vast amount of relevant research. Connected to those search-related questions, we invited the interviewees to sketch any kind of dependencies and connections that may exist between two generic research projects. We then asked how a service that was aware of such relations could impact research work. Finally, we invited participants to reflect on their concerns, hopes, and expectations regarding such a RDM tool.


\subsection{Data Analysis}
We recorded a total of 11 hours during the 15 interviews (Mean = 43.8 minutes, SD = 8 minutes). All recordings were transcribed non-verbatim. We used Atlas.ti to organize and analyze the transcriptions. We used Grounded Theory Method \cite{muller2014curiosity} to explore the data. Two of the authors independently performed open coding of two transcriptions. We repeatedly discussed the open codes, also in the presence of a third author who moderated the discussions. We recorded open codes and rules for those codes in memos. Those helped to further reason about the data. Next, we performed axial coding based on the open codes recorded in Atlas.ti. We represented axial codes as code groups in the data analysis software. The Atlas.ti code group report that we made available as supplementary material captures this state of the data analysis. It refers to 28 axial codes and 379 open codes. As Atlas.ti did not provide means to further support the data analysis, we focused on creating and expanding memos in common word processors. Based on our continuous comparison of data to data, we described and tested categories and four core concepts: \textsc{Practice}, \textsc{Education}, \textsc{Adoption}, \textsc{Barriers}, and \textsc{Impact}.

\section{Findings}
\label{section:cross_findings}

We present findings from our study based on five core concepts: \textsc{Practice}, \textsc{Education}, \textsc{Adoption}, \textsc{Barriers}, and \textsc{Impact}.

\subsection{Practice}

Interviewees extensively discussed the role of RDM in the data and research life cycle. In fact, data management already plays an important role in the production and collection of data. The interviewees stressed that they acquire data in four different ways: analysts produce the data themselves; they order data from companies; they use publicly accessible data (open data); and they request them from data providers within their collaborative frameworks.

\begin{quote}
In general, I would say that most of them is generally produced by themselves, but there's also cases where for genomics or metabolomic studies, it's all produced by external vendors. (P4) \newline
You can access the data on the webnet pages of the ESA, of NASA. [...] you click on satellites, you get the product which you want, you select your time range, your special ranges, and then you download them. (P2)
\end{quote}

Severals participants highlighted the role of technology in collaboration-internal data distribution. P1 described that she requests data from so called wet labs that are part of the research collaboration. She pointed out the role technology plays in sharing those data: ``\textit{I'm usually asking for the data. [...] We're having an online sharing platform which is provided from Heidelberg, which is especially for researchers to share our data. They're uploading it and tell us that it's there.}'' P1 further described experiences with commercial vendors. She highlighted that they do not necessarily share all data they have. In fact, she was wondering ``\textit{if they even send their data analysis protocol with the data. [...] Maybe it's even because they're keeping it secret for their purposes, maybe.}''

Informants described a wide variety of data analysis and modeling approaches and the role of technology in this crucial step in the research life cycle:

\begin{quote}
Based on the analysis, they write a plan and they execute that plan. The treatment plans and the reports on the treatment, they print and they sign and they sent to us to keep. At the moment that's still a highly analog process. We don't have systems in place yet to record those data digitally. (P3)
\newline
Again, it's a spectrum. Keep in mind, at one end of the user community, we've got people who are Excel spreadsheet people. [...] At the other end of it, then you've got people who are on the JASMIN system. They are writing large traces of software which may be stored in a GitHub repository. (P15)
\end{quote}



Independent of the level of technology use in the data analysis, most participants described the value of comprehensive data management as part of the research life cycle. Probed by the generic preservation service mockup, several researchers stressed the value of having such a tool in their environment. However, informants showed concern for the effort needed to document and preserve their work on such a service: \textit{``The only thing that comes into my mind is how much time does it cost the user to fill this out? [...] They like to write it down and basically they think it's done.''} (P7) Most study participants stressed that the extra effort needed must be met by strong benefits and use cases which need to be communicated to the researchers. The following statements reflect this common notion:

\begin{quote}
My concerns would be that it wouldn't be taken up by scientists because they think it's too much work on top of their normal work. [...] If it's made clear that it doesn't cost extra time and that it saves time in the end, I don't know, by presenting a good use case or so, then it should be fine. Otherwise, people may remain skeptic. (P2) \newline
Those formats, the file formats, that contain the metadata-- It would be very helpful to have some kind of service to extract those metadata from those files and to put it in a structured way. (P3)  
\end{quote}


In collaborative environments, sharing of data and analysis scripts is common. Access to research resources depends on the state of the research and visibility within an organization. P4 described this: \textit{``The private ones are in the development stage of any project or code before it's published.''} Given restricted access to resources and the overall difficulty to locate them, most researchers described that they would ask for data and resources directly through personal communication: \textit{``I would contact people, but by email, and say: I've heard that you have this and this data or portrait, do you have results or so which you can share?''} (P2)

\subsection{Education}

Education in RDM practices has proven to be a core concept. One that is subject to change in computational data-intensive science. There is a notion of lacking awareness:

\begin{quote}
There was no awareness that there is a thing called data management and that it is important. There was no motivation to do it actually because the benefit wasn't clear. I think it was mainly not knowing how things work. For example, the concept of data workflow was never ever presented or not aware. (P3)
\end{quote}

The data managers highlighted the value of engaging in communication with the researchers and to provide support. Here, P6 emphasized the difference between helping and educating: \textit{``I think the way we can help them best is just to make the help that we offer as concrete as possible and not go around a lot of business and tell them things they would not ever use. No, I don't think the majority of the researchers would want to be educated. \textit{Help}, that's the word I think.''} P15 stressed that this supportive process requires efforts to adapt to the individual situation of the researcher: 

\begin{quote}
We'll then sit down with them and then say: Okay, all right we recommend that you do this. [...] They may turn around and say, 'Well I can't do that, I'm an excel spreadsheet user.' Okay, so you're going to have to take a step back and say, 'Right, well what can we do?'
\end{quote}

Knowing about and finding suitable infrastructure is a key challenge that most participants referred to. P14 reflected on willingness and ability of research sharing: \textit{``In my experience, they either don't have the time to publish the materials, the code and the data. I think even more often they just don't know how to do it and where.''} There is consensus that coordinated efforts are needed to train researchers, both as part of formal education and afterwards: \textit{``And we are not trained as physicists to the good practices, even though there are efforts ongoing. So, ok, more education about the good practices is a recommendation.''} (P9)

P2 stressed that RDM practices need to become part of formal curricula. He further emphasized that to start educating oneself, there are numerous learning resources available: \textit{``There's a lot of literature out there, web courses, webinars, on data management.''} The need for training in good practices and the provision of suitable and easily accessible infrastructure becomes evident by the descriptions of our study participants. P7 referred to a \textit{"haphazard way"} of storing resource. P4 described an experience of his own that is based on the personal use of general-purpose storage drives: 

\begin{quote}
I discovered that there was a directory of work that I hadn't touched for five or so years. The two other people were actually - because I was about to go to delete it, I noticed that the file dates were actually quite recent. Turned out that inadvertently two other people had been using it to share the data on that thing, but it was actually my hosted shared drive.
\end{quote}

In this context, P3 stressed the value of a research preservation service. He emphasized that as they currently have no suitable tools available, titles of folders and the folder structure become the meta-data description: \textit{``We have a simple file server and we have a manual structure for researchers to use when they use certain kinds of techniques. [...] you have to call your file folders like this, et cetera. The metadata is really mostly in the structure that is used on the file server firm."}

Besides actually storing and sharing large data volumes in a structured way, computational and data-intensive science pose further challenges related to RDM practices. In particular, citing of data and analysis software is not yet a common practice. P15 referred to data citations as being ``\textit{still in its infancy to some extent.}'' P13 added: ``\textit{Usually, people are not aware that this is something you should also cite. [...] Actually, you should cite data and software usage as you should cite papers, but people are not aware of that.}''

Finally, a crucial part of RDM education relates to changing perceptions regarding quality and judgement of shared resources. Several researchers referred to a fear of judgement for shared resources that they do not perceive to be \textit{perfect}: \textit{``Yeah I think some people are kind of shy to... that people might judge them for their code being a mess or something and so they tend to, want to keep / hide it away.''} (P12)


\subsection{Adoption}

While P3 highlighted that they do not have developers available to create much needed data management tools, several participants noted that they had already undergone such developments. Adoption may be based on support and enforcement within smaller teams and institutes: ``\textit{R is the programming language that is used here in our institute, because the professor of the lab is also very keen to use open scientific tools and open source tools.}'' (P14) Another participant mentioned the development of a tool that provides similar functionalities to the generic preservation service. However, he stressed that the adoption of such tools may not just be a question of acceptance among the researchers and institutes, but the whole administration:

\begin{quote}
My hope for the pilot is that it is taken up as a supported tool by the IT department. Because this is not something you can do easily on a group level or even an institute level because you have to make guarantees of 10 years storage retention times if you start having data. (P14)
\end{quote}

P2 further reflected on this aspect of adoption, considering potential enforcement: \textit{``But if there's only like a handful people of say, the institution or the community using it, then it's useless. It has to become a common tool [...] or say it has to grow over a certain critical point, then it's useful. Or it has to be imposed from top down saying: We use this tool, period.''} Yet, most participants stated that there are no clear policies in place in their institution. Or that the policies are too generic:

\begin{quote}
There is a policy concerning the practice, which just says that you have to work clean and make your work reproducible but it doesn't mention any tools. You read it and you say: 'Yes, I comply!' But, really, do you comply? In most cases, no, you don't. (P2)
\end{quote}

More pressure might be exercised by publishers and funding agencies. P4 pointed out that changes in their adoption of practices and technology lead back to recommendations from an institutional review. And P6 stressed that funders prescribe the use of certain platforms: ``\textit{the researcher funder can say, 'We want you to publish or to store it on Easy'.}'' Most informants referred to the role of publishers. They stressed that publishers increasingly require submission and sharing of additional resources and meta-data. However, technologies used to describe the data put an extra burden on researchers:

\begin{quote}
This is usually at the point where the vast corrections of the article are just done and people are stressed out because the deadline is really near then you still need this one little code to say I uploaded my data there than to spend- this tool makes life hard and people are reluctant to use it. (P7)
\end{quote}

Finally, all data managers and some researchers referred to a benefit of documenting and sharing research resources, namely the impact on citations. P8 stated: \textit{``To enable their research be more visible. So they can get citations. That's the best known to encourage them to submit.''}

\subsection{Barriers}

We already hinted towards barriers in the adoption of comprehensive RDM practices in the previous three sections. In this section, we expand on the notion of barriers with particular regard to challenges imposed by increasing data volumes and computational processing. The concept of data preservation is --- although not necessarily well adopted --- well understood. Yet, in computational research the notion of data can become ambiguous. P2 emphasized that in environmental research, the data is just the output of a model. The actual information, the part that needs to be preserved, is the model itself: 
\begin{quote}
Now, the point is, do you keep all that data or do you keep the model code including all the settings and the environment which it was run in? And then just save that and have someone, who wants to have the data, actually run the model again. Or, there is a consent and communication of what is data because some people don't work in this field, they don't see the model output as data. It's just model output.    
\end{quote}

An important motivation for our study is that science is becoming increasingly data-intensive and that data is analyzed through computational processing. But this does not mean that all data is ready to be processed and to be treated as part of a modern comprehensive RDM: \textit{``Curators have gathered a lot of heterogeneous documents on the collection of individual objects [...] That's about 800 meters of paper information. We are in the process of digitizing, but maybe that will come up later.''} (P3) While analogue data represent the far end of the data science spectrum, challenges connected to digital data formats are more common. P11 provided an example for this. Her research focuses on the extraction and public distribution of annotated research images which were published in open access journals: \textit{``Many articles are in XML format. That's easy to pass and so we decided only to collect XML articles. But, a lot of articles are in the PDF format.''}

Data formats were addressed by most informants. Closed formats provoke issues, as scientists either can't use them at all or as they need to convert them for automated processing. In order to make them inter-operable, researchers and / or facilitators invest efforts:

\begin{quote}
We built a big database for translating all the different stuff, because we generally stopped requesting in a specific format. We're just asking what they have and if it's in our database and then we're translating it for ourselves. (P11) \newline
We have a lot of problems with software specific data. [...] You get the data in the specific format for the software. Well, we cannot do anything. Well, we cannot even open it. (P1)
\end{quote}

Besides proprietary formats, also commercial software and closed repository services pose challenges to the management and sharing of data. Commenting on the service mockups, P10 stressed that generic, public and well-known services are likely more attractive to researchers than internal, tailored systems that create less visibility. However, once researchers opt for such an approach, data will likely stay there: \textit{``This ResearchGate is not an inter-operable system, it's closed. You will never be able to export the information from there so you need to balance the two.''} (P10)


\begin{quote}
First of all, it should be an open format. In my idea, research is not open and reproducible if the code is only usable with a licensed software, so MATLAB or something like that. It should be something that is open which everyone can download and use, for example, R or Python, whatever, but licensed software is, in my opinion, not open and reproducible. (P14)
\end{quote}

Participants stressed that increasing data volumes pose challenges related to storage capacity, processing and validation of (meta)data, and increased data noise and waste. As they need to find ways to deal with big data, data managers learn from others: \textit{``A project which basically is a landscape analysis of seeing how we can manage big sets of data better within the organization. We are really looking for good examples of organizations where they also deal with large amounts of data.''} (P3)

\textit{Sharing} of data is also impacted by the big data challenge. Not only because sharing requires suitable infrastructure, but also because data must be validated. In particular, researchers and data managers need to ensure that they have the rights to share data and that they respect privacy regulations:

\begin{quote}
We will not be able to expose because we don't have the capacity to curate such data. [...] Our policy doesn't allow to expose raw data unless it's fully curated, except remove the personality identifiers, name of farmers, so we cannot just take dumped and put it for the use. That's when it's a loss because we could have observed certain behavior. (P10) \newline
They share their work but not in the entire community and actually openly. At least in the metereology and climate sciences, that's not a normal thing to do and I know that sometimes people are very anxious about that because they are not sure about licenses of the data. (P2)
\end{quote}

RDM tools commonly enable researchers to restrict access to resources within the system. However, P1 noted that a system that is in principal more visible and prone to attacks or accidental data publication might bear serious concerns:

\begin{quote}
We're working with industry. A big fear of industry is that data is getting, not hacked, but accidentally made public. I don't care if my thing is accidentally public. Well, it's public anyways at some future point, but we signed the contracts with the industry, and they will kill us if we accidentally make something public. It needs to be 100 percent secure that --- hacked is a completely different level --- but it needs to be in a very closed system.
\end{quote}

Another major barrier that was discussed by the participants relates to the growing complexity and novel practices in computing. Several participants pointed out that analysts are usually domain experts, but no professional programmers: \textit{``Either a complex analysis code or just a visualization macro, the challenge is the same. Someone who is not a professional programmer, who writes code, as if this code is going to be used only once. So, doesn't care about writing meaningful comments, or naming the variables in a meaningful way. I mean it's intended to be private and to be used once. Then it gets used by ten people across ten years.''} (P9)

But not only the quality of the code and its documentation suffers from this. Informants discussed that even if analyses are accessible, it is usually not possible to re-execute them, as the authors are not trained in such computing practices:

\begin{quote}
This, of course, also means that it's not always executable. So, maybe with Docker, for example, ... But, this usually takes a lot of effort for authors, and they have to learn the new technologies. Most researchers have no idea about Docker. (P14) \newline
Computationally, it was almost always a disaster. Reusing code, I don't know why, but it seems to be just impossible. [...] You're not interested in doing exactly the same thing, you want to reinvent something, use new data for it. (P1)
\end{quote}

As P14 further pointed out, there is not only a need for different skills, but also for suitable infrastructure: \textit{``Ideally, they would use some kind of online infrastructure. There are, for example, infrastructure such as MyBinder, where you can submit your Python notebooks from your GitHub repository and then you can execute it online.''} But, \textit{``many, many publishers don't support that. They just say you should attach the code and the data. So at least it should be somehow possible to attach it in a folder.''}


\subsection{Impact}

Regarding the challenges that researchers face, our informants also discussed potential benefits from comprehensive RDM. Those relate, in particular, to an increased efficiency based on the ability to re-execute and reproduce existing work. But, also to a decrease in frustration that is caused by current practices. In addition, participants pointed out new opportunities for the discovery and navigation of relevant data and analyses. When we asked them to imagine what they would want to search for if a service like the mockup preservation service existed, participants told us about various uses, including navigating people, finding examples to learn from, fostering collaboration, and reviewing administrative status overviews of ongoing research projects.

Several participants also discussed additional effects of reuse --- and especially reuse tracking. In particular, they referred to the impact on the recognition of an institute's or organization's work. Proving their value for the wider scientific community can provide strong arguments in the interaction with funding agencies. This might be particularly important for globally operating organizations that need to convince funders from different nations:

\begin{quote}
[...] otherwise, they will be decided to fund only national institution instead of international institution may be located as we are, for example, in Lebanon as headquarter but in main office. Why I would say the British government should fund as a way to do something if we are not able to demonstrate that our data is being used by a British student to downloading from our server. In this way, we are just putting the data out there open-access but nobody's able to trace it. (P10)

When we've got people who have made the data available, and then they need to give some information about their funder. What type of communities they are supporting, what proportion of the users of the data are actually commercial users, for example, or personal users or now outside the core research domain. (P15)
\end{quote}

P10 further stressed that the reuse tracking is also important as feedback to improve scientific processes: \textit{``From the research side, obviously, if you are able to study the behavior of a specific researcher that is downloading a specific stream of data, you are also able over time to influence some data quality process, some engagement, and so forth. That one is obviously a feeding back the results of behavioral analysis of interest to the research cycle to improve on the specific aspect.''} P11 added: \textit{``Because this is a (publicly) funded project, of course when we write the report, we want to say: 'This many people used our images.' Also, of course, to see when we change something, if that had an impact or not.''}

Finally, few of the informants discussed that adopting comprehensive RDM practices --- and openly showing their efforts --- is a responsible way of dealing with unique objects. For example, P3 referred to a historical obligation: \textit{``We have a huge colonial past. [...] We have this whole, this big responsibility to show those people that we take care of that collection in a proper way because it's their heritage [...]''} And also in the natural sciences, experiments exist that are unique. P13 referred to particle accelerators that are ``working like if you have a data collision, it's a once in a lifetime event. [...] it's super important that people can share this data and can reuse data.''

\section{Discussion}

The findings of the cross-domain study relate to several of the challenges we characterized in the HEP study in Chapter \ref{ch:requirements}. In particular, issues concerning the accessibility and re-usability of research artifacts were described extensively by study participants across diverse scientific domains. We learned that data formats, growing data volumes, and fast-changing requirements and practices in computational science are some of the main barriers for effective RDM and reuse in science. Thorough education, as well as meaningful incentives, policies, and encouragement are key in the adoption of comprehensive RDM. Based on our findings, we introduce the \CommitmentModel, which closely maps those considerations in four stages: Non-Reproducible Practices, Overcoming Barriers, Sustained Commitment, and Reward. We describe and discuss the model in Chapter \ref{ch:hci_role} --- The Role of HCI in Motivating Reproducible Science --- as its development profits from further reflection on findings of all studies presented in this thesis. 

%
%







\subsection{Limitations and Future Work}

We strive to foster the replicability of our work and to provide a base for future research. To do so, we made several of the study's resources available as supplementary material. Those resources include the semi-structured interview guide, the Atlas.ti data analysis code group report, and the paper resources that depict the generic research preservation service mockups.

We want to note our recruitment strategy as both a limitation and strength of this study. The recruitment represents a limitation, as one third of the participants was recruited based on their participation in the Open Repositories 2019 conference. Almost half of the participants had shared research / RDM responsibilities. While we consider the mix of different perspectives a strength of our study, we find that it does not suit the systematic mapping of \textit{secondary usage forms} of RDM tools. The professional background of the study participants makes it likely that they voice issues and concerns around RDM and reuse more loudly. While we have to understand that in the interpretation of our findings, we argue that this unique perspective of scientists and data managers across a wider set of scientific fields presents a great opportunity to learn about needs and requirements of RDM and reuse in science. 

\section{Conclusion}

This chapter presented a systematic study of motivations and practices around RDM and reuse across a wide variety of scientific domains. The findings from our interview study with 15 researchers and research data managers highlighted the delicate balance between researchers' frustration about bad data practices, lack of knowledge and ambiguity in RDM practices, and hesitation to commit to comprehensive RDM. Based on our data analysis, we mapped practices around RDM and reuse across multiple scientific domains and described five core concepts: \textsc{Practice}, \textsc{Education}, \textsc{Adoption}, \textsc{Barriers}, and \textsc{Impact}. Based on those, we present a \CommitmentModel, which we describe in detail in Chapter \ref{ch:hci_role}.
	
	\part{Gamification: Motivating Reproducible Practices}\label{part:gamification}

\chapter{Gamification Design Requirements for Reproducible Science }
\label{ch:gamification_requirements}

In Part \ref{part:requirements}, we reported on practices around RDM, reproducibility, and reuse in HEP and beyond. We investigated requirements for cyberinfrastructure design and found that supportive tools need to incentivize contributions. Based on those findings, we further investigated the application of motivational design tools in the context of reproducible science. In particular, we focused on the application of gamification in the science context. We acknowledge that gamification is a powerful design tool that has proven to create motivation and engaging interaction with tools and practices in a wide variety of applications (see Section \ref{section:gamification_background}). But, we argue that gamification in the science context has not only received less research attention. It is likely subject to different design requirements that necessitate dedicated requirements research.

Gamification in the workplace has been subject of extensive research studies \cite{Swacha2016, Oprescu2014}, indicating that gamification mechanisms increase the motivation of employees to collaborate with colleagues, to document project-related knowledge \cite{Schacht2014}, and to engage more enterprise users in the pursuit of business goals \cite{Dale2014}. However, little focus has been placed on scientific work environments, even though questions on the role of gamification in research have been raised \cite{Deterding:2015:GRS:2702613.2702646}. In particular, we are missing systematic design processes for tools employed in the scientific workplace. So far, gamified interaction in science mostly focused on supporting the learning process of students \cite{Ibanez2014}, and designing engaging experiences in citizen science where the general public is motivated to contribute to scientific knowledge through micro tasks \cite{Eveleigh2013,Bowser2014}.

Scientists often underlie a less stringent organizational hierarchy than corporate employees. Merali \cite{Merali2010} reported on practices within the LHC experiments. She highlighted that those are different from other complex organizations, typically encountered in industry or government. Merali referred to Karin Knorr Cetina, a sociologist who studied the collaborations at CERN for almost 30 years. Knorr Cetina agreed that ``the industrial model cannot work.'' Top-down decision making is given up in favor of numerous highly specialized teams. As Merali's work shows, the common practice of cooperation and inclusion of various different institutes plays a role in the employment framework of scientists. Dozens and hundreds of institutes are involved within the various LHC collaborations\footnote{LHC Research Programme: Institutes. https://greybook.cern.ch/greybook/researchProgram/detail?id=LHC}. A spokesperson of one of the two biggest collaborations noted that ``in industry, if people don't agree with you and refuse to carry out their tasks, they can be fired, but the same is not true in the LHC collaborations.'' That is because ``physicists are often employed by universities, not by us.'' This absence of a strong and enforcing command structure also establishes a special need for motivational design.

Studying findings from gamification research in corporate environments, we find that suggested approaches might not directly apply to scientific workplaces. For example, Swacha and Muszy{\'{n}}ska \cite{Swacha2016} proposed several patterns for gamification of work, one of which they call \textit{Sense of progress}. They stated that when an ``employee sees no direct result of his/her actions (and considers) them futile and fruitless'', we have to make him/her ``aware that every action he/she performs is a step in progress.'' While this is certainly as true for researchers as for any other professionals, the proposed solutions are difficult to map to researchers' workflows that are characterized by novelty and creativity. The authors proposed to reward ``points even for simple routine tasks, define point levels marking stages of progress (and to) visualize progress bars showing the distance to the next level.'' Of course, we have similar mechanisms in academia: students have to attend lectures and pass exams to get credit points. In HEP, researchers have to earn points, for example for their community work within the big research collaborations. Yet, such simple extrinsic rewards cannot evaluate the process of scientific knowledge creation as a whole. Scoring a highscore or advancing to a certain level does not earn a PhD. Scientific progress includes demonstrating failure, postulating hypotheses, and preparing research data for reuse in their community. Those are advancements in science that are hard to quantify by an algorithm.

It becomes increasingly evident that gamification is much more than the application of point-based rewards, leaderboards and badges, but instead profits from a holistic design process that appeals to the intrinsic motivation of the players \cite{Brito2015, Dale2014}. If we think about a design model for gamification in science, we must keep in mind that meaningless game elements not only lack motivational benefits, but rather alienate users \cite{Nicholson2015}. This certainly applies as well to scientists who are trained to think critically. Meaningful gamification design requires a deep understanding of the users, their contexts, practices and needs \cite{kumar2013gamification, werbach2012win}. Proposed gamification design models reflect the need for extensive user --- or player --- research. For example, in their six step design process, Werbach and Hunter \cite{werbach2012win} devoted one step to: \textit{Describe your players}. Kumar and Herger \cite{kumar2013gamification} described the Player Centered Design model that requires designers to \textit{Know your player} and \textit{Understand human motivation}. Design processes for scientific tools might particularly profit from reflecting scientists' practices and motivations within this layer.

In this chapter, we present our research on requirements of gamification in highly skilled science. This approach is in concert with recent calls to investigate the uses and effects of gamification beyond classic application areas. In \textit{The Maturing of Gamification Research}, published by Nacke and Deterding \cite{Nacke2017}, the authors highlighted that gamification's early research focused on few contexts like education. As not all contexts and desired behaviors are equally suited for gamification, ``extending the use of gamification beyond these contexts, and systematically studying the moderating effects of different individual and situational contexts is thus very much in need today.'' The authors argued that ``we are just at the beginning of understanding what gamification design elements and methods best map onto what application domains.''

In \textit{Rethinking Gamification} \cite{fuchs2014rethinking}, Deterding stressed that ``motivational design should revolve around designing whole systems for motivational affordances, not adding elements with presumed-determined motivational effects.'' Recent work from Orji, Tondello and Nacke \cite{orji2018personalizing} represents a good example of context-specific gamification research, as they mapped the impact of persuasive strategies on gamification user types for persuasive gameful health systems. Basing their study on storyboards, they illustrated how gamification research profits from novel methods. This approach also inspired our prototype-centered study design, mapping moderating effects of game design elements in science.

In this chapter, we first reflect on work related to gamification in science. We stress that while most research focused on citizen science, the success of Open Science Badges motivates research on gamification in scientific environments. Next, we present our study design. In particular, we illustrate the user-centered design and evaluation process of two contrasting gamified preservation service prototypes. We present findings from our evaluation with CERN physicists and present design implications for gamification in science. Finally, we discuss how our findings map to the success of Open Science Badges.

\begin{tcolorbox}[title = The introduction to this chapter is based on the following publication.]
Sebastian Feger, Sünje Dallmeier-Tiessen, Paweł Woźniak, and Albrecht Schmidt. 2018. Just Not The Usual Workplace: Meaningful Gamification in Science. Mensch und Computer 2018 -- Workshopband (2018).
\end{tcolorbox}

\begin{tcolorbox}[title = This chapter is based on the following publication.]
Sebastian S. Feger, Sünje Dallmeier-Tiessen, Paweł W. Woźniak,
and Albrecht Schmidt. 2019. Gamification in Science: A Study of Requirements in the Context of Reproducible Research. In CHI Conference on Human Factors in Computing Systems Proceedings (CHI 2019), May 4–9, 2019, Glasgow, Scotland Uk. ACM, New York, NY, USA, 14 pages.
\newline\url{https://doi.org/10.1145/3290605.3300690}

\color{darkgray}
------------------------------------------------------------------------------------------------------\newline
Several of the study's resources are openly available as supplementary material in the ACM Digital Library.
\end{tcolorbox}

\section{Related Work}

Studying gamification in a research setting represents an opportunity to extend our knowledge of the applicability and constraints of gamification beyond traditional contexts. So far, gamification in science focused on designing engaging experiences in citizen science, motivating the general public to contribute to scientific knowledge through micro tasks \cite{Eveleigh2013,Bowser2014}. The CHI workshop summary from Deterding et al. \cite{Deterding:2015:GRS:2702613.2702646} raised questions on the role of gamification in research. Still, they focused on citizen science, as they tried to encourage users to provide self-tracking data and to participate in research activities.

The reproducibility crisis represents a strong example of a scientific challenge that motivates the study of needs and constraints of gamification in research settings. Documenting and sharing research data and resources are key requirements of reproducible research \cite{Bechhofer2013, Wilkinson2016}. But, the efforts required to prepare, document and share experimental data ~\cite{Borgman:1297241} are often not matched by the perceived gain. Persuasive gamification design might provide motivation for scientists to conduct reproducible research. Kidwell et al. \cite{Kidwell2016} studied adoption of OSB in the \textit{Psychological Science} journal. Authors who made their data and / or materials openly available received corresponding badges, displayed on top of their paper. In their quantitative study, they found that badges increased reported and actual sharing rates significantly, both in comparison to previous sharing behaviors in the same journal and other journals in the same discipline. Yet, despite this indication that game elements can significantly impact open sharing practices, empirical studies on the moderating effects of gamification in science were still missing.

\section{Study Design}

\begin{figure}
  \centering
  \includegraphics[width=0.8\columnwidth]{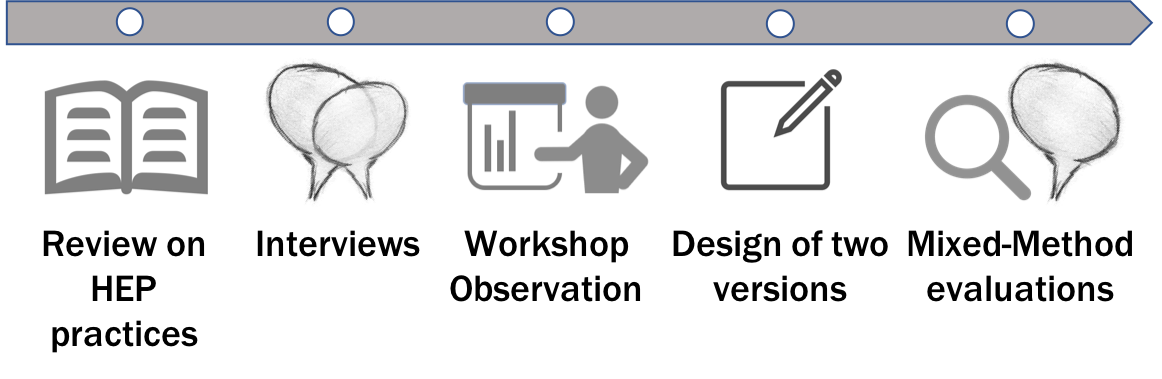}
  \caption{Schematic representation of the design and evaluation process.}~\label{fig:gamification_process}
\end{figure}

Figure \ref{fig:gamification_process} provides a schematic representation of the human-centered design process. In this section, we detail our study design, with particular regards to the prototype development and evaluation.

\subsection{Gamification Designs}

In line with existing gamification design models that emphasize studying needs, practices and motivations of target users, we set out to learn about HEP data analysts. We started by reviewing published studies on field practices, field differences and scholarly communication in HEP. Next, we conducted semi-structured interviews with 12 HEP data analysts to learn about their workflow practices and perceptions of CAP (see Chapter \ref{ch:requirements}). Finally, we observed a one-day workshop that was attended by representatives of the four major LHC collaborations. The service developers presented the latest features and collaboration representatives discussed their needs and wishes for the future service development. We gained full access to the workshop notes and presentations.

To stimulate feedback, we decided to create two prototypes that are based on our researcher- and service-centered insights. Following our initial expectation that gamification in a professional scientific context is most likely to profit from a serious, informative, and rule-based design language, we created the Rational-Informative Design (RID). The RID was designed to make little use of most common game elements like points and leaderboards. Instead, it uses elements of `Social networks', `Social discovery', `Signposting', and `Challenges', as suggested by Tondello et al. \cite{Tondello2017}. This enables an exploration of gameful design elements in the HEP context. Yet, as scientists are already subjected to a high degree of competition, we also created a contrasting Simple Game Elements Design (SGED) version that focuses on point-based rewards and competitive elements. The basic UI design rules (color schemes, arrangements, etc.) are the same for both versions and are inspired by the actual service design. We built interactive prototypes with a high level of detail using the prototyping tool Balsamiq. This approach is also motivated by recent, novel research methods, mapping persuasive strategies to gamification user types based on storyboards \cite{orji2018personalizing}. 

\subsection{Evaluation}

We conducted mixed-method, within-subjects evaluations with 10 HEP data analysts. As indicated in Table \ref{tab:table_participants_gamification}, researchers from CMS and LHCb were primarily recruited, as their CAP submission forms are most complex and time demanding. We particularly considered recruiting participants with a very diverse set of professional experiences and roles. Participants who had finished their PhD three or more years ago were considered \textit{senior.} We further identified current or previous \textit{conveners}, physicists who have a particular project management role within a collaboration. The 10 participants included 3 female researchers, reflecting the employment structure of research physicists at CERN \cite{CERN-HR-STAFF-STAT-2017}. The analysts were 27 to 53 years old (Avg = 36, SD = 8.2). No remuneration was provided, as all evaluation sessions were conducted during normal working hours and all participants were employed by CERN or an associated institute.

\begin{table}
  \centering
  \begin{tabular}{l c c c r}
    {\small\textit{Ref}}
    & {\small \textit{Affiliation}}
    & {\small \textit{Gender}}
    & {\small \textit{Experience}}
    & {\small \textit{Order}}\\
    \midrule
    P1 & CMS & Male & Senior & SGED-RID \\ 
    P2 & LHCb & Male & Postdoc & RID-SGED \\ 
    P3 & CMS & Female & Senior & SGED-RID \\
    P4 & LHCb & Female & PhD student & RID-SGED \\
    P5 & CMS & Female & Senior & SGED-RID \\
    P6 & LHCb & Male & PhD student & RID-SGED \\
    P7 & CMS & Male & Senior, Convener & SGED-RID \\
    P8 & ATLAS & Male & Senior, Professor & RID-SGED \\
    P9 & CMS & Male & Senior & SGED-RID \\
    P10 & CMS & Male & Postdoc & RID-SGED \\
    \bottomrule
  \end{tabular}
  \caption{Overview of the study participants indicating the order of prototype use.}~\label{tab:table_participants_gamification}
\end{table}

\subsubsection{Structure}

First, participants were introduced to CAP. They were shown the analysis submission form of their corresponding collaboration, in order to get familiar with the context. Afterwards, half of the participants started exploring the RID prototype, the other half the SGED one. They started with the dashboard and explored the various views on their own. We prepared a few questions for every principal view that aimed to stimulate feedback. Following the exploration, we asked the physicists to respond to a 7-point Likert scale questionnaire, structured as follows:

\begin{itemize}
    \item We used two subscales of an abbreviated \ac{IMI}. We considered assessing the perceived \textbf{Value / Usefulness} (5 items) to be of key importance for gamification in science, as well as \textbf{Interest / Enjoyment} (4 items). Enjoyment has also been used to characterize user preferences of game design elements by Tondello et al. \cite{Tondello2017}. The interest / enjoyment subscale assesses intrinsic motivation per se, while task meaningfulness appeals to the innate need for autonomy \cite{sailer2017gamification}.
    \item We further asked to rate a statement that targets the \textbf{suitability} of the design: \textit{The system is NOT suitable for a research preservation service}. Finally, \textit{The system would influence me to document my analyses}, targets the \textbf{persuasiveness} of the design, also core to the study of Orji et al. \cite{orji2018personalizing}.
\end{itemize}

Afterwards, participants explored the other prototype and the process was repeated. In the following, we asked analysts to discuss the two versions. Finally, analysts were invited to fill in a short questionnaire with six items, assessing the validity of our underlying design assumptions.

\subsubsection{Data Analysis}
We collected 5.2 hours of recording during the evaluation sessions. All recordings were transcribed non-verbatim and Atlas.ti data analysis software was used to organize, analyze and code the transcriptions. We performed Thematic Analysis \cite{Blandford:2222613} to identify emerging themes. Two authors independently performed open coding of the first two transcriptions. They discussed and merged their codes. The resulting code tree was used in coding the remaining transcriptions. A hundred and three codes and 10 code groups resulted from this combined effort. The code tree was used as reference in coding the remaining transcriptions. In total, 124 codes were created through 287 quotations (1 -- n associated codes). Code groups were adapted and merged, resulting in 9 code groups. We constructed the four high-level themes based on those code groups. For example, the theme `Scientific practice' is based on `Speaking scientists' language' and `Known mechanisms'.

%
%

\section{Design}

In this section, we reflect on findings of our researcher-centered design process. We present the two gamified research preservation service prototypes.

\subsection{Researcher-Centered Design}

We first detail the insights gathered from our research activities, studying practices and motivations of HEP data analysts, as well as perceptions towards research preservation. Based on those, we present target behaviors for the gamification design.

\subsubsection{Literature Review: HEP Field Practices}

Various studies report on the role researchers play within the huge collaborations. In her article \textit{The Large Human Collider} \cite{Merali2010}, Merali documented the high level of identification with the detector. She devoted an entire section to researchers sacrificing their identity to their respective LHC collaboration. Merali referred to Karin Knorr Cetina, a sociologist who studied CERN's collaborations for almost three decades. Knorr Cetina confirmed that CERN ``functions as a commune, where particle physicists gladly leave their homes and give up their individuality to work for the greater whole.'' In her earlier work, she even described ``the erasure of the individual epistemic subject in HEP experiments.'' \cite{cetina2009epistemic} 

\subsubsection{Interview: Practices, Needs and Perceptions}

In our first interviews study (Chapter \ref{ch:requirements}), participants reported commonly sharing their analysis resources (codes, datasets and configurations) with their colleagues. Yet, we realized that despite the very early invention and adoption of collaborative technologies, the information and communication architecture is shaped by traditional forms of communication. Searching for resources is hindered by challenges imposed by the databases or unstructured presentation of materials. Analysts are in high demand for information, but rely heavily on e-mail communication, trying to satisfy their information and resource needs through personal networks.

The communication architecture results in a high level of uncertainty. Participants highlighted that reliably informing all dependent analysts about issues in a common analysis resource is difficult, if not impossible. E-mail communication in collaborations with several thousand members and highly distributed institutes is not sufficient. The interviews revealed the value of collaboration in HEP, as well as challenges of engaging in collaborative behavior. Analysts cannot know all relevant colleagues in their highly distributed collaborations. We envisioned rich analysis documentation as a strategy to increase the visibility of researchers, thereby improving chances to engage in useful collaboration.

Finally, analysts reinforced the value of centralized and automated workflow execution. They highlighted the efforts of setting up their own environments and acquiring computing time. Some of the analysts described to run their analyses on their institute's servers, as there is less competition for computing resources. However, doing so hinders sharing and collaboration with researchers outside their institute and requires substantial efforts when changing institutes. Thus, automated analysis re-execution on a centralized preservation service represents a strong incentive to keep documented analyses up-to-date.

\subsubsection{Workshop Observation}

The service developers presented for the first time the full-cycle execution from CAP to REANA, a partner project that aims to re-execute analysis workflows on a centralized computing framework. Matching our interview analysis, this functionality was acknowledged very positively by the attending researchers. It confirmed our initial thoughts of promoting execution of appropriately documented and structured analyses on REANA. As we learned in the workshop, there is a second dimension to analysis structure and automation which relates to the use of workflow management systems. Making use of such tools fosters scalable, machine-readable, and executable analysis designs, representing an important step towards automated re-execution. 

\subsubsection{Target Behaviors (TB)}
Based on the findings of our research activities, we developed four target behaviors that we want to encourage through our gamification designs:

\begin{itemize}
    \item \textbf{TB\#1 -- Document and provide rich descriptions. } Primarily, we want to encourage data analysts to document and preserve their work thoroughly.
    \item \textbf{TB\#2 -- Communicate relevant information. } Analysts who discover issues related to collaboration-shared analysis fragments should be encouraged to share their findings. In turn, we expect to create awareness that documenting analysis resources can be a strategy to cope with uncertainty.
    \item \textbf{TB\#3 -- Use automation capabilities. } Structuring and documenting analyses for central re-execution represents an opportunity to speed up analysis workflows. We expect physicists who follow this target behavior to experience benefits through automated and more efficient workflows that provide motivation to keep the documentation up-to-date.
    \item \textbf{TB\#4 -- Embrace collaborative opportunities. } A central research preservation service provides opportunities for analysts to increase their visibility and the visibility of their work. This likely leads to valuable collaboration.
\end{itemize}

\subsection{Prototypes}

We designed two interactive prototypes of a gamified research preservation service. They are based on two principal views: dashboard and analysis page. Those are inspired by views that already exist in CAP. In addition, we created a profile page to list activities and achievements. In this section, we depict the dashboard and analysis pages of both versions. The complete, interactive, high-resolution Balsamiq archives are provided as supplementary material.

\subsubsection{Simple Game Elements Design (SGED)}

\begin{figure}
  \centering
  \includegraphics[width=1.0\columnwidth]{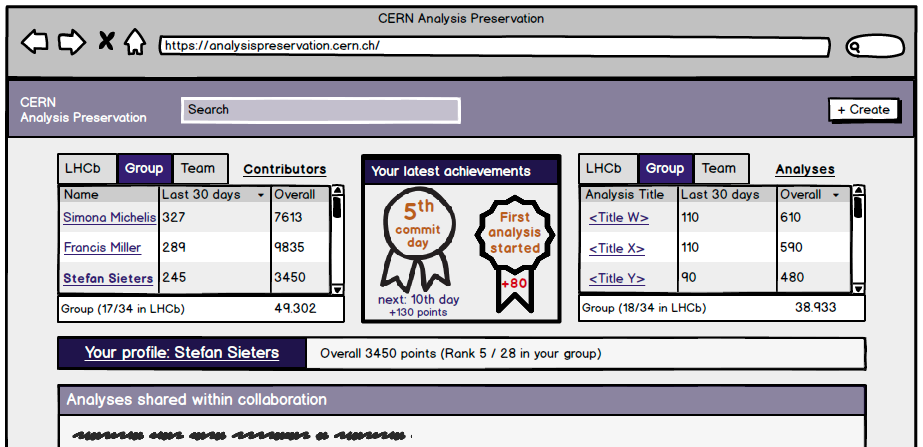}
  \caption{Dashboard of the SGED prototype.}~\label{fig:sged_dashboard}
\end{figure}

\begin{figure}
  \centering
  \includegraphics[width=0.8\columnwidth]{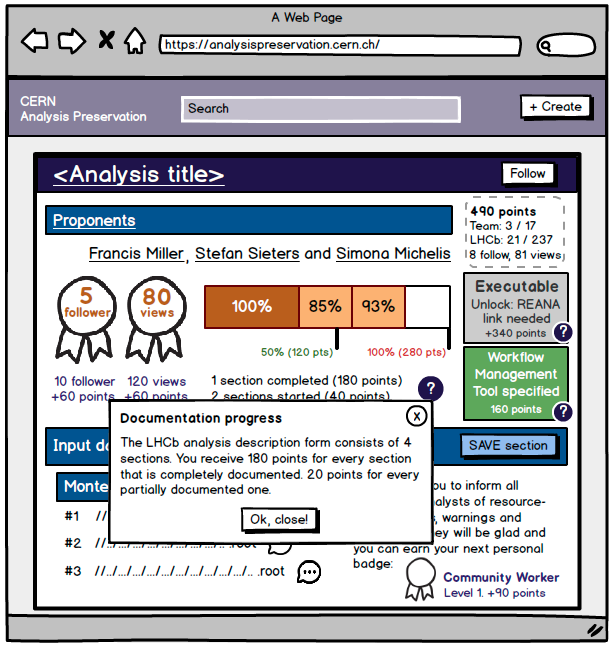}
  \caption{Analysis page of the SGED prototype.}~\label{fig:sged_analysis}
\end{figure}

\begin{figure}
  \centering
  \includegraphics[width=0.6\columnwidth]{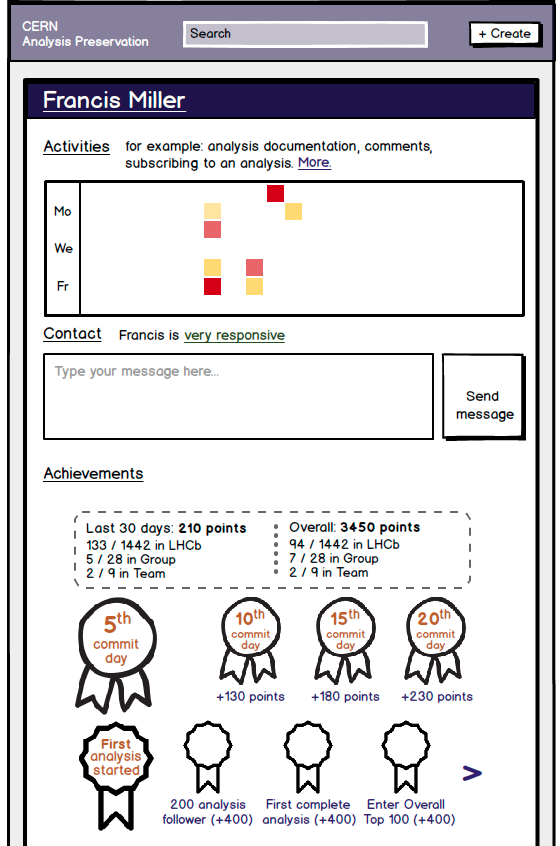}
  \caption{SGED prototype profile page.}~\label{fig:sged_profile}
\end{figure}

As shown in Figure \ref{fig:sged_dashboard}, the dashboard design of the SGED focuses on achievements and competition. Latest rewards are displayed in the center, together with awarded points and upcoming challenges. As indicated by the shown badges, they primarily attempt to stimulate documentation (TB\#1). Two leaderboards depict the performance of contributors and of analyses. In order to foster collaboration (TB\#4), leaderboards can be set to show the entire LHC collaboration or single groups or teams. Listed contributors link to corresponding profile pages (see Figure \ref{fig:sged_profile}) and analysis titles to analysis pages. 

The analysis page, depicted in Figure \ref{fig:sged_analysis}, educates and stimulates researchers towards using central computing resources for automated (re-)execution of analyses (TB\#3). Badges are awarded both for establishing a link with the REANA project and the integration of a workflow management system. Analysis points are awarded to the analysis, as well as to all proponents. Having learned about the importance of visibility and collaboration (TB\#4), we added rewards and challenges that target analysis impact (views / followers). The documentation progress bar gives a visible overview of the completeness of the analysis and incentivizes further contributions (TB\#1). Finally, the importance of sending relevant resource-related information is highlighted and compliance incentivized (TB\#2).

\subsubsection{Rational-Informative Design (RID)}

\begin{figure}
  \centering
  \includegraphics[width=1.0\columnwidth]{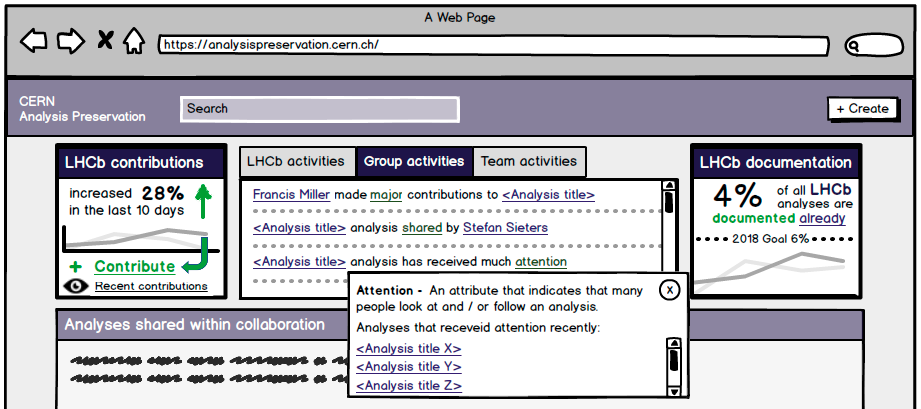}
  \caption{Dashboard of the RID prototype.}~\label{fig:rid_dashboard}
\end{figure}

\begin{figure}
  \centering
  \includegraphics[width=0.7\columnwidth]{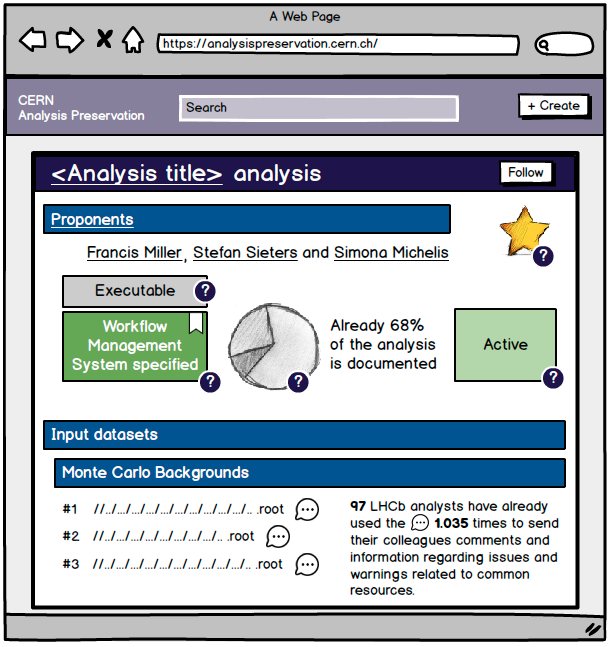}
  \caption{Analysis page of the RID prototype.}~\label{fig:rid_analysis}
\end{figure}

The dashboard in this version displays an activity stream. As depicted in Figure \ref{fig:rid_dashboard}, researchers can again control the desired granularity (TB\#4). Entries in the stream refer to a researcher and / or analysis, and a \textit{rule} that needs to be fulfilled (TB\#1). When selected, we display further information concerning the rule, as well as analyses that comply with it (TB\#4). Having learned about the particularly strong identification of HEP researchers with their collaboration, we decided to depict the collaboration's preservation status and a community goal. Thereby, we expect to trigger researchers' sense of identification to stimulate contributions (TB\#1) that impact a common good of the collaboration.

The analysis page, shown in Figure \ref{fig:rid_analysis}, is designed to report on statuses and does not make use of point-based rewards. Badges are used to educate and indicate use of automated, centralized analysis (re-)execution and workflow tools (TB\#3). A pie chart indicates the number of blocks that are fully, partially or not at all documented. Depending on the level of documentation, we show encouraging messages (TB\#1). Analyses that continue to receive contributions (TB\#1) are indicated as \textit{active}. 
Based on our previous research, we expect this to be a meaningful attribution, as active analyses are more likely to be of interest to other collaboration members (TB\#4). A star marks popular analyses that have many followers and views. Finally, we show information related to the usage of the resource-related communication features (TB\#2). Detailing the number of analysts who have used the feature, we aim to stimulate the identification of analysts with their collaboration and provide an opportunity to directly impact those collaboration-related statistics.

%
%
\section{Results}

\begin{figure}
  \centering
  \includegraphics[width=0.9\columnwidth]{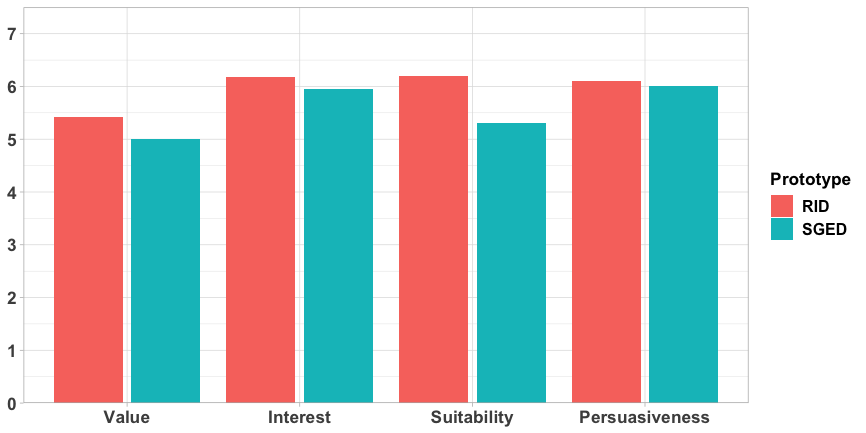}
  \caption{Value, Enjoyment, Suitability and Persuasiveness of the prototypes.}~\label{fig:rid_sged_results}
\end{figure}

The results of the IMI scales (\textit{Value / Usefulness} and \textit{Interest / Enjoyment}) and the statements regarding \textit{Suitability} and \textit{Persuasiveness} are shown in Figure \ref{fig:rid_sged_results}. The results are as follows [Mean (SD)]:

\begin{itemize}
    \item \textit{RID}: Value 5.42 (0.95), Interest 6.18 (0.75), Suitability 6.2 (0.75), Persuasiveness 6.1 (1.04).
    \item \textit{SGED}: Value 5.0 (0.45), Interest 5.95 (1.06), Suitability 5.3 (1.55), Persuasiveness 6.0 (1.26).
\end{itemize}
 
As depicted, the RID consistently scores better, although RID and SGED stimulate almost identical enjoyment / interest and persuasion. The most pronounced difference between the two designs concerns the \textit{suitability} in research preservation. While the RID scores as well as in the previous subscales, the SGED is considered less suitable. Ordering effects with more than a one-point difference were observed only for \textit{SGED suitability} (SGED first: 6.4, SGED second: 4.2) and \textit{SGED persuasiveness} (SGED first: 6.8; SGED second: 5.2). This suggests that participants more critically reflected on controversial elements in the SGED after exploring the overall suitable RID. We focus on explaining this effect through our extensive qualitative findings.

Overall, participants confirmed our underlying design assumptions, rating the following statements on a 5-point likert scale (Strongly disagree : -2\textbf{;} Strongly agree: 2):

\begin{itemize}
    \item I am willing to document and share my analyses as a contribution to the research quality of my collaboration: \textbf{1.8}
    \item I do \textbf{NOT} expect my career opportunities to profit from an increased visibility within my collaboration \textbf{( R )}: \textbf{-1.4}
    \item My analysis work would profit from access to the sum of well-documented <collaboration> analyses: \textbf{1.9}
    \item I would hope to engage in more collaboration, if I managed to increase my visibility within <collaboration>: \textbf{1.3}
    \item I think that I would want to frequently check the <collaboration> or group activities on the service dashboard: \textbf{0.9}
\end{itemize}

\section{Findings}

Four themes emerged from our qualitative data analysis: \textsc{Contribution}, \textsc{Metrics}, \textsc{Applications}, and \textsc{Scientific Practice}. We present each theme and our understanding of opportunities and constraints for gamification in science.

\subsection{Contribution}

Most participants (P1, P2, P5, P6, P8, P10) referred to improved career opportunities, resulting from game elements that reflect their contributions. To this end, a variety of mechanisms --- from rankings to badges --- seem valuable, as long as they can increase visibility within the huge collaborations:

\begin{quote} 
We are so many people in the collaborations, of course. Especially if we want to continue in our field, we have to get some visibility somehow. (P6) \newline
And if it's known that you were one of the first users of a particular technique, this can really help get your name out there. (P2)
\end{quote}

In this context, P1, P2, P6, and P10 explicitly mentioned their desire to refer to service achievements and scores in job applications. But, the resulting competition also triggers concerns. In particular P2, P4, P6, P7, and P9 warned about unhealthy competition: \textit{``Imagine that my two PhD students had rank number 1 and rank number 2 and they compete with each other. I would find it a potential for a toxic situation.''} (P7)

\subsubsection{Reflecting Contribution and Quality}

Given the potential impact of scores and achievements, all analysts discussed concerns related to the accurate mapping of research contributions in the gamification layer. P1 -- P4 and P9 pointed to different roles within an analysis team. Concerning the preservation on the service, P3 noted that \textit{``it may be, for example, there is one person in a group who is taking care of this.''} Thus, mechanisms are needed to reflect contributions:

\begin{quote} 
Maybe you can split the points for several people. Because if you give a big amount of points and only one person is allowed to push the button, this probably is a bit unfair. [...] You should find the means of splitting the awards or something. (P1)
\end{quote}
 
One physicist, P3, further worried that difficult tasks with low visibility might not be fully recognized, referring to the example of someone who struggles to solve an issue in programming code. P4 added that metrics need to consider analysis complexity, because \textit{``if I preserve my shitty analysis 100 percent and someone else who actually was published in Nature preserves 60 percent, that does not really tell that my analysis is better than the other analysis.''}

\subsubsection{Team Rather Than Individual Contributions}

Given the challenges that result from recognizing contributions, researchers (P1, P3, P7, P9, P10) strongly advocated promotion of team contributions, rather than personal ones. In fact, our analysis suggests that while competition on an individual level is controversial, comparison between teams and analyses is generally accepted.

\begin{quote} 
Any comparison between analyses, or everything you say about analyses I think it's very good. [...] I think people like to play that. But when you go inside one analysis things might get complicated. (P9) \newline
To boast that we do gracious things as a team. That would look less silly if it's at a team level. Rather than the individual that are gaining one more price. (P7)
\end{quote}

\subsection{Metrics}

A major theme that emerged from our data analysis relates to the selection of meaningful metrics in gamification design. Analysts described four dimensions: Frequency, accessibility, discouragement, and social implications.

\subsubsection{Frequency}

A core dimension that has to be considered in the design of game elements is \textit{frequency} of contributions and activities. Most analysts referred to an expected unbalanced distribution of activities on the research preservation service. In particular P4 stressed that \textit{``it's just I feel there is this peak activity. People preserve in 3 days and then they stop.''} Our data analysis revealed that the impact of frequency needs to be considered in various design elements. For example, both P2 and P4 commented on the SGED ribbon \textit{5th commit day}:

\begin{quote} 
So, I feel like there is a peak activity... that's why I feel that this 5th commit day is not so applicable. (P4) \newline
Fifth means just like the fifth day in general, I think that's fine, but I would not want to encourage like continuous streaks of days, because I really don't like encouraging people to work outside their working hours. [...] And I also... At least when I work, I sort of dip in and out of projects quite frequently. So, I wouldn't want to have like any pressure to have to work on something for continuous block. (P2)
\end{quote}

P8 further depicted the effect of elements that are not frequently updated: \textit{``Then there are other things that stays, like yellow or red for you know a year. Then everyone just kind of stops paying attention. It turns to be more depressing than anything.''}

\subsubsection{Accessibility}

The previous statement also highlights requirements connected to the accessibility of goals and achievements. Although all participants acknowledged the analysis badges \textit{Executable} and \textit{Workflow Management Tool} to be important and valuable, analysts warned that the goals might be too high. P4 proposed to add more levels to provide intermediate and accessible goals:

\begin{quote} 
I think this maybe a too high goal to strive for. Because, I said that the biggest obstacle is probably that people know that it is going to take a lot of time. So, if you set them a very high standard [...] it's like immediately judging them. It's not executable! So, I'm thinking maybe there should be more statuses. (P4)
\end{quote}

Both P7 and P8 referred to \textit{binary} mechanisms and highlighted that they are not likely to map reality. Concerning the analysis documentation progress bar (SGED) and pie chart (RID), P7 stated that \textit{``things are never binary. There is always partial completion. And one can think also about more than three categories.''}

\subsubsection{Discouragement}

Participants highlighted adverse motivational effects resulting from discouraging statistics. Those are expected to be most pronounced in the early stages of the service operation where they expect few activities and few preserved analyses. Looking at the low documentation statistic of their collaboration on the RID prototype, P1 and P4 expressed their disappointment. P1, P4, and P7 proposed to only show encouraging and positive information. For example:

\begin{quote}
You want the star. [...] I guess it's an element that appears if you over-perform and does not appear otherwise." (P7) \newline
It's good for the preservation coordinator to show momentum. Of course, you would only show it if it's actually full. (P1)
\end{quote}

\subsubsection{Social implications}

Besides increasing visibility and improving career prospects, metrics also have social implications. P2, P4, P5, and P9 commented on perceptions of the activity stream and collaboration documentation overview (RID). Looking at low numbers, P2 stated: \textit{``If I saw that, I'd be like: Maybe I can help get that number up.''} The analysts described their close identification with their collaboration. P4 even introduced the term \textit{tribalism}, to better illustrate the strong group feeling. Shown metrics can thus provoke social pressure:

\begin{quote} 
I think is cool, is to have the total goal for 2018, for instance. Like you really feel that you are contributing to the whole project, right. (P5) \newline
[...] there are 20 people in this group. And then there is like higher probability that someone is going to make some activity. And then you are again going to feel like: oh my god, my peers are preserving. I should probably do the same thing. (P4)
\end{quote}

\subsection{Applications}

Our data analysis revealed that the gamification layer not only provides incentives and benefits to individual researchers. Instead, it can play an important role in two application areas: \textit{Education / Training} and \textit{Administration}.

\subsubsection{Education / Training}

Most analysts (P2, P3, P4, P7, P8, P9, P10) indicated that game elements can educate researchers about best practices. For example, P2 highlighted \textit{``that an analysis that's like well documented, that's very reproducible, and does all the best practices, does probably end up with more views.''} Thus, the researcher would like to sort by analysis views, to take inspiration from reproducible analyses. P2, P3, P4, and P8 highlighted that those mechanisms can be most beneficial at the start of a new project and as learning material for new students. P8 even sees opportunities to change current practice:

\begin{quote} 
There is people who are doing things in an old way and then there is a new way of doing it where things are more reproducible etc. And what I think what we largely want to do is get signals to people of which ones are like doing it best practice way and which ones aren't.
\end{quote}

In this context, P4 and P7 cautioned about potential issues. P4 worried that the rank of analyses might not necessarily reflect suitability for teaching. Less complex analyses could score high, while more sophisticated ones might not. Yet, innovative, more complex analyses might set a better example.

Concerning the connection between point-based awards and elements that simulate best practices (SGED), P7 cautioned about patronizing researchers. The convener also highlighted that generally suitable practices might not always apply in novelty-based research. Seeing the RID analysis page with the same workflow elements later, the convener judged the mechanisms to be suitable because analysts are not forced to comply with a certain practice.

\subsubsection{Administration}

Senior researchers (P1, P3, P7, P10) described how the transparency that is created by the gamification layer can be used in administrative tasks. The analysts indicated that the status transparency allows to more easily detect barriers. They described detecting issues based on percentage-based documentation overview on the analysis level. In addition, P7 saw it as an opportunity to assess performance on a higher level:

\begin{quote}
And maybe I can navigate in the front part. [...] To check who is over-performing or under-performing. To see what are the weak links and where to act. So, that's definitely the manager view and this sounds like the right thing to do in fact.
\end{quote}

In addition, P2 referred to the role of transparency and achievement in formal reviews. The analyst indicated that particularly the \textit{workflow management tool} and \textit{executable} badges would influence his perceived trust in an analysis.

\subsection{Scientific practice}

As our data analysis shows, the impact and requirements for game elements and mechanisms are manifold. Yet, a common denominator is the use of well-known scientific mechanisms.

\subsubsection{Speaking scientists' language}

Most participants (P3, P4, P6, P7, P8, P9) explicitly referred to the impact of design language on perceptions in a scientific environment. They highlighted that design needs to adapt to scientific practice in order to be well-perceived: 

\begin{quote} 
It's probably for me -- as a scientist... I'm disturbed, because it's sort of... I may be happy with gamification, but I don't want it to look like it. (P3) \newline
The central part (RID activity stream) is professional. While the previous (SGED leaderboards) looks like something to engage a certain kind of people. [...] This is really professional and it's... Maybe it's less fun, but looks more useful. (P7)
\end{quote}

There is little controversy about game elements that use scientific language. While P3 considered community goals in collaboration statistics (SGED) to represent a \textit{``certain balance between the pure game type gamification elements and something which is sort of easily acceptable in a scientific domain''}, P4 argued that \textit{percentages} are already a strong and familiar metric for analysis completion, making points obsolete. P3 further highlighted the strength of familiar language:

\begin{quote} 
Well, this gives sort of a scientific view of... Probably is more attractive to scientists because it gives you graphs. It's the language we speak, rather than points and awards and that kind of things. Which is something which is not our language in that sense. But it still gives you a scale. (P3)
\end{quote}

\subsubsection{Known mechanisms}

Besides familiar language, almost all analysts pointed to the suitability of known mechanisms. Concerning the analysis star in the RID, P3 and P4 described parallels to GitHub and GitLab mechanisms, commonly used code repositories. P1 compared achievement overviews and personal goals to mechanisms on Stackoverflow, a popular online developer community, and indicated that he would appreciate similar mechanisms in this context. P5 illustrated how points on the preservation service could potentially map to formally required collaboration service points. The researcher described that analysts need to fulfill certain tasks as part of their obligations to the collaboration. Yet, \textit{``there are not so many opportunities to get the service points. And they are taken. So, somehow if you are able to arrive to some kind of agreement with the collaboration, for example CMS, and you can say like: I am going to change this many points in the analysis preservation page. I am going to exchange them by one service point.''} Finally, P8 highlighted the value of design elements that are more than just status elements, but rather provide a meaningful entry point:

\begin{quote}
And also when you do that it gives you a little badge. Which says \textit{launch binder}. Which in some sense is more like a button that looks like it does something. It's not just like collecting stars. It's an actionable something, you know. It also looks similar in terms of being you know a badge. (P8)
\end{quote}

\section{Discussion}

In the following, we discuss how our findings can be used to design engaging interactions through gamification in science. As our results suggest, a variety of game elements and mechanisms can provide value and enjoyment, while still being persuasive and suitable to the professional context. The overall low difference between RID and SGED in the quantitative assessment is not surprising, considering the work of Tondello, Mora and Nacke \cite{Tondello2017}. In their paper, they mapped 49 of the most frequently used gameful design elements, assigned to 8 groups, to gamification user types and personality traits and found that the overall difference ``is not extraordinary but still pronounced, with approximately 20\% difference between the lowest and the highest scoring groups.'' In addition, we see the overall low differences in our evaluation as evidence of the success of our extensive researcher-centered design process. Although our qualitative findings highlight constraints and requirements of individual game mechanisms, researchers appreciated the underlying target behaviors and best practices that we aimed to stimulate.

We consider the qualitative focus of our study to be a key strength. It allowed us to better understand the impact, opportunities, and requirements of individual game mechanisms. In the following discussion, we see how prevalent challenges in scientific work need to be reflected in design requirements for gamification in science. We postulate that we have to consider \textit{controversial elements} very carefully in this competitive environment. We conclude with \textit{design recommendations} and a note on how they relate to the success of Open Science Badges \cite{Kidwell2016}.

\subsection{Reflect the Scientific Environment and Contribution}

Scientists were particularly concerned about the reflection of research quality and personal contribution on the gamification layer. This means that designers \textbf{need to provide mechanisms that allow to distribute awards and visibility based on individuals' contributions.} While this is a core requirement that applies to all game elements, it applies in particular to point-based rewards and rankings. In addition, it is important to \textbf{enable promotion of work based on quality, impact, and purpose instead of relying solely on general and static service mechanisms.} This is particularly important as ranking and promotion of work has significant implications on education and training. Promoting work that does not fit these purposes risks providing misguided references for researchers that aim to learn about techniques or best practices. Thus, administrative or community mechanisms need to be created that allow to adapt ranking and visibility of work depending on the desired application.

Given multiple applications and uses of the gamification layer, systems should \textbf{allow to adapt presentation to desired use cases}. For example, the system could provide filter mechanisms in analysis rankings which are tailored to training efforts. As imagined also by one of the participants, presentation could be adapted based on the role of the user. Logging in to a system, senior researchers could profit from more visible performance overviews, while early-career researchers would most likely profit from relevant research activities.

Given that studies and analyses in science are often conducted over a long period of time, \textbf{it is crucial to provide accessible goals}. This applies particularly to research-related achievements. Awards that promote best practices should not only target the ultimate goal which requires months and years of effort, but intermediate steps. Whenever possible, \textbf{binary reward mechanisms should be replaced by more multifaceted structures}. This likely prevents discouragement through goals that are very hard. Instead, it might provide a \textit{sense of progress}, one of the design pattern for gamification of work by Swacha and Muszy{\'{n}}ska \cite{Swacha2016}, making an ``employee aware that every action he/she performs is a step in progress.'' Yet, doing so might become more challenging in a scientific context which is characterized by novelty and creativity.

\subsection{Find Potential Breaking Points}

Our results suggest that both prototypes are likely to be well-received. Still, our qualitative analysis pointed to a fine line between particularly valuable and suitable design elements and those with a potential for controversy. Concerns were particularly pronounced for explicit personal rankings and point-based incentives. Some participants feared to patronize researchers and to limit them in their choice. Yet, others pointed to those mechanisms as their favorite design elements, allowing them to compete and aggressively promote necessary best practices. Given our findings, we consider those mechanisms to be highly \textit{controversial} and \textbf{system designers should weight potential costs and benefits employing controversial mechanisms}. 

Our findings suggest that independent of individual design elements, \textbf{mechanisms that promote team or analysis achievements are overall accepted, while personal promotion is controversial.} This can be seen particularly in statements referring to leaderboards and activity streams. While some researchers saw personal metrics as a particularly strong opportunity to compete and to gain visibility, others worried about creating an unhealthy environment and a potentially \textit{toxic situation.} Yet, promotion of collaborative achievements is overall accepted and desired, even if they employ the same design elements like leaderboards and activity streams.

There is little controversy regarding mechanisms and language known from established scientific practice. Our results suggest that \textbf{studying and integrating community-specific language profits perceived value and suitability}. Similarly, \textbf{the use of well-known mechanisms that are employed in common research tools} seems to create acceptance.

\subsection{Create Active and Social Environment}

Our findings indicate that several dimensions need to be considered in designing game elements for a research setting. In particular, \textbf{the design of game mechanisms should consider expected \textit{frequency} of status changes and activities}. Introducing intermediate and accessible goals allows to communicate progress for elements that are otherwise expected to stay in the same condition for a long time. Related to \textit{frequency}, our findings suggest that \textbf{design needs to deal with potentially \textit{discouraging} statistics and messages.} This concerns both collaboration-wide statistics as well as elements that depict the status of individual analyses. This applies especially in the early stages of a service or analysis. In response, we can provide explanations of why statistics are less promising than most would expect, pointing for example to the fact that a service became operational only a short while ago.

Finally, we encourage to systematically \textbf{consider social factors resulting from design}. Not only did our findings show that researchers agree on mechanisms that foster cooperation and that they find individual promotion controversial. We also perceived indications of positive social pressure. Statistics and elements that depict activities are likely to create positive peer pressure, in particular if researchers have a strong identification with their research collaboration.

\subsection{Role of Open Science Badges}

A recent systematic literature review concluded that Open Science Badges are the only evidence-based incentive \cite{Rowhani-Farid2017} that promotes data sharing of research in the health and medical domain. In fact, in their quantitative study, Kidwell et al. \cite{Kidwell2016} found a significant increase in data sharing of submissions to the \textit{Psychological Science} journal that adopted those badges. Based on our findings and design implications, we discuss five aspects explaining why those mechanisms had a positive impact. First, the badges allow promoting best practices that are considered highly important in the community. We employed similar mechanisms in our study that were very well received by participants. Second, while badges are visibly placed on the paper and in the digital library of participating journals, \textit{no} adverse indication is given, highlighting that a paper has not yet received those awards. Third, promotion of rewarded papers increases their visibility, as well as the visibility of authors. This is especially true if search engines of digital libraries highlight corresponding search results. Through increased visibility, researchers can expect increasing citations and improved career prospects. Fourth, the fact that badges are assigned to papers instead of individual researchers certainly fosters acceptance, as we have previously discussed. Finally, the badges provide accessible goals, a first step towards reproducibility. ACM takes this notion even further, introducing fine-grained badges that focus on very accessible goals \cite{boisvert2016incentivizing, acmbadgesweb}.

\section{Limitations and Future Work}

The findings and design implications of our study are based on evaluations with HEP researchers. We discussed how they relate to the underlying mechanisms of Open Science Badges. Those represent a strong example of successful game elements in (reproducible) science. We envision potential for future research, mapping requirements of gamification in diverse scientific domains to gather additional requirements resulting from differing practices and needs. To foster future work, we released relevant resources, in particular the interactive Balsamiq archives, the questionnaires, questionnaire responses, and the semi-structure interview guide. Based on our findings, we envision future work to study the effects of controversial design mechanisms, in particular related to personality types of scientists. This might allow to provide individual and personality-based experiences that prevent use of alienating mechanisms for some researchers, while providing stimulating ones to others.

The rudimentary nature of the prototypes and the lack of deployment represent both a limitation to this study, as well as a necessary step in the systematic study of requirements for gamification in a highly skilled scientific environment. The prototypes allowed confronting researchers with a wide variety of very different game elements, without the risk of deploying a design that may have negative consequences. We also consider the number of participants suitable for the qualitative focus of this study. The ratings of the questionnaire concerning Value, Interest, Suitability, and Persuasiveness represent a valuable indicator for the potential of gamification in science. Yet, we recognize that the information value of the questionnaire would profit from a greater number of participants. In this context, we further envision implementation and evaluation of our design implications in production research tools. Although we have perceived the interactive prototypes to be suitable for evaluation with our participants, production systems would allow mapping researchers' behaviors and perceptions over a longer period of time.

\section{Conclusion}

This chapter presented a systematic study of perceptions and requirements of gamification in science. We conducted our study in the context of research reproducibility, one of the most prevalent challenges in science today that suffers from motivating researchers to document and preserve their work. Through several research activities, we learned about opportunities in designing gamification for research preservation in data-intensive experimental physics. Based on our researcher-centered design, we created two interactive prototypes of a preservation service that make use of contrasting gamification strategies.

Our evaluation showed that both the rational-informative and the openly competitive prototypes were considered valuable, enjoyable, suitable, and persuasive. Through thematic analysis of our interviews, we identified four themes that inform about design requirements for gamification in science: \textsc{Contribution}, \textsc{Metrics}, \textsc{Applications}, and \textsc{Scientific Practice}. Our data analysis revealed that gamification needs to address well-known challenges of the research process, including the fair reflection of quality and individual contribution. Our findings point to a high level of controversy related to the promotion of individual achievements and suggest that team and analysis-related rewards are generally accepted and desired.

Finally, we discussed implications designing for gamification in science that we expect to impact prevalent scientific challenges. We further discussed how already existing Open Science Badges relate to our design implications.

\chapter{Tailored Science Badges: Enabling New Forms of Research Interaction }
\label{ch:tailored_badges}

Gamification is a promising tool in motivating reproducible research practices. Our previous study has shown that prototypes of a gamified preservation service created a persuasive, valuable, and enjoyable interaction that motivates documentation and sharing (see Chapter \ref{ch:gamification_requirements}). In the discussion of our findings, we placed particular emphasize on the success of Open Science Badges (OSB) which acknowledge and promote open research practices. OSB have proven to significantly impact data and material sharing of submissions to a psychological science journal \cite{Kidwell2016}. The general nature of those badges led to their adoption in 67 journals across a variety of scientific domains \cite{osfbadges}. ACM introduced a set of even more fine-grained badges that promote sharing and reproducibility in experimental computer science \cite{boisvert2016incentivizing}. Their badge design showcases the balance between more directed support of research conducted within the ACM's scientific scope and the desire to remain applicable to all threads of research within ACM's diverse scientific landscape. In this chapter, we explore designs and uses of science badges that are even \textit{more directed} to specific practices and needs of individual scientific fields and organizations. We introduce the notion of \textit{tailored science badges}. Those are badges \textit{closely tailored to a target community, scientific domain, and infrastructure}. We consider them as a complement to generic science badges like the OSB and ACM badges, as we expect tailored science badges to enable a more targeted promotion and acknowledgement of reproducible research practices.

Besides promoting and motivating scientific practices, science badges are expected to improve the discoverability of building blocks of research \cite{nustguerrilla}. Through \textit{tailored} promotion of research content and uses, we aim to support navigation and discovery of preserved research. The targeted exposure of scientific building blocks is expected to provide an effective means to increase the visibility of scientists and research work, which in turn motivates contributions. Given the impact on the visibility of research and scientists' careers, we acknowledge that the provenance of science badges plays a crucial role in establishing researchers' trust \cite{nustguerrilla}.

Based on the findings from our user-centered gamification design and research study (see Chapter \ref{ch:gamification_requirements}), we designed and implemented six tailored science badges in CAP. In this chapter, we first reflect on gamification design in science and position tailored science badges in the context of related research. Next, we detail the design and implementation of the tailored science badges in CAP. We then describe the study design and present results and findings from the evaluation. Finally, we discuss design implications and put our findings into perspective for the wider gamification research. In particular, we stress that tailored science badges enable new uses and more targeted interaction with research repositories that open up new applications for gamification beyond motivation.

\begin{tcolorbox}[title = This chapter is based on the following manuscript currently being prepared for submission.]
Sebastian S. Feger, Paweł W. Woźniak, Jasmin Niess, and Albrecht Schmidt. Tailored Science Badges: Enabling New Forms of Research Interaction.

\color{darkgray}
------------------------------------------------------------------------------------------------------\newline
We made several of the study's resources available to the reviewers of this PhD thesis and will make those resources available to the conference / journal reviewers of our manuscript.
\end{tcolorbox}

\section{Related Work}

Badges, one of the most common elements in gamification design, are already being used to promote and motivate documentation and sharing of research artifacts. Open Science Badges (OSB) have proven to significantly increase sharing of data and material in a psychological science journal \cite{Kidwell2016}. Rowhani-Farid et al. \cite{Rowhani-Farid2017} conducted a systematic review of incentives for data sharing in medical research and concluded that OSB ``is the only tested incentive that motivated researchers to share data.'' There are three OSB, acknowledging \textit{open data}, \textit{open materials}, and \textit{preregistration}. ACM introduced even more fine-grained badges for their digital library, including \textit{artifacts reusable} and \textit{results reproduced} \cite{boisvert2016incentivizing}. 
The \textit{general} nature of those science badges allows conferences and journals across a diverse scientific landscape to adopt and award them. OSB are already issued by 67 journals in various scientific domains, from geoscience to neurochemistry \cite{osfbadges}. While the design and implementation of tailored science badges requires significantly more efforts, we expect them to enable a more focused interaction with research repositories.

In our previous study on gamified preservation service prototypes, we discussed how our findings relate to the impact of science badges (see Chapter \ref{ch:gamification_requirements}). We reasoned about the underlying factors that contribute to the success and acceptance of the badges: 1) They allow to promote valuable and accepted best practices; 2) Badges create incentives, but do not punish; 3) Badges increase visibility; 4) They acknowledge papers, not individuals; and 5) They provide accessible goals. We argued that the overall acceptance of badges makes them particularly suitable for scientific environments, where other game design elements are more controversial. We further stressed that particle physics researchers wanted to explicitly find analyses on the service that are directly \textit{executable (reusable)} or considered \textit{educational} or \textit{innovative}. Researchers further wanted to navigate analyses based on \textit{popularity}, \textit{completeness}, and number of forks (\textit{fundamental)}. Based on our findings, we chose to implement those six badges in CAP. We considered that this approach allows for a most effective evaluation of the impact of \textit{tailored science badges}, as it is based on extensive user-centered requirements research.

N\"ust et al. \cite{nustguerrilla} expect badges to play a key role in exposing ``building blocks of research.'' They argued that in today's computational and data-driven science, the links between publications and underlying digital material are often not sufficiently transparent. Thus, they investigated the ``concept of badges to expose, not only advertise, the building blocks of scholarship.'' The authors described the implementation of a badge server and stressed that further research is needed to ``investigate potential effects on willingness to publish research compendia and elaborate on trust.'' Given the potential impact of science badges, they argued that ``the provenance of badges (i.e. who awarded it, to what, using which criteria) would be crucial in a scholarly setting to establish trust.'' As we also consider trust to be key to the acceptance of tailored science badges, we evaluate trust and commitment for each of the six badges that we introduce.

\section{Tailored Science Badges Implementation}

Based on previous requirements research, we designed and implemented six tailored science badges in CAP. Here, we detail the design and implementation.

\subsection{Design of the Badges}

We decided to base the design of the tailored science badges on the findings from our previous study on gamification requirements (see Chapter \ref{ch:gamification_requirements}). Our evaluation of a gamified physics preservation service pointed to six applications, uses, and characteristics of preserved research that can be exposed through game design elements. In Table \ref{tab:badges}, we list the six corresponding badges and their descriptions.

\begin{table}[h]
  \centering
  \begin{tabular}{Sc Sc}
  
     \rowincludegraphics[scale=0.3]{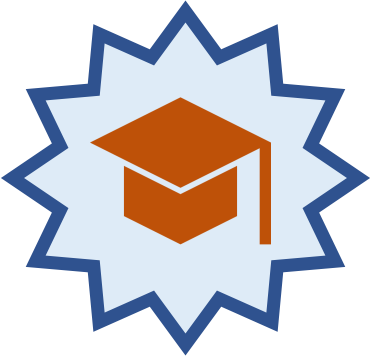} &
     \makecell{\textbf{Educational}\\Work that is particularly educational.\\The award is directly based on the feedback\\of members of your collaboration.} \\
     
     \rowincludegraphics[scale=0.3]{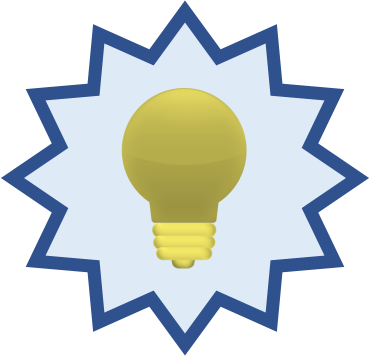} &
     \makecell{\textbf{Innovative}\\Rewards work that is innovative.\\The award is directly based on the feedback\\of members of your collaboration.} \\
     
     \rowincludegraphics[scale=0.3]{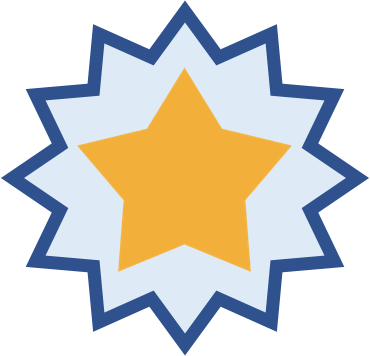} &
     \makecell{\textbf{Popular}\\Popular analyses in your collaboration.\\Popularity is based on the number of\\researchers viewing an analysis.} \\
    
     \rowincludegraphics[scale=0.3]{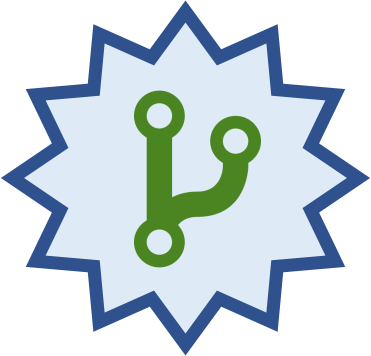} &
     \makecell{\textbf{Fundamental}\\Refers to work that is fundamental:\\Analyses published on CAP can be cloned.\\Cloned research provides a foundation for\\future research. Frequently cloned work\\receives this award.} \\
     
     \rowincludegraphics[scale=0.3]{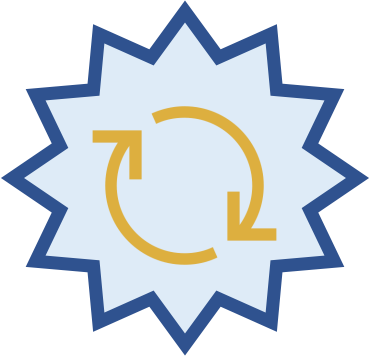} &
     \makecell{\textbf{Reusable}\\Award goes to work that is reusable:\\ Analyses which can be re-executed \\ on ReAna receive this award.} \\
     
     \rowincludegraphics[scale=0.3]{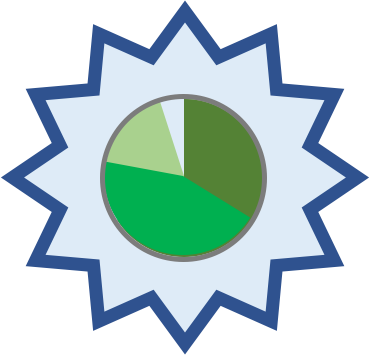} &
     \makecell{\textbf{Thorough} \\ Awarded to analyses which have \\more than 90\% of the fields documented.} \\
     
  \end{tabular}
  \caption{Overview of the six tailored science badges and their descriptions.}
  \label{tab:badges}
\end{table}

\begin{figure}
  \centering
  \includegraphics[width=1.0\columnwidth]{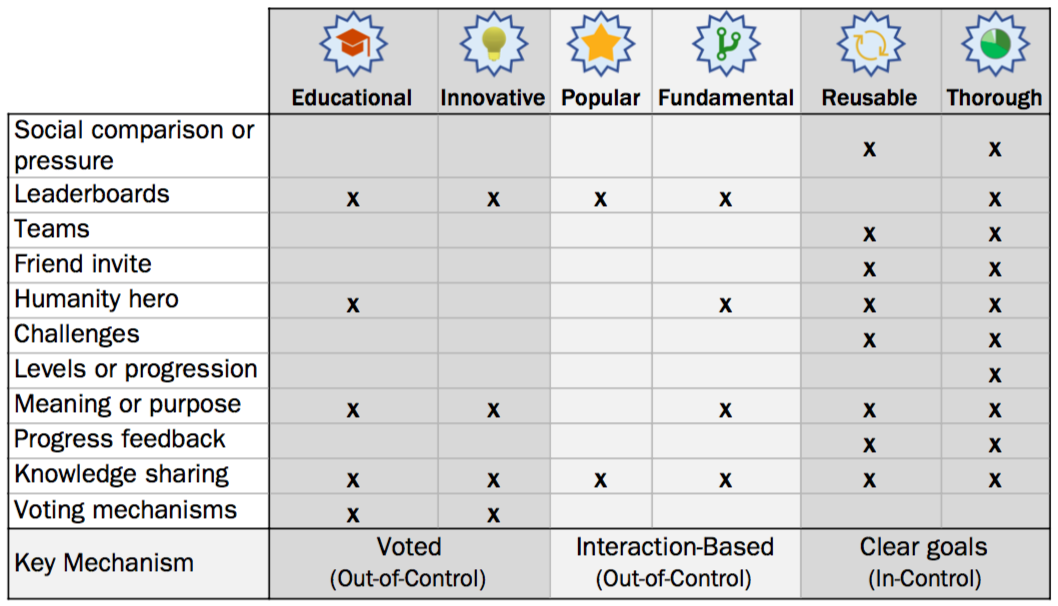}
  \caption{Mapping of the six badges to gameful design elements.
  }~\label{fig:badges_elements}
\end{figure}

\begin{figure}
  \centering
  \includegraphics[width=0.35\columnwidth]{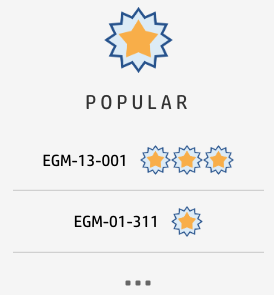}
  \caption{An overview of popular analyses on the service dashboard.
  }~\label{fig:dashboard_popular}
\end{figure}

\begin{figure}
  \centering
  \includegraphics[width=0.5\columnwidth]{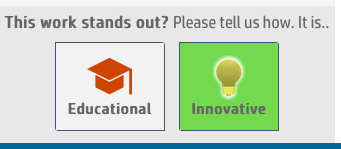}
  \caption{The educational and innovative badges are awarded based on community votes.}~\label{fig:voting}
\end{figure}

As part of the design process, we related the science badges to established game mechanisms. Figure \ref{fig:badges_elements} shows their connection to applicable gameful design elements, as listed by Tondello et al. \cite{Tondello2017}. For example, most of the badges provide means to apply an ordinal measurement scale which enables representation in \textit{leaderboards}. Figure \ref{fig:dashboard_popular} depicts an overview of popular work on the service dashboard. Based on the number of analysis views, popular analyses can receive one, two, or three popular badges. Instead, the reusable badge does not support comparisons, as work is either executable on ReAna or not. \textit{Progress feedback} is another example of a gameful design element which is applicable mostly to the reusable and thorough badges. These badges are based on clear goals and a system can easily measure the progress towards those goals. In contrast, we designed \textit{voting} mechanisms to characterize educational and innovative work (see Figure \ref{fig:voting}).

The process of relating our tailored science badges to applicable game design elements supported us in identifying and describing the key mechanisms and criteria of the badges. We identified three mechanisms: Community votes, community interaction, and clear goals. As depicted in Figure \ref{fig:badges_elements}, the educational and innovative badges are based on community votes. The popular and fundamental badge are based on user interactions (number of views and number of forks / clones). The reusable and thorough badges are based on clear rules. Here, the researchers are in full control of reaching the badge criteria on their own. In the cases of voted and interaction-based badges, analysts have to trust their colleagues and the system to make a fair judgement of their work. However, they can expect that a thorough documentation of high-quality research is likely to increase their chances to earn those badges. Besides that, contributors have no direct control.

\begin{figure}
  \centering
  \includegraphics[width=0.6\columnwidth]{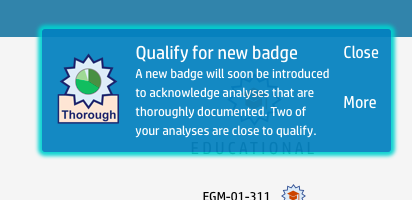}
  \caption{A notification informs about the introduction of a new science badge.}~\label{fig:badges_notification}
\end{figure}

\begin{figure}
  \centering
  \includegraphics[width=0.55\columnwidth]{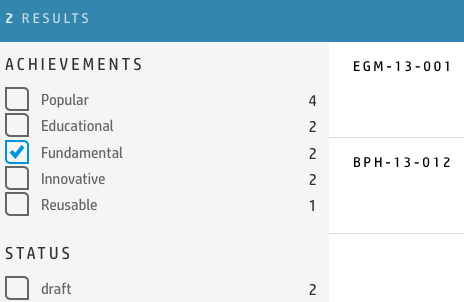}
  \caption{Dedicated facets for badge achievements were integrated on the search page.}~\label{fig:badges_search}
\end{figure}

\begin{figure}
  \centering
  \includegraphics[width=0.6\columnwidth]{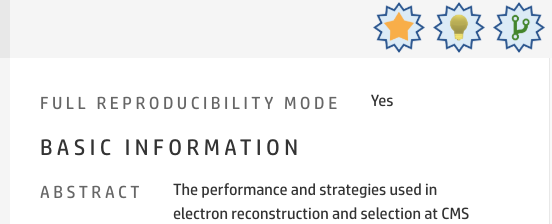}
  \caption{The badge banner promotes analysis achievements.}~\label{fig:preview}
\end{figure}

\subsection{Service Implementation}

Our implementation of tailored science badges is based on the production CAP service in the state of early 2019. We added several badge-related views to existing pages of CAP and created several new views. Screenshots of all those pages are available as supplementary material. In the following, we briefly describe the major changes:

\textbf{Dashboard}: We added overviews and leaderboards for each badge. Figure \ref{fig:dashboard_popular} shows a selection of popular badges on the dashboard. Each list contains up to four references to analyses. The bottom element references the search page.
    
\textbf{Search page}: As shown in Figure \ref{fig:badges_search}, we implemented dedicated achievement facets on the search results page.
    
\textbf{Analysis page}: As previously mentioned and depicted in Figure \ref{fig:voting}, we implemented a voting mechanism to promote educational and innovative work. Furthermore, we added a banner to analysis pages which displays achievements. Figure \ref{fig:preview} shows an analysis with three awarded badges. Finally, we added a printable banner that opens when one of the badges is selected. This banner is designed to export key information about the analysis, including title, authors, abstract, and all awarded badges. 

\begin{figure}
  \centering
  \includegraphics[width=1.0\columnwidth]{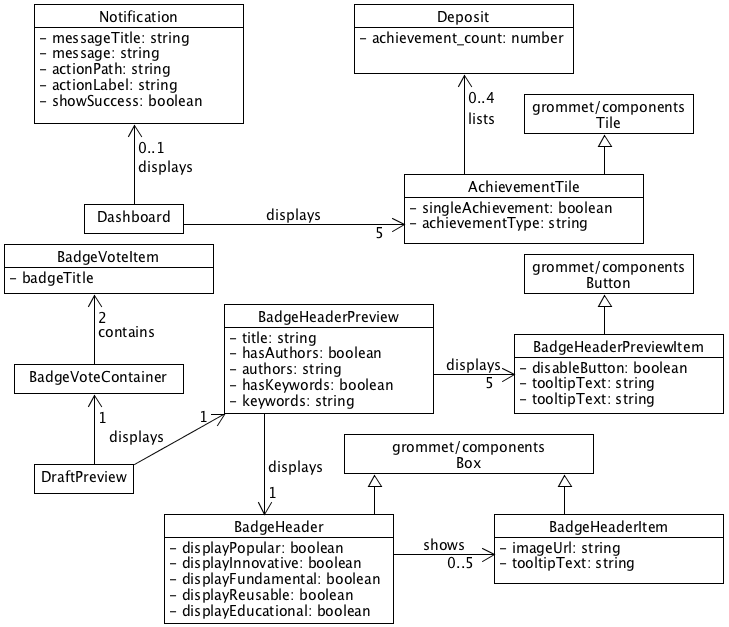}
  \caption{Simplified UML class diagram of the science badges layer.}~\label{fig:uml_badges}
\end{figure}

The implementation branch is openly accessible on GitHub\footnote{The science badges implementation is publicly available on GitHub: \url{https://github.com/sefeg/analysispreservation.cern.ch/tree/gamification\_feb19}}. It should be noted that the main goal of this implementation was to conduct this study. As such, it implements the study design that is based on multiple events. The gamified service creates the illusion of a seamless integration into the production system. However, it does not currently support data manipulations (e.g. the vote buttons do not actually store the information) as we encourage participants to explore and discuss all mechanisms without worrying about the immediate consequences of their actions. We further want to stress that any party interested in using the code should consider the findings and design implications that we discuss in this chapter, in order to introduce a most meaningful implementation of tailored science badges in RDM tools. Figure \ref{fig:uml_badges} shows a simplified UML class diagram of the tailored science badges implementation in CAP. It should be noted that for the purpose of simplicity in communicating the overall structure, we considered both React \textit{classes} that extend \textit{React.component} and JavaScript \textit{functions} as \textit{classes} in this diagram. For the functions, we added parent relationships to the root components they return. This simplified UML class diagram is designed to help any party interested in re-using and adapting the implementation to quickly understand the overall structure and relationships. In particular, we see that the \textit{Dashboard} and \textit{DraftPreview} display most of the containers that hold information about the science badges.

%
%
\section{Study Design}

We carried out 11 mixed-method evaluations, to establish an empirical understanding of the impact of tailored science badges on researchers' motivation, and ability to navigate and discover research repositories. Here, we describe the recruitment of participants, the structure of the evaluation sessions and the data analysis.

\subsection{Participants}

We recruited 11 research physicists working at CERN. None of the research analysts participated in any previous study related to gamification in the scientific context. The participants were 29 to 48 years old (average = 35 years, SD = 6.6 years). We assured participants that we would not disclose the age of individual research analysts. The 11 interviewees were all male. This partially reflects CERN's employment structure: according to 2018 personnel statistics, between 80\% and 91\% (depending on the contract type) of research physicists working at CERN were male ~\cite{CERN-HR-STAFF-STAT-2018}. All interviewees were employed by CERN or by an institute that is collaborating with CERN. As all interviews were conducted during regular working hours, they became part of an analyst's regular work day. Thus, participants received no extra remuneration for the study participation.

\begin{table}
  \centering
  \begin{tabular}{c c c r}
    {\small\textit{Reference}}
    & {\small \textit{Affiliation}}
    & {\small \textit{Experience}} \\
    \midrule
    P1 & ATLAS & Postdoc \\
    P2 & ATLAS & Postdoc \\
    P3 & LHCb & Upper Management \\
    P4 & ATLAS & Postdoc \\
    P5 & FCC & Postdoc \\
    P6 & CMS & Convener \\
    P7 & CMS & Convener \\
    P8 & CMS & Postdoc \\
    P9 & CMS & Postdoc \\
    P10 & ATLAS & Upper Management \\
    P11 & ATLAS & Postdoc \\
    \bottomrule
  \end{tabular}
  \caption{Overview of the researchers recruited for the study on tailored science badges.
  }~\label{tab:table_participants_badges}
\end{table}

Table \ref{tab:table_participants_badges} provides an overview of the 11 participants. We recruited physics data analysts with a diverse set of experiences and roles within the LHC collaborations. In order to create a most complete understanding of perceptions, requirements, needs, and impact of tailored science badges in particle physics research preservation, we made sure to recruit both early-career and senior researchers. We recruited two conveners. Although conveners have a project management role within a collaboration, they are often still involved in technical analysis work. In addition, we recruited two active or former members of the upper management of two of the collaborations. We asked those two participants to rate only a subset of the questionnaire, as they are unlikely to preserve analyses themselves. However, we consider their participation a strength of our study, as they provide an administrative perspective that related work has not profited from.

None of the interviewees had any hierarchical connection to any of the authors. And none of the participants had previously taken part in any other research conducted by any author of this work. The participants reflect the cultural diversity at CERN. We did not list the nationalities of individual participants, as this might allow to identify some of the researchers based on the information already provided in Table \ref{tab:table_participants_badges}. However, we can report the nationalities involved in alphabetical order: Austrian, British, German, Japanese, Portuguese, Spanish, Swiss. We conducted all interviews in English, which all interviewees spoke fluently. English is the predominant language in research at CERN.


\subsection{Evaluation Structure}

In this section, we describe the structure of the evaluation sessions. The complete evaluation material, including the interview protocol and questionnaire are available as supplementary material.

First, we introduced the participants to CAP. As they had not used CAP before, we asked them to explore the current production version \textbf{without badges}. In particular, they reviewed some of the available analyses, the analysis description template and the search page. We then asked them to respond to a questionnaire designed to evaluate the value, enjoyment, identified regulation, external regulation, suitability and persuasiveness of this service. For the value, enjoyment, suitability and persuasiveness subscales, we re-used the questionnaire items from our gamification in science requirements research. We decided to reuse them, as we aim to relate our findings on tailored science badges to previous research on game design elements in science. The value and enjoyment subscales are based on the Intrinsic Motivation Inventory (IMI). Besides assessing intrinsic motivation, we also wanted to assess the impact of extrinsic motivation on contributions to preservation technology. To do so, we slightly adapted the Identified Regulation and External Regulation subscales of the Situational
Motivation Scale (SIMS). Finally, we asked participants to provide a list of keywords or short sentences that describe for which reasons they would want to use this service.

Next, we switched to our version of CAP \textbf{with badges}. The participants were immediately directed to the dashboard. As depicted in Figure \ref{fig:dashboard_popular}, they saw an overview of preserved analyses that were published by their colleagues and that had been awarded a badge. Here, it should be noted that we populated the database with a set of actual physics analyses. Next, the participant received a notification, referring to an analysis of their own that just got awarded the popular badge. As the participants had not used CAP before, we asked them to imagine pre-populated physics analyses as being their own. Participants were invited to open the analysis and comment on the different badge-related mechanisms on the analysis page (e.g. the exportable badge banner or the badge preview, as depicted in Figure \ref{fig:preview}). Here, we asked about the value of badges on their own analyses and on analyses preserved by their colleagues.

Back on the dashboard, another notification appeared. As depicted in Figure \ref{fig:badges_notification}, the participants were informed about the upcoming introduction of a new badge: the \textit{thorough} badge. Analysts were asked to get more information about this badge by following the link. On the referenced page, the key criteria for the thorough badge was described: more than 90\% of the analysis fields have to be documented. Two analyses owned by the participant were listed as close to reaching this goal. We then asked about the value of thoroughness in research preservation and the importance of such a badge in navigating the research repository. Finally, we invited the participants to use and review the vote mechanisms (Figure \ref{fig:voting}) and the badge-related search facets. We asked them corresponding questions and concluded the practical exercises on this CAP version.

We then invited the researchers to answer the same questionnaire of before, assessing the value, enjoyment, identified regulation, external regulation, suitability and persuasiveness of this service. We also asked them, again, to provide a list of keywords or short sentences that describe for which reasons they would want to use this service. Doing so, we aimed to record and compare potential changes in the perceived uses of the service versions.

Finally, we asked the analysts to rate the suitability, trust and goal commitment for each of the six badges. To assess \textit{suitability}, we re-used a slightly adapted statement from our previous study: The [title] badge is NOT suitable for a research preservation service ( R ). We used the following statements regarding \textit{trust} in innovative and educational badges: I trust the \textit{research community} to make a fair assessment of [innovative / educational] work. \textit{Trust} statements for the other badges were constructed as follows: I trust that \textit{the system} will calculate and award the [title] badge fairly. It should be noted that the two participants from upper management (see Table \ref{tab:table_participants_badges}) rated agreement only to those two scales. To assess \textit{goal commitment}, we employed the five-item goal commitment scale by Klein et al. \cite{klein2001assessment}.

\subsection{Qualitative Data Analysis}

We recorded a total of 6.2 hours during the evaluation sessions. We transcribed the recordings non-verbatim and used Atlas.ti data analysis software to analyze and code the transcriptions. We performed Thematic Analysis \cite{Blandford:2222613} to identify themes. Two authors performed open coding of the first two interviews. They discussed and merged their codes and assigned them to code groups. This code tree was used in coding the remaining transcriptions. In total, we created 153 codes. We further discussed the resulting code groups and adapted and merged some of them. Fourteen code groups resulted from this highly iterative and collaborative process. Out of those, we constructed three high-level themes: \textsc{Effects}, \textsc{Content Interaction}, and \textsc{Criteria}. The theme \textsc{Effects}, for example, is based on the code groups `Visibility', `Career', `Feedback', and `Motivation'.

\section{Results}

\begin{figure}
  \centering
  \includegraphics[width=0.9\columnwidth]{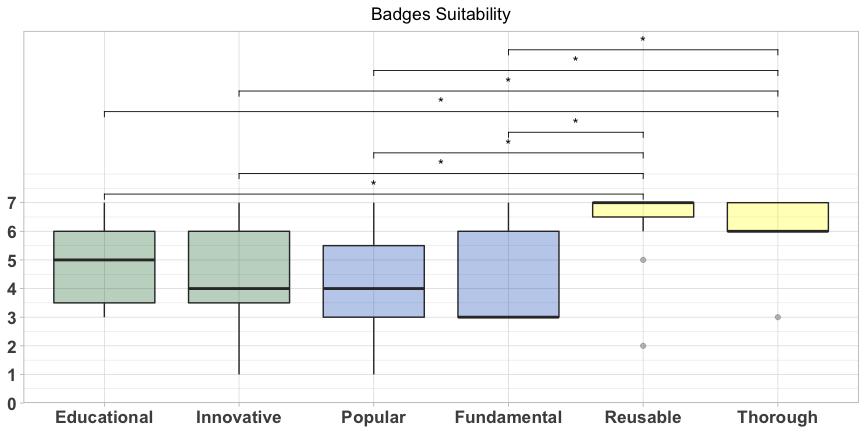}
  \caption{Box plot for badges \textit{suitability}. Significant differences are marked (*).}~\label{fig:suitability}
\end{figure}

We performed pairwise Wilcoxon comparisons with Holm p-adjustment for badges \textit{suitability}, \textit{trust}, and \textit{commitment}. Our analysis of the questionnaire responses showed that participating physicists found both the \textit{reusable} and \textit{thorough} badges significantly more \textit{suitable} than all other badges (Figure \ref{fig:suitability}). This means that the badges in the \textit{clear goals (in-control)} group were considered significantly more suitable than those based on different key mechanisms. The color schemes of the badge plots in this section relate to the underlying core mechanisms: community votes (green), community interaction (blue), and clear goals (yellow).

\begin{figure}
  \centering
  \includegraphics[width=0.9\columnwidth]{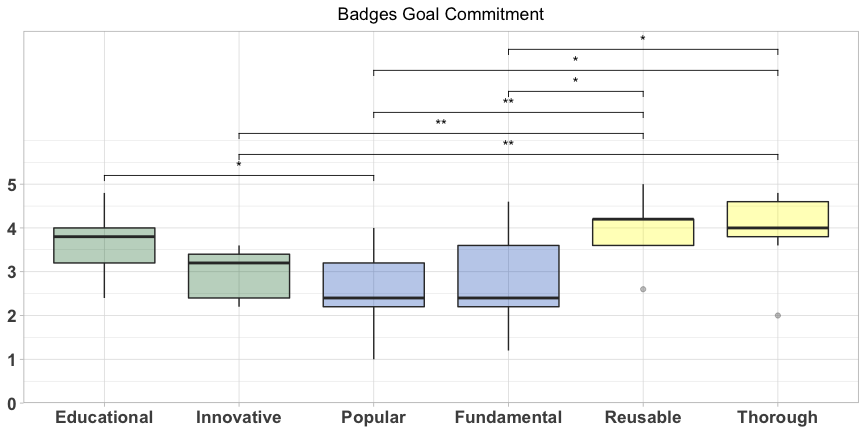}
  \caption{Badges goal commitment (5-point scale). Significant differences are marked (*).}~\label{fig:commitment}
\end{figure}

\begin{figure}
  \centering
  \includegraphics[width=0.9\columnwidth]{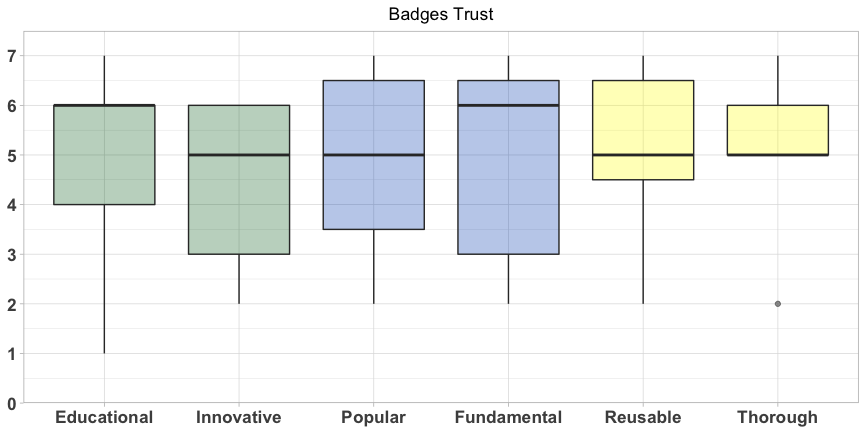}
  \caption{Box plot for \textit{trust} towards the badges. No significant differences.}~\label{fig:trust}
\end{figure}

The pair-wise comparison of rated \textit{goal commitment} showed significant differences in participants' commitment towards the \textit{reusable} badge, as compared to the \textit{innovative}, \textit{popular}, and \textit{fundamental} badges. Differences in goal commitment towards the \textit{thorough} badge, as compared to the \textit{innovative}, \textit{popular}, and \textit{fundamental} badges are also significant. All significant differences are marked in Figure \ref{fig:commitment}. The analysis of \textit{trust} towards the badges (Figure \ref{fig:trust}) showed no significant differences.

\begin{figure}
  \centering
  \includegraphics[width=0.9\columnwidth]{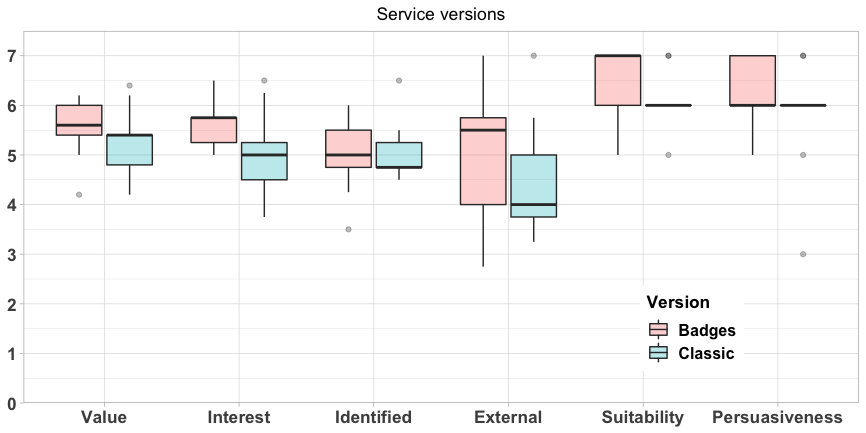}
  \caption{Box plot concerning the service versions.}~\label{fig:versions}
\end{figure}

\begin{figure}
  \centering
  \includegraphics[width=0.9\columnwidth]{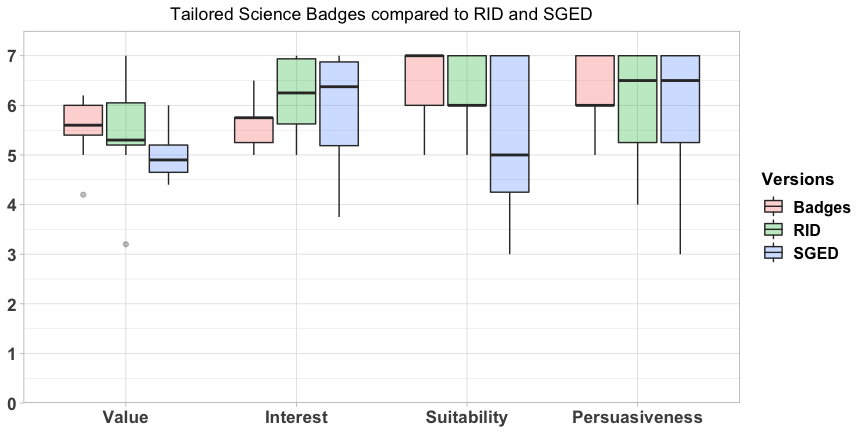}
  \caption{Box plot comparing tailored science badges with RID and SGED.}~\label{fig:study_comparison}
\end{figure}

As illustrated in Figure \ref{fig:versions}, we analysed the questionnaire responses related to the two service versions. We found no significant differences in rated value, interest, identified / external regulation, suitability, and persuasiveness between CAP versions with (Badges) and without (Classic) badges. However, median and mean\footnote{Except for identified regulation} scores of the Badge version are consistently higher than those of the Classic CAP version. 
Finally, we compared value, interest, suitability, and persuasiveness between CAP with Badges and the RID and SGED prototypes from our previous study (Figure \ref{fig:study_comparison}). We find that rated \textit{value} of CAP Badges differs significantly from value of the SGED prototype. There are no further significant differences between the three service versions.

\section{Findings}

Our qualitative data analysis provides further insights into researchers' assessment of the badges. In particular, regarding the uses and the value of tailored science badges, and the constraints and requirement of their implementation. We present those findings based on three themes:  \textsc{Effects}, \textsc{Content Interaction}, and \textsc{Criteria}.

\subsection{Effects}


Researchers perceived the suitability of individual badges differently. Still, they rated the service as suitable, persuasive, and valuable overall. Based on our qualitative data analysis, this is mainly due to the badges' generally positive effects. In fact, most participants referred to an increase in visibility. Both for research analyses and researchers:

\begin{quote} 
I mean if it shows up on the main page, people will have to look for it, I guess. Top analysis more people will have to look for it, I guess. Which makes sense, I suppose. (P9) \newline
So, fundamental is I think getting exactly at that. Because then you have some master student who forks it and they do a lot of work on their masters thesis and never publish it in a peer-reviewed journal and never gets cited. But it's still work. It's still interesting science. And that would capture that. (P4)
\end{quote}

In addition, the badges are likely to provide an opportunity for smaller groups or smaller experiments to get visibility: ``\textit{I am thinking more to smaller experiments. Because they are completely invisible. So, yes, this could be nice.}'' (P2)

A participant further discussed multiplication effects enabled by the increased visibility:

\begin{quote} 
It would give me some insurance that my analysis is interesting. Would probably also tell others that this is interesting and it would make it more likely for others to actually look at the analysis. Again, boosting the popularity. Yeah, so I mean it would be nice if you got this if this was available. (P7)
\end{quote}

Ultimately, analysts expected that the increased visibility impacts career opportunities. P6 imagined that researchers would add the exportable badges banner to their CV. And P7 thought about an official mechanism where the number of awarded badges are considered as criteria in the promotion of employees. The convener discussed this as an approach to improve the transparency within the organization, as current processes are thought of as rather intransparent. 

Related to visibility and career opportunities, researchers discussed the role of presentations. They imagined that the exportable badge banner might be a valuable resource in presentations. P4 even asked to provide badges tailored to preserved presentations:

\begin{quote} 
When people make talks, the whole point is you are presenting yourself. This is different then a publication which has a thousand authors and isn't actually attached to you. You know it's your publication, whereas on a talk it's a name, maybe on behalf of, but you are giving the talk. (P4)
\end{quote}

Finally, most participants discussed feedback as an important driver that is enabled by the badges:

\begin{quote} 
So, my very first reaction compared to the first version is much more positive. Specifically, the notification I think it's good. So that you get positive feedback. [...] And there is some abstract later gain, but you often don't get notified normally. And so if you get this notification I think it's very useful. (P1)
\end{quote}

P5 asked for the possibility to provide short comments as part of the vote mechanism:

\begin{quote} 
That actually seems interesting. Because I wonder if I can then look at the discussion and can learn a bit more and get more views on more opinions of this analysis. So, this is... If there is actually a discussion there to be viewed, this seems kind of like an interesting thing. (P5)
\end{quote}

\subsection{Content Interaction}


Most participants described badges as a tool that enables new forms of interaction with preserved content --- and with research work in general. Foremost, they provide a mechanism to navigate large research repositories.

\begin{quote} 
I like this too. (Educational) Exactly the same as the innovative tag. [...] Yeah this one is I think, it's good to have a few of the analyses of the big pool stand out in certain aspects. (P8) \newline
I think the biggest problem at the moment, it's just that we are beyond 900 papers [...] you basically try to look into the details of the individual analyses, you know the thoroughness badge would probably be very good to have. (P7)
\end{quote}

The participants pointed out that the badges are likely to provoke browsing on the service and aid in discovery that would currently rely on unstructured forms of direct communication. In this context, P2 referred to structured and collaboration-wide visible feedback provided by the vote mechanism:

\begin{quote} 
Now, I put my editorial hat on. I like the concept. [...] Feedback of people that's the kind of things that you hear around coffee-discussions. Oh yeah, go to this analysis. It's nicely done. It's nicely documented. You can start from there and learn from it. But, it's never written anywhere. So, that is useful. (P2)
\end{quote}

P1, P4, P5, and P8 referred to mechanisms of serendipitous discovery that is likely to result from researchers browsing the content promoted by the badges:

\begin{quote} 
You don't really find their work and it's difficult to discover like this. Unless you work with them and you know where they put their stuff. Can be very nice to kind of like discover analysis and like that to get an overview of stuff like that. (P5)
\end{quote}

Most researchers discussed re-execution of preserved analyses as most desirable goal. In this context, convener P7 discussed the re-usable badge as a mechanism to filter noise: ``\textit{Because most of them there is no information that goes beyond the very basics. I could at least filter all the noise. That's something important.}''

P4 expanded upon the notion of improved navigation:

\begin{quote} 
I think the main thing is attaching the badges is important. Because it gives you a different way to query the database. [...] I think the main things is you attached new information, that the current way we archive science doesn't afford. (P4)
\end{quote}

\subsection{Criteria}


Given that researchers rated the individual badges differently, they extensively commented on requirements for designing tailored science badges. The initial contact with the gamified service proved to be a critical moment, as also the strong initial reactions of the following two researchers show:

\begin{quote} 
I see the gamification already, there. So, I am not sure about the achievements being used. I do think that having something where people can say, I have used this piece or this has been visited these many times, or people have left a star, or something like this. So, I don't know what popular is-- (P6) \newline
So, you basically rate the analysis or somehow like this right... Ok, I mean then the question would probably be how you rate something. Or what is more interesting than others. (P11)
\end{quote}

These quotes refer to two major challenges that resurfaced during the entire exploration of the service by every participant: inter-badge comparison and the need to understand the rules of the badges.

The tooltips placed on the badges proved effective in communicating the individual mechanisms and rules of the badges. Based on those descriptions, the physicists stressed that sophisticated protection mechanisms for most of the badges would need to be implemented. This implementation should be communicated to the users to establish trust. The use of those protections is twofold. First, they protect from \textit{unintended} side effects:

\begin{quote} If it's forks, then it could have the nasty effect that- if there is a problem with some particular analysis, people try to fork it several times. [...] Fork again. But still doesn't work. You see? Oh it's very re-usable! We forked a lot. No, it's not. But you can get around that. But you would need to put protections at the number of unique forks by unique people. (P3)
\end{quote}

Second, it protects from any attempts to game the system. P3 refers to such concerns related to the thorough badge: ``\textit{If it is automatically calculated by the computer, it would tend to encourage people to just add some meaningless words everywhere. Or some minimal, just to have something in all fields. While a documented analysis is something different.}'' (P3)

Physicists also referred to adoption within their collaboration as criteria for use of certain elements. For example, P5 commented on the exportable badge banner:

\begin{quote} 
It depends a lot on how like collaboration or colleagues would use it. I think I wouldn't go ahead using this kind of thing, because people would kind of wonder, why is this like a popular analysis. And who gives out badges and stuff.
\end{quote}

Participants further discussed the role of the administration in awarding some badges. Regarding the vote mechanisms, the browsing on the service might not be sufficient to provide strong and reliable data for the community feedback:

\begin{quote} 
So, here my question kind of is: When does this happen that I am on this page of a different analysis and think `Ok, I want to vote on this'. (P2)
\end{quote}

Instead, P2, P3, and P7 imagined that feedback for the innovative and educational badges could be based on the \textit{``decision of some sort of experts.''} (P2) Although here, as P2 continued to state, \textit{``the main worry is that these experts then be overloaded and then the quality of their work may not be that high.''}

Most participants reflected on differences in complexity of the individual badges. Foremost, they distinguished between the complexity in terms of awarding them:

\begin{quote}
Like the popular. Definitely, it's just counting. This is easy I would say. And educational and innovative. I mean this is how other people see the analysis. Ok, that's also fine. And then fundamental, reusable, yes, there I have a bit more doubts I would say. This is a fair thing and it would work. (P10)
\end{quote}

This reflects a common observation we made during the sessions: Participants tried to imagine examples of analyses that might qualify for individual badges. Finding examples proved to be a crucial step in being able to evaluate the usefulness and suitability of a badge. This was especially true for the re-usable badge that aims for a goal that only few particle physics analyses qualified for at the time of the interviews. Here, several participants explicitly asked for a finer granularity. P7 provided examples of more accessible steps towards the reusable badge:

\begin{quote} 
So, I think there could actually be smaller steps towards this, so you know basically your code is available via the portal or something. It's like the first thing. Then it also compiles. [...] There should be more granularity there. (P7)
\end{quote}

The discussions regarding complexity also relate to common scientific challenges. P10 had concerns with the fundamental badge, as \textit{``basically (...) all what we are doing is fundamental.''} P5 wondered about the meaning of the innovative badge, as \textit{``research is supposed to be innovative by definition.''}


\section{Discussion}

We discuss our findings from the evaluation of tailored science badges that we implemented in a particle physics research preservation service. First, we describe how the scope of tailored science badges differs from other generic game design elements in science. Next, we discuss design implications for the implementation and adoption of tailored science badges. Finally, we stress how tailored science badges move the design goal from \textit{motivating} practices, to \textit{supporting} research practices and content \textit{interaction}. We expect and wish that our findings and discussions will spark a debate within the SIGCHI community on meaningful implementations and adoption of science badges.

\subsection{Scope of Tailored Science Badges}

This study presented some of the first empirical findings on the design and evaluation of game design elements that are specifically tailored to a science tool and research community. With this tailored design approach, those badges target a different scope than Open Science Badges (OSB) \cite{osfbadges} and ACM badges \cite{acmbadgesweb}. While OSB and ACM badges can be easily adopted by a wide variety of journals and conferences, \textbf{tailored science badges enable a more focused support of scientific practices. They also differ in terms of underlying mechanisms.} OSB badges are awarded based on the review of committees and experts. The same mechanism applies for most of the ACM badges. However, ACM foresees a form of community interaction related to the \textit{Results Reproduced} badge: The reward can be claimed once other researchers report that they successfully reproduced findings from an ACM publication.

We found that participants were concerned about overloading committees or experts with tasks of reviewing content and awarding badges. And that this might impact the quality of the reviews. Here, we particularly profited from the assessments of two members of the upper management of the particle physics collaborations. Notably, researchers recorded no significant differences in \textit{trust} towards the six badges. Provided that the badges are based on strong protection mechanisms, \textbf{researchers stressed that they overall trust the system and their research community to make fair assessments --- independent of the underlying mechanism}. This is a valuable finding, as it also provides a different perspective on reward mechanisms among the more general science badges like OSB and ACM badges. 


\subsection{Adoption}

We observed that \textbf{the initial contact with the gamified service is crucial} in the process of assessing the value of science badges. While most physicists directly commented that the CAP Badges version is more attractive and appealing than the Classic version, most researchers immediately started to compare and reason about the individual badges. They often stopped at the first badge that was not clear to them or that they found troubling. At this point, they showed initial concerns for the badge implementation in general. It is reasonable to imagine that many researchers at this stage would lose interest in the badges or even the service if they had no motive to further reason about the badges. In this study, we explicitly asked the participants to further explore the service and to review the mechanisms of the individual badges, at which point the initially concerned researchers stressed that they considered most of the badges useful. In conclusion, \textbf{we need to guide scientists who are experiencing a gamified research service for the first time through the initial exploration process}. This guidance might be provided through notifications that inform about the introduction of a badge or through helpful tips displayed during the first use.

In general, participants found the tooltips useful that appeared once they hovered the mouse over badges. While the information was helpful in communicating the basic concept and reward mechanism of a badge, researchers often started thinking about good examples of analyses that would qualify for a particular badge. Occasionally, this proved difficult, as some of the badges promoted mechanisms that were not yet applicable for the majority of the research work conducted within the collaborations. This was particularly true for the reusable badge. Thus, \textbf{providing strong examples and justifications in the tooltips can foster understanding and assessment of badges}. In addition, researchers repeatedly asked for strong protection mechanisms to prevent deliberate or accidental manipulation. \textbf{Service and tool designers do not only need to implement protection mechanisms, but also communicate their implementation to the users}. We argue that communicating badge motivation, strong examples, and protective mechanisms is essential to justify and explain ``the provenance of badges (i.e. who awarded it, to what, using which criteria), (which) would be crucial in a scholarly setting to establish trust'' \cite{nustguerrilla}.


Our findings showed that individual badges can be controversial. All researchers mentioned concerns related to the implementation of at least one badge. However, their \textbf{perception of the overall service seemed to be informed by the most suitable and useful elements}. Still, we need to stress that this might not necessarily be the case if a tool implemented game design elements that provoked most serious concerns. In our study, no researcher mentioned that any of the implemented badges would represent a major barrier. This is likely due to the fact that our badge design was informed by previous and extensive research \cite{Feger2019Gamification}. 

Related to adoption, we find that \textbf{the design of \textit{tailored} game design elements that promote scientific practices needs to explore mechanisms that reflect achievements outside the original application context}. Research data management tools that are tailored to organizations, institutes, or scientific fields are likely to restrict access to the corresponding research community. While this is not an issue for scientists who stay within the original research area, it becomes challenging for those who change their academic framework or move to industry. Thus, designers should \textbf{consider the implementation of exportable formats, as well as forms of communicating achievements that are comprehensible outside the original research context}.

\subsection{Beyond Motivation}

Gamification is commonly used to motivate actions and practices \cite{Ibanez2014,knaving2018understanding}. In our previous gamification study, we also stressed \textit{creating motivation} as primary aim for our study on two gamified research preservation service prototypes. Yet, most participants in this study on tailored science badges did not explicitly discuss motivation. Instead, they discussed the \textsc{Effects} and uses that the implementation of the science badges enable. Uses related to the impact on content discovery and repository navigation even emerged as part of a dedicated theme: \textsc{Content Interaction}. Improved content interaction foremost profits those who want to find and use information within the research repository. But, the participants stressed that this also provides a strong incentive to contribute to the preservation service and to follow certain practices which will likely result in rewards and more visibility within the research collaboration. Given that participants discussed increased visibility as a driver in the career development, \textbf{we argue that the tailored science badges provide an implicit form of motivation that is tied to new forms of interaction with preserved research.} That way, they also differ from the more generic OSB and ACM badges. OSB badges appear on corresponding publications. However, adopting journals are not mandated to implement facets within their digital libraries. To date, ACM only added one badge (Artifact Badge) as search criteria in the \textit{Advanced Search} of their digital library\footnote{Retrieved March 3, 2020. \url{https://dl.acm.org/advsearch.cfm?coll=DL&dl=ACM}}.
\textbf{We recommend that designers and adopters of science badges --- tailored and general --- explore means to systematically make the sum of additional meta-data collected on research artefacts accessible to the research community.}

To be clear, we still consider \textit{acknowledging} and \textit{motivating} open science practices as key design rationales in the implementation and adoption of tailored science badges. However, we find notable that researchers' perceptions of tailored science badges shifted from \textit{motivational drivers} towards \textit{tools that provide new forms of interaction with preserved research}. In particular, as the desire to integrate one's own research into this cyberinfrastructure framework promises to provide implicit forms of motivation for researchers to follow comprehensive RDM practices. Further exploration of the relationship between meaningful forms of content interaction and implicit motivation might pave new ways for design thinking in gamification, which could be closely connected to the exploration of new application contexts~\cite{Nacke2017}.

\subsection{Limitations and Future Work}

We aim to foster the replicability of our work and to provide a base for future research in the context of tailored science badges and gamification in science. Thus, we make several of the study resources available as supplementary material. Those include the study protocol, Atlas.ti code group report, the questionnaire, questionnaire responses, plots, and screenshots from the service implementation.

We presented findings from the first implementation and evaluation of tailored science badges in a fully functional particle physics research preservation service. Implementing the badges in this open source preservation service is a limitation of the study, as we previously presented findings on gamification design in this environment. However, our findings were limited to the design and evaluation of two gamified preservation service \textit{mockups}. Evaluating a fully functional implementation of game design elements in a research tool represents a novel contribution. In fact, we argue that this study represents a necessary second step in the systematic development of gamified RDM tools that must precede long-term evaluations in production environments. This is largely due to the fact that such an evaluation would need to involve researchers whom we have to convince about the value of RDM in the first place. As service designers, we must not risk deploying gamified services into the scientific cyberinfrastructure without having an empirical understanding of the effects. We simply cannot risk to alienate researchers who commit to open science practices. Based on our findings, we envision opportunities for future work to explore and evaluate tailored science badges in long-term studies across a larger sample. In particle physics and beyond. It would be particularly interesting to map commonalities and differences between requirements for gamification in general, and tailored science badges in particular, between distinct fields of science.

\section{Conclusion}

This chapter presented a systematic study on the design and evaluation of \textit{tailored science badges} in a particle physics research preservation service. We evaluated the science badges implementation with 11 research physicists. The participants were postdocs, group leaders, and members of the upper management of the physics collaborations. Our findings showed that the badges enable new forms of research discovery and navigation within research repositories. We presented researchers' perceptions, as well as the discussed uses, requirements and needs related to the design of tailored science badges in three themes: \textsc{Effects}, \textsc{Content Interaction}, and \textsc{Criteria}. Based on our findings, we related the mechanisms and uses of tailored science badges to the wider concept of gamification in science. In particular, we discussed how design rationales behind tailored science badges differ from generic science badges. Finally, we presented design implications for the implementation and adoption of tailored game design elements, and discussed gamification beyond motivation.

	\part{Conclusion and Future Work}\label{part:conclusion}
	\section*{Outline}

The focus of our work at CERN SIS over the past three years was to conduct research on requirements and opportunities for designing interactive tools that support and motivate reproducible science practices. As a result, we reported findings from four empirical studies involving 42 researchers and data managers in HEP and beyond. In addition to our research activities, I interfaced with physics users and software developers, implemented tailored science badges in CAP, and supported the project management of CAP and COD. Furthermore, I conducted usability tests of several versions of the CAP prototype that was accessible to all members of the ALICE, ATLAS, CMS, and LHCb collaborations.

To adequately reflect the dual character of the work and research conducted over the past 36 months, we first provide an extensive overview of the role of HCI in reproducible science in Chapter \ref{ch:hci_role}. That chapter is conceptually based on the CHI 2019 publication ``\textit{The Role of HCI in Reproducible Science: Understanding, Supporting and Motivating Core Practices}'' \cite{Feger2019RoleHCI} which informed the subtitle of this thesis. In particular, we introduce and describe two models that increase and detail our understanding of how to design systems effective in supporting and motivating reproducible science practices: 1) a \CommitmentModel; and 2) a conceptual model of components and interactions involved in RDM tools. Those models reflect and relate to the findings presented in Parts \ref{part:requirements} and \ref{part:gamification}, as well as related work (Chapter \ref{ch:background}). In addition, we discuss implications of the models, design guidelines for HCI practitioners, and emerging research challenges. Finally, we present our vision of how HCI could help introduce \ac{URP}. While it is important to note that our empirical research and design recommendations stem largely from research in HEP, we argue that the general nature of most of our findings makes them likely to profit science beyond experimental physics.

In Chapter \ref{ch:conclusion}, we summarize our research contributions with regards to our four key RQs and the four stages of the \CommitmentModel. We comment on the role of replication in HCI and discuss limitations of our work. Finally, in Chapter \ref{ch:future_work}, we discuss opportunities for future work, with a focus on supporting the transitions in the commitment evolution model.

\chapter{The Role of HCI in Understanding, Supporting, and Motivating Reproducible Science}
\label{ch:hci_role}

Our work over the past three years focused on the study of interactive tools for reproducible science. It further involved components of practice, as I organized and conducted usability tests, implemented tailored science badges, provided feedback to software developers, and supported the project management of CAP and COD. In this chapter, we aggregate findings from our various activities and provide guidance for HCI practitioners and researchers in designing effective tools for comprehensive RDM. 

First, we introduce a \CommitmentModel. The model increases our understanding of how to design systems effective at supporting and motivating comprehensive RDM. We further present a conceptual model of components and interactions involved in RDM tools that illustrates the interplay of our findings and research threads. 

Next, we summarize and depict the role of both HCI practitioners and researchers through design recommendations and emerging research challenges. Defining and establishing those roles is particularly important as we recognize that today's availability of online technologies enables institutes, libraries, and service providers to develop platforms that support scientists in preserving and sharing their research \cite{Pasquetto:2016:ODS:2858036.2858543, Worden2017}. We argue that HCI methods are valuable assets in the systematic study and design of interactive tools for reproducible science. 

Finally, we depict our vision of \ac{URP}, which we envision to transform interaction with RDM tools. We describe URP and the role that HCI plays in its study and implementation.

\begin{tcolorbox}[title = {This chapter in general, and Sections 7.3 and 7.4 in particular, are conceptually based on the following publication. Sections 7.1, 7.2, and 7.5 are based on publications referred to in the corresponding sections.}]

Sebastian S. Feger, Sünje Dallmeier-Tiessen, Paweł W. Woźniak, and Albrecht Schmidt. 2019. The Role of HCI in Reproducible Science: Understanding, Supporting and Motivating Core Practices. In Extended Abstracts of the 2019 CHI Conference on Human Factors in Computing Systems (CHI EA '19). ACM, New York, NY, USA, Paper LBW0246, 6 pages.
    \newline\url{https://doi.org/10.1145/3290607.3312905}

\end{tcolorbox}

\section{The Personal RDM Commitment Evolution Model}
\label{section:evolution_model}

We presented our cross-domain study on practices around RDM and reuse in Chapter \ref{ch:cross_domain}. Based on the findings from this study, we introduce and discuss the \CommitmentModel. We decided to introduce the model in this concluding Part \ref{part:conclusion}, as it ultimately relates to findings from all studies presented in this thesis. In addition, the model underlines the important role of HCI in supporting and motivating reproducible science practices. 

\begin{figure}
\centering
  \includegraphics[width=1.0\columnwidth]{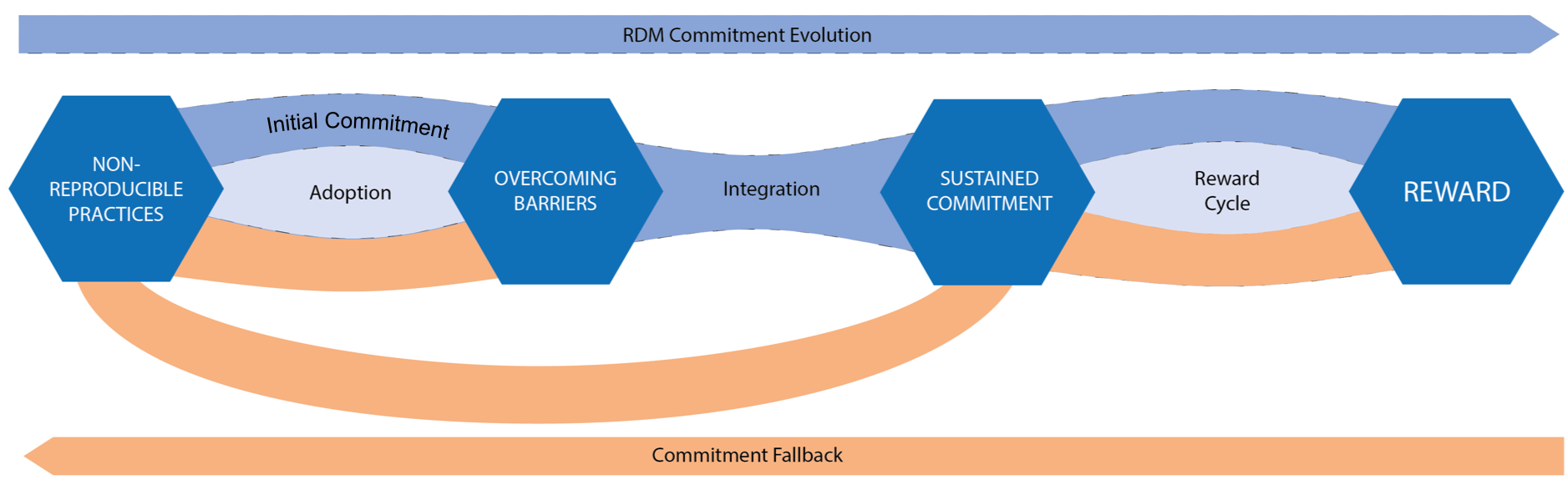}
  \caption{The \CommitmentModel.}~\label{fig:rdm_model}
\end{figure}

The \CommitmentModel, depicted in Figure \ref{fig:rdm_model}, improves our understanding of how researchers transition from \textbf{non-reproducible practices} to \textbf{sustained commitment} for comprehensive RDM. It emphasizes the role that institutional and scientific frameworks play in the \textbf{adoption} of initial commitment. The model further reflects growing complexity and demands for RDM and effective reuse in data-intensive computational science and highlights the value of suitable cyberinfrastructure and education to \textbf{overcome barriers}. The \CommitmentModelTrailingSpace places particular emphasis on \textit{motivation} in RDM. Commitment must be met by meaningful \textbf{rewards}. The continuous stimulation of this \textbf{reward cycle}, as well as a steady support related to adoption and barriers, contributes to researchers' \textit{commitment evolution}. Instead, a lack of encouraging and supporting socio-technical frameworks, and meaningful incentives, likely leads to a \textit{commitment fallback}. 
We argue that the model provides an additional dimension to our understanding of ``human interventions in relation to data'', as described by Muller et al. \cite{Muller:2019:DSW:3290605.3300356}. In particular, it suggests that additional interventions (i.e. documentation, preservation, and sharing) must be \textit{incorporated} into the data and analysis lifecycle, rather than added \textit{retrospectively}. One of the key differences to the interventions described by Muller et al. lies in the components of \textit{motivation}: Muller et al. described practices that are part of the analysis process, and thus become part of a reputation or reward economy --- be it a scientific or industrial one. As our research shows, the same is not true for interventions that fall within RDM. Our findings and the RDM commitment model are not limited to data-intensive computational analyses. Rather, they are based on studies with researchers from numerous fields of science, including biology, chemistry, meteorology, geology, and foremost physics. Based on the study of practices and requirements with scientists from key organizations and branches of science, we expect to provide guidance to the wider scientific community.

In the following, we define each stage of the \CommitmentModelTrailingSpace and describe infrastructure components involved. References to \textsc{themes} and statements from study participants refer to the Findings section (Section \ref{section:cross_findings}) of the cross-domain study (Chapter \ref{ch:cross_domain}). We describe the role of scholars and practitioners in facilitating and stimulating transition between individual stages and discuss them in the context of findings from related work.

\begin{tcolorbox}[title = This chapter is based on the following publication.]
Sebastian S. Feger, Paweł W. Woźniak, Lars Lischke, and Albrecht Schmidt. 2020. ‘Yes, I comply!’: Motivations and Practices around Research Data Management and Reuse across Scientific Fields. In Proceedings of the ACM on Human-Computer Interaction, Vol. 4, CSCW2, Article 141 (October 2020). ACM, New York, NY. 26 pages.
\newline\url{https://doi.org/10.1145/3415212} 

\color{darkgray}
------------------------------------------------------------------------------------------------------\newline
Several of the study's resources are openly available as supplementary material in the ACM Digital Library.
\end{tcolorbox}

\subsection{Non-Reproducible Practices}

Researchers in a wide variety of scientific domains are unable to reproduce published work, including their own \cite{Baker2016}. This points to a universal challenge related to comprehensive RDM. In our HEP study, reported in Chapter \ref{ch:requirements}, we found that researchers perceive a mismatch between the effort needed to commit to effective RDM and the personal gain for doing so. The cross-domain study in Chapter \ref{ch:cross_domain} reported on qualitative findings from a wide variety of scientific fields. Findings related to \textsc{Practice} described non-reproducible practices in a wider scientific context. We referred to exemplary and representative statements of participants from Biology (P1) and Meteorology (P2):

\begin{quote}
My concerns would be that it wouldn't be taken up by scientists because they think it’s too much work on top of their normal work. [...] If it's made clear that it doesn't cost extra time and that it saves time in the end, I don't know, by presenting a good use case or so, then it should be fine. Otherwise, people may remain skeptic. (P2)

[...] if they even send their data analysis protocol with the data. [...] Maybe it's even because they're keeping it secret for their purposes, maybe.
\end{quote}

Researchers and research data managers described various drivers for commitment towards more open and reproducible practices. As our findings show, \textbf{initial commitment} is stimulated by three primary factors:

1) The initial commitment can be a direct result of \textbf{policies and organizational rules}. Policies may be issued by different stakeholders and even provide a motivation for institutes to aim for transparent and open RDM, as statements from P10 (Agricultural Science) and P15 (Environmental Science) in \textsc{Impact} underline. Conferences and journals also started to encourage and even enforce sharing of research resources \cite{Belhajjame2014, Stodden2014}. Participants stressed that publishers and funders often prescribe specific tools. We referred to statements from a policy offer (P6) and a Biology / Chemistry researcher (P7) in \textsc{Adoption}. P7 contrasted time pressure with efforts imposed by such tools.

2) Researchers reported that the unavailability of research material, the need for direct and unstructured communication with colleagues to exchange resources, and the inability to reproduce research lead to frustration. As a result, researchers develop an \textbf{intrinsic motivation to make a positive contribution} to the research community by following important RDM practices. This is particularly reflected in our HEP studies. In our first study on HEP practices and requirements (Chapter \ref{ch:requirements}), P2 exemplarily stated: \textit{``I want everyone else's analyses to be there and equally that means that they might want my analysis to be there.''} In the later study on two gamification prototypes and design requirements, another participant referred to as P2 summarized this notion very well: \textit{``I have a strong philosophical interest in sharing things with the collaboration anyway.''}

3) Finally, researchers commit to open and reproducible science practices when they \textbf{expect to be rewarded} within the academic reputation economy. Most interviewees referred to increased visibility and impact on the citation count as core motivators. We referred to a statement from P8 who studied effects of opening up an academic repository to a wider audience and who concluded that visibility and citations are ``the best known to encourage them to submit.''

%
%
%

These three drivers of initial commitment are not exclusive. In fact, it is likely that the initial commitment is based on a mix. In the model, we refer to those drivers as \textit{Initial Commitment} in the transition from Non-Reproducible Practices to Overcoming Barriers. While we might have little control over policies, we expect to lower data management efforts through the design of tailored and supportive data management tools. By doing so, we can lower the overall threshold for the initial commitment to RDM.

\subsection{Overcoming Barriers}

Researchers and data managers described barriers in the adoption of comprehensive RDM practices. As we described in the \textsc{Education} and \textsc{Barriers} core concepts, many of them are deeply rooted in the interaction with technical infrastructure or the lack of suitable infrastructure. In this context, most informants stressed that researchers often do not know where to start and which tools are available. In \textsc{Education}, we referred to a representative statement of a Geoinformatics researcher / data manager (P14) who focused on the reuse of computational workflows. As a result of lacking education and awareness of suitable infrastructure, researchers adopt \textit{haphazard} practices (P4 and P7 in \textsc{Education}) that lead to unstructured archives on storage drives (P3 in \textsc{Education}). Overcoming those barriers represents a serious obstacle in the RDM commitment evolution.

Data managers and service providers can support researchers overcome barriers. They need to make sure that tools are designed and available that enable researchers to cope with common challenges. As findings in the \textsc{Barriers} core concept showed, one thread of common challenges relates to formats and interoperability. We selected representative statements that hint to conflicts between analogue and digital data (P3, Arts and Curation), the treatment of difficult-to-process formats (P11, Research Image Reuse), closed data (P10, Agricultural Science), and proprietary standards (P1, Biology). Tools need to be able to process and translate between formats that are common in the target domain. This includes data at all stages of the data and research cycle. Given the challenges that growing data volumes and computational reusabilty pose, RDM tools should provide widely accessible mechanisms that enable preservation of reproducible computational research. Tools like the REANA analysis platform provide accessible starting points.

The effectiveness of the support --- both in terms of human and technology support --- in overcoming the initial barriers impacts the commitment evolution. In case drivers for commitment are not strong enough, or the barriers prove to be insuperable, researchers are likely to give up their attempts to overcome the barriers and return to the first stage. When they overcome the barriers, they integrate the adopted practices into their research workflows.

\subsection{Sustained Commitment and Rewards}

Researchers successfully integrated RDM practices into their workflows when they arrive at the Sustained Commitment and Reward stages. Yet, the drivers of commitment need to be maintained, in order to sustain commitment. As we discussed earlier, participants described citations as one of the key motivators for adopting open practices. A clear impact on a researcher's citation count is likely to confirm the initial motivation and stimulate transition to the last stage: the reward stage. Study participants referred to additional types of rewards in \textsc{Impact}. For example, the ability to track and demonstrate reuse can provide strong arguments in the interaction with funding agencies. However, we note that those are benefits that participants discussed with a focus on institutions, rather than individuals.

In our HEP requirements study (Chapter \ref{ch:requirements}), we discussed \textit{secondary usage forms} of preservation technology. We described those as uses that are not part of the core mission of such tools. Instead, they provide contributors with meaningful benefits. We found that in HEP, secondary usage forms relate to uncertainty coping, fostering of collaboration, and the integration of automated and structured workflow processes. We stressed that the secondary uses benefit foremost researchers who actively contribute their work ad-hoc during the research lifecycle. In terms of the transition between the sustained commitment stage and the reward stage, this is an important consideration. Services and RDM tools that offer meaningful \textit{secondary usage forms} are likely to more frequently stimulate this transition.

In \textsc{Practice}, we noted that most participants contrasted the extra effort required to meaningful benefits and use cases. We related to statements from two interviewees who advocated ``presenting a good use case'' (P2, Meteorology) and supporting the meta-data extraction of preserved documents in a structured way (P3, Arts and Curation). However, we were not able to map common \textit{secondary uses} across scientific domains. In contrast to our HEP study, the cross-domain study was not tailored to a single branch of science, but included participants from a wide variety of scientific domains. While this allowed to present a more universal understanding of practices, needs, and requirements around RDM and reuse in science, the findings do not strongly contribute to the systematic description of secondary usage forms. We argue that the character of such uses is strongly dependent on individual domains, thus necessitating focused studies with a greater number of participants from within a specific domain of interest. 

\subsection{Model Implications}

Understanding the stages and transitions on the road to sustained open science poses a number of challenges and requirements for future systems. In the following, we provide an overview of model implications.

\textbf{We propose to further explore meaningful forms of motivation and rewards.} We stressed before that we recommend systematic investigations of \textit{secondary usage forms} across a wide set of scientific branches. Our model suggests that receiving continuous rewards is more likely to provide sustained commitment for RDM than abstract and long-term future rewards. \textbf{We further advocate the study of \textit{game} design elements in the context of the reward cycle.} Badges, in particular Open Science Badges, have shown to encourage sharing of research material \cite{Kidwell2016}. ACM introduced a set of badges that respond even better to challenges in computational research \cite{boisvert2016incentivizing, acmbadgesweb}. Findings from our research on gamification in reproducible science showed that game design elements can enable new forms of interaction with preserved research. Future work on gamification in the context of RDM is likely to address findings from Huang et al. \cite{huang2013meanings}. They investigated meanings and boundaries of scientific software sharing and found that ``what is important is not simply making more software available, but addressing issues of navigation, selection and awareness.'' Related to the motivational component of RDM, finally, we consider it important to \textbf{study differences and commonalities of motivational drivers involved in the \textit{Adoption} and \textit{Reward Cycle} transitions}.

\textbf{Improve communication around research artefacts.} We found that enabling communication around preserved material is a key driver of secondary usage forms (Chapter \ref{ch:requirements}). Birnholtz and Bietz \cite{birnholtz2003data} described similar findings related to the design of systems that support science data sharing. They noted that ``[...] the sharing of data follows the paths established by existing social networks. Thus, one possible way to encourage data sharing behaviour may be to provide facilities for communication around shared data abstractions.'' The need for information is also reflected in our cross-domain study. We argue that \textbf{communication facilities should be implemented that integrate the wider research ecosystem}. Such communication strategies go beyond information exchange of researchers and should, where appropriate, integrate the wider framework of stakeholders, including collaborating research groups, industry partners, and commercial vendors.

Finally, our data and model suggest that service designers and research data managers must \textbf{find a balance between enforcement and initial drivers of motivation}. We described enforcement at various levels, from personal instruction by supervisors, to general funding policies. In combination with meaningful goals, they can stimulate initial commitment. However, the model further implies that the impact of regulations decreases as researchers continue to commit to RDM practices. One can easily think about practical applications of this implication. For example, organisational policies could mandate preservation of a set of clearly defined resources that make up certain fields of a tailored preservation service like CAP. The policy would likely be effective at enforcing researchers to preserve and share common denominators of research at a given organisation. But, as a single instrument it is prone to fail in adapting to \textit{novelty and creativity} in science. Policies might provide researchers with little motivation to encourage and advise implementation of highly specific features in the preservation service, or promote documentation and sharing beyond what is required. Thus, we suggest that not only do policies shape technologies, and technologies shape policies \cite{Pasquetto:2016:ODS:2858036.2858543}, but a mix of motivational drivers and policies shape technologies and RDM commitment at different stages in the commitment evolution.

In conclusion, we propose to use the \CommitmentModelTrailingSpace to understand \textit{why} researchers commit to open and reproducible science. We argue that the model provides valuable guidance in assessing and providing drivers of commitment, from early adoption to sustained RDM practices. In the following section, we describe a conceptual model that explains \textit{how} interactive tools for reproducible science impact drivers of commitment and rewards.

\section{Towards a Conceptual Model of User-Centered Design in Reproducible Science}

\label{section:ucd_model}

\begin{tcolorbox}[title = This chapter is based on the following publications.]

\begin{itemize}
    \item Sebastian Feger and Paweł W. Woźniak. 2019. More Than Preservation: A Researcher-Centered Approach to Reproducibility in Data Science. Accepted and presented at the \textbf{CHI 2019} Workshop on \textit{Human-Centered Study of Data Science Work Practices}. Published on CERN CDS. 4 pages. \newline \url{http://cds.cern.ch/record/2677268}
    
    \item Sebastian Feger, Sünje Dallmeier-Tiessen, Pamfilos Fokianos, Dinos Kousidis, et al. More than preservation: Creating motivational designs and tailored incentives in research data repositories. 2019. Peer-reviewed, accepted presentation proposal for a full talk at \textbf{Open Repositories 2019}. Published on CERN CDS. 5 pages. \url{https://cds.cern.ch/record/2691945}
\end{itemize}
\end{tcolorbox}

\begin{figure}
  \centering
  \includegraphics[width=1.0\columnwidth]{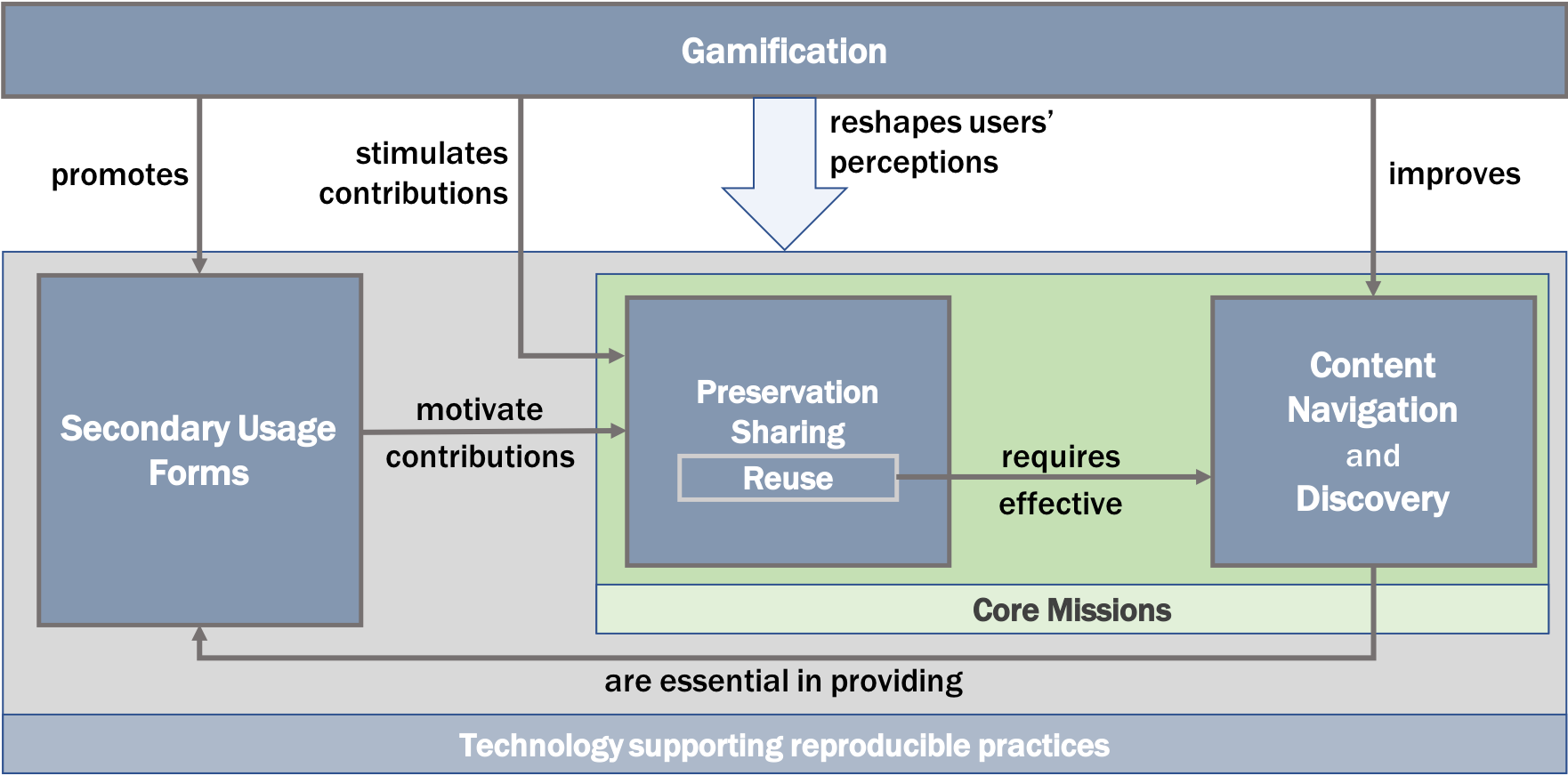}
  \caption{Conceptual model of components and interactions involved in RDM tools: Gamification and Secondary Uses support and motivate RDM.}~\label{fig:bigger_picture}
\end{figure}

Our work shows that UCD for research expands the design space of RDM tools. Figure \ref{fig:bigger_picture} shows a conceptual model of components and interactions involved in RDM tools. This model is based on the research we presented in this thesis. It depicts the interplay and mutual dependencies between the core missions of RDM tools, secondary usage forms, and gamification.

The data management tool and its core missions is at the center of the model. Tools like CAP need to enable effective and efficient preservation, sharing, and reuse of research (see Sections \ref{section:needs_req_repro_background} and \ref{section:cap_framework}). The ability to reuse artefacts depends on suitable navigation and discovery mechanisms. This is especially true in large research environments. Thus, we consider \textit{providing strong navigation and discovery components} core missions of RDM tools.

Our model references two additional key elements that illustrate the interplay and dependencies between components of RDM tools: secondary usage forms and gamification. In Chapter \ref{ch:requirements}, we learned that \textit{secondary usage forms} motivate contributions to preservation tools. In particular, uses related to coping with uncertainty, providing collaboration-stimulating mechanisms, and automation rely on thorough and structured documentation, preservation, and sharing. Thereby, they directly support core missions of RDM tools. This is a bi-directional dependency, as depicted in our model in Figure \ref{fig:bigger_picture}. Secondary uses rely on means to navigate and discover content. Secondary uses related to uncertainty coping and collaboration-stimulation require effective means for content interaction, as they rely on discovery and communication within the research community. Hence, the mutual dependency between core missions and secondary usage forms.

We extensively studied gamification in the context of RDM. In Chapter \ref{ch:gamification_requirements}, we reported on our study of two gamified RDM service prototypes. We found that HEP researchers considered a variety of game design elements suitable to increase their visibility within the large research collaborations. The participants stressed that the gamification layer could positively impact their careers. Thus, we expect that the prospect of increased visibility stimulates contributions to RDM tools. Our mockups and findings also demonstrated how game design elements can promote communication, with particular regard to secondary usage forms that rely on information exchange. Based on those findings, we implemented and evaluated tailored science badges in CAP (see Chapter \ref{ch:tailored_badges}). Our findings showed that tailored badges not only stimulate contributions. Rather, they enable new forms of interaction with research repository content. Study participants expected them to represent primary means for discovering educational, innovative, and reusable work. Thus, game design elements improve content navigation and discovery. They further reshape users' perceptions of RDM tools, as they are effective at illustrating uses of research content. Cases in point are the badges and corresponding lists of analyses on the service dashboard. 

The conceptual model of UCD in reproducible science shows that the findings and research topics in this thesis are closely interconnected. Based on this model, we suggest that researchers and designers investigate how components and interactions involved in RDM tools depend upon and enable each other. In this context, we need to stress that our findings on gamification and secondary uses stem primarily from HEP. We do not claim that the model represents a complete analysis of all components involved in the design of RDM tools. Rather, we encourage designers and researchers to build upon and expand this model based on findings beyond HEP. In the following section, we present accessible starting points for HCI practitioners.

\section{The Role of Practitioners}
\label{section:role_practitioners}

The design and operation of tools that support reproducibility profits from the involvement of HCI practitioners at all stages of the technologies' life cycle. 
Based on our involvement in the development and support of preservation technology in HEP, we present recommendations for the design of RDM tools that we consider particularly relevant for HCI practitioners:

\textbf{\textit{Map Practices.}} As part of a platform's design, research workflows need to be well understood. Applicable to both platform types, this is particularly important for \textit{tailored} tools. Service developers need to involve target communities in the design process, to map submission, search, and reuse needs.
    
\textbf{\textit{Lower Efforts.}} Given that the effort to document and share research is a main barrier, data description mechanisms must be supportive. Well-designed submission forms, as well as auto-suggest and auto-complete mechanisms that build on knowledge of research workflows are essential.

\begin{figure}
  \centering
  \includegraphics[width=0.7\columnwidth]{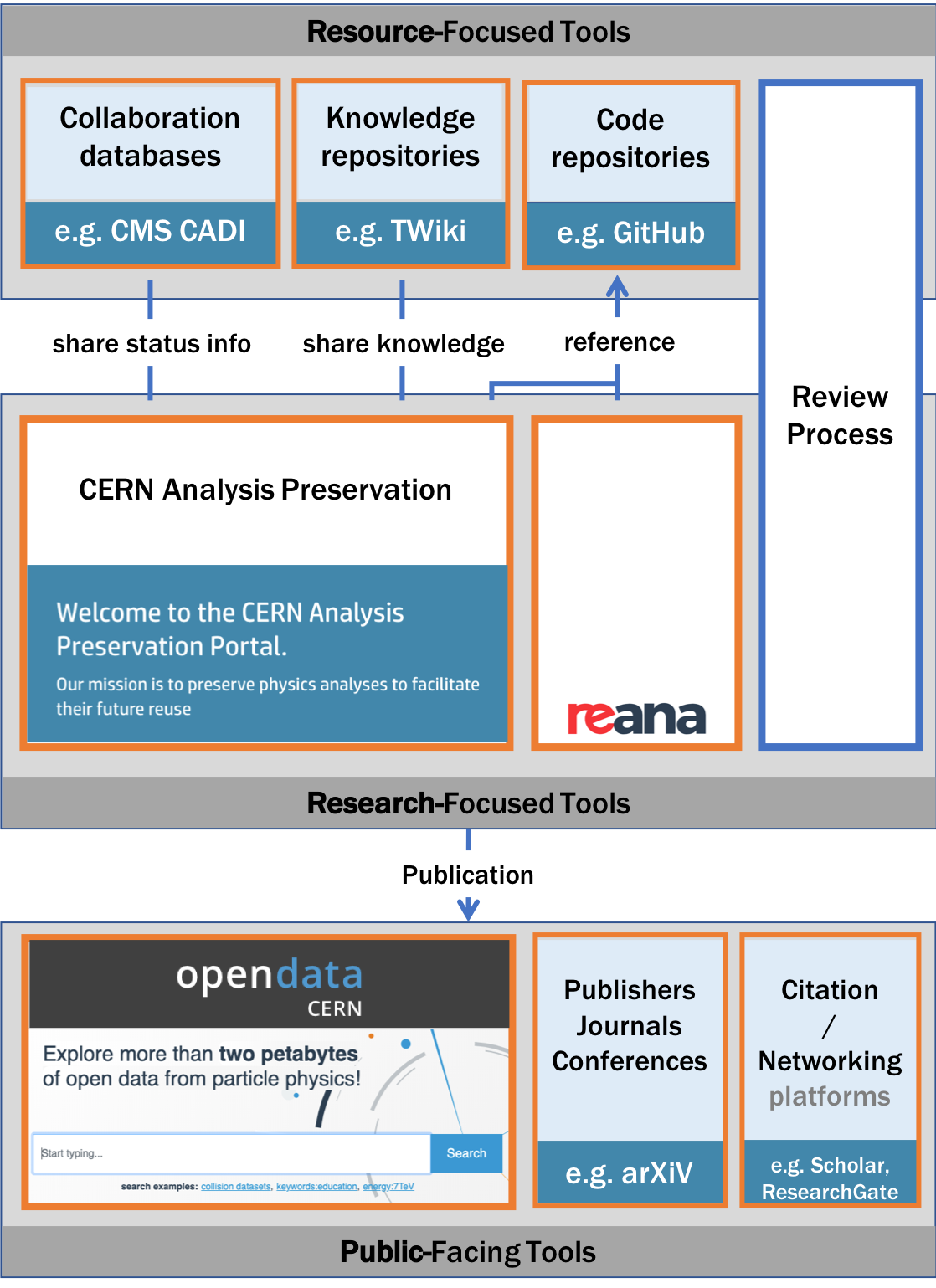}
  \caption{Illustration of the wider ecosystem of science infrastructure. Based on three types of tools: Resource-Focused, Research-Focused, Public-Facing.}~\label{fig:ecosystem}
\end{figure}

\textbf{\textit{Integrate with existing tools.}} Understand the architecture and interplay of existing tools across the research, preservation, and publication layers in the target domain. Develop tools and services that integrate into this wider ecosystem of research tools, in order to provide meaningful and seamless interaction. Based on our empirical studies, Figure \ref{fig:ecosystem} provides a systematic mapping of tools and connections in HEP. We distinguish between tools on three different layers. First, \textit{resource}-focused tools manage a limited scope of research artefacts. Examples include code repositories like GitHub and knowledge repositories like TWiki. Second, \textit{research}-focused tools like CAP and REANA manage analyses in their entirety. They reference and make use of research artefacts managed by \textit{resource}-focused tools. Our research showed that internal review processes concern information managed within both layers. Finally, \textit{public}-facing tools include open data repositories like COD, as well as scientific networking and publication services. We found that implementing a \textit{research}-focused service layer that interfaces with \textit{public}-facing tools can profit researchers, as this connection provides opportunities to lower efforts and increase transparency. In conclusion, it is crucial to integrate RDM tools into the wider ecosystem of science infrastructure. Interfaces between the various services and layers enable effective and efficient sharing of research artefacts that benefit researchers and scientific communities.

\textbf{\textit{Think beyond data management.}} Enabling effective RDM is the core mission of the tools that practitioners design. However, investing time and effort into understanding how the interaction with RDM tools can be motivating and rewarding is crucial. While mapping practices in the target domain, place particular emphasis on understanding \textit{secondary usage forms}. Those will be essential to engage the scientific community.

\textbf{\textit{Ensure Usability.}} Services need to be tested with users as part of the design process, to improve their usability and to detect barriers. This has to be a continuous process, as research description templates on tailored services need to be adapted to novelty and creativity in science.
    
\textbf{\textit{Provide an Interface.}} Given the close involvement of practitioners with the research community, HCI practitioners are in an ideal position to interface between technology developers and researchers. Main responsibilities should include: promotion of infrastructure developments to the research community; and feedback communication to the development team.

We provided an overview of how HCI practitioners can and need to systematically support the design of supportive RDM tools. In the following section, we complete our discussion of responsibilities and roles of HCI in reproducible science by presenting emerging challenges for HCI researchers.

\section{Emerging Research Challenges}
\label{section:role_researchers}

The involvement of HCI practitioners in the design and operation of services for science reproducibility allows to create a most supportive and efficient interaction with technology. Yet, minimizing the effort is not necessarily enough to engage scientists at large~\cite{Borgman2006}. HCI research has a unique opportunity to impact reproducible practices through the systematic study of requirements and incentive structures:

\textbf{\textit{Understanding Requirements.}} A pillar of HCI research is based on studying requirements of user groups and populations. In the context of reproducible research, we need to further understand the requirements and connections of scientific communities and individual researchers to preservation, sharing, infrastructure, and knowledge lifecycles \cite{Jirotka2006}. \textit{What new forms of community interaction do platforms provide? What role do policies play in the design and interaction with technology for reproducible research? \cite{Pasquetto:2016:ODS:2858036.2858543} What common requirements apply to diverse forms of research, including computational, qualitative, and descriptive research? And how do requirements differ between these different research forms?} Jackson and Barbrow \cite{Jackson2013} stressed the value of studying requirements in field-specific investigations, pointing out the ``need to supplement or replace generic, tool-centered, and aspirational accounts of cyberinfrastructure development with approaches that start from the individual histories of practice and value in specific scientific fields.''

\textbf{\textit{Incentives / Rewards.}} Research repositories often advertise opportunities to increase citation counts. They emphasize that ``the potential of data citations can affect researchers' data sharing preferences from private to more open'' \cite{Garza:2015:FCD:2783446.2783605}. 
Rowhani-Farid et al. \cite{Rowhani-Farid2017} reported a lack of incentives in their systematic literature review in the medical domain. They noted that even though ``data is the foundation of evidence-based health and medical research, it is paradoxical that there is only one evidence-based incentive to promote data sharing.'' They referred to open science badges \cite{Kidwell2016} and concluded that ``more well-designed studies are needed in order to increase the currently low rates of data sharing.''

A fundamental understanding of researchers' needs enables description and implementation of new incentive structures. In our research in HEP, we found that technology can create meaningful incentives that profit contributing scientists in their work \cite{Feger2019Requirements}. Our first interview study points to \textit{secondary usage forms} of preservation technology that can support coping with uncertainty and stimulating useful collaboration. As our research shows, tailored systems can particularly profit from a systematic study of incentives. They are adapted to a particular scientific domain and require a thorough design process in order to provide tailored submission and reuse mechanisms. Extending the research and design process to study and implement also tailored incentives and rewards seems desirable. However, not all institutions and (specialized) research domains can afford implementation and support of highly tailored systems. Thus, future research needs to study more general frameworks for incentives that can be applied across different research forms.

\textbf{\textit{Motivational Design.}} Badges, one of the most common game design elements, have shown to encourage research data openness in the Psychological Science journal \cite{Kidwell2016}. ACM announced the introduction of an even larger set of badges that aim to incentivize reproducible practices \cite{boisvert2016incentivizing}. 
Yet, we have limited knowledge about needs and constraints of gamification in highly skilled scientific environments \cite{feger2018just}. Our research on gamified prototypes of a preservation service in HEP showed that gamification can provide motivation if scientific practices are reflected in the design \cite{Feger2019Gamification}. We contrasted two prototypes in a mixed-method study. While one made use of most common game design elements (including points and leaderboards), the other used a more informative language. Both were rated persuasive and suitable by the experimental physicists. They highlighted how game mechanisms can provide motivation through a fair representation of contributions and best practice efforts. Our research on tailored science badges showed that game design elements can further provide meaningful new forms of interaction in research repositories. We consider large-scale and long-term studies of gamification in reproducible science as promising direction of research.

In this chapter, we introduced and described two models that are based on findings from four empirical studies. We further described the roles and responsibilities of both HCI researchers and practitioners in designing supportive and motivating tools for reproducible science. We expect that our models and descriptions of responsibilities and challenges will lead to the design of more supportive generic services and stimulate the development of tailored RDM tools. To conclude this chapter, we venture an outlook into ubiquitous knowledge preservation strategies. We discuss our vision of how future RDM tools could be directly integrated into research workflows in the next section.

%
%
\section{Making Digital Research Preservation Ubiquitous}
\label{section:urp}

To conclude this chapter and to push the boundaries of interaction with RDM tools, we 
introduce Ubiquitous Research Preservation (URP), which we envision to automate preservation in computational science. Based on our research, we contribute a characterization of preservation processes, illustrate the spectrum of technology interventions, and describe research challenges and opportunities for HCI in the implementation of URP in computation-based scientific domains.


\begin{tcolorbox}[title = This section is based on the following publication.]
Sebastian S. Feger, Sünje Dallmeier-Tiessen, Pascal Knierim, Passant El.Agroudy, Paweł W. Woźniak, and Albrecht Schmidt. 2020. Ubiquitous Research Preservation: Transforming Knowledge Preservation in Computational Science. MetaArXiv Preprint. 4 pages.
\newline \url{https://doi.org/10.31222/osf.io/qmkc9}
\end{tcolorbox}

\subsection{Motivation and Background}

Oleksik et al. \cite{Oleksik2014} reported on their observational study of electronic lab notebooks (ELN) in a research organization. They found that the flexibility of digital media can lead to much less precision during experiment recording and that \textit{`freezing'} parts of the record might be necessary. The authors stressed that \textit{``ELN environments need to incorporate automatic or semi- automatic features that are supported by sophisticated technologies [...].''} 

Studying the use of a hybrid laboratory notebook, Tabard et al. \cite{tabard2008individual} found that \textit{``users clearly do not want to focus on the process of capturing information.''} Yet, they also noted that automated mechanisms can be intrusive and that users need to be in control of the recording and sharing. They illustrated the importance of reflection in the scientific process and highlighted how access to preserved, redundant information supports reflection, as \textit{``scientists understand how their thoughts have evolved over time.''} 


Kery et al. \cite{Kery2018} asked scientists to think about ``a magical perfect record'' in their study of literate programming tools. Participants created queries referring to ``many kinds of contextual details, including libraries used, outputs, plots,  [...].'' Participants described their inability to find prior analyses and illustrated consequences. The authors found that in literate programming tools, ``version control is currently poor enough that records of prior iterations often do not exist.''

\subsection{Technology Interventions for Research Preservation}

To describe the spectrum of technology intervention in the preservation of machine-processed research, we characterize, based on our empirical findings, preservation efforts from a researcher point of view. Researchers commonly document, preserve, and share information and resources in lab notebooks, cloud services, or dedicated research preservation services (e.g. Figshare and Zenodo). Or, they decide to commit assets to repositories (e.g. GitHub). In either case, those actions are mostly \textsc{user-initiated}. Scientists who --- for any reason --- decide to preserve or share their research make a \textsc{conscious} selection of their study data and materials. In Figure \ref{fig:researcher_interaction}, we describe how we assigned those characteristics to the dimensions \textsc{Initiative} and \textsc{Resource Awareness}. 

\subsubsection{Towards Ubiquitous Research Preservation}

\begin{figure}
    \centering
    \includegraphics[width=1.0\columnwidth]{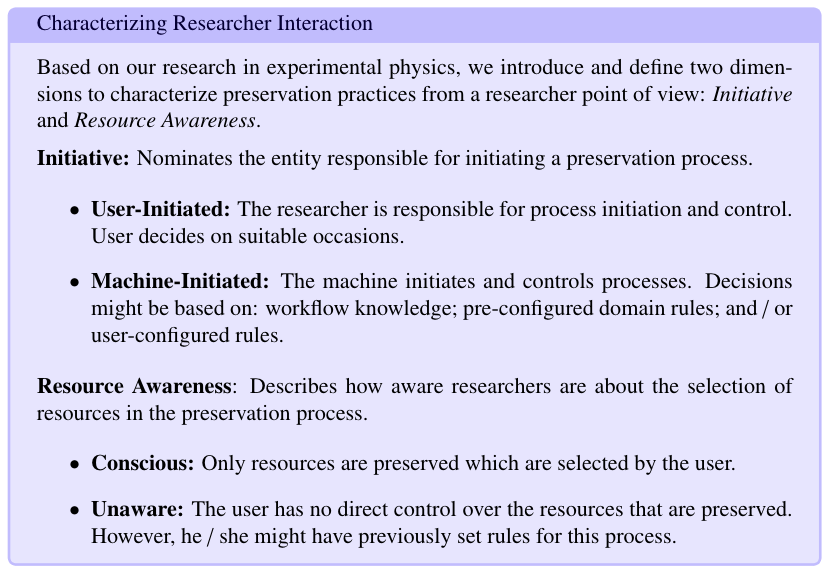}
    \caption{Researcher interaction based on Initiative and Resource Awareness.}
    \label{fig:researcher_interaction}
\end{figure}

In our first study on sharing practices in HEP, reported in Chapter \ref{ch:requirements}, we found that HEP data analysis work is based on common building blocks that foster implementation of automated recording and processing strategies. Related work presented similar notions of automated features \cite{Oleksik2014, tabard2008individual} and perfect records \cite{Kery2018}. We consider that the dimensions \textsc{Initiative} and \textsc{Resource Awareness} are suitable to develop a more formal description of automated preservation strategies. In contrast to current user-initiated preservation efforts, automated workflow recording could be entirely \textsc{Machine-Initiated}. Here, researchers might be \textsc{Unaware} of continuous background preservation efforts. Figure \ref{fig:researcher_interaction} provides a complete overview of the dimensions and described characteristics.

\begin{figure}
  \centering
  \includegraphics[width=1.0\columnwidth]{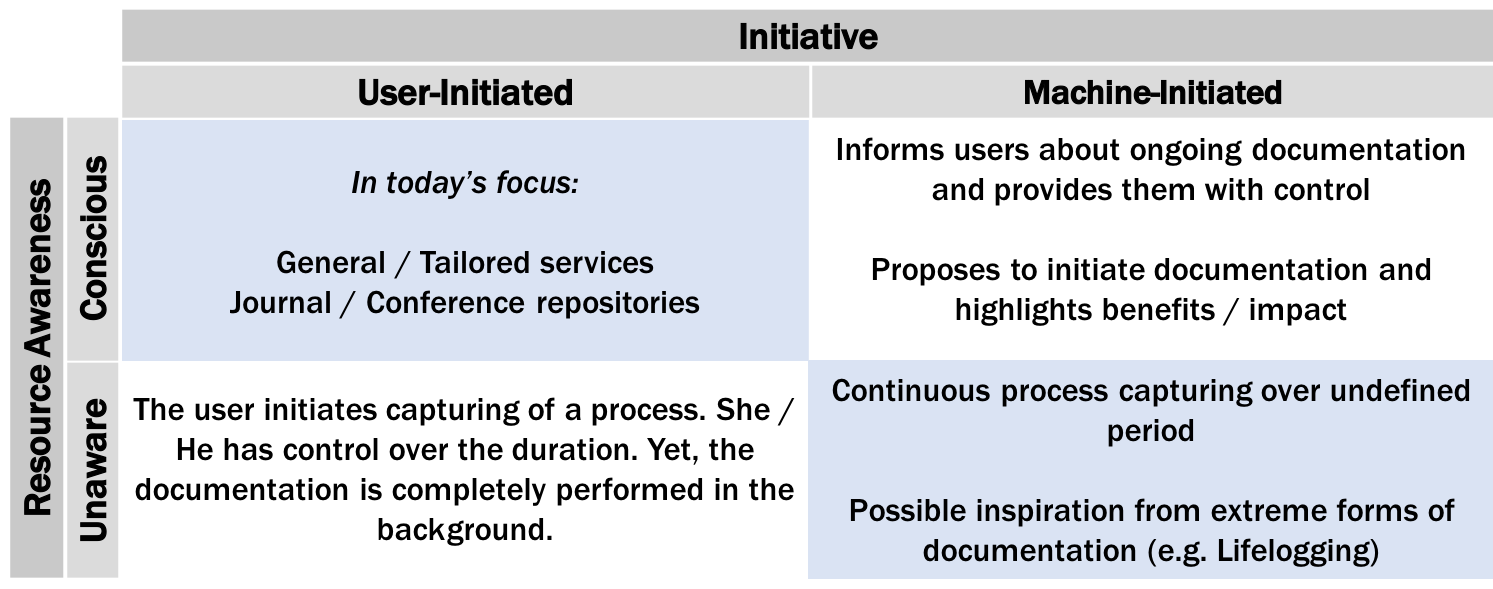}
  \caption{Spectrum of ubiquitous preservation technologies.}~\label{fig:dimensions}
\end{figure}

Described dimensions and characteristics enable a wide spectrum of technology interventions, as depicted in Figure \ref{fig:dimensions}. For example, technology could implement completely \textsc{machine-initiated/unaware} preservation of computational processes. Such an approach could guarantee (near-) continuous workflow recording, possibly taking inspiration from extreme forms of documentation like lifelogging. 

\begin{figure}
  \centering
  \includegraphics[width=0.45\columnwidth]{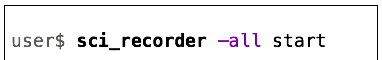}
  \caption{Speculative prototype of \textsc{User-Initiated / Unaware} interactions.}~\label{fig:terminal}
\end{figure}

Tabard et al. \cite{tabard2008individual} emphasized that \textit{control} is an important factor in research preservation. Technology supporting \textsc{user-initiated/unaware} interactions might make an important contribution towards acceptance. For example, a researcher who considers that a process could become relevant in the future could start an application or execute a command that initiates recording of computational states and changes (see Figure \ref{fig:terminal}). The researcher should be able to stop this process at any time.

\begin{figure}
  \centering
  \includegraphics[width=0.65\columnwidth]{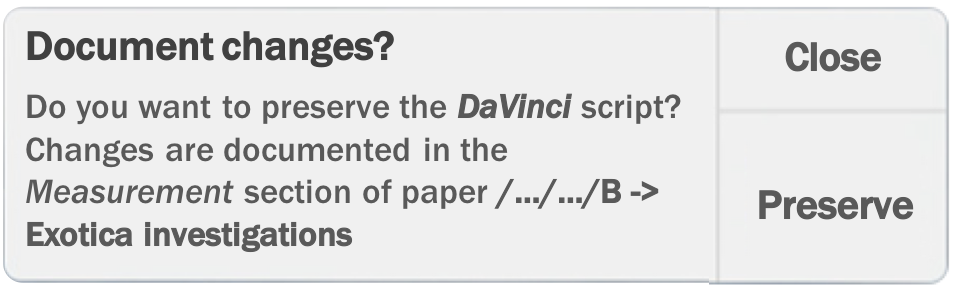}
  \caption{\textsc{Machine-Initiated / Conscious} interaction might provide needed control.}~\label{fig:notification}
\end{figure}

\textsc{Machine-initiated/conscious} interaction could provide researchers with control. Here, the machine might actively propose users to preserve certain processes. This decision would need to be based on pre-defined triggers or in-depth workflow knowledge. A researcher might receive a notification detailing the proposed initiation of a preservation process or activity (see Figure \ref{fig:notification}).

We refer to this spectrum of technology interventions for machine-supported recording of computation-based research workflows as \textbf{\textsc{Ubiquitous Research Preservation (URP)}}.

 \begin{tcolorbox}[
colframe=gray!70,
colback=blue!1,
coltitle=gray!10!black,
title = Definitions] 
   
  \textbf{Ubiquitous Research Preservation (URP)} refers to the machine-supported scientific knowledge recording and preservation process of computational workflows.
   
  \smallskip
  \smallskip
  
  \textbf{URP technology} \textit{initiates} and/or \textit{controls} partial or complete preservation.
  \end{tcolorbox} 
  
In the following section, we present key research challenges that need to be addressed to enable the design of URP technology.
  
\subsection{Research Challenges}
Our research and related studies hinted towards various challenges resulting from automated recording strategies. Here, we expand on challenges and opportunities for research on URP technology:

\smallskip

\textbf{Usefulness.} To create complete `magical records', preserved data need to be annotated, searchable, and suitable for desired use cases. It will be important to manage the signal-to-noise ratio, as well as to find suitable ways for information discovery and presentation.

\smallskip

\textbf{Generalizability.} As URP technology profits from knowledge about research practices for the recording and presentation of information, development of assistive technology across heterogeneous environments needs to be further researched. Research questions include: How can technology assess researchers' practices and needs and integrate into their workflows? Can we create accessible templates based on learned and confirmed structures? How does technology adapt to scientific novelty and creativity?


\smallskip

\begin{figure}
  \centering
  \includegraphics[width=0.65\columnwidth]{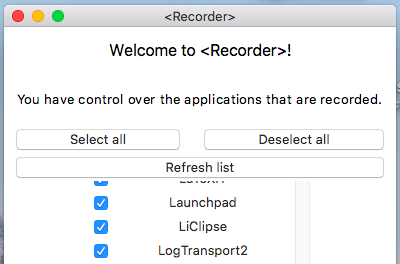}
  \caption{The <Recorder> captures screens and titles of selected applications.}~\label{fig:recorder}
\end{figure}

\textbf{Control.} Acceptance of URP technology will depend on researchers' perceived control over the preservation process. Figure \ref{fig:recorder} shows our \textit{<Recorder>} that continuously captures the screen and title of applications that the user selected for recording. Though we need to further evaluate the \textit{<Recorder>}, it is clear that researchers want to control capturing and sharing. This conflict between exercising control over the preservation process and desired automated preservation requires further study.

\smallskip

\begin{figure}
  \centering
  \includegraphics[width=0.65\columnwidth]{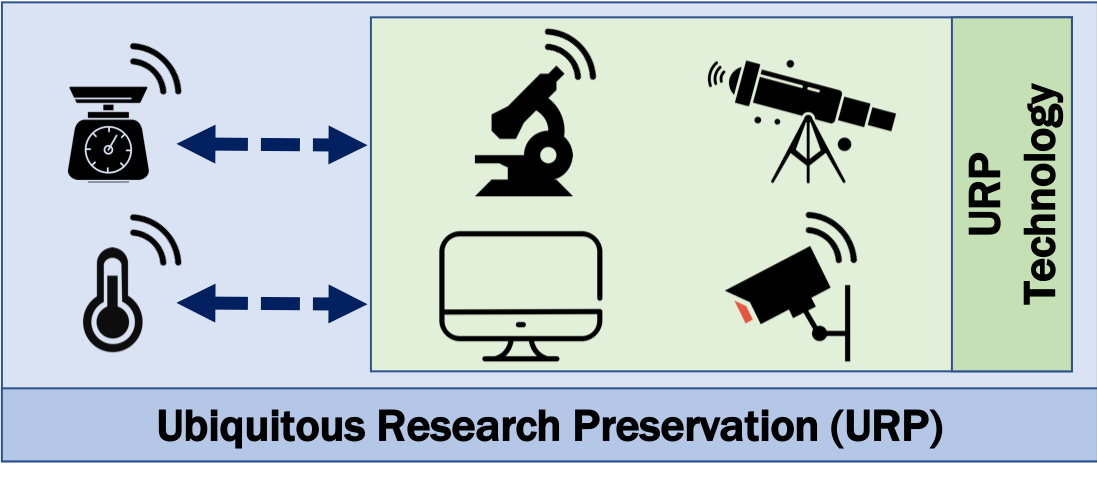}
  \caption{URP technology interaction architecture.}~\label{fig:landscape}
\end{figure}

\textbf{Integration.} The landscape of connected devices that measure, generate, or process scientific data is large and diverse. Devices range from desktop computers to microscopes and sensors. Integrating all those data sources into the preservation process poses further challenges regarding user control, network safety, and system architectures. As depicted in Figure \ref{fig:landscape}, some devices will implement URP strategies. And even though our examples and developments are mostly limited to computer applications, a wide variety of connected devices can offer URP by directly communicating with repository servers. Other devices can be connected to URP technology which acts as a proxy in the preservation process.

\subsection{Discussion and Conclusion}

We described our past and current efforts aiming to spark discussions and further research on machine-automated preservation in computation-based science. We illustrated a broad spectrum of technology interventions that we refer to as \textit{Ubiquitous Research Preservation (URP).} We expect URP to make a positive impact on researchers' ability to reflect on past processes, to provide training material, and to improve the reproducibility of their work. Yet, we do not intent to oversimplify complex use cases. Preservation is a first step towards supporting those, but it is not the only requirement. The decision to share resources does not only depend on the effort to preserve data, but on various other factors, including competition, fear of judgement, and privacy policies.

We described four major research challenges for the design and acceptance of URP technology. \textit{Usefulness} and \textit{control} will be important for the acceptance and use of URP systems. \textit{Generalizability} needs to be considered, to provide fast and wide access to URP tools and to include even branches of science and organizations that find it challenging to spend considerable resources on the development and adaptation of URP systems. Finally, the diverse landscape of connected, data-producing, or data-processing devices needs to be \textit{integrated} into URP systems. Developments and URP architectures must not be limited to computer applications.

We expect our vision of URP to profit computational science. It might bring us closer to ``a magical perfect record'' \cite{tabard2008individual}. Clearly, this would benefit HEP researchers who often spend years working on a particular analysis. Yet, benefits of URP will not be limited to data-intensive natural science. We argue that URP will likely benefit reporting, understanding, and transparency in all fields of science. Research today relies on computation. Scientists in all fields, including the humanities and social sciences, use computers to reason about related findings, prepare interview protocols, analysis data, and report their results. Similarly, URP is likely to profit computer users well beyond science.

\chapter{Conclusion }
\label{ch:conclusion}

In this thesis, we presented our research on interactive tools for reproducible science. This thesis represents the first systematic application of HCI methods and tools in studying, supporting, and motivating reproducible research practices. In this chapter, we first summarize our research contributions. We place particular emphasis on our researcher-centered definition of reproducibility and illustrate how our findings relate to the four key research questions. Next, we comment on how our findings inform discussions on the role of replication in HCI. Finally, we discuss limitations of our research.

\section{Research Contributions}

In Part \ref{part:intro}, we introduced our researcher-centered definition of reproducibility that reflects research practices and criteria relevant to scientists: \textit{\ReproDefinition} We expect this definition to be applicable across a wide variety of scientific domains, as it does not limit the purpose for (re-)using scientific work, nor does it prescribe constraints related to the users interacting with research artefacts. Instead, we argue, based on our empirical research, that the definition reflects what scientists care about most: having to overcome as little resistance as possible in sharing, accessing, and re-using scientific work. By reflecting interests and practices of researchers, the definition further contrasts common definitions of reproducibility and related terms that have ambiguous meanings across the sciences --- and sometimes even within the same field. We expect that adoption of our researcher-centered definition enables more focused and meaningful designs of interactive tools for reproducible science. In the following, we summarize our contributions and illustrate how they impact design thinking in science reproducibility. In particular, we illustrate contributions related to the four research questions: 1) Role of Technology in Supporting Reproducibility; 2) Practices and Design Requirements; 3) Stimulating and Motivating Reproducible Practices; and 4) Role of HCI in Reproducible Science. We conclude this chapter by commenting on the value of replication in HCI and by discussing limitations of our work.

\subsection{RQ1 --- Role of Technology in Supporting Reproducibility}
\label{ssection:rq1}

We investigated and articulated the role of technology in supporting reproducible research practices in various forms. In Section \ref{section:open}, we reflected on the value of openness in reproducible science and concluded that \textit{open is not enough}. We argued that in order to support core RDM practices, the sharing of resources must be systematically supported through appropriate tools. We related this reflection to the service infrastructure at CERN, underlining the motivation for our research on interactive tools for reproducible science. 

We dedicated Part \ref{ch:requirements} to the study of RDM practices and the role of technology in supporting those practices. First, we focused on HEP. In Chapter \ref{ch:requirements}, we reported on our study related to CAP at CERN. The participating physics data analysts stressed that technology plays a central role in sharing and managing analysis resources. In particular, they emphasized the role of e-mail communication and code repositories in their current research workflows. However, the interview participants highlighted that those tools do not sufficiently meet today's RDM requirements that are heavily impacted by growing data volumes and collaborations. We found that CAP is perceived as a tool that has the potential to address those challenges. Yet, adoption is subject to design requirements that we discuss in the context of RQ2.

To develop a wider understanding of the role of technology in reproducible science, we expanded the HEP study to a wide variety of diverse scientific fields, including biology, art and museum sciences, chemistry, geology, and agricultural research. Our findings showed that technology is instrumental in supporting RDM practices. Both, on the level of the individual researcher, as well as institutes and organizations. Similar to our findings in HEP, we found that current infrastructure is often inadequate for RDM tasks. However, this depends on individual institutes and fields of science. While some do not have the resources to develop adequate tools, others are in the process of reflecting practices, demands, and policies in their infrastructure design.

Based on the findings of our studies, we introduced the \CommitmentModel. This model underlines the importance of developing suitable RDM tools that help in \textit{Overcoming Barriers} (Stage 2) and creating \textit{Sustained Commitment} (Stage 3). 

\subsection{RQ2 --- Practices and Design Requirements}
\label{ssection:rq2}

Related to RQ1, we learned that tools play an important role in supporting reproducible practices. In our HEP study on RDM practices and design requirements in HEP (Chapter \ref{ch:requirements}), we found that researchers welcome a dedicated knowledge and analysis preservation tool like CAP. Still, they are worried about adoption within the community. The analysts stressed that the research community would be happy to use resources on CAP, but most scientists would be reluctant to contribute information as well. The participants stressed that supportive mechanisms provided by a tailored tool ease analysis preservation. Yet, only lowering efforts is not enough. Instead, we found that providing meaningful incentives has to be a key consideration in design thinking. In the context of CAP and HEP, we noted that analysts asked for support in overcoming some of the challenges they face in their research work. In particular, the communication and information architecture described by the study participants leads to uncertainty. That uncertainty relates to updates within the large LHC collaborations, the permanent preservation of datasets, and communication of data-related warnings and issues. We found that especially those researchers who contribute documentation and resources to services like CAP could profit from support in overcoming those challenges. Besides uncertainty, we characterized meaningful rewards related to automation, structured designs, and collaboration-stimulation. We referred to \textit{secondary usage forms}, to describe uses of RDM tools that, while not part of the core missions of a preservation tool, provide contributors with meaningful benefits. The ability to provide secondary uses appears directly connected to researchers' contributions. Thus, researchers benefit in a meaningful way from their efforts. Once secondary uses of RDM tools are well understood by the community, they might factor into the initial decision to contribute analysis resources and to transition from the first stage of our \CommitmentModelTrailingSpace (Non-Reproducible Practices) to the next. Profiting frequently from secondary uses likely contributes to sustained commitment for comprehensive RDM.

\subsection{RQ3 --- Stimulating and Motivating Reproducible Practices}
\label{ssection:rq3}

In terms of RQ2, we found that providing meaningful incentives and rewards is key in the design and implementation of supportive RDM tools. Based on this understanding, we investigated requirements for interaction tools that encourage and motivate behaviours and practices. In particular, we studied requirements for gamification in highly skilled science. We related to gamification as motivational tool in work environments and argued that differences in the socio-technical frameworks between enterprise employees and scientists necessitate dedicated research on gamification in science. We further recognized the need to base any developments on a thorough user-centered design process, as we cannot risk to alienate scientists with game mechanisms that are not perceived meaningful.

In Chapter \ref{ch:gamification_requirements}, we reported on the design of two contrasting prototypes of gamified research preservation services that are inspired by CAP. The prototypes make use of very different game mechanisms. The RID prototype focuses on a rational-informative communication that is based on activity overviews, contribution statistics, and progress bars. The SGED prototype uses most common competitive game design elements, including points, badges, and leaderboards. Our evaluation showed that both are considered valuable, enjoyable, suitable, and persuasive. We found that the physics researchers rated both prototypes positively overall. They considered that all mechanisms provide means to increase the visibility of their work and consequently their career prospects. However, given that impact, we found that the gamification layer needs to carefully reflect individual contributions and scientific practices. Based on our findings, we outlined mechanisms that explain the success of OSB in adopting and motivating sharing practices. We argue that this study represents the first systematic research on requirements of gamification in highly skilled science. 

Based on the findings from our gamification requirements research, we designed and implemented six badges in CAP (see Section \ref{ch:tailored_badges}). Those badges are closely connected to the CAP service and the HEP community. We introduced the notion of `tailored science badges' to reflect this special character and contrasted them with generic science badges like OSB and ACM badges. In the evaluation, researchers stressed that the tailored science badges foremost enabled new forms of interaction with preserved research. In particular, researchers expected the badges to impact content navigation and discovery. That way, the badges allow to increase the visibility of research which represents a motivation for scientists to thoroughly document their work on the service. The study participants did not strongly perceive the tailored science badges as a form of explicit motivation, but described motivation in an implicit form based on the ability to show relevant research more effectively. We described this effect as `beyond motivation' and argued that our findings pave new ways for gamification research in general.

Overall, findings related to RQ3 showed the potential for motivating reproducible research practices. Rewarding scientists with meaningful badges or similar achievements relates to stage four of the \CommitmentModelTrailingSpace (Rewards) and is key to create Sustained Commitment (Stage 3).


\subsection{RQ4 --- Role of HCI in Reproducible Science}
\label{ssection:rq4}

We reported on four empirical studies in this thesis. In total, 42 researchers and research data managers informed our findings. We consider our research process to represent the first systematic application of HCI methods in designing interactive tools for reproducible science. In addition, I took part in the regular CAP service design, verification, and adoption process over the course of three years. Based on the findings and experiences of this systematic research and design process, we illustrated how HCI can support and transform interaction with interactive tools for reproducible science. We contributed a \CommitmentModelTrailingSpace (Section \ref{section:evolution_model}) and a conceptual model of UCD in reproducible science (Section \ref{section:ucd_model}) that provide guidance in the design of interactive RDM tools. In Section \ref{section:role_practitioners}, we detailed the role of practitioners in mapping practices, lowering efforts, ensuring usability, and providing an interface between the research community and service developers. We described emerging research challenges for HCI scholars in Section \ref{section:role_researchers}. In particular, we emphasized research opportunities related to universal design requirements, incentives and rewards, and motivational design. Finally, we presented our vision of URP (Section \ref{section:urp}) and illustrated how HCI can take a lead in the development of tools that integrate RDM seamlessly into the research process. Based on the sum of experiences, findings, and contributions, we argue that HCI plays a crucial role in the design, development, and adoption of reproducible practices in science.

\subsection{Summary of Contributions}

We presented findings from four empirical studies that followed a systematic research process. We conducted 45 interviews with 42 distinct researchers and research data managers. Out of those, 30 researchers worked as HEP data analysts at CERN, a leading research organization in one of the most data-intensive branches of science. We were not allowed to offer reimbursement for study participation. Still, we managed to recruit highly skilled participants: out of the 42, a total of 30 participants held a doctoral degree, and seven were PhD students. Seven of the researchers who held a PhD had a particularly senior role as professor, team leader, or member of the upper management of the LHC collaborations. In total, we recorded 29 hours of interview and evaluation sessions and transcribed 34 sessions without external support.

Based on the sum of our findings, we contributed a researcher-centered definition of reproducibility in Chapter \ref{ch:background}. We expect that this definition will provide a common ground for discussing root causes of and solutions to irreproducibility across the sciences. In Chapter \ref{ch:requirements}, we contributed a systematic mapping of communication and sharing practices, interaction with a tailored preservation tool, and secondary usage forms in HEP. We expanded our findings through a cross-domain study which we reported in Chapter \ref{ch:cross_domain}. We involved researchers and data managers from a variety of scientific domains, including biology, chemistry, agricultural science, and meteorology. Based on our findings, we introduced and described the \CommitmentModelTrailingSpace in Section \ref{section:evolution_model}. The model details how and why researchers internalize RDM practices and depicts stimuli and requirements in the transition from non-reproducible practices to sustained commitment for RDM.

We are committed to make most of our study resources available as supplementary material. This includes the interview protocols of all studies and the two prototypes of the requirements study on gamification in science (see Chapter \ref{ch:gamification_requirements}). In this study, we provided a first systematic account of how highly skilled scientists perceive gamification in professional research tools. We expect that researchers and designers can reuse our prototypes to inform about perceptions and requirements of gamification in diverse fields of science. In addition, we made the GitHub branch of our implementation of tailored science badges in CAP publicly available. In Chapter \ref{ch:tailored_badges}, we presented the evaluation of the tailored badges and discussed how they can provide new forms of research interaction.

In Chapter \ref{ch:hci_role}, we described and advocated the role of HCI in systematically supporting and transforming open and reproducible science practices. We introduced a conceptual model of UCD in reproducible science and described challenges for both HCI practitioners and researchers. We concluded with our vision of URP that we expect to transform knowledge preservation in science. Finally, in Chapter \ref{ch:future_work}, we contribute an extensive account of opportunities for future work that we expect to support the various transitions of the \CommitmentModel.

\section{Replication in HCI}

In this thesis, we described and advocated HCI's role in reproducible science. Ultimately, this raises the question whether and how our findings also inform about the role of \textit{replication in HCI}. Greiffenhagen and Reeves \cite{Greiffenhagen2013} stressed that ``to focus the discussion of replication in HCI, it would be very helpful if one could gather more examples from different disciplines, from biology to physics, to see whether and how replications are valued in these.''

Our work focused on understanding practices and needs of researchers, rather than values of reproducibility in scientific fields. Thus, we are hardly in a position to contribute directly to discussions on the value of replication in HCI as a field of science. Instead, we argue that HCI scholars and practitioners will profit from systematic sharing, as reflected in our researcher-centered definition of reproducibility. Our findings and related work \cite{hutson2018artificial, gundersen2018state} further suggest that computational reusability will become increasingly important in HCI, as the field continues to adopt methods around big data science, \ac{ML}, and \ac{AI}.

We know that sharing of computational resources in HCI research is hindered by some of the same issues that we described in this thesis \cite{echtler2018open, wacharamanotham2019transparency}. Thus, we consider our models suitable to guide general design thinking in the development of computational research repositories and RDM tools in HCI. In particular, our findings related to science badges provide accessible starting points for promoting, acknowledging, and motivating OS practices in our field. We hope that our work inspires a discussion around the design and adoption of science badges (e.g. the ACM badges) within the SIGCHI community and its conferences.

\section{Limitations}

Our research focused on practices and requirements of RDM and reproducible science in HEP. This is a strength and limitation of our work. HEP represents a unique environment that deals with challenges unmatched by most other branches of science. We argue that this is a strength, as findings in this environment are likely to become relevant to increasingly data-intensive branches of science in the future. The invention of the WWW in 1989 at CERN represents a good example. Still, future research needs to apply our methods and designs in diverse scientific fields, in order to expand and generalize our findings through field-specific investigations that account for diverse research practices.

Our empirical findings are mostly based on qualitative research. We conducted two semi-structured interview studies and two mixed-method evaluation studies with a strong qualitative focus. Again, the qualitative focus represents a strength and limitation. We argue that a qualitative approach was needed to provide a first systematic account of practices and requirements around RDM and tailored technology interaction in reproducible science. We should not risk to employ RDM tools without having studied their integration into research workflows. Doing so could alienate scientists and further jeopardize commitment for open science. The same applies for gamification. We know that meaningless ad-hoc implementations of game design elements alienate users. We expected that the same is true for highly skilled scientists who are trained to think critically. For that reason, we first studied perceptions towards two prototypes that made use of a variety of different game design elements. Based on our findings, we implemented tailored science badges in CAP. Here again, we decided to focus on single evaluation sessions with a qualitative focus. The science badges represented the first actual implementation of game design elements in CAP. We needed to study how researchers perceive and interact with them, before we could decide on a wider deployment across the LHC collaborations. We are confident that the last iteration of our gamification research matured to a level that enables implementation and wider dissemination within the LHC experiments. This will open the door to quantitative research based on long-term usage behaviour.

\chapter{Future Work }
\label{ch:future_work}

In Chapter \ref{ch:conclusion}, we summarized our research contributions based on the four principal RQs. Ultimately, those contributions relate to the four stages of the \CommitmentModel, as introduced in Figure \ref{fig:rdm_model}: Non-Reproducible Practices, Overcoming Barriers, Sustained Commitment, and Reward. In this chapter, we illustrate challenges and opportunities for future work, with particular regard to the transition between those stages. First, we depict the three transitions: Adoption, Integration, and Reward Cycle. We relate those to four key areas of future work: 1) Generalize findings, methods, and uses; 2) Support impactful secondary usage forms; 3) Reflect and integrate internal contributions; and 4) Advance gamification in science. We detail those areas and describe opportunities for HCI researchers and practitioners in designing interactive tools for reproducible science.

\section{Commitment Transitions}

In this section, we depict the three transitions of the \CommitmentModelTrailingSpace and relate those to areas of future work.

\subsection{Adoption}

\begin{figure}
\centering
  \includegraphics[width=0.5\columnwidth]{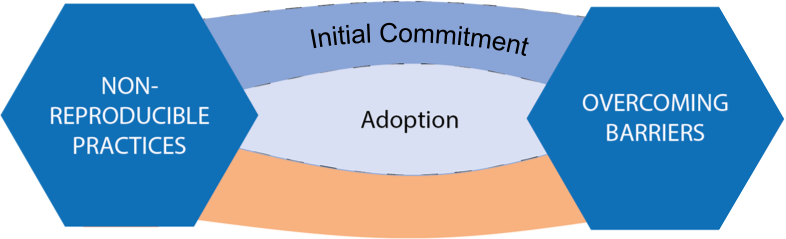}
  \caption{Adoption describes the transition between Non-Reproducible Practices and Overcoming Barriers.}~\label{fig:adoption}
\end{figure}

We refer to \textit{Adoption} as the transition between Non-Reproducible Practices and Overcoming Barriers (see Figure \ref{fig:adoption}). We found and described different drivers of initial commitment for entering the Adoption phase. Those include internal enforcement (e.g. supervisor enforcement, institute policy), general policies (e.g. conference / journal submissions, funding agencies), motivation to follow good scientific practice, and expectations to receive rewards. To further our understanding of how practices in a specific field or institute impact willingness for adoption, we need to systematically expand our knowledge of practices and requirements in the interaction with supportive RDM tools and existing practice. We describe this area of future work related to the generalization of our findings in Section \ref{section:fw_generalize}. Studying the wider applicability is also relevant for future work on secondary usage forms, as discussed in Section \ref{section:fw_suf}. We discussed secondary uses mainly in the context of rewards. Still, we consider that the prospect of profiting from those uses represents a driver of initial commitment. In fact, we argue that initial adoption of comprehensive RDM is based on a mix of enforcement and the prospect of rewards. In Chapter \ref{ch:hci_role}, we argued ``that not only do policies shape technologies, and technologies shape policies [141], but a mix of motivational drivers and policies shape technologies and RDM commitment at different stages in the commitment evolution.'' We consider the study of commitment drivers in the adoption phase and across the entire personal RDM commitment evolution as a promising area of future research.

\subsection{Integration}

\begin{figure}
\centering
  \includegraphics[width=0.5\columnwidth]{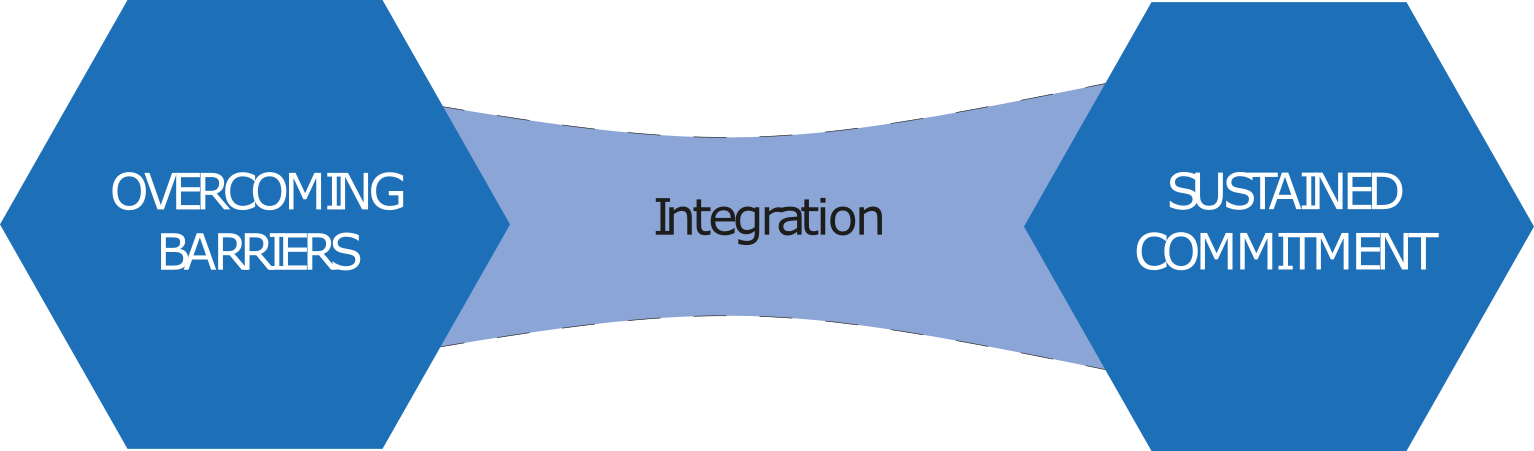}
  \caption{Integration describes the transition between Overcoming Barriers and Sustained Commitment.}~\label{fig:integration}
\end{figure}

Integration refers to the transition between Overcoming Barriers and Sustained Commitment. Providing a suitable and supportive environment is a key prerequisite of Integration. At the very least, this means that an appropriate socio-technical infrastructure must enable researchers to follow core RDM practices. Personal attitudes and the experiences during this integration process will be crucial in transitioning to Sustained Commitment. Related to areas of future work, we envision the development of tools that integrate into research workflows for the purpose of supporting documentation and preservation. Our vision refers to Ubiquitous Research Preservation (see Section \ref{section:urp}). We further recognize that integration will depend on the scientific community \textit{relating} to one's achievements. Fulfilling the psychological need \textit{relatedness} can be particularly challenging in tailored systems that restrict access to members of experiments or institutes. In Section \ref{section:fw_internal_external}, we present challenges related to the reflection and integration of internal contributions. We envision future RDM systems to seamlessly integrate into scientific practice and to reflect good practice efforts and achievements across scientific, industrial, and organizational frameworks.

\subsection{Reward Cycle}

\begin{figure}
\centering
  \includegraphics[width=0.5\columnwidth]{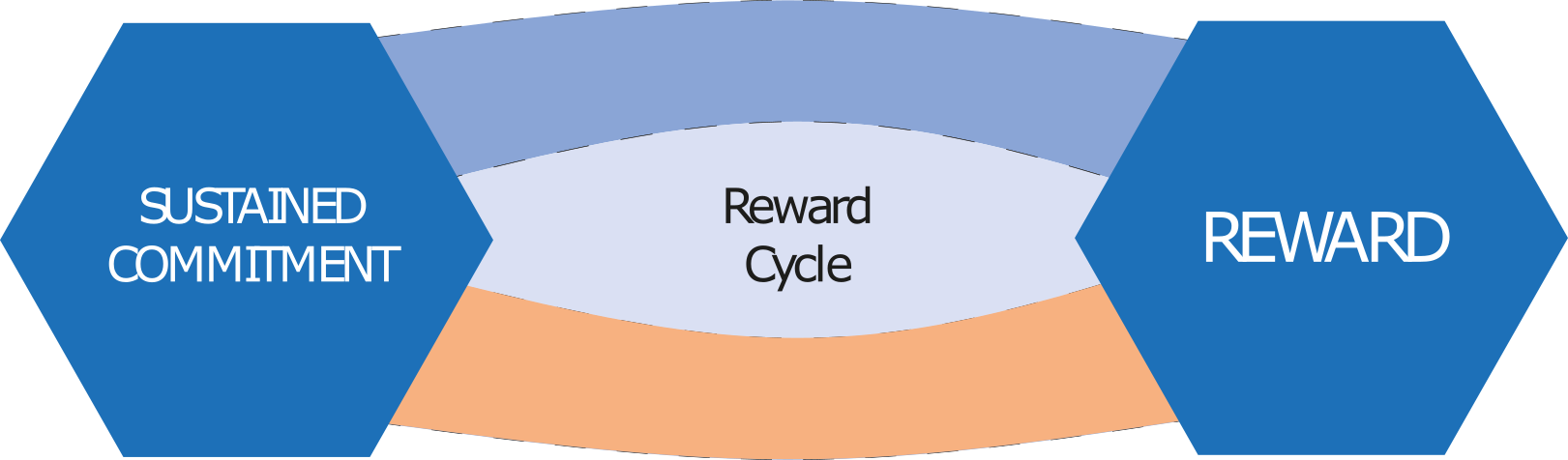}
  \caption{Reward Cycle describes the transition between Sustained Commitment and Rewards.}~\label{fig:reward_cycle}
\end{figure}

Reward Cycle is the bi-directional transition between Sustained Commitment and Reward. It refers to the process of receiving, acknowledging, and internalizing meaningful rewards. In the context of future work, we describe two key areas of study. First, in Section \ref{section:fw_suf}, we illustrate research opportunities related to secondary usage forms. Second, we describe required next steps to advance gamification in (reproducible) science in Section \ref{section:gamification_fw}. We expect that future research in this direction generally advances our understanding of motivation in HCI which has been described as focusing narrowly on extrinsic motivations \cite{knaving2015flow}, lacking deeper engagement with theories \cite{tyackself}, and leaving open questions related to conception, abstraction, and measurement of user engagement \cite{doherty2018engagement}.

%
%
\section{Generalize Findings, Methods, and Uses}
\label{section:fw_generalize}

Our research findings are largely based on practices in HEP and CERN's largest experiments. As we emphasized previously, we consider this focus on one of the largest scientific data producers beneficial to inform the design of supportive tools in science overall. As big data processing becomes prevalent across the sciences, challenges observed in particle physics become relevant to other fields. However, to verify applicability of our findings and to refine them to specific practices in diverse branches of science, we stress that future work should systematically map practices and requirements around RDM and adoption of supportive technology. We posit that our researcher-centered definition of reproducibility (see Section \ref{section:definition}) is suitable to guide future research and design thinking. We further expect that future work will contribute to a refined understanding of the researcher-centered perspective on reproducibility.

We trust that the wider study of practices and requirements will enable the design and adoption of more widely usable tools that still provide close support in the RDM process. Finding a good balance between the wide applicability of generic services and community-tailored tools is a crucial future step. CAP might provide a great starting point. It is based on solid and advanced data storage and search layers and places powerful templates at its core. Integration of easy-to-configure form builders\footnote{Open source libraries provide an accessible starting point: https://github.com/formio} might make an important step towards integration into communities beyond particle physics. Studying how institutes and service providers adapt and adopt such a RDM tool can make a positive impact on science reproducibility. Future work should particularly focus on understanding the role of policies in the design of RDM tools, as well as their adoption. While we have to consider policies in driving compliance, we expect that policies will ultimately limit the development of tailored services. Considering the LHC experiments, we noted that practices across working groups and teams differed. Collaboration and institute policies can only demand a minimum common denominator. That way, policies will likely prevent a fast adaption of tailored templates to novel research practices. It would be interesting to see if and how scientific communities and tools reflect contributions beyond the required minimum.

We acknowledge that requirements for encouraging reproducible research practices are likely to differ between distinct organizations and fields of science. We envision the development and validation of a standardized instrument to systematically assess the need for policies and components of motivation and reward. An open science motivation scale could be constructed within the framework of SDT. In particular, around intrinsic motivation, amotivation, and different regulatory styles of extrinsic motivation. Examples of such instruments include the Sport Motivation Scale \cite{pelletier1995toward} and the Academic Motivation Scale \cite{vallerand1992academic} that measures motivation in high school education. In the OS context, we can imagine a motivation scale that asks the following basic question: \textit{Why do you document and share your research?} Based on our findings, some of the statements might include: \textit{Because my supervisors demand it} (External Regulation); \textit{I profit from my colleagues' resources. I would look bad if I did not return anything} (Introjected Regulation); and \textit{I don't know anymore, it feels like a waste of time as no one acknowledges my efforts} (Amotivation). Clearly, the construction of such a scale would require a systematic analysis of motivations across the diverse scientific landscape and large-scale validation of the scale items. We argue that those efforts would be well invested as such an instrument could be a valuable tool in the design of interactive tools and policies for reproducible science. An OS motivation scale would allow to compare motivation across and beyond science, systems, and incentives. Applied to an organization or scientific field, designers and policy makers could take informed decisions about requirements and effects of initial and sustained drivers of commitment.


%
%
\section{Support Impactful Secondary Usage Forms}
\label{section:fw_suf}

In Chapter \ref{ch:requirements}, we introduced the notion of `secondary usage forms' of RDM tools. Based on our findings, we argued that those will be essential in the adoption of tools like CAP. Secondary usage forms are uses of RDM tools that, while not central to the core mission of the tools, are key in providing contributors with meaningful rewards. We found that in HEP those secondary uses relate to coping with uncertainty, providing structured designs and automation, and stimulating collaboration. Future work should focus on the systematic design and implementation of secondary uses.

\begin{figure}
\centering
  \includegraphics[width=0.6\columnwidth]{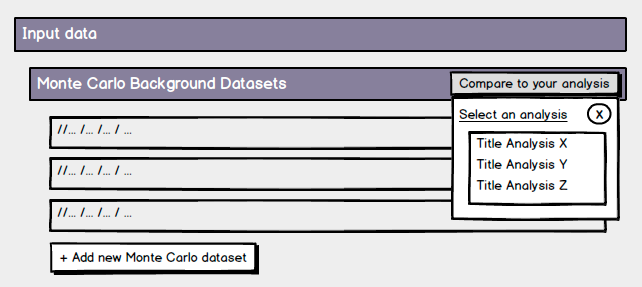}
  \caption{Concept for supporting structured and automated analysis comparisons.}~\label{fig:automated_comparison}
\end{figure}

\begin{figure}
\centering
  \includegraphics[width=0.6\columnwidth]{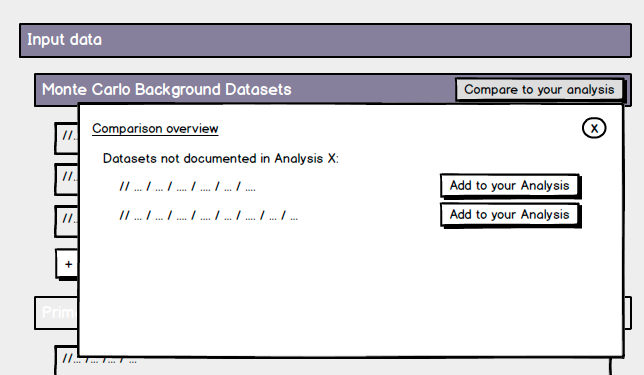}
  \caption{Automated analysis comparisons can support researchers and prevent errors.}~\label{fig:dataset_comparison}
\end{figure}

In this section, we provide starting points related to computational research in particle physics. Based on our findings in HEP, we sketched prototypes that support structured and automated analysis comparison. In the prototype in Figure \ref{fig:automated_comparison}, we added the button `Compare to your analysis' in the dataset section of a colleague's analysis. Researchers can select amongst analyses that they documented in the service, in order to display a comparison (see Figure \ref{fig:dataset_comparison}). This prototype corresponds directly to the expectation of one of our study participants who liked to see input dataset mismatches amongst similar analyses. 

\begin{figure}
\centering
  \includegraphics[width=0.7\columnwidth]{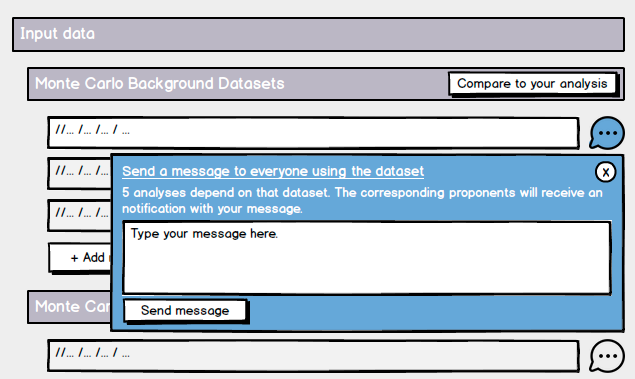}
  \caption{Design concept for resource-based communication.}~\label{fig:dataset_communication}
\end{figure}

Figure \ref{fig:dataset_communication} envisions a resource-based communication feature that allows to reliably communicate information, warnings, and errors to anyone who depends on the specific resource. This concept is strongly related to uncertainty coping, as described in Chapter \ref{ch:requirements}. It further relates to the call for ``facilities for communication around shared data abstractions'' by Birnholtz and Bietz \cite{birnholtz2003data}. We expect that researchers will acknowledge benefits provided by those secondary uses as they explore RDM tools. We further expect that they will find the secondary uses most beneficial for analyses that are documented and preserved during the research process, rather than retrospectively. Thus, we imagine that the implementation of secondary usage forms will increase researchers' willingness to contribute resources to RDM tools early in the research lifecycle. Future work should test those hypotheses. Future research should further investigate how the implementation of secondary uses can be communicated effectively to the research community.

Understanding meaningful secondary uses beyond HEP will play an important role in the adoption of RDM tools beyond physics. We argued that the general nature of the described secondary usage forms make them likely applicable in other scientific domains. However, we concluded in Chapter \ref{ch:cross_domain}, that cross-domain studies with few participants make a clear description of those uses challenging. Thus, we advertise field-specific investigations of secondary uses. Their goal is to collect data and findings that enable us to build models for the design of secondary usage forms across scientific fields.

%
%
\section{Reflect and Integrate Internal Contributions}
\label{section:fw_internal_external}

The findings from our empirical studies showed that RDM tools like CAP integrate into a wider ecosystem of established tools and processes. Future research should systematically investigate components of this ecosystem and examine how tailored and general RDM tools integrate, complement, replace, or interact with those components. To provide a starting point, we mapped the HEP ecosystem of science infrastructure in Figure \ref{fig:ecosystem}. We aggregated findings from our interview studies, related to tools and processes, into eight components and three layers. We distinguished between tools that focus on complete research and execution cycles (\textbf{Research}-Focused Tools), services that are designed to manage a very specific set of resources (\textbf{Resource}-Focused Tools), and services that provide a public interface for research (\textbf{Public}-Facing Tools). We expect that acceptance of research-focused tools will depend on the meaningful integration of pre-existing resource-focused tools. It would be valuable to test this hypothesis in future work and to systematically map the interplay between those two layers. Similarly, we expect that the further study and implementation of interfaces between research-focused tools and public-facing services will benefit adoption and might provide new forms of secondary usage forms.

The study participants repeatedly noted that RDM tools provide means to demonstrate and check compliance with important scientific practice. While preserved resources on CAP enable an implicit communication of efforts, future research should focus on investigating opportunities to express, summarize, and communicate contributions more explicitly. This is probably easier in tailored tools like CAP where a homogeneous research community understands and applies common metrics. Yet, such tailored tools pose challenges with regards to the fair reflection of contributions in the outside world. For example, researchers in HEP will want to show their contributions when they transition away from particle physics. Exploring trusted and exportable formats that communicate efforts and contributions will be crucial.

%
%

\section{Advance Gamification in Science}
\label{section:gamification_fw}

The gamification research we presented in Part \ref{part:gamification} represents the first systematic study of requirements and effects of gamification in highly skilled science. We emphasized the importance of a systematic design process to ensure acceptance of game design elements amongst scientists. To that end, we first conducted a prototype-based requirements study. Next, we designed, implemented, and evaluated tailored science badges based on our previous findings. Future work should explore long-term usage behaviour of gamified tools in science. In particular, it will be valuable to study and contrast short-term adoption and long-term commitment.

Future work should further explore effects of individual game elements in highly skilled (scientific) environments. Our requirements research in Chapter \ref{ch:gamification_requirements} showed that many different types of elements and mechanisms seem suitable, even though some elements were controversial. Our implementation (see Chapter \ref{ch:tailored_badges}) focused on tailored science badges. Implementing and comparing additional game design elements in the scientific environment reflects current gamification research challenges \cite{Tondello2017} and will likely provide new opportunities to encourage good scientific practices. In addition, it will allow further exploration of new forms of interaction with scientific content through the gamification layer. In this context, we expect to further our understanding of gamification `beyond motivation'.

We motivated dedicated gamification research for science by emphasizing that the socio-technical frameworks of scientists likely differ from those of industry employees. However, frameworks, practices, and expectations further differ between researchers across different scientific fields. Thus, we need to study gamification requirements and effects in diverse branches of science to build a wider understanding of how gamification motivates scientific practice in particular, and impacts highly skilled environments in general. Here, future work should relate to current gamification research investigating the effects of persuasive strategies on different personality traits and gamification user types \cite{orji2018personalizing}.

Finally, it would be important to explore how achievements can be communicated effectively. We found that achievements on the gamification layer can impact human resource decisions, e.g. hiring and promotion. This also relates to the previous section on \textit{reflecting internal contributions}. To foster acceptance amongst scientists, it would be valuable to understand how those achievements can be communicated to parties of interest outside of tailored and closed tools.

On a concluding note, we want to stress once more that the advent of widely used research cyberinfrastructure and tailored RDM tools provides new opportunities for openness, transparency, and reproducibility in science. We expect that future work on research topics described in this chapter will contribute to a positive change in science that benefits research communities, scientists, decision makers, and the general public.
	
	\cleardoublepage

	\backmatter
	
	
	\part{Bibliography}
	\bibliographystyle{abbrvnat}
	\bibliography{bibliography/bibliography}

\begin{thebibliography}{205}
\providecommand{\natexlab}[1]{#1}
\providecommand{\url}[1]{\texttt{#1}}
\expandafter\ifx\csname urlstyle\endcsname\relax
  \providecommand{\doi}[1]{doi: #1}\else
  \providecommand{\doi}{doi: \begingroup \urlstyle{rm}\Url}\fi

\bibitem[Abras et~al.(2004)Abras, Maloney-Krichmar, and Preece]{abras2004user}
C.~Abras, D.~Maloney-Krichmar, and J.~Preece.
\newblock User-centered design.
\newblock \emph{Bainbridge, W. Encyclopedia of Human-Computer Interaction.
  Thousand Oaks: Sage Publications}, 37\penalty0 (4):\penalty0 445--456, 2004.

\bibitem[ACM(2018)]{acmbadgesweb}
ACM.
\newblock {Artifact Review and Badging}, April 2018.
\newblock URL
  \url{https://www.acm.org/publications/policies/artifact-review-badging}.
\newblock Retrieved September 10, 2018.

\bibitem[Baker(2016)]{Baker2016}
M.~Baker.
\newblock 1,500 scientists lift the lid on reproducibility.
\newblock \emph{Nature}, 533\penalty0 (7604):\penalty0 452--454, 2016.
\newblock ISSN 0028-0836.
\newblock \doi{10.1038/533452a}.

\bibitem[Ball(2015)]{ball2015carl}
P.~Ball.
\newblock {Carl Djerassi (1923--2015)}.
\newblock \emph{Nature}, 519\penalty0 (7541):\penalty0 34--34, 2015.
\newblock \doi{10.1038/519034a}.

\bibitem[Barba(2018)]{barba2018terminologies}
L.~A. Barba.
\newblock {Terminologies for Reproducible Research}.
\newblock \emph{arXiv preprint arXiv:1802.03311}, 2018.

\bibitem[Bechhofer et~al.(2013)Bechhofer, Buchan, {De Roure}, Missier,
  Ainsworth, Bhagat, Couch, Cruickshank, Delderfield, Dunlop, Gamble,
  Michaelides, Owen, Newman, Sufi, and Goble]{Bechhofer2013}
S.~Bechhofer, I.~Buchan, D.~{De Roure}, P.~Missier, J.~Ainsworth, J.~Bhagat,
  P.~Couch, D.~Cruickshank, M.~Delderfield, I.~Dunlop, M.~Gamble,
  D.~Michaelides, S.~Owen, D.~Newman, S.~Sufi, and C.~Goble.
\newblock {Why linked data is not enough for scientists}.
\newblock \emph{Future Generation Computer Systems}, 29\penalty0 (2):\penalty0
  599--611, 2013.
\newblock ISSN 0167739X.
\newblock \doi{10.1016/j.future.2011.08.004}.

\bibitem[Begley and Ellis(2012)]{Begley2012}
C.~G. Begley and L.~M. Ellis.
\newblock {Drug development: Raise standards for preclinical cancer research.}
\newblock \emph{Nature}, 483\penalty0 (7391):\penalty0 531--3, 2012.
\newblock ISSN 1476-4687.
\newblock \doi{10.1038/483531a}.

\bibitem[Belhajjame et~al.(2014)Belhajjame, Zhao, Garijo, Hettne, Palma,
  Corcho, G{\'{o}}mez-P{\'{e}}rez, Bechhofer, Klyne, and Goble]{Belhajjame2014}
K.~Belhajjame, J.~Zhao, D.~Garijo, K.~Hettne, R.~Palma, {\'{O}}.~Corcho, J.-M.
  G{\'{o}}mez-P{\'{e}}rez, S.~Bechhofer, G.~Klyne, and C.~Goble.
\newblock {The Research Object Suite of Ontologies: Sharing and Exchanging
  Research Data and Methods on the Open Web}.
\newblock \emph{arXiv preprint arXiv: 1401.4307}, \penalty0 (February
  2014):\penalty0 20, 2014.
\newblock URL \url{http://arxiv.org/abs/1401.4307}.

\bibitem[Bell et~al.(2009)Bell, Hey, and Szalay]{bell2009beyond}
G.~Bell, T.~Hey, and A.~Szalay.
\newblock {Beyond the Data Deluge}.
\newblock \emph{Science}, 323\penalty0 (5919):\penalty0 1297--1298, 2009.
\newblock ISSN 0036-8075.
\newblock \doi{10.1126/science.1170411}.

\bibitem[Bentley et~al.(1995)Bentley, Horstmann, Sikkel, and
  Trevor]{bentley1995supporting}
R.~Bentley, T.~Horstmann, K.~Sikkel, and J.~Trevor.
\newblock {Supporting Collaborative Information Sharing with the World Wide
  Web: The BSCW Shared Workspace System}.
\newblock In \emph{Proceedings of the 4th International WWW Conference},
  volume~1, pages 63--74, 1995.

\bibitem[Berners-Lee et~al.(1992)Berners-Lee, Cailliau, Groff, and
  Pollermann]{Berners-Lee1992}
T.~Berners-Lee, R.~Cailliau, J.-F. Groff, and B.~Pollermann.
\newblock {World-Wide Web: The Information Universe}.
\newblock \emph{Internet Research}, 2:\penalty0 52--58, 1992.
\newblock \doi{10.1108/eb047254}.

\bibitem[Birnholtz and Bietz(2003)]{birnholtz2003data}
J.~P. Birnholtz and M.~J. Bietz.
\newblock Data at work: supporting sharing in science and engineering.
\newblock In \emph{Proceedings of the 2003 international ACM SIGGROUP
  conference on Supporting group work}, pages 339--348. ACM, 2003.
\newblock \doi{10.1145/958160.958215}.

\bibitem[Blandford et~al.(2016)Blandford, Furniss, and
  Makri]{Blandford:2222613}
A.~Blandford, D.~Furniss, and S.~Makri.
\newblock \emph{Qualitative HCI Research: Going Behind the Scenes}, pages
  51--60.
\newblock Synthesis Lectures on Human-Centered Informatics. Morgan \& Claypool
  Publishers, 2016.
\newblock ISBN 9781627057608.
\newblock \doi{10.2200/S00706ED1V01Y201602HCI034}.

\bibitem[Bohle(2013)]{bohle2013science}
S.~Bohle.
\newblock What is e-science and how should it be managed.
\newblock \emph{Nature, Spektrum der Wissenschaft (Scientific American)}, 2013.

\bibitem[Boisvert(2016)]{boisvert2016incentivizing}
R.~F. Boisvert.
\newblock Incentivizing reproducibility.
\newblock \emph{Communications of the ACM}, 59\penalty0 (10):\penalty0 5--5,
  2016.
\newblock \doi{10.1145/2994031}.

\bibitem[Borgman(2006)]{Borgman2006}
C.~L. Borgman.
\newblock {What can Studies of e-Learning Teach us about Collaboration in
  e-Research? Some Findings from Digital Library Studies}.
\newblock \emph{Computer Supported Cooperative Work}, 15\penalty0 (4):\penalty0
  359--383, 2006.
\newblock ISSN 15737551.
\newblock \doi{10.1007/s10606-006-9024-1}.

\bibitem[Borgman(2007)]{Borgman:1297241}
C.~L. Borgman.
\newblock \emph{{Scholarship in the digital age: information, infrastructure,
  and the internet}}.
\newblock MIT Press, Cambridge, MA, 2007.

\bibitem[Borgman et~al.(2013)Borgman, Edwards, Jackson, Chalmers, Bowker,
  Ribes, Burton, and Calvert]{borgman2013knowledge}
C.~L. Borgman, P.~N. Edwards, S.~J. Jackson, M.~K. Chalmers, G.~C. Bowker,
  D.~Ribes, M.~Burton, and S.~Calvert.
\newblock {Knowledge Infrastructures: Intellectual Frameworks and Research
  Challenges}.
\newblock 2013.

\bibitem[Boukhelifa et~al.(2017)Boukhelifa, Perrin, Huron, and
  Eagan]{Boukhelifa2017}
N.~Boukhelifa, M.-E. Perrin, S.~Huron, and J.~Eagan.
\newblock {How Data Workers Cope with Uncertainty: A Task Characterisation
  Study}.
\newblock \emph{In Proceedings of the 2017 CHI Conference on Human Factors in
  Computing Systems}, pages 3645--3656, 2017.
\newblock \doi{10.1145/3025453.3025738}.

\bibitem[Boulton et~al.(2011)Boulton, Rawlins, Vallance, and
  Walport]{boulton2011science}
G.~Boulton, M.~Rawlins, P.~Vallance, and M.~Walport.
\newblock Science as a public enterprise: the case for open data.
\newblock \emph{The Lancet}, 377\penalty0 (9778):\penalty0 1633--1635, 2011.

\bibitem[Bowser et~al.(2014)Bowser, Hansen, Preece, He, Boston, and
  Hammock]{Bowser2014}
A.~Bowser, D.~Hansen, J.~Preece, Y.~He, C.~Boston, and J.~Hammock.
\newblock {Gamifying citizen science: A study of two user groups}.
\newblock In \emph{17th ACM Conference on Computer Supported Cooperative Work
  and Social Computing, CSCW 2014}, pages 137--140, 2014.
\newblock ISBN 9781450325417.
\newblock \doi{10.1145/2556420.2556502}.

\bibitem[Brito et~al.(2015)Brito, Vieira, and Duran]{Brito2015}
J.~Brito, V.~Vieira, and A.~Duran.
\newblock {Towards a Framework for Gamification Design on Crowdsourcing
  Systems: The G.A.M.E. Approach}.
\newblock \emph{2015 12th International Conference on Information Technology -
  New Generations}, pages 445--450, 2015.
\newblock ISSN 978-1-4799-8828-0.
\newblock \doi{10.1109/ITNG.2015.78}.

\bibitem[Br{\"u}hlmann et~al.(2013)Br{\"u}hlmann, Mekler, and
  Opwis]{bruhlmann2013gamification}
F.~Br{\"u}hlmann, E.~Mekler, and K.~Opwis.
\newblock Gamification from the perspective of self-determination theory and
  flow.
\newblock \emph{University of Basel}, 2013.

\bibitem[Brun et~al.(2012)Brun, Carminati, and Carminati]{brun2012web}
R.~Brun, F.~Carminati, and G.~G. Carminati.
\newblock \emph{From the Web to the Grid and Beyond: Computing Paradigms Driven
  by High-Energy Physics}.
\newblock Springer Science \& Business Media, 2012.
\newblock \doi{10.1007/978-3-642-23157-5}.

\bibitem[Campbell et~al.(2003)Campbell, Maglio, Cozzi, and Dom]{Campbell2003}
C.~S. Campbell, P.~P. Maglio, A.~Cozzi, and B.~Dom.
\newblock {Expertise identification using email communications}.
\newblock In \emph{CIKM '03: Proceedings of the twelfth international
  conference on Information and knowledge management}, pages 528--531, January
  2003.
\newblock ISBN 1581137230.
\newblock \doi{10.1145/956863.956965}.

\bibitem[Cavusoglu et~al.(2015)Cavusoglu, Li, and Huang]{Cavusoglu2015}
H.~Cavusoglu, Z.~Li, and K.-W. Huang.
\newblock {Can Gamification Motivate Voluntary Contributions?}
\newblock \emph{Proceedings of the 18th ACM Conference Companion on Computer
  Supported Cooperative Work {\&} Social Computing - CSCW'15 Companion}, pages
  171--174, 2015.
\newblock ISSN 9781450329460.
\newblock \doi{10.1145/2685553.2698999}.
\newblock URL \url{http://dl.acm.org/citation.cfm?doid=2685553.2698999}.

\bibitem[CERN(2013)]{birth:1998446}
CERN.
\newblock {The birth of the web}.
\newblock Dec 2013.
\newblock URL \url{http://cds.cern.ch/record/1998446}.
\newblock Retrieved March 15, 2018.

\bibitem[CERN(2017)]{CERN-HR-STAFF-STAT-2017}
CERN.
\newblock {CERN Annual Personnel Statistics 2017}.
\newblock 2017.
\newblock URL \url{https://cds.cern.ch/record/2317058}.

\bibitem[CERN(2018)]{CERN-HR-STAFF-STAT-2018}
CERN.
\newblock {CERN Annual Personnel Statistics 2018}.
\newblock 2018.
\newblock URL \url{http://cds.cern.ch/record/2677223}.

\bibitem[Cetina(2009)]{cetina2009epistemic}
K.~K. Cetina.
\newblock \emph{Epistemic cultures: How the sciences make knowledge}.
\newblock Harvard University Press, 2009.

\bibitem[Chard et~al.(2015)Chard, Pruyne, Blaiszik, Ananthakrishnan, Tuecke,
  and Foster]{Chard2015}
K.~Chard, J.~Pruyne, B.~Blaiszik, R.~Ananthakrishnan, S.~Tuecke, and I.~Foster.
\newblock {Globus Data Publication as a Service: Lowering Barriers to
  Reproducible Science}.
\newblock In \emph{Proceedings - 11th IEEE International Conference on
  eScience, eScience 2015}, pages 401--410, 2015.
\newblock ISBN 9781467393256.
\newblock \doi{10.1109/eScience.2015.68}.

\bibitem[Chen et~al.(2016)Chen, Dallmeier-Tiessen, Dani, Dasler, Fern{\'a}ndez,
  Fokianos, Herterich, and {\v{S}}imko]{chen2016cern}
X.~Chen, S.~Dallmeier-Tiessen, A.~Dani, R.~Dasler, J.~D. Fern{\'a}ndez,
  P.~Fokianos, P.~Herterich, and T.~{\v{S}}imko.
\newblock {CERN Analysis Preservation: A Novel Digital Library Service to
  Enable Reusable and Reproducible Research}.
\newblock In N.~Fuhr, L.~Kov{\'a}cs, T.~Risse, and W.~Nejdl, editors,
  \emph{Research and Advanced Technology for Digital Libraries}, pages
  347--356, Cham, 2016. Springer International Publishing.
\newblock ISBN 978-3-319-43997-6.
\newblock \doi{10.1007/978-3-319-43997-6_27}.

\bibitem[Chen et~al.(2018)Chen, Dallmeier-Tiessen, Dasler, Feger, Fokianos,
  Gonzalez, Hirvonsalo, Kousidis, Lavasa, Mele, et~al.]{chen2018open}
X.~Chen, S.~Dallmeier-Tiessen, R.~Dasler, S.~Feger, P.~Fokianos, J.~B.
  Gonzalez, H.~Hirvonsalo, D.~Kousidis, A.~Lavasa, S.~Mele, et~al.
\newblock Open is not enough.
\newblock \emph{Nature Physics}, 2018.
\newblock \doi{10.1038/s41567-018-0342-2}.

\bibitem[Cho(2011)]{Cho1564}
A.~Cho.
\newblock {Particle Physicists{\textquoteright} New Extreme Teams}.
\newblock \emph{Science}, 333\penalty0 (6049):\penalty0 1564--1567, 2011.
\newblock ISSN 0036-8075.
\newblock \doi{10.1126/science.333.6049.1564}.

\bibitem[Chuang and Pfeil(2018)]{chuang2018transparency}
L.~L. Chuang and U.~Pfeil.
\newblock {Transparency and Openness Promotion Guidelines for HCI}.
\newblock In \emph{Extended Abstracts of the 2018 CHI Conference on Human
  Factors in Computing Systems}, page SIG04. ACM, 2018.
\newblock \doi{10.1145/3170427.3185377}.

\bibitem[Cockburn et~al.(2018)Cockburn, Gutwin, and
  Dix]{Cockburn:2018:HNM:3173574.3173715}
A.~Cockburn, C.~Gutwin, and A.~Dix.
\newblock {HARK No More: On the Preregistration of CHI Experiments}.
\newblock In \emph{Proceedings of the 2018 CHI Conference on Human Factors in
  Computing Systems}, CHI '18, pages 141:1--141:12, New York, NY, USA, 2018.
  ACM.
\newblock ISBN 978-1-4503-5620-6.
\newblock \doi{10.1145/3173574.3173715}.

\bibitem[Cockburn(2019)]{cockburn_2019}
H.~Cockburn.
\newblock Scientists may have discovered fifth force of nature, laboratory
  announces.
\newblock \emph{Independent}, Nov 2019.

\bibitem[Collaboration(2012)]{Collaboration2012}
O.~S. Collaboration.
\newblock {An Open, Large-Scale, Collaborative Effort to Estimate the
  Reproducibility of Psychological Science}.
\newblock \emph{Perspectives on Psychological Science}, 7\penalty0
  (6):\penalty0 657--660, 2012.
\newblock ISSN 1745-6916.
\newblock \doi{10.1177/1745691612462588}.

\bibitem[COS(2019)]{osfbadges}
COS.
\newblock {Open Science Badges}.
\newblock 2019.
\newblock URL \url{https://cos.io/our-services/open-science-badges}.
\newblock Retrieved February 5, 2020.

\bibitem[Cragin et~al.(2010)Cragin, Palmer, Carlson, and Witt]{cragin2010data}
M.~H. Cragin, C.~L. Palmer, J.~R. Carlson, and M.~Witt.
\newblock Data sharing, small science and institutional repositories.
\newblock \emph{Philosophical Transactions of the Royal Society A:
  Mathematical, Physical and Engineering Sciences}, 368\penalty0
  (1926):\penalty0 4023--4038, 2010.
\newblock \doi{10.1098/rsta.2010.0165}.

\bibitem[Cross and Cummings(2004)]{Cross2004}
R.~Cross and J.~N. Cummings.
\newblock {Tie and network correlates of individual performance in
  knowledge-intensive work}.
\newblock \emph{Academy of Management Journal}, 47\penalty0 (6):\penalty0
  928--937, 2004.
\newblock ISSN 00014273.

\bibitem[Crowston et~al.(2012)Crowston, Wei, Howison, and
  Wiggins]{crowston2012free}
K.~Crowston, K.~Wei, J.~Howison, and A.~Wiggins.
\newblock Free/libre open-source software development: What we know and what we
  do not know.
\newblock \emph{ACM Computing Surveys (CSUR)}, 44\penalty0 (2):\penalty0 7,
  2012.
\newblock \doi{10.1145/2089125.2089127}.

\bibitem[Dale(2014)]{Dale2014}
S.~Dale.
\newblock {Gamification: Making work fun, or making fun of work?}
\newblock \emph{Business Information Review}, 31\penalty0 (2):\penalty0 82--90,
  2014.
\newblock ISSN 17416450.
\newblock \doi{10.1177/0266382114538350}.

\bibitem[Dallmeier~Tiessen et~al.(2015)Dallmeier~Tiessen, Herterich,
  Igo-Kemenes, Šimko, and Smith]{dallmeier_tiessen_use_cases}
S.~Dallmeier~Tiessen, P.~Herterich, P.~Igo-Kemenes, T.~Šimko, and T.~Smith.
\newblock {CERN analysis preservation (CAP) - Use Cases}.
\newblock Nov. 2015.
\newblock \doi{10.5281/zenodo.33693}.

\bibitem[Darlington et~al.(2010)Darlington, Ball, Howard, McMahon, and
  Culley]{darlington2010principles}
M.~Darlington, A.~Ball, T.~Howard, C.~McMahon, and S.~Culley.
\newblock Principles for engineering research data management.
\newblock \emph{ERIM Project Document, erim6rep101028mjd. Bath, UK: University
  of Bath. Accessed}, 8, 2010.

\bibitem[De~Waard et~al.(2015)De~Waard, Cousijn, and Aalbersberg]{de10aspects}
A.~De~Waard, H.~Cousijn, and I.~Aalbersberg.
\newblock 10 aspects of highly effective research data: Good research data
  management makes data reusable.
\newblock December 2015.
\newblock URL
  \url{https://www.elsevier.com/connect/10-aspects-of-highly-effective-research-data}.

\bibitem[Deci and Ryan(1985)]{deci1985toward}
E.~L. Deci and R.~M. Ryan.
\newblock Toward an organismic integration theory.
\newblock In \emph{Intrinsic motivation and self-determination in human
  behavior}, pages 113--148. Springer, 1985.

\bibitem[Delfanti(2016)]{doi:10.1177/0306312716659373}
A.~Delfanti.
\newblock Beams of particles and papers: How digital preprint archives shape
  authorship and credit.
\newblock \emph{Social Studies of Science}, 46\penalty0 (4):\penalty0 629--645,
  2016.
\newblock \doi{10.1177/0306312716659373}.

\bibitem[Denny(2013)]{Denny:2013:EVA:2470654.2470763}
P.~Denny.
\newblock {The Effect of Virtual Achievements on Student Engagement}.
\newblock In \emph{Proceedings of the SIGCHI Conference on Human Factors in
  Computing Systems}, CHI '13, pages 763--772, New York, NY, USA, 2013. ACM.
\newblock ISBN 978-1-4503-1899-0.
\newblock \doi{10.1145/2470654.2470763}.

\bibitem[Deterding et~al.(2011)Deterding, Khaled, Nacke, and
  Dixon]{Deterding2011}
S.~Deterding, R.~Khaled, L.~E. Nacke, and D.~Dixon.
\newblock Gamification: Toward a definition.
\newblock In \emph{CHI 2011 gamification workshop proceedings}, volume~12.
  Vancouver BC, Canada, 2011.

\bibitem[Deterding et~al.(2015)Deterding, Canossa, Harteveld, Cooper, Nacke,
  and Whitson]{Deterding:2015:GRS:2702613.2702646}
S.~Deterding, A.~Canossa, C.~Harteveld, S.~Cooper, L.~E. Nacke, and J.~R.
  Whitson.
\newblock {Gamifying Research: Strategies, Opportunities, Challenges, Ethics}.
\newblock In \emph{Proceedings of the 33rd Annual ACM Conference Extended
  Abstracts on Human Factors in Computing Systems}, CHI EA '15, pages
  2421--2424, New York, NY, USA, 2015. ACM.
\newblock ISBN 978-1-4503-3146-3.
\newblock \doi{10.1145/2702613.2702646}.

\bibitem[Djerassi(2012)]{djerassi2012cantor}
C.~Djerassi.
\newblock \emph{Cantor's dilemma}.
\newblock Doubleday, 2012.

\bibitem[Doherty and Doherty(2018)]{doherty2018engagement}
K.~Doherty and G.~Doherty.
\newblock {Engagement in HCI: Conception, Theory and Measurement}.
\newblock \emph{ACM Computing Surveys (CSUR)}, 51\penalty0 (5):\penalty0 1--39,
  2018.
\newblock \doi{10.1145/3234149}.

\bibitem[Echtler and H\"{a}ussler(2018)]{echtler2018open}
F.~Echtler and M.~H\"{a}ussler.
\newblock {Open Source, Open Science, and the Replication Crisis in HCI}.
\newblock In \emph{Extended Abstracts of the 2018 CHI Conference on Human
  Factors in Computing Systems}, CHI EA '18, pages alt02:1--alt02:8, New York,
  NY, USA, 2018. ACM.
\newblock ISBN 978-1-4503-5621-3.
\newblock \doi{10.1145/3170427.3188395}.

\bibitem[Ehrlich et~al.(2007)Ehrlich, Lin, and
  Griffiths-Fisher]{ehrlich2007searching}
K.~Ehrlich, C.-Y. Lin, and V.~Griffiths-Fisher.
\newblock Searching for experts in the enterprise: combining text and social
  network analysis.
\newblock In \emph{Proceedings of the 2007 international ACM conference on
  Supporting group work}, pages 117--126. ACM, 2007.
\newblock \doi{10.1145/1316624.1316642}.

\bibitem[Eiselmayer et~al.(2019)Eiselmayer, Wacharamanotham, Beaudouin-Lafon,
  and Mackay]{Eiselmayer:2019:TIE:3290605.3300447}
A.~Eiselmayer, C.~Wacharamanotham, M.~Beaudouin-Lafon, and W.~E. Mackay.
\newblock {Touchstone2: An Interactive Environment for Exploring Trade-offs in
  HCI Experiment Design}.
\newblock In \emph{Proceedings of the 2019 CHI Conference on Human Factors in
  Computing Systems}, CHI '19, pages 217:1--217:11, New York, NY, USA, 2019.
  ACM.
\newblock ISBN 978-1-4503-5970-2.
\newblock \doi{10.1145/3290605.3300447}.

\bibitem[Elmer et~al.(2017)Elmer, Neubauer, and Sokoloff]{elmer2017strategic}
P.~Elmer, M.~Neubauer, and M.~D. Sokoloff.
\newblock Strategic plan for a scientific software innovation institute (s2i2)
  for high energy physics.
\newblock \emph{arXiv preprint arXiv:1712.06592}, 2017.

\bibitem[Esteva et~al.(2019)Esteva, Jansen, Arduino, Sharifi-Mood, Dawson, and
  Balandrano-Coronel]{esteva2019curation}
M.~Esteva, C.~Jansen, P.~Arduino, M.~Sharifi-Mood, C.~N. Dawson, and
  J.~Balandrano-Coronel.
\newblock Curation and publication of simulation data in designsafe, a natural
  hazards engineering open platform and repository.
\newblock \emph{Publications}, 7\penalty0 (3):\penalty0 51, 2019.

\bibitem[Evans and Bryant(2008)]{Evans2008}
L.~Evans and P.~Bryant.
\newblock {LHC Machine}.
\newblock \emph{Journal of Instrumentation}, 3\penalty0 (08):\penalty0 S08001,
  2008.
\newblock ISSN 1748-0221.
\newblock \doi{10.1088/1748-0221/3/08/S08001}.

\bibitem[Eveleigh et~al.(2013)Eveleigh, Jennett, Lynn, and Cox]{Eveleigh2013}
A.~Eveleigh, C.~Jennett, S.~Lynn, and A.~L. Cox.
\newblock {"I want to be a captain! I want to be a captain!": gamification in
  the old weather citizen science project}.
\newblock \emph{Proceedings of the First International Conference on Gameful
  Design, Research, and Applications - Gamification '13}, pages 79--82, 2013.
\newblock ISSN 9781450328159.
\newblock \doi{10.1145/2583008.2583019}.

\bibitem[Falessi et~al.(2006)Falessi, Cantone, and
  Becker]{falessi2006documenting}
D.~Falessi, G.~Cantone, and M.~Becker.
\newblock Documenting design decision rationale to improve individual and team
  design decision making: an experimental evaluation.
\newblock In \emph{Proceedings of the 2006 ACM/IEEE international symposium on
  Empirical software engineering}, pages 134--143. ACM, 2006.
\newblock \doi{10.1145/1159733.1159755}.

\bibitem[Faniel and Jacobsen(2010)]{faniel2010reusing}
I.~M. Faniel and T.~E. Jacobsen.
\newblock Reusing scientific data: How earthquake engineering researchers
  assess the reusability of colleagues’ data.
\newblock \emph{Computer Supported Cooperative Work (CSCW)}, 19\penalty0
  (3-4):\penalty0 355--375, 2010.
\newblock \doi{10.1007/s10606-010-9117-8}.

\bibitem[Fecher et~al.(2017)Fecher, Friesike, Hebing, and
  Linek]{fecher2017reputation}
B.~Fecher, S.~Friesike, M.~Hebing, and S.~Linek.
\newblock A reputation economy: how individual reward considerations trump
  systemic arguments for open access to data.
\newblock \emph{Palgrave Communications}, 3:\penalty0 17051, 2017.
\newblock \doi{10.1057/palcomms.2017.51}.

\bibitem[Feger(2019)]{Feger:2691945}
S.~S. Feger.
\newblock {{More than preservation: Creating motivational designs and tailored
  incentives in research data repositories}}.
\newblock \penalty0 (CERN-OPEN-2019-007):\penalty0 5 p, Jan 2019.
\newblock URL \url{https://cds.cern.ch/record/2691945}.

\bibitem[Feger and Woźniak(2019)]{Feger:2677268}
S.~S. Feger and P.~W. Woźniak.
\newblock {More Than Preservation: A Researcher-Centered Approach to
  Reproducibility in Data Science}.
\newblock \penalty0 (CERN-OPEN-2019-003), Jan 2019.
\newblock URL \url{http://cds.cern.ch/record/2677268}.

\bibitem[Feger et~al.(2018)Feger, Dallmeier-Tiessen, Wo{\'z}niak, and
  Schmidt]{feger2018just}
S.~S. Feger, S.~Dallmeier-Tiessen, P.~Wo{\'z}niak, and A.~Schmidt.
\newblock {Just Not The Usual Workplace: Meaningful Gamification in Science}.
\newblock \emph{Mensch und Computer 2018-Workshopband}, 2018.
\newblock \doi{10.18420/muc2018-ws03-0366}.

\bibitem[Feger et~al.(2019{\natexlab{a}})Feger, Dallmeier-Tiessen, Schmidt, and
  Wo{\'z}niak]{Feger2019Requirements}
S.~S. Feger, S.~Dallmeier-Tiessen, A.~Schmidt, and P.~W. Wo{\'z}niak.
\newblock {Designing for Reproducibility: A Qualitative Study of Challenges and
  Opportunities in High Energy Physics}.
\newblock In \emph{Proceedings of the SIGCHI Conference on Human Factors in
  Computing Systems - CHI'19}, 2019{\natexlab{a}}.
\newblock ISBN 978-1-4503-5970-2/19/05.
\newblock \doi{10.1145/3290605.3300685}.

\bibitem[Feger et~al.(2019{\natexlab{b}})Feger, Dallmeier-Tiessen, Wo{\'z}niak,
  and Schmidt]{Feger2019Gamification}
S.~S. Feger, S.~Dallmeier-Tiessen, P.~W. Wo{\'z}niak, and A.~Schmidt.
\newblock {Gamification in Science: A Study of Requirements in the Context of
  Reproducible Research}.
\newblock In \emph{Proceedings of the SIGCHI Conference on Human Factors in
  Computing Systems - CHI'19}, 2019{\natexlab{b}}.
\newblock ISBN 978-1-4503-5970-2/19/05.
\newblock \doi{10.1145/3290605.3300690}.

\bibitem[Feger et~al.(2019{\natexlab{c}})Feger, Dallmeier-Tiessen, Wo{\'z}niak,
  and Schmidt]{Feger2019RoleHCI}
S.~S. Feger, S.~Dallmeier-Tiessen, P.~W. Wo{\'z}niak, and A.~Schmidt.
\newblock {The Role of HCI in Reproducible Science: Understanding, Supporting
  and Motivating Core Practices}.
\newblock \emph{Proceedings of the SIGCHI Conference on Human Factors in
  Computing Systems - CHI'19}, 2019{\natexlab{c}}.
\newblock \doi{10.1145/3290607.3312905}.

\bibitem[Feger et~al.(2020{\natexlab{a}})Feger, Dallmeier-Tiessen, Knierim,
  El.Agroudy, Wo{\'z}niak, and Schmidt]{feger2020urp}
S.~S. Feger, S.~Dallmeier-Tiessen, P.~Knierim, P.~El.Agroudy, P.~W.
  Wo{\'z}niak, and A.~Schmidt.
\newblock {Ubiquitous Research Preservation: Transforming Knowledge
  Preservation in Computational Science}.
\newblock \emph{MetaArXiv}, March 2020{\natexlab{a}}.
\newblock \doi{10.31222/osf.io/qmkc9}.

\bibitem[Feger et~al.(2020{\natexlab{b}})Feger, Wo{\'z}niak, Lischke, and
  Schmidt]{Feger2020Comply}
S.~S. Feger, P.~W. Wo{\'z}niak, L.~Lischke, and A.~Schmidt.
\newblock {`Yes, I comply!': Motivations and Practices around Research Data
  Management and Reuse across Scientific Fields}.
\newblock In \emph{Proceedings of the ACM on Human-Computer Interaction, Vol.
  4, CSCW2, Article 141 (October 2020)}, 2020{\natexlab{b}}.
\newblock \doi{10.1145/3415212}.

\bibitem[Feitelson(2015)]{Feitelson2015}
D.~G. Feitelson.
\newblock {From Repeatability to Reproducibility and Corroboration}.
\newblock \emph{ACM SIGOPS Operating Systems Review}, 49\penalty0 (1):\penalty0
  3--11, 2015.
\newblock ISSN 01635980.
\newblock \doi{10.1145/2723872.2723875}.

\bibitem[Fontijn and Hoonhout(2007)]{fontijn2007functional}
W.~Fontijn and J.~Hoonhout.
\newblock {Functional Fun with Tangible User Interfaces}.
\newblock In \emph{2007 First IEEE International Workshop on Digital Game and
  Intelligent Toy Enhanced Learning (DIGITEL'07)}, pages 119--123. IEEE, 2007.
\newblock \doi{10.1109/DIGITEL.2007.26}.

\bibitem[FORCE11(2014)]{FAIR}
FORCE11.
\newblock {The FAIR data principles}.
\newblock Website, 2014.
\newblock Retrieved August 8, 2017 from
  \url{https://www.force11.org/group/fairgroup/fairprinciples}.

\bibitem[Foundation et~al.(2017)Foundation, Albrecht, Alves~Jr, Amadio,
  Andronico, Anh-Ky, Aphecetche, Apostolakis, Asai, Atzori,
  et~al.]{foundation2017roadmap}
H.~Foundation, J.~Albrecht, A.~A. Alves~Jr, G.~Amadio, G.~Andronico, N.~Anh-Ky,
  L.~Aphecetche, J.~Apostolakis, M.~Asai, L.~Atzori, et~al.
\newblock {A Roadmap for HEP Software and Computing R\&D for the 2020s}.
\newblock \emph{arXiv preprint arXiv:1712.06982}, 2017.
\newblock \doi{10.1007/s41781-018-0018-8}.

\bibitem[Fuchs et~al.(2014)Fuchs, Fizek, Ruffino, Schrape,
  et~al.]{fuchs2014rethinking}
M.~Fuchs, S.~Fizek, P.~Ruffino, N.~Schrape, et~al.
\newblock \emph{Rethinking gamification}.
\newblock meson press, 2014.
\newblock \doi{10.14619/001}.

\bibitem[Gaillard and Pandolfi(2017)]{Melissa:2276551}
M.~Gaillard and S.~Pandolfi.
\newblock {CERN Data Centre passes the 200-petabyte milestone}.
\newblock Jul 2017.
\newblock URL \url{http://cds.cern.ch/record/2276551}.

\bibitem[Garza et~al.(2015)Garza, Goble, Brooke, and
  Jay]{Garza:2015:FCD:2783446.2783605}
K.~Garza, C.~Goble, J.~Brooke, and C.~Jay.
\newblock Framing the community data system interface.
\newblock In \emph{Proceedings of the 2015 British HCI Conference}, British HCI
  '15, pages 269--270, New York, NY, USA, 2015. ACM.
\newblock ISBN 978-1-4503-3643-7.
\newblock \doi{10.1145/2783446.2783605}.

\bibitem[Gentil-Beccot et~al.(2010)Gentil-Beccot, Mele, and
  Brooks]{Gentil-Beccot2010}
A.~Gentil-Beccot, S.~Mele, and T.~C. Brooks.
\newblock {Citing and reading behaviours in high-energy physics}.
\newblock \emph{Scientometrics}, 84\penalty0 (2):\penalty0 345--355, 2010.
\newblock ISSN 01389130.
\newblock \doi{10.1007/s11192-009-0111-1}.

\bibitem[Goble(2016)]{goble_repro}
C.~Goble.
\newblock {What is reproducibility}.
\newblock April 2016.
\newblock URL
  \url{https://www.slideshare.net/carolegoble/what-is-reproducibility-gobleclean}.
\newblock Retrieved October 8, 2019.

\bibitem[Gooch et~al.(2016)Gooch, Vasalou, Benton, and
  Khaled]{Gooch:2016:UGM:2858036.2858231}
D.~Gooch, A.~Vasalou, L.~Benton, and R.~Khaled.
\newblock Using gamification to motivate students with dyslexia.
\newblock In \emph{Proceedings of the 2016 CHI Conference on Human Factors in
  Computing Systems}, CHI '16, pages 969--980, New York, NY, USA, 2016. ACM.
\newblock ISBN 978-1-4503-3362-7.
\newblock \doi{10.1145/2858036.2858231}.

\bibitem[Gopalakrishnan et~al.(2017)Gopalakrishnan, Benhur, Kaushik, and
  Passala]{Gopalakrishnan2017}
G.~Gopalakrishnan, K.~Benhur, A.~Kaushik, and A.~Passala.
\newblock Professional network analytics platform for enterprise collaboration.
\newblock In \emph{Companion of the 2017 ACM Conference on Computer Supported
  Cooperative Work and Social Computing}, pages 5--8, 2017.
\newblock \doi{10.1145/3022198.3023264}.

\bibitem[Greiffenhagen and Reeves(2013)]{Greiffenhagen2013}
C.~Greiffenhagen and S.~Reeves.
\newblock {Is Replication important for HCI?}
\newblock \emph{CEUR Workshop Proceedings}, 976:\penalty0 8--13, 2013.
\newblock ISSN 16130073.

\bibitem[Gundersen and Kjensmo(2018)]{gundersen2018state}
O.~E. Gundersen and S.~Kjensmo.
\newblock State of the art: Reproducibility in artificial intelligence.
\newblock In \emph{Thirty-second AAAI conference on artificial intelligence},
  2018.

\bibitem[Gustafsson(2006)]{Gustafsson2006}
H.~A. Gustafsson.
\newblock {LHC experiments}.
\newblock \emph{Nuclear Physics A}, 774\penalty0 (1-4):\penalty0 361--368,
  2006.
\newblock ISSN 03759474.
\newblock \doi{10.1016/j.nuclphysa.2006.06.056}.

\bibitem[Hamari(2013)]{hamari2013transforming}
J.~Hamari.
\newblock Transforming homo economicus into homo ludens: A field experiment on
  gamification in a utilitarian peer-to-peer trading service.
\newblock \emph{Electronic commerce research and applications}, 12\penalty0
  (4):\penalty0 236--245, 2013.
\newblock \doi{10.1016/j.elerap.2013.01.004}.

\bibitem[Hamari and Koivisto(2014)]{hamari2014measuring}
J.~Hamari and J.~Koivisto.
\newblock Measuring flow in gamification: Dispositional flow scale-2.
\newblock \emph{Computers in Human Behavior}, 40:\penalty0 133--143, 2014.
\newblock \doi{10.1016/j.chb.2014.07.048}.

\bibitem[Hamari et~al.(2014)Hamari, Koivisto, and Sarsa]{hamari2014does}
J.~Hamari, J.~Koivisto, and H.~Sarsa.
\newblock {Does Gamification Work? -- A Literature Review of Empirical Studies
  on Gamification}.
\newblock In \emph{2014 47th Hawaii international conference on system sciences
  (HICSS)}, pages 3025--3034. IEEE, 2014.
\newblock \doi{10.1109/HICSS.2014.377}.

\bibitem[Harper et~al.(2008)Harper, Rodden, Rogers, Sellen, Human,
  et~al.]{harper2008human}
E.~R. Harper, T.~Rodden, Y.~Rogers, A.~Sellen, B.~Human, et~al.
\newblock Human-computer interaction in the year 2020.
\newblock 2008.

\bibitem[Howison and Herbsleb(2011)]{howison2011scientific}
J.~Howison and J.~D. Herbsleb.
\newblock Scientific software production: incentives and collaboration.
\newblock In \emph{Proceedings of the ACM 2011 conference on Computer supported
  cooperative work}, pages 513--522. ACM, 2011.
\newblock \doi{10.1145/1958824.1958904}.

\bibitem[Howison and Herbsleb(2013)]{howison2013}
J.~Howison and J.~D. Herbsleb.
\newblock Incentives and integration in scientific software production.
\newblock In \emph{Proceedings of the 2013 Conference on Computer Supported
  Cooperative Work}, CSCW ’13, page 459–470, New York, NY, USA, 2013.
  Association for Computing Machinery.
\newblock ISBN 9781450313315.
\newblock \doi{10.1145/2441776.2441828}.
\newblock URL \url{https://doi.org/10.1145/2441776.2441828}.

\bibitem[Huang et~al.(2013)Huang, Ding, Lee, Lu, and Gu]{huang2013meanings}
X.~Huang, X.~Ding, C.~P. Lee, T.~Lu, and N.~Gu.
\newblock Meanings and boundaries of scientific software sharing.
\newblock In \emph{Proceedings of the 2013 conference on Computer supported
  cooperative work}, pages 423--434. ACM, 2013.
\newblock \doi{10.1145/2441776.2441825}.

\bibitem[Hunicke et~al.(2004)Hunicke, LeBlanc, and Zubek]{Hunicke2004}
R.~Hunicke, M.~LeBlanc, and R.~Zubek.
\newblock {MDA: A Formal Approach to Game Design and Game Research}.
\newblock \emph{Workshop on Challenges in Game AI}, pages 1--4, 2004.
\newblock ISSN 03772217.
\newblock \doi{10.1.1.79.4561}.

\bibitem[Hurlin et~al.(2014)Hurlin, P{\'e}rignon, Stodden, Leisch, and
  Peng]{hurlin2014runmycode}
C.~Hurlin, C.~P{\'e}rignon, V.~Stodden, F.~Leisch, and R.~Peng.
\newblock Runmycode. org: A research-reproducibility tool for computational
  sciences.
\newblock \emph{Implementing reproducible research. CRC Press, Boca Raton, FL},
  pages 367--381, 2014.

\bibitem[Hutson(2018)]{hutson2018artificial}
M.~Hutson.
\newblock Artificial intelligence faces reproducibility crisis.
\newblock \emph{Science}, 359\penalty0 (6377):\penalty0 725--726, 2018.
\newblock ISSN 0036-8075.
\newblock \doi{10.1126/science.359.6377.725}.

\bibitem[Ibanez et~al.(2014)Ibanez, Di-Serio, and Delgado-Kloos]{Ibanez2014}
M.-B. Ibanez, A.~Di-Serio, and C.~Delgado-Kloos.
\newblock {Gamification for Engaging Computer Science Students in Learning
  Activities: A Case Study}.
\newblock \emph{IEEE Transactions on Learning Technologies}, 7\penalty0
  (3):\penalty0 291--301, 2014.
\newblock ISSN 1939-1382.
\newblock \doi{10.1109/TLT.2014.2329293}.

\bibitem[Jackson and Barbrow(2013)]{Jackson2013}
S.~J. Jackson and S.~Barbrow.
\newblock Infrastructure and vocation: field, calling and computation in
  ecology.
\newblock In \emph{Proceedings of the SIGCHI conference on Human Factors in
  Computing Systems}, pages 2873--2882, 2013.
\newblock \doi{10.1145/2470654.2481397}.

\bibitem[Jianu and Laidlaw(2012)]{Jianu:2012:ESU:2207676.2208704}
R.~Jianu and D.~Laidlaw.
\newblock An evaluation of how small user interface changes can improve
  scientists' analytic strategies.
\newblock In \emph{Proceedings of the SIGCHI Conference on Human Factors in
  Computing Systems}, CHI '12, pages 2953--2962, New York, NY, USA, 2012. ACM.
\newblock ISBN 978-1-4503-1015-4.
\newblock \doi{10.1145/2207676.2208704}.

\bibitem[Jirotka et~al.(2006)Jirotka, Procter, Rodden, and Bowker]{Jirotka2006}
M.~Jirotka, R.~Procter, T.~Rodden, and G.~C. Bowker.
\newblock Special issue: Collaboration in e-research.
\newblock \emph{Computer Supported Cooperative Work (CSCW)}, 15\penalty0
  (4):\penalty0 251--255, Aug 2006.
\newblock ISSN 1573-7551.
\newblock \doi{10.1007/s10606-006-9028-x}.

\bibitem[Jirotka et~al.(2013)Jirotka, Lee, and Olson]{jirotka2013supporting}
M.~Jirotka, C.~P. Lee, and G.~M. Olson.
\newblock {Supporting Scientific Collaboration: Methods, Tools and Concepts}.
\newblock \emph{Computer Supported Cooperative Work (CSCW)}, 22\penalty0
  (4-6):\penalty0 667--715, 2013.
\newblock \doi{10.1007/s10606-012-9184-0}.

\bibitem[Kaltenbrunner(2017)]{kaltenbrunner2017digital}
W.~Kaltenbrunner.
\newblock {Digital Infrastructure for the Humanities in Europe and the US:
  Governing Scholarship through Coordinated Tool Development}.
\newblock \emph{Computer Supported Cooperative Work (CSCW)}, 26\penalty0
  (3):\penalty0 275--308, 2017.
\newblock \doi{10.1007/s10606-017-9272-2}.

\bibitem[Kappen et~al.(2017)Kappen, Mirza-Babaei, and
  Nacke]{Kappen:2017:GTA:3116595.3116604}
D.~L. Kappen, P.~Mirza-Babaei, and L.~E. Nacke.
\newblock {Gamification Through the Application of Motivational Affordances for
  Physical Activity Technology}.
\newblock In \emph{Proceedings of the Annual Symposium on Computer-Human
  Interaction in Play}, CHI PLAY '17, pages 5--18, New York, NY, USA, 2017.
  ACM.
\newblock ISBN 978-1-4503-4898-0.
\newblock \doi{10.1145/3116595.3116604}.

\bibitem[Karasti et~al.(2006)Karasti, Baker, and Halkola]{Karasti2006}
H.~Karasti, K.~S. Baker, and E.~Halkola.
\newblock {Enriching the notion of data curation in e-Science: Data managing
  and information infrastructuring in the Long Term Ecological Research (LTER)
  network}.
\newblock \emph{Computer Supported Cooperative Work}, 15\penalty0 (4):\penalty0
  321--358, 2006.
\newblock ISSN 15737551.
\newblock \doi{10.1007/s10606-006-9023-2}.

\bibitem[Kay et~al.(2016)Kay, Haroz, Guha, and Dragicevic]{kay2016special}
M.~Kay, S.~Haroz, S.~Guha, and P.~Dragicevic.
\newblock {Special Interest Group on Transparent Statistics Guidelines}.
\newblock In \emph{Proceedings of the 2016 CHI Conference Extended Abstracts on
  Human Factors in Computing Systems}, pages 1081--1084, 2016.
\newblock \doi{10.1145/3170427.3185374}.

\bibitem[Kay et~al.(2017)Kay, Haroz, Guha, Dragicevic, and
  Wacharamanotham]{Kay:2017:MTS:3027063.3027084}
M.~Kay, S.~Haroz, S.~Guha, P.~Dragicevic, and C.~Wacharamanotham.
\newblock {Moving Transparent Statistics Forward at CHI}.
\newblock In \emph{Proceedings of the 2017 CHI Conference Extended Abstracts on
  Human Factors in Computing Systems}, CHI EA '17, pages 534--541, New York,
  NY, USA, 2017. ACM.
\newblock ISBN 978-1-4503-4656-6.
\newblock \doi{10.1145/3027063.3027084}.

\bibitem[Kervin and Hedstrom(2012)]{kervin2012research}
K.~Kervin and M.~Hedstrom.
\newblock How research funding affects data sharing.
\newblock In \emph{Proceedings of the ACM 2012 Conference on Computer Supported
  Cooperative Work Companion}, pages 131--134. ACM, 2012.
\newblock \doi{10.1145/2141512.2141560}.

\bibitem[Kery et~al.(2018)Kery, Radensky, Arya, John, and Myers]{Kery2018}
M.~B. Kery, M.~Radensky, M.~Arya, B.~E. John, and B.~A. Myers.
\newblock {The Story in the Notebook: Exploratory Data Science using a Literate
  Programming Tool}.
\newblock In \emph{Proceedings of the SIGCHI Conference on Human Factors in
  Computing Systems - CHI'18}, pages 1--11, 2018.
\newblock ISBN 9781450356206.
\newblock \doi{10.1145/3173574.3173748}.

\bibitem[Kidwell et~al.(2016)Kidwell, Lazarevi{\'{c}}, Baranski, Hardwicke,
  Piechowski, Falkenberg, Kennett, Slowik, Sonnleitner, Hess-Holden, Errington,
  Fiedler, and Nosek]{Kidwell2016}
M.~C. Kidwell, L.~B. Lazarevi{\'{c}}, E.~Baranski, T.~E. Hardwicke,
  S.~Piechowski, L.~S. Falkenberg, C.~Kennett, A.~Slowik, C.~Sonnleitner,
  C.~Hess-Holden, T.~M. Errington, S.~Fiedler, and B.~A. Nosek.
\newblock {Badges to Acknowledge Open Practices: A Simple, Low-Cost, Effective
  Method for Increasing Transparency}.
\newblock \emph{PLoS Biology}, 2016.
\newblock ISSN 15457885.
\newblock \doi{10.1371/journal.pbio.1002456}.

\bibitem[Klein et~al.(2001)Klein, Wesson, Hollenbeck, Wright, and
  DeShon]{klein2001assessment}
H.~J. Klein, M.~J. Wesson, J.~R. Hollenbeck, P.~M. Wright, and R.~P. DeShon.
\newblock The assessment of goal commitment: A measurement model meta-analysis.
\newblock \emph{Organizational behavior and human decision processes},
  85\penalty0 (1):\penalty0 32--55, 2001.
\newblock \doi{10.1006/obhd.2000.2931}.

\bibitem[Knaving and Bj{\"o}rk(2013)]{knaving2013designing}
K.~Knaving and S.~Bj{\"o}rk.
\newblock Designing for fun and play: exploring possibilities in design for
  gamification.
\newblock In \emph{Proceedings of the first International conference on gameful
  design, research, and applications}, pages 131--134, 2013.
\newblock \doi{10.1145/2583008.2583032}.

\bibitem[Knaving et~al.(2015)Knaving, Wo{\'z}niak, Fjeld, and
  Bj{\"o}rk]{knaving2015flow}
K.~Knaving, P.~Wo{\'z}niak, M.~Fjeld, and S.~Bj{\"o}rk.
\newblock {Flow is Not Enough: Understanding the Needs of Advanced Amateur
  Runners to Design Motivation Technology}.
\newblock In \emph{Proceedings of the 33rd Annual ACM Conference on Human
  Factors in Computing Systems}, pages 2013--2022, 2015.
\newblock \doi{10.1145/2702123.2702542}.

\bibitem[Knaving et~al.(2018)Knaving, Wo{\'z}niak, Niess, Poguntke, Fjeld, and
  Bj{\"o}rk]{knaving2018understanding}
K.~Knaving, P.~W. Wo{\'z}niak, J.~Niess, R.~Poguntke, M.~Fjeld, and
  S.~Bj{\"o}rk.
\newblock Understanding grassroots sports gamification in the wild.
\newblock In \emph{Proceedings of the 10th Nordic Conference on Human-Computer
  Interaction}, pages 102--113. ACM, 2018.
\newblock \doi{10.1145/3240167.3240220}.

\bibitem[Konkol et~al.(2019)Konkol, Kray, and
  Pfeiffer]{konkol2019computational}
M.~Konkol, C.~Kray, and M.~Pfeiffer.
\newblock Computational reproducibility in geoscientific papers: Insights from
  a series of studies with geoscientists and a reproduction study.
\newblock \emph{International Journal of Geographical Information Science},
  33\penalty0 (2):\penalty0 408--429, 2019.
\newblock \doi{10.1080/13658816.2018.1508687}.

\bibitem[Kouzes et~al.(2009)Kouzes, Anderson, Elbert, Gorton, and
  Gracio]{kouzes2009changing}
R.~T. Kouzes, G.~A. Anderson, S.~T. Elbert, I.~Gorton, and D.~K. Gracio.
\newblock The changing paradigm of data-intensive computing.
\newblock \emph{Computer}, 42\penalty0 (1):\penalty0 26--34, 2009.
\newblock \doi{10.1109/MC.2009.26}.

\bibitem[Krasznahorkay et~al.(2019)Krasznahorkay, Csatlos, Csige, Gulyas,
  Koszta, Szihalmi, Timar, Firak, Nagy, Sas, and
  Krasznahorkay]{krasznahorkay2019new}
A.~J. Krasznahorkay, M.~Csatlos, L.~Csige, J.~Gulyas, M.~Koszta, B.~Szihalmi,
  J.~Timar, D.~S. Firak, A.~Nagy, N.~J. Sas, and A.~Krasznahorkay.
\newblock New evidence supporting the existence of the hypothetic x17 particle,
  2019.

\bibitem[Kumar and Herger(2013)]{kumar2013gamification}
J.~Kumar and M.~Herger.
\newblock Gamification at work: Designing engaging business software.
\newblock In \emph{International Conference of Design, User Experience, and
  Usability}, pages 528--537. Springer, 2013.
\newblock \doi{10.1007/978-3-642-39241-2_58}.

\bibitem[Larkoski et~al.(2017)Larkoski, Marzani, Thaler, Tripathee, and
  Xue]{larkoski2017exposing}
A.~Larkoski, S.~Marzani, J.~Thaler, A.~Tripathee, and W.~Xue.
\newblock {Exposing the QCD splitting function with CMS open data}.
\newblock \emph{Physical review letters}, 119\penalty0 (13):\penalty0 132003,
  2017.
\newblock \doi{10.1103/PhysRevLett.119.132003}.

\bibitem[Leek and Peng(2015)]{Leek2015}
J.~T. Leek and R.~D. Peng.
\newblock {Reproducible research can still be wrong: Adopting a prevention
  approach.}
\newblock In \emph{Proceedings of the National Academy of Sciences of the
  United States of America}, volume 112, pages 1645--6, 2015.
\newblock \doi{10.1073/pnas.1421412111}.

\bibitem[Leonelli(2013)]{leonelli2013current}
S.~Leonelli.
\newblock {Why the Current Insistence on Open Access to Scientific Data? Big
  Data, Knowledge Production, and the Political Economy of Contemporary
  Biology}.
\newblock \emph{Bulletin of Science, Technology \& Society}, 33\penalty0
  (1-2):\penalty0 6--11, 2013.
\newblock \doi{10.1177/0270467613496768}.

\bibitem[Mackay et~al.(2007)Mackay, Appert, Beaudouin-Lafon, Chapuis, Du,
  Fekete, and Guiard]{Mackay2007}
W.~E. Mackay, C.~Appert, M.~Beaudouin-Lafon, O.~Chapuis, Y.~Du, J.-D. Fekete,
  and Y.~Guiard.
\newblock {Touchstone: exploratory design of experiments}.
\newblock In \emph{CHI '07 Proceedings of the SIGCHI Conference on Human
  Factors in Computing System}, pages 1425--1434, 2007.
\newblock ISBN 9781595935939.
\newblock \doi{10.1145/1240624.1240840}.

\bibitem[Mayernik et~al.(2008)Mayernik, Wallis, Pepe, and
  Borgman]{mayernik2008whose}
M.~S. Mayernik, J.~C. Wallis, A.~Pepe, and C.~L. Borgman.
\newblock Whose data do you trust? integrity issues in the preservation of
  scientific data.
\newblock 2008.
\newblock Presented at the iConference, Los Angeles, CA.

\bibitem[Mays and Pope(2020)]{mays2020quality}
N.~Mays and C.~Pope.
\newblock Quality in qualitative research.
\newblock \emph{Qualitative research in health care}, pages 211--233, 2020.

\bibitem[Merali(2010)]{Merali2010}
Z.~Merali.
\newblock {The Large Human Collider}.
\newblock \emph{Nature}, 464\penalty0 (7288):\penalty0 482--484, 2010.
\newblock ISSN 00280836.
\newblock \doi{10.1038/464482a}.

\bibitem[Meyrick(2006)]{meyrick2006good}
J.~Meyrick.
\newblock What is good qualitative research? a first step towards a
  comprehensive approach to judging rigour/quality.
\newblock \emph{Journal of health psychology}, 11\penalty0 (5):\penalty0
  799--808, 2006.

\bibitem[Mihaly(1990)]{mihaly1990flow}
C.~Mihaly.
\newblock Flow: The psychology of optimal performance.
\newblock 1990.

\bibitem[Molin et~al.(2016)Molin, Wo\'{z}niak, Lundstr\"{o}m, Treanor, and
  Fjeld]{Molin:2016:UDA:2971485.2971561}
J.~Molin, P.~W. Wo\'{z}niak, C.~Lundstr\"{o}m, D.~Treanor, and M.~Fjeld.
\newblock {Understanding Design for Automated Image Analysis in Digital
  Pathology}.
\newblock In \emph{Proceedings of the 9th Nordic Conference on Human-Computer
  Interaction}, NordiCHI '16, pages 58:1--58:10, New York, NY, USA, 2016. ACM.
\newblock ISBN 978-1-4503-4763-1.
\newblock \doi{10.1145/2971485.2971561}.

\bibitem[Muller(2014)]{muller2014curiosity}
M.~Muller.
\newblock {Curiosity, Creativity, and Surprise as Analytic Tools: Grounded
  Theory Method}.
\newblock In \emph{Ways of Knowing in HCI}, pages 25--48. Springer, 2014.
\newblock \doi{10.1007/978-1-4939-0378-8\_2}.

\bibitem[Muller et~al.(2019)Muller, Lange, Wang, Piorkowski, Tsay, Liao, Dugan,
  and Erickson]{Muller:2019:DSW:3290605.3300356}
M.~Muller, I.~Lange, D.~Wang, D.~Piorkowski, J.~Tsay, Q.~V. Liao, C.~Dugan, and
  T.~Erickson.
\newblock {How Data Science Workers Work with Data: Discovery, Capture,
  Curation, Design, Creation}.
\newblock In \emph{Proceedings of the 2019 CHI Conference on Human Factors in
  Computing Systems}, CHI '19, pages 126:1--126:15, New York, NY, USA, 2019.
  ACM.
\newblock ISBN 978-1-4503-5970-2.
\newblock \doi{10.1145/3290605.3300356}.

\bibitem[Mullet et~al.(1997)Mullet, Fry, and Schiano]{mullet1997your}
K.~Mullet, C.~Fry, and D.~Schiano.
\newblock On your marks, get set, browse!
\newblock In \emph{CHI'97 Extended Abstracts on Human Factors in Computing
  Systems}, pages 113--114. ACM, 1997.
\newblock \doi{10.1145/1120212.1120285}.

\bibitem[Munaf{\`o} et~al.(2017)Munaf{\`o}, Nosek, Bishop, Button, Chambers,
  Du~Sert, Simonsohn, Wagenmakers, Ware, and Ioannidis]{munafo2017manifesto}
M.~R. Munaf{\`o}, B.~A. Nosek, D.~V. Bishop, K.~S. Button, C.~D. Chambers,
  N.~P. Du~Sert, U.~Simonsohn, E.-J. Wagenmakers, J.~J. Ware, and J.~P.
  Ioannidis.
\newblock A manifesto for reproducible science.
\newblock \emph{Nature human behaviour}, 1\penalty0 (1):\penalty0 0021, 2017.
\newblock \doi{10.1038/s41562-016-0021}.

\bibitem[Nacke and Deterding(2017)]{Nacke2017}
L.~E. Nacke and S.~Deterding.
\newblock {The maturing of gamification research}.
\newblock \emph{Computers in Human Behavior}, 2017.
\newblock ISSN 07475632.
\newblock \doi{10.1016/j.chb.2016.11.062}.

\bibitem[Nakamura and Csikszentmihalyi(2009)]{nakamura2009flow}
J.~Nakamura and M.~Csikszentmihalyi.
\newblock {Flow Theory and Research}.
\newblock \emph{{The Oxford Handbook of Positive Psychology}}, pages 195--206,
  2009.
\newblock \doi{10.1093/oxfordhb/9780195187243.013.0018}.

\bibitem[Newman et~al.(2003)Newman, Ellisman, and Orcutt]{newman2003data}
H.~B. Newman, M.~H. Ellisman, and J.~A. Orcutt.
\newblock Data-intensive e-science frontier research.
\newblock \emph{{Communications of the ACM}}, 46\penalty0 (11):\penalty0
  68--77, 2003.
\newblock \doi{10.1145/948383.948411}.

\bibitem[Nicholson(2015)]{Nicholson2015}
S.~Nicholson.
\newblock \emph{A RECIPE for Meaningful Gamification}, pages 1--20.
\newblock Springer International Publishing, Cham, 2015.
\newblock ISBN 978-3-319-10208-5.
\newblock \doi{10.1007/978-3-319-10208-5_1}.

\bibitem[Norman and Draper(1986)]{norman1986user}
D.~A. Norman and S.~W. Draper.
\newblock \emph{{User Centered System Design; New Perspectives on
  Human-Computer Interaction}}.
\newblock CRC Press, 1986.
\newblock ISBN 978-0-89859-781-3.

\bibitem[Nosek et~al.(2016)Nosek, Alter, Banks, Borsboom, Bowman, Breckler,
  Buck, Chambers, Chin, Christensen, et~al.]{nosek2016transparency}
B.~A. Nosek, G.~Alter, G.~C. Banks, D.~Borsboom, S.~Bowman, S.~Breckler,
  S.~Buck, C.~Chambers, G.~Chin, G.~Christensen, et~al.
\newblock Transparency and openness promotion (top) guidelines.
\newblock 2016.
\newblock Retrieved from osf.io/9f6gx.

\bibitem[Nosek et~al.(2018)Nosek, Ebersole, DeHaven, and
  Mellor]{nosek2018preregistration}
B.~A. Nosek, C.~R. Ebersole, A.~C. DeHaven, and D.~T. Mellor.
\newblock The preregistration revolution.
\newblock \emph{Proceedings of the National Academy of Sciences}, 115\penalty0
  (11):\penalty0 2600--2606, 2018.
\newblock \doi{10.1073/pnas.1708274114}.

\bibitem[N{\"u}st et~al.(2017)N{\"u}st, Konkol, Pebesma, Kray, Schutzeichel,
  Przibytzin, and Lorenz]{nust2017opening}
D.~N{\"u}st, M.~Konkol, E.~Pebesma, C.~Kray, M.~Schutzeichel, H.~Przibytzin,
  and J.~Lorenz.
\newblock {Opening the Publication Process with Executable Research Compendia}.
\newblock \emph{D-Lib Magazine}, 23\penalty0 (1/2), 2017.
\newblock \doi{10.1045/january2017-nuest}.

\bibitem[N{\"u}st et~al.(2019)N{\"u}st, Lohoff, Einfeldt, Gavish, G{\"o}tza,
  Jaswal, Khalid, Meierkort, Mohr, Rendel, et~al.]{nustguerrilla}
D.~N{\"u}st, L.~Lohoff, L.~Einfeldt, N.~Gavish, M.~G{\"o}tza, S.~T. Jaswal,
  S.~Khalid, L.~Meierkort, M.~Mohr, C.~Rendel, et~al.
\newblock {Guerrilla Badges for Reproducible Geospatial Data Science}.
\newblock \emph{AGILE 2019}, 2019.
\newblock \doi{10.31223/osf.io/xtsqh}.

\bibitem[O'Carroll et~al.(2017)O'Carroll, Rentier, Cabello~Vald{\`e}s,
  Esposito, Kaunismaa, Maas, Metcalfe, Vandevelde, Halleux, Kamerlin,
  et~al.]{o2017evaluation}
C.~O'Carroll, B.~Rentier, C.~Cabello~Vald{\`e}s, F.~Esposito, E.~Kaunismaa,
  K.~Maas, J.~Metcalfe, K.~Vandevelde, I.~Halleux, C.~L. Kamerlin, et~al.
\newblock {Evaluation of Research Careers fully acknowledging Open Science
  Practices -- Rewards, incentives and/or recognition for researchers
  practicing Open Science}.
\newblock Technical report, Publication Office of the Europen Union, 2017.

\bibitem[Oleksik et~al.(2012)Oleksik, Milic-Frayling, and
  Jones]{Oleksik:2012:BDS:2145204.2145376}
G.~Oleksik, N.~Milic-Frayling, and R.~Jones.
\newblock Beyond data sharing: Artifact ecology of a collaborative
  nanophotonics research centre.
\newblock In \emph{Proceedings of the ACM 2012 Conference on Computer Supported
  Cooperative Work}, CSCW '12, pages 1165--1174, New York, NY, USA, 2012. ACM.
\newblock ISBN 978-1-4503-1086-4.
\newblock \doi{10.1145/2145204.2145376}.

\bibitem[Oleksik et~al.(2014)Oleksik, Milic-Frayling, and Jones]{Oleksik2014}
G.~Oleksik, N.~Milic-Frayling, and R.~Jones.
\newblock {Study of electronic lab notebook design and practices that emerged
  in a collaborative scientific environment}.
\newblock In \emph{Proceedings of the 17th ACM conference on Computer supported
  cooperative work {\&} social computing - CSCW '14}, pages 120--133, 2014.
\newblock ISBN 9781450325400.
\newblock \doi{10.1145/2531602.2531709}.

\bibitem[Oprescu et~al.(2014)Oprescu, Jones, and Katsikitis]{Oprescu2014}
F.~Oprescu, C.~Jones, and M.~Katsikitis.
\newblock {I PLAY AT WORK-ten principles for transforming work processes
  through gamification}.
\newblock \emph{Frontiers in Psychology}, 5\penalty0 (JAN), 2014.
\newblock ISSN 16641078.
\newblock \doi{10.3389/fpsyg.2014.00014}.

\bibitem[Orji et~al.(2018)Orji, Tondello, and Nacke]{orji2018personalizing}
R.~Orji, G.~F. Tondello, and L.~E. Nacke.
\newblock {Personalizing Persuasive Strategies in Gameful Systems to
  Gamification User Types}.
\newblock In \emph{Proceedings of the 2018 CHI Conference on Human Factors in
  Computing Systems}, page 435. ACM, 2018.
\newblock \doi{10.1145/3173574.3174009}.

\bibitem[Paine et~al.(2015)Paine, Sy, Piell, and Lee]{paine2015examining}
D.~Paine, E.~Sy, R.~Piell, and C.~P. Lee.
\newblock Examining data processing work as part of the scientific data
  lifecycle: Comparing practices across four scientific research groups.
\newblock \emph{iConference 2015 Proceedings}, 2015.

\bibitem[Pasquetto et~al.(2016)Pasquetto, Sands, Darch, and
  Borgman]{Pasquetto:2016:ODS:2858036.2858543}
I.~V. Pasquetto, A.~E. Sands, P.~T. Darch, and C.~L. Borgman.
\newblock {Open Data in Scientific Settings: From Policy to Practice}.
\newblock In \emph{Proceedings of the 2016 CHI Conference on Human Factors in
  Computing Systems}, CHI '16, pages 1585--1596, New York, NY, USA, 2016. ACM.
\newblock ISBN 978-1-4503-3362-7.
\newblock \doi{10.1145/2858036.2858543}.

\bibitem[Pasquier et~al.(2017)Pasquier, Lau, Trisovic, Boose, Couturier,
  Crosas, Ellison, Gibson, Jones, and Seltzer]{pasquier2017if}
T.~Pasquier, M.~K. Lau, A.~Trisovic, E.~R. Boose, B.~Couturier, M.~Crosas,
  A.~M. Ellison, V.~Gibson, C.~R. Jones, and M.~Seltzer.
\newblock If these data could talk.
\newblock \emph{Scientific data}, 4, 2017.
\newblock \doi{10.1038/sdata.2017.114}.

\bibitem[Pelletier et~al.(1995)Pelletier, Tuson, Fortier, Vallerand, Briere,
  and Blais]{pelletier1995toward}
L.~G. Pelletier, K.~M. Tuson, M.~S. Fortier, R.~J. Vallerand, N.~M. Briere, and
  M.~R. Blais.
\newblock {Toward a New Measure of Intrinsic Motivation, Extrinsic Motivation,
  and Amotivation in Sports: The Sport Motivation Scale (SMS)}.
\newblock \emph{Journal of sport and Exercise Psychology}, 17\penalty0
  (1):\penalty0 35--53, 1995.
\newblock \doi{10.1123/jsep.17.1.35}.

\bibitem[Pilat and Fukasaku(2007)]{pilat2007oecd}
D.~Pilat and Y.~Fukasaku.
\newblock {OECD Principles and Guidelines for Access to Research Data from
  Public Funding}.
\newblock \emph{Data Science Journal}, 6:\penalty0 OD4--OD11, 2007.
\newblock \doi{10.1787/9789264034020-en-fr}.

\bibitem[Piwowar and Vision(2013)]{10.7717/peerj.175}
H.~A. Piwowar and T.~J. Vision.
\newblock Data reuse and the open data citation advantage.
\newblock \emph{PeerJ}, 1:\penalty0 e175, Oct. 2013.
\newblock ISSN 2167-8359.
\newblock \doi{10.7717/peerj.175}.

\bibitem[Ponti et~al.(2015)Ponti, Hillman, and
  Stankovic]{Ponti:2015:SGO:2793107.2810293}
M.~Ponti, T.~Hillman, and I.~Stankovic.
\newblock {Science and Gamification: The Odd Couple?}
\newblock In \emph{Proceedings of the 2015 Annual Symposium on Computer-Human
  Interaction in Play}, CHI PLAY '15, pages 679--684, New York, NY, USA, 2015.
  ACM.
\newblock ISBN 978-1-4503-3466-2.
\newblock \doi{10.1145/2793107.2810293}.

\bibitem[Pontika et~al.(2015)Pontika, Knoth, Cancellieri, and
  Pearce]{pontika2015fostering}
N.~Pontika, P.~Knoth, M.~Cancellieri, and S.~Pearce.
\newblock {Fostering open science to research using a taxonomy and an eLearning
  portal}.
\newblock In \emph{Proceedings of the 15th international conference on
  knowledge technologies and data-driven business}, page~11. ACM, 2015.
\newblock \doi{10.1145/2809563.2809571}.

\bibitem[Prior(2019)]{prior_2019}
R.~Prior.
\newblock A 'no-brainer nobel prize': Hungarian scientists may have found a
  fifth force of nature.
\newblock \emph{CNN}, Nov 2019.
\newblock URL
  \url{https://edition.cnn.com/2019/11/22/world/fifth-force-of-nature-scn-trnd/index.html}.

\bibitem[Qin(2016)]{qin2016metadata}
J.~Qin.
\newblock Metadata and reproducibility: A case study of gravitational wave data
  management.
\newblock \emph{International Journal of Digital Curation}, 11\penalty0
  (1):\penalty0 218--231, 2016.
\newblock \doi{10.2218/ijdc.v11i1.399}.

\bibitem[Reichling and Wulf(2009)]{Reichling2009}
T.~Reichling and V.~Wulf.
\newblock {Expert Recommender Systems in Practice : Evaluating Semi-automatic
  Profile Generation}.
\newblock In \emph{Proceedings of the SIGCHI Conference on Human Factors in
  Computing Systems}, pages 59--68, 2009.
\newblock ISBN 9781605582467.
\newblock \doi{10.1145/1518701.1518712}.

\bibitem[Richards et~al.(2014)Richards, Thompson, and
  Graham]{Richards:2014:BDM:2658537.2658683}
C.~Richards, C.~W. Thompson, and N.~Graham.
\newblock Beyond designing for motivation: The importance of context in
  gamification.
\newblock In \emph{Proceedings of the First ACM SIGCHI Annual Symposium on
  Computer-human Interaction in Play}, CHI PLAY '14, pages 217--226, New York,
  NY, USA, 2014. ACM.
\newblock ISBN 978-1-4503-3014-5.
\newblock \doi{10.1145/2658537.2658683}.

\bibitem[Robson et~al.(2015)Robson, Plangger, Kietzmann, McCarthy, and
  Pitt]{Robson2015}
K.~Robson, K.~Plangger, J.~H. Kietzmann, I.~McCarthy, and L.~Pitt.
\newblock {Is it all a game? Understanding the principles of gamification}.
\newblock \emph{Business Horizons}, 58\penalty0 (4):\penalty0 411--420, 2015.
\newblock ISSN 00076813.
\newblock \doi{10.1016/j.bushor.2015.03.006}.

\bibitem[Rolland and Lee(2013)]{rolland2013beyond}
B.~Rolland and C.~P. Lee.
\newblock Beyond trust and reliability: reusing data in collaborative cancer
  epidemiology research.
\newblock In \emph{Proceedings of the 2013 conference on Computer supported
  cooperative work}, pages 435--444. ACM, 2013.
\newblock \doi{10.1145/2441776.2441826}.

\bibitem[Rosenblatt(2016)]{Rosenblatt336ed5}
M.~Rosenblatt.
\newblock An incentive-based approach for improving data reproducibility.
\newblock \emph{Science Translational Medicine}, 8\penalty0 (336):\penalty0
  336ed5--336ed5, 2016.
\newblock ISSN 1946-6234.
\newblock \doi{10.1126/scitranslmed.aaf5003}.

\bibitem[Rowhani-Farid et~al.(2017)Rowhani-Farid, Allen, and
  Barnett]{Rowhani-Farid2017}
A.~Rowhani-Farid, M.~Allen, and A.~G. Barnett.
\newblock {What incentives increase data sharing in health and medical
  research? A systematic review}.
\newblock \emph{Research Integrity and Peer Review}, 2\penalty0 (1):\penalty0
  4, 2017.
\newblock ISSN 2058-8615.
\newblock \doi{10.1186/s41073-017-0028-9}.

\bibitem[Ruhi(2015)]{ruhi2015level}
U.~Ruhi.
\newblock {Level Up Your Strategy: Towards a Descriptive Framework for
  Meaningful Enterprise Gamification}.
\newblock \emph{Technology Innovation Management Review}, 2015.
\newblock \doi{10.22215/timreview/918}.

\bibitem[Rule et~al.(2018)Rule, Tabard, and Hollan]{rule2018exploration}
A.~Rule, A.~Tabard, and J.~D. Hollan.
\newblock {Exploration and Explanation in Computational Notebooks}.
\newblock In \emph{Proceedings of the 2018 CHI Conference on Human Factors in
  Computing Systems}, page~32. ACM, 2018.
\newblock \doi{10.1145/3173574.3173606}.

\bibitem[Russell(2013)]{russell2013if}
J.~F. Russell.
\newblock If a job is worth doing, it is worth doing twice: researchers and
  funding agencies need to put a premium on ensuring that results are
  reproducible.
\newblock \emph{Nature}, 496\penalty0 (7443):\penalty0 7--8, 2013.
\newblock \doi{10.1038/496007a}.

\bibitem[Ryan and Deci(2000)]{ryan2000self}
R.~M. Ryan and E.~L. Deci.
\newblock Self-determination theory and the facilitation of intrinsic
  motivation, social development, and well-being.
\newblock \emph{American psychologist}, 55\penalty0 (1):\penalty0 68, 2000.

\bibitem[Ryan and Patrick(2009)]{ryan2009self}
R.~M. Ryan and H.~Patrick.
\newblock Self-determination theory and physical.
\newblock \emph{Hellenic journal of psychology}, 6:\penalty0 107--124, 2009.

\bibitem[Sailer et~al.(2017)Sailer, Hense, Mayr, and
  Mandl]{sailer2017gamification}
M.~Sailer, J.~U. Hense, S.~K. Mayr, and H.~Mandl.
\newblock How gamification motivates: An experimental study of the effects of
  specific game design elements on psychological need satisfaction.
\newblock \emph{Computers in Human Behavior}, 69:\penalty0 371--380, 2017.
\newblock \doi{10.1016/j.chb.2016.12.033}.

\bibitem[Schacht et~al.(2014)Schacht, Morana, and Maedche]{Schacht2014}
S.~Schacht, S.~Morana, and A.~Maedche.
\newblock {The Project World: Gamification in Project Knowledge Management}.
\newblock In \emph{Proceedings of the 22nd European Conference on Information
  Systems (ECIS)}, number June, pages 1--10, 2014.
\newblock ISBN 9780991556700 (ISBN).

\bibitem[Schmidt(2009)]{Schmidt2009}
S.~Schmidt.
\newblock {Shall we Really do it Again? The Powerful Concept of Replication is
  Neglected in the Social Sciences}.
\newblock \emph{Review of General Psychology}, 13\penalty0 (2):\penalty0
  90--100, 2009.
\newblock ISSN 1089-2680.
\newblock \doi{10.1037/a0015108}.

\bibitem[Schwartz et~al.(2010)Schwartz, Pappas, and Sandlow]{schwartz2010data}
A.~Schwartz, C.~Pappas, and L.~J. Sandlow.
\newblock Data repositories for medical education research: issues and
  recommendations.
\newblock \emph{Academic Medicine}, 85\penalty0 (5):\penalty0 837--843, 2010.
\newblock \doi{10.1097/ACM.0b013e3181d74562}.

\bibitem[Seaborn and Fels(2015)]{seaborn2015gamification}
K.~Seaborn and D.~I. Fels.
\newblock Gamification in theory and action: A survey.
\newblock \emph{International Journal of human-computer studies}, 74:\penalty0
  14--31, 2015.
\newblock \doi{10.1016/j.ijhcs.2014.09.006}.

\bibitem[{Sears}(2011)]{2011AGUFMIN53B1628S}
J.~R.~L. {Sears}.
\newblock {Data Sharing Effect on Article Citation Rate in Paleoceanography}.
\newblock \emph{AGU Fall Meeting Abstracts}, Dec. 2011.

\bibitem[{Segal} et~al.(2000){Segal}, {Robertson}, {Gagliardi}, and
  {Carminati}]{segal2000grid}
B.~{Segal}, L.~{Robertson}, F.~{Gagliardi}, and F.~{Carminati}.
\newblock {Grid computing: the European Data Grid Project}.
\newblock In \emph{2000 IEEE Nuclear Science Symposium. Conference Record (Cat.
  No.00CH37149)}, volume~1, pages 2/1 vol.1--, Oct 2000.
\newblock \doi{10.1109/NSSMIC.2000.948988}.

\bibitem[Shami et~al.(2007)Shami, Yuan, Cosley, Xia, and Gay]{Shami2007}
N.~S. Shami, Y.~C. Yuan, D.~Cosley, L.~Xia, and G.~Gay.
\newblock {That's what friends are for: facilitating 'who knows what' across
  group boundaries}.
\newblock In \emph{Proceedings of the 2007 international ACM conference on
  Supporting group work}, pages 379--382, 2007.
\newblock ISBN 978-1-59593-845-9.
\newblock \doi{10.1145/1316624.1316681}.

\bibitem[Shami et~al.(2011)Shami, Muller, and Millen]{shami2011browse}
N.~S. Shami, M.~Muller, and D.~Millen.
\newblock Browse and discover: social file sharing in the enterprise.
\newblock In \emph{Proceedings of the ACM 2011 conference on Computer supported
  cooperative work}, pages 295--304. ACM, 2011.
\newblock \doi{10.1145/1958824.1958868}.

\bibitem[Stanculescu et~al.(2016)Stanculescu, Bozzon, Sips, and
  Houben]{Stanculescu:2016:WPE:2818048.2820061}
L.~C. Stanculescu, A.~Bozzon, R.-J. Sips, and G.-J. Houben.
\newblock {Work and Play: An Experiment in Enterprise Gamification}.
\newblock In \emph{Proceedings of the 19th ACM Conference on Computer-Supported
  Cooperative Work \& Social Computing}, CSCW '16, pages 346--358, New York,
  NY, USA, 2016. ACM.
\newblock ISBN 978-1-4503-3592-8.
\newblock \doi{10.1145/2818048.2820061}.

\bibitem[Stodden and Miguez(2014)]{Stodden2014}
V.~Stodden and S.~Miguez.
\newblock {Best Practices for Computational Science: Software Infrastructure
  and Environments for Reproducible and Extensible Research}.
\newblock \emph{Journal of Open Research Software}, 2\penalty0 (1):\penalty0
  21, 2014.
\newblock ISSN 2049-9647.
\newblock \doi{10.5334/jors.ay}.

\bibitem[Strasser(2015)]{strasser2015research}
C.~Strasser.
\newblock {Research Data Management: A Primer Publication of the National
  Information Standards Organization}.
\newblock \emph{National Information Standards Organization}, 2015.

\bibitem[Sufi et~al.(2014)Sufi, Hong, Hettrick, Antonioletti, Crouch, Hay,
  Inupakutika, Jackson, Pawlik, Peru, et~al.]{sufi2014software}
S.~Sufi, N.~C. Hong, S.~Hettrick, M.~Antonioletti, S.~Crouch, A.~Hay,
  D.~Inupakutika, M.~Jackson, A.~Pawlik, G.~Peru, et~al.
\newblock Software in reproducible research: advice and best practice collected
  from experiences at the collaborations workshop.
\newblock In \emph{{Proceedings of the 1st ACM SIGPLAN Workshop on Reproducible
  Research Methodologies and New Publication Models in Computer Engineering}},
  page~2. ACM, 2014.
\newblock \doi{10.1145/2618137.2618140}.

\bibitem[Swacha and Muszy{\'{n}}ska(2016)]{Swacha2016}
J.~Swacha and K.~Muszy{\'{n}}ska.
\newblock {Design patterns for gamification of work}.
\newblock \emph{Proceedings of the Fourth International Conference on
  Technological Ecosystems for Enhancing Multiculturality - TEEM '16}, pages
  763--769, 2016.
\newblock \doi{10.1145/3012430.3012604}.

\bibitem[Sy(2007)]{sy2007adapting}
D.~Sy.
\newblock Adapting usability investigations for agile user-centered design.
\newblock \emph{Journal of usability Studies}, 2\penalty0 (3):\penalty0
  112--132, 2007.

\bibitem[Tabard et~al.(2008)Tabard, Mackay, and Eastmond]{tabard2008individual}
A.~Tabard, W.~E. Mackay, and E.~Eastmond.
\newblock From individual to collaborative: the evolution of prism, a hybrid
  laboratory notebook.
\newblock In \emph{Proceedings of the 2008 ACM conference on Computer supported
  cooperative work}, 2008.
\newblock \doi{10.1145/1460563.1460653}.

\bibitem[Thomer et~al.(2016)Thomer, Twidale, Guo, and
  Yoder]{Thomer:2016:CSS:2851581.2892549}
A.~K. Thomer, M.~B. Twidale, J.~Guo, and M.~J. Yoder.
\newblock {Co-designing Scientific Software: Hackathons for Participatory
  Interface Design}.
\newblock In \emph{Proceedings of the 2016 CHI Conference Extended Abstracts on
  Human Factors in Computing Systems}, CHI EA '16, pages 3219--3226, New York,
  NY, USA, 2016. ACM.
\newblock ISBN 978-1-4503-4082-3.
\newblock \doi{10.1145/2851581.2892549}.

\bibitem[Tondello et~al.(2017)Tondello, Mora, and Nacke]{Tondello2017}
G.~F. Tondello, A.~Mora, and L.~E. Nacke.
\newblock {Elements of Gameful Design Emerging from User Preferences}.
\newblock In \emph{Proceedings of the Annual Symposium on Computer-Human
  Interaction in Play - CHI PLAY '17}, pages 129--142, 2017.
\newblock ISBN 9781450348980.
\newblock \doi{10.1145/3116595.3116627}.

\bibitem[Tripathee et~al.(2017)Tripathee, Xue, Larkoski, Marzani, and
  Thaler]{tripathee2017jet}
A.~Tripathee, W.~Xue, A.~Larkoski, S.~Marzani, and J.~Thaler.
\newblock {Jet substructure studies with CMS open data}.
\newblock \emph{Physical Review D}, 96\penalty0 (7):\penalty0 074003, 2017.
\newblock \doi{10.1103/PhysRevD.96.074003}.

\bibitem[Tsai et~al.(2016)Tsai, Kohrt, Matthews, Betancourt, Lee, Papachristos,
  Weiser, and Dworkin]{tsai2016promises}
A.~C. Tsai, B.~A. Kohrt, L.~T. Matthews, T.~S. Betancourt, J.~K. Lee, A.~V.
  Papachristos, S.~D. Weiser, and S.~L. Dworkin.
\newblock Promises and pitfalls of data sharing in qualitative research.
\newblock \emph{Social Science \& Medicine}, 169:\penalty0 191--198, 2016.

\bibitem[Tyack and Mekler(2020)]{tyackself}
A.~Tyack and E.~D. Mekler.
\newblock {Self-Determination Theory in HCI Games Research: Current Uses and
  Open Questions}.
\newblock In \emph{{Proceedings of the 2020 CHI Conference on Human Factors in
  Computing Systems}}, 2020.
\newblock ISBN 978-1-4503-6708-0.
\newblock \doi{10.1145/3313831.3376723}.

\bibitem[Vallerand et~al.(1992)Vallerand, Pelletier, Blais, Briere, Senecal,
  and Vallieres]{vallerand1992academic}
R.~J. Vallerand, L.~G. Pelletier, M.~R. Blais, N.~M. Briere, C.~Senecal, and
  E.~F. Vallieres.
\newblock The academic motivation scale: A measure of intrinsic, extrinsic, and
  amotivation in education.
\newblock \emph{Educational and psychological measurement}, 52\penalty0
  (4):\penalty0 1003--1017, 1992.
\newblock \doi{10.1177/0013164492052004025}.

\bibitem[van~de Sandt et~al.(2019)van~de Sandt, Lavasa, Dallmeier-Tiessen, and
  Petras]{van2019definition}
S.~van~de Sandt, A.~Lavasa, S.~Dallmeier-Tiessen, and V.~Petras.
\newblock The definition of reuse.
\newblock \emph{Data Science Journal}, 18:\penalty0 22, 2019.
\newblock \doi{10.5334/dsj-2019-022}.

\bibitem[Velden(2013)]{Velden2013}
T.~Velden.
\newblock {Explaining Field Differences in Openness and Sharing in Scientific
  Communities}.
\newblock \emph{In Proceedings of the ACM Conference on Computer-Supported
  Cooperative Work and Social Computing (CSCW'13)}, pages 445--457, 2013.
\newblock \doi{10.1145/2441776.2441827}.

\bibitem[Vertesi and Dourish(2011)]{vertesi2011value}
J.~Vertesi and P.~Dourish.
\newblock The value of data: considering the context of production in data
  economies.
\newblock In \emph{Proceedings of the ACM 2011 conference on Computer supported
  cooperative work}, pages 533--542. ACM, 2011.
\newblock \doi{10.1145/1958824.1958906}.

\bibitem[Vicente-S{\'a}ez and Mart{\'\i}nez-Fuentes(2018)]{vicente2018open}
R.~Vicente-S{\'a}ez and C.~Mart{\'\i}nez-Fuentes.
\newblock Open science now: A systematic literature review for an integrated
  definition.
\newblock \emph{Journal of business research}, 88:\penalty0 428--436, 2018.
\newblock \doi{10.1016/j.jbusres.2017.12.043}.

\bibitem[Vines et~al.(2014)Vines, Albert, Andrew, D{\'e}barre, Bock, Franklin,
  Gilbert, Moore, Renaut, and Rennison]{vines2014availability}
T.~H. Vines, A.~Y. Albert, R.~L. Andrew, F.~D{\'e}barre, D.~G. Bock, M.~T.
  Franklin, K.~J. Gilbert, J.-S. Moore, S.~Renaut, and D.~J. Rennison.
\newblock {The Availability of Research Data Declines Rapidly with Article
  Age}.
\newblock \emph{Current biology}, 24\penalty0 (1):\penalty0 94--97, 2014.
\newblock \doi{10.1016/j.cub.2013.11.014}.

\bibitem[Wacharamanotham et~al.(2018)Wacharamanotham, Kay, Haroz, Guha, and
  Dragicevic]{Wacharamanotham:2018:SIG:3170427.3185374}
C.~Wacharamanotham, M.~Kay, S.~Haroz, S.~Guha, and P.~Dragicevic.
\newblock {Special Interest Group on Transparent Statistics Guidelines}.
\newblock In \emph{Extended Abstracts of the 2018 CHI Conference on Human
  Factors in Computing Systems}, CHI EA '18, pages SIG08:1--SIG08:4, New York,
  NY, USA, 2018. ACM.
\newblock ISBN 978-1-4503-5621-3.
\newblock \doi{10.1145/3170427.3185374}.

\bibitem[Wacharamanotham et~al.(2019)Wacharamanotham, Eisenring, Haroz, and
  Echtler]{wacharamanotham2019transparency}
C.~Wacharamanotham, L.~Eisenring, S.~Haroz, and F.~Echtler.
\newblock {Transparency of CHI Research Artifacts: Results of a Self-Reported
  Survey}.
\newblock 2019.
\newblock \doi{10.31219/osf.io/3bu6t}.

\bibitem[Wallis et~al.(2013)Wallis, Rolando, and Borgman]{Wallis2013}
J.~C. Wallis, E.~Rolando, and C.~L. Borgman.
\newblock {If We Share Data, Will Anyone Use Them? Data Sharing and Reuse in
  the Long Tail of Science and Technology}.
\newblock \emph{PLoS ONE}, 8\penalty0 (7), 2013.
\newblock ISSN 19326203.
\newblock \doi{10.1371/journal.pone.0067332}.

\bibitem[Wegner(1987)]{wegner1987transactive}
D.~M. Wegner.
\newblock {Transactive Memory: A Contemporary Analysis of the Group Mind}.
\newblock In \emph{Theories of group behavior}, pages 185--208. Springer, 1987.
\newblock \doi{10.1007/978-1-4612-4634-3_9}.

\bibitem[Werbach and Hunter(2012)]{werbach2012win}
K.~Werbach and D.~Hunter.
\newblock \emph{For the Win: How Game Thinking Can Revolutionize Your
  Business}.
\newblock Wharton Digital Press, 2012.
\newblock ISBN 9781613630235.

\bibitem[Whyte and Tedds(2011)]{whyte2011making}
A.~Whyte and J.~Tedds.
\newblock \emph{Making the case for research data management}.
\newblock Digital Curation Centre, 2011.

\bibitem[Wilkinson et~al.(2016)Wilkinson, Dumontier, Aalbersberg, Appleton,
  Axton, Baak, Blomberg, Boiten, {da Silva Santos}, Bourne, Bouwman, Brookes,
  Clark, Crosas, Dillo, Dumon, Edmunds, Evelo, Finkers, Gonzalez-Beltran, Gray,
  Groth, Goble, Grethe, Heringa, Hoen, Hooft, Kuhn, Kok, Kok, Lusher, Martone,
  Mons, Packer, Persson, Rocca-Serra, Roos, van Schaik, Sansone, Schultes,
  Sengstag, Slater, Strawn, Swertz, Thompson, van~der Lei, van Mulligen,
  Velterop, Waagmeester, Wittenburg, Wolstencroft, Zhao, and
  Mons]{Wilkinson2016}
M.~D. Wilkinson, M.~Dumontier, I.~J. Aalbersberg, G.~Appleton, M.~Axton,
  A.~Baak, N.~Blomberg, J.-W. Boiten, L.~B. {da Silva Santos}, P.~E. Bourne,
  J.~Bouwman, A.~J. Brookes, T.~Clark, M.~Crosas, I.~Dillo, O.~Dumon,
  S.~Edmunds, C.~T. Evelo, R.~Finkers, A.~Gonzalez-Beltran, A.~J. Gray,
  P.~Groth, C.~Goble, J.~S. Grethe, J.~Heringa, P.~A.~t. Hoen, R.~Hooft,
  T.~Kuhn, R.~Kok, J.~Kok, S.~J. Lusher, M.~E. Martone, A.~Mons, A.~L. Packer,
  B.~Persson, P.~Rocca-Serra, M.~Roos, R.~van Schaik, S.-A. Sansone,
  E.~Schultes, T.~Sengstag, T.~Slater, G.~Strawn, M.~a. Swertz, M.~Thompson,
  J.~van~der Lei, E.~van Mulligen, J.~Velterop, A.~Waagmeester, P.~Wittenburg,
  K.~Wolstencroft, J.~Zhao, and B.~Mons.
\newblock {The FAIR Guiding Principles for scientific data management and
  stewardship}.
\newblock \emph{Scientific Data}, 3:\penalty0 160018, 2016.
\newblock ISSN 2052-4463.
\newblock \doi{10.1038/sdata.2016.18}.

\bibitem[Wilson et~al.(2014)Wilson, Chi, Reeves, and
  Coyle]{Wilson:2014:RWI:2559206.2559233}
M.~L. Wilson, E.~H. Chi, S.~Reeves, and D.~Coyle.
\newblock {RepliCHI: The Workshop II}.
\newblock In \emph{CHI '14 Extended Abstracts on Human Factors in Computing
  Systems}, CHI EA '14, pages 33--36, New York, NY, USA, 2014. ACM.
\newblock ISBN 978-1-4503-2474-8.
\newblock \doi{10.1145/2559206.2559233}.

\bibitem[Wilson et~al.(2013)Wilson, Resnick, Coyle, and Chi]{Wilson2013}
M.~L.~L. Wilson, P.~Resnick, D.~Coyle, and E.~H. Chi.
\newblock {RepliCHI}.
\newblock \emph{CHI '13 Extended Abstracts on Human Factors in Computing
  Systems on - CHI EA '13}, page 3159, 2013.
\newblock \doi{10.1145/2468356.2479636}.
\newblock URL \url{http://dl.acm.org/citation.cfm?doid=2468356.2479636}.

\bibitem[Wittenburg et~al.(2010)Wittenburg, Van~de Sompel, Vigen, Bachem,
  Romary, Marinucci, Andersson, Genova, Best, Los,
  et~al.]{wittenburg2010riding}
P.~Wittenburg, H.~Van~de Sompel, J.~Vigen, A.~Bachem, L.~Romary, M.~Marinucci,
  T.~Andersson, F.~Genova, C.~Best, W.~Los, et~al.
\newblock Riding the wave: How europe can gain from the rising tide of
  scientific data.
\newblock 2010.
\newblock {Final report of the High Level Expert Group on Scientific Data. A
  submission to the European Commission}.

\bibitem[Worden(2017)]{Worden2017}
D.~J. Worden.
\newblock {Emerging Technologies for Data Research: Implications for Bias,
  Curation, and Reproducible Results}.
\newblock In \emph{Human Capital and Assets in the Networked World}. 2017.
\newblock \doi{10.1108/978-1-78714-827-720171003}.

\bibitem[Zhao et~al.(2017)Zhao, Arya, Whitehead, Chan, and
  Etemad]{Zhao:2017:KUE:3025453.3025982}
Z.~Zhao, A.~Arya, A.~Whitehead, G.~Chan, and S.~A. Etemad.
\newblock {Keeping Users Engaged Through Feature Updates: A Long-Term Study of
  Using Wearable-Based Exergames}.
\newblock In \emph{Proceedings of the 2017 CHI Conference on Human Factors in
  Computing Systems}, CHI '17, pages 1053--1064, New York, NY, USA, 2017. ACM.
\newblock ISBN 978-1-4503-4655-9.
\newblock \doi{10.1145/3025453.3025982}.

\bibitem[Zimmerman(2007)]{zimmerman2007not}
A.~Zimmerman.
\newblock Not by metadata alone: the use of diverse forms of knowledge to
  locate data for reuse.
\newblock \emph{International Journal on Digital Libraries}, 7\penalty0
  (1-2):\penalty0 5--16, 2007.
\newblock \doi{10.1007/s00799-007-0015-8}.

\end{thebibliography}

	

	\cleardoublepage
	\phantomsection
	
	\listoffigures	
	\cleardoublepage
	\listoftables
	\cleardoublepage
	%
%

\chapter*{List of Acronyms}\markboth{LIST OF ACRONYMS}{}
\addcontentsline{toc}{chapter}{List of Acronyms}

\begin{acronym}[MyAbbMyAbb] 
	\setlength{\itemsep}{-0.06cm}
	
	\acro{ACM}{Association for Computing Machinery}
	\acro{AI}{Artificial Intelligence}
	\acro{BMBF}{German Federal Ministry of Education and Research}
	\acro{BPNT}{Basic psychological needs theory}
	
	\acro{CAP}{CERN Analysis Preservation}
	\acro{CERN}{European Organization for Nuclear Research}
	\acro{CET}{Cognitive Evaluation Theory}
	
	\acro{COD}{CERN Open Data}
	\acro{COS}{Center for Open Science}
	\acro{CSCW}{Computer-Supported Cooperative Work}
	
	\acro{EC}{European Commission}
	\acro{ELN}{Electronic Lab Notebook}
	\acro{EU}{European Union}
	\acro{ERC}{Executable Research Compendium}
	
	\acro{GCT}{Goal Contents Theory}
	
	\acro{HCI}{Human-Computer Interaction}
	\acro{HEP}{High Energy Physics}
	
	\acro{IMI}{Intrinsic Motivation Inventory}
	
	\acro{JSON}{JavaScript Object Notation}
	
	\acro{LHC}{Large Hadron Collider}
	
	\acro{ML}{Machine Learning}
	
	\acro{OIT}{Organismic Integration Theory}
	\acro{OR}{Open Repositories}
	\acro{OS}{Open Science}
	\acro{OSF}{Open Science Framework}
	\acro{OSB}{Open Science Badges}

    \acro{REANA}{REusable ANAlysis}
	\acro{RDM}{Research Data Management}
	\acro{RMT}{Relationship Motivation Theory}
	\acro{RQ}{Research Question}
	
	\acro{SDT}{Self-determination theory}
	\acro{SIG}{Special Interest Group}
	\acro{SIGCHI}{Special Interest Group on Computer–Human Interaction}
	\acro{SIS}{Scientific Information Service}
	
	\acro{TMS}{Transactive Memory Systems}
	\acro{TOP}{Transparency and Openness Promotion}

	\acro{UCD}{User-Centered Design}
	\acro{URP}{Ubiquitous Research Preservation}
	\acro{US}{United States}
	\acro{UX}{User Experience} 

	\acro{WWW}{World Wide Web}
\end{acronym}


	\cleardoublepage
	
	\selectlanguage{english}
\markboth{Declaration}{Declaration}

\selectlanguage{ngerman}

~\vfill

\begin{Large}Eidesstattliche Versicherung\end{Large}

\begin{small}(Siehe Promotionsordnung vom 12.07.11, § 8, Abs. 2 Pkt. 5)\end{small}

Hiermit erkl\"{a}re ich an Eidesstatt, dass die Dissertation von mir selbstst\"{a}ndig und ohne unerlaubte Beihilfe angefertigt wurde.

München, den 12. März 2020

\vspace{2cm}

\hfill \myname
~\vfill

\selectlanguage{english} 
	
\end{document}